\documentclass[a4paper,10pt,twoside]{cpc-hepnp}
\pdfoutput=1 

\usepackage{CJK,upgreek,fancyhdr}
\usepackage{diagbox}
\usepackage{lineno}
\usepackage{xspace}
\usepackage{upgreek}
\usepackage{lmodern}

\usepackage{subfigure}
\usepackage{multicol}
\usepackage{booktabs}
\usepackage{amssymb,bm,mathrsfs,bbm,amscd}
\usepackage[tbtags]{amsmath}
\usepackage{lastpage}
\usepackage{graphicx}
\usepackage[english]{babel}
\usepackage[colorlinks=false,pdfpagemode=FullScreen,,setpagesize=off,pdfborder={0 0 0}]{hyperref}
\usepackage{overpic}
\usepackage{lineno}
\usepackage{color}

\newcommand\myfigure[1]{%
\medskip\noindent\begin{minipage}{\columnwidth}
\centering%
#1%
\end{minipage}\medskip}

\newcommand{\boldm}[1] {\mathversion{bold}#1\mathversion{normal}}

\def\KS{{K^0_{\mathrm{\scriptscriptstyle S}}}}
\def\CP{{\ensuremath{C\!P}}\xspace}
\def\invfb{\ensuremath{\mbox{\,fb}^{-1}}\xspace}

\def\Dbar{{\kern 0.2em\overline{\kern -0.2em D}{}}\xspace}
\def\Dz{{\ensuremath{D^0}}\xspace}
\def\Dtz{{\ensuremath{\tilde{D}^0}}\xspace}
\def\Dzb{{\ensuremath{\Dbar{}^0}}\xspace}
\def\Dstz{{\ensuremath{D^{*0}}}\xspace}
\def\Dtstz{{\ensuremath{\tilde{D}^{*0}}}\xspace}
\def\Dstzb{{\ensuremath{\Dbar{}^{*0}}}\xspace}

\def\Bbar{{\ensuremath{\kern 0.18em\overline{\kern -0.18em B}{}}}\xspace}
\def\Bsb{{\ensuremath{\Bbar{}^0_s}}\xspace}
\def\Bs{{\ensuremath{B^0_s}}\xspace}
\def\Bz{{\ensuremath{B^0}}\xspace}
\def\Bp{{\ensuremath{B^+}}\xspace}

\newcommand{\BsDtphi}{\Bs\rightarrow \tilde{D}^{(*)0}\phi}
\newcommand{\BsDphi}{\Bs\rightarrow D^{(*)0}\phi}
\newcommand{\BsDbphi}{\Bs\rightarrow \Dbar^{(*)0}\phi}
\newcommand{\BsDstbphi}{\Bs\rightarrow \Dbar^{*0}\phi}

\newcommand{\BstoDbphi}{\Bs\rightarrow \Dzb \phi}

\def\pip{{\ensuremath{\pi^{+}}}\xspace}
\def\pim{{\ensuremath{\pi^{-}}}\xspace}
\def\piz{{\ensuremath{\pi^{0}}}\xspace}
\def\Kp{{\ensuremath{K^{+}}}\xspace}
\def\Km{{\ensuremath{K^{-}}}\xspace}

\def\AT{{\ensuremath{\cal A}}\xspace}
\def\BT{{\ensuremath{\cal B}}\xspace}

\def\rB{{\ensuremath{r_{B}}}}
\def\rstB{{\ensuremath{r^*_{B}}}}
\def\deltaB{{\ensuremath{\delta_{B}}}}
\def\deltastB{{\ensuremath{\delta^*_{B}}}}
\def\rBst{{\ensuremath{r^{(*)}_{B}}}}
\def\deltaBst{{\ensuremath{\delta^{(*)}_{B}}}}

\begin{document}
\begin{CJK*}{UTF8}{gkai}


\fancyhead[c]{\small Chinese Physics C~~~Vol. xx, No. x (202x) xxxxxx}
\fancyfoot[C]{\small 010201-\thepage}

\footnotetext[0]{Received xxxx June xxxx}

\title{Study of the CKM angle $\gamma$ sensitivity using flavor untagged $\BsDtphi$ decays
}

\maketitle

\begin{center}
\author{D.~Ao$^{1}$, D.~Decamp$^{2}$, W.~B.~Qian$^{1}$, S.~Ricciardi$^{3}$, H.~Sazak$^{4}$, S.~T'Jampens$^{2}$, V.~Tisserand$^{4}$, Z.~R.~Wang$^{5}$, Z.~W.~Yang$^{5}$, S.~N.~Zhang$^{6}$, X.~K.~Zhou$^{1}$
 \small
 \\{$1$: University of Chinese Academy of Sciences, Beijing, China,}
 \\{$2$: Univ. Grenoble Alpes, Univ. Savoie Mont Blanc, CNRS, IN2P3-LAPP, Annecy, France,}
 \\{$3$: STFC Rutherford Appleton Laboratory, Didcot, United Kingdom,}
 \\{$4$: Universit{\'e} Clermont Auvergne, CNRS/IN2P3, LPC, Clermont-Ferrand, France,}
 \\{$5$: Center for High Energy Physics, Tsinghua University, Beijing, China,}
 \\{$6$: School of Physics State Key Laboratory of Nuclear Physics and Technology, Peking University, Beijing, China,}
}

\renewcommand{\thefootnote}{\fnsymbol{footnote}}

\end{center}
\linenumbers
\begin{abstract}
A sensitivity study for the measurement of the CKM angle $\gamma$ from $\BsDtphi$ decays is performed using $D$ meson reconstructed in the quasi flavour-specific modes $K\pi$, $K3\pi$, $K\pi\piz$, and \CP-eigenstate modes $KK$ and $\pi\pi$, where the notation $\tilde{D}^0$ corresponds to a $\Dz$ or a $\Dzb$ meson. The LHCb experiment is taken as a use case. A statistical uncertainty of about $8-19^{\circ}$ can be achieved with the $pp$ collision data collected by the LHCb experiment from year 2011 to 2018. The sensitivity to $\gamma$ should be of the order $3-8^{\circ}$ after accumulating
23\invfb of $pp$ collision data by 2025, while it is expected to further improve with 300\invfb by the second half of the 2030 decade. The accuracy depends on the strong parameters $\rBst$ and $\deltaBst$, describing, together with $\gamma$, the interference between the leading amplitudes of the $\BsDtphi$ decays.
\end{abstract}



 \begin{figure}[t]
\centering
\includegraphics[width=0.45\textwidth,height=0.2\textheight]{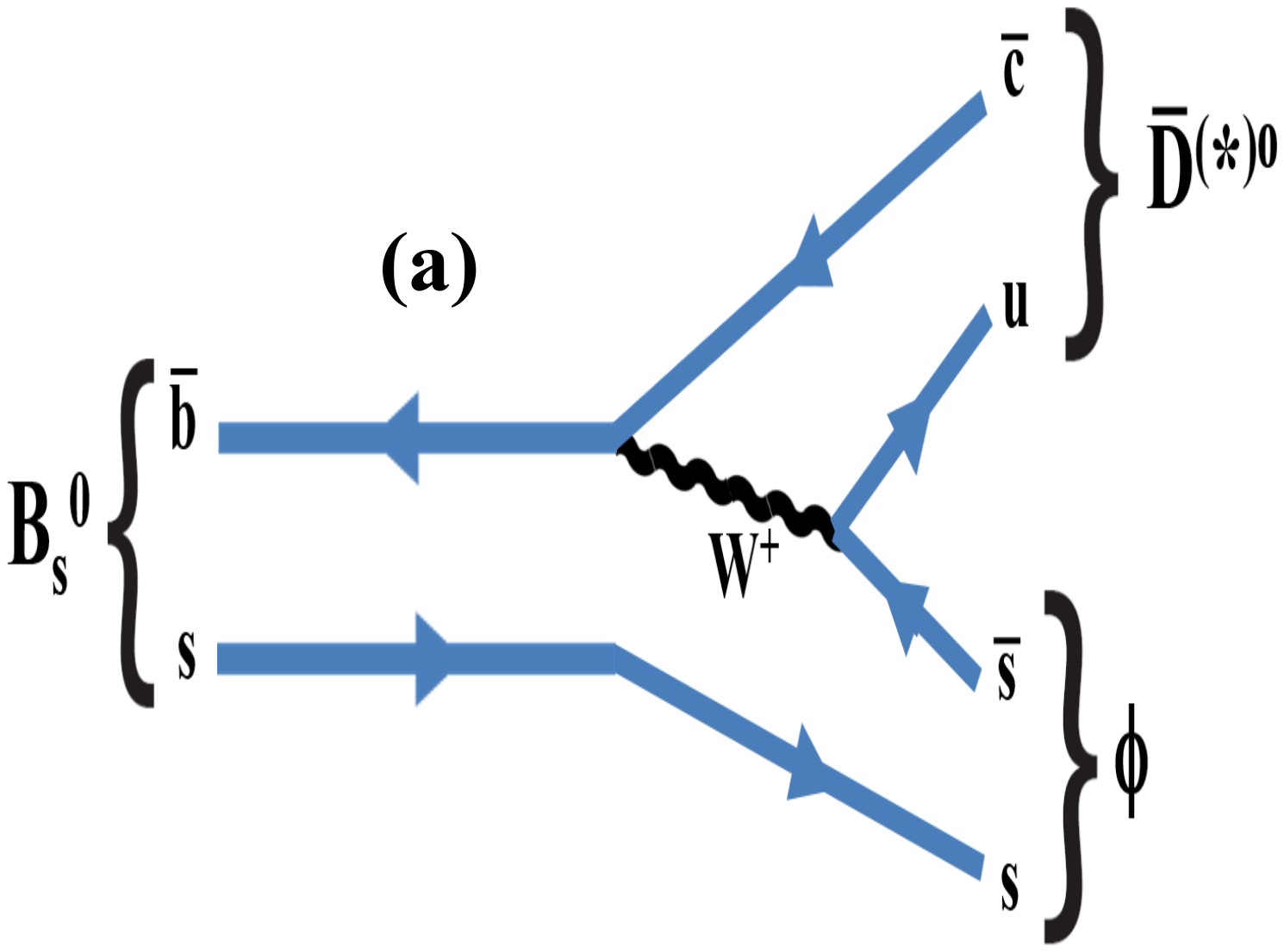}
\includegraphics[width=0.45\textwidth,height=0.2\textheight]{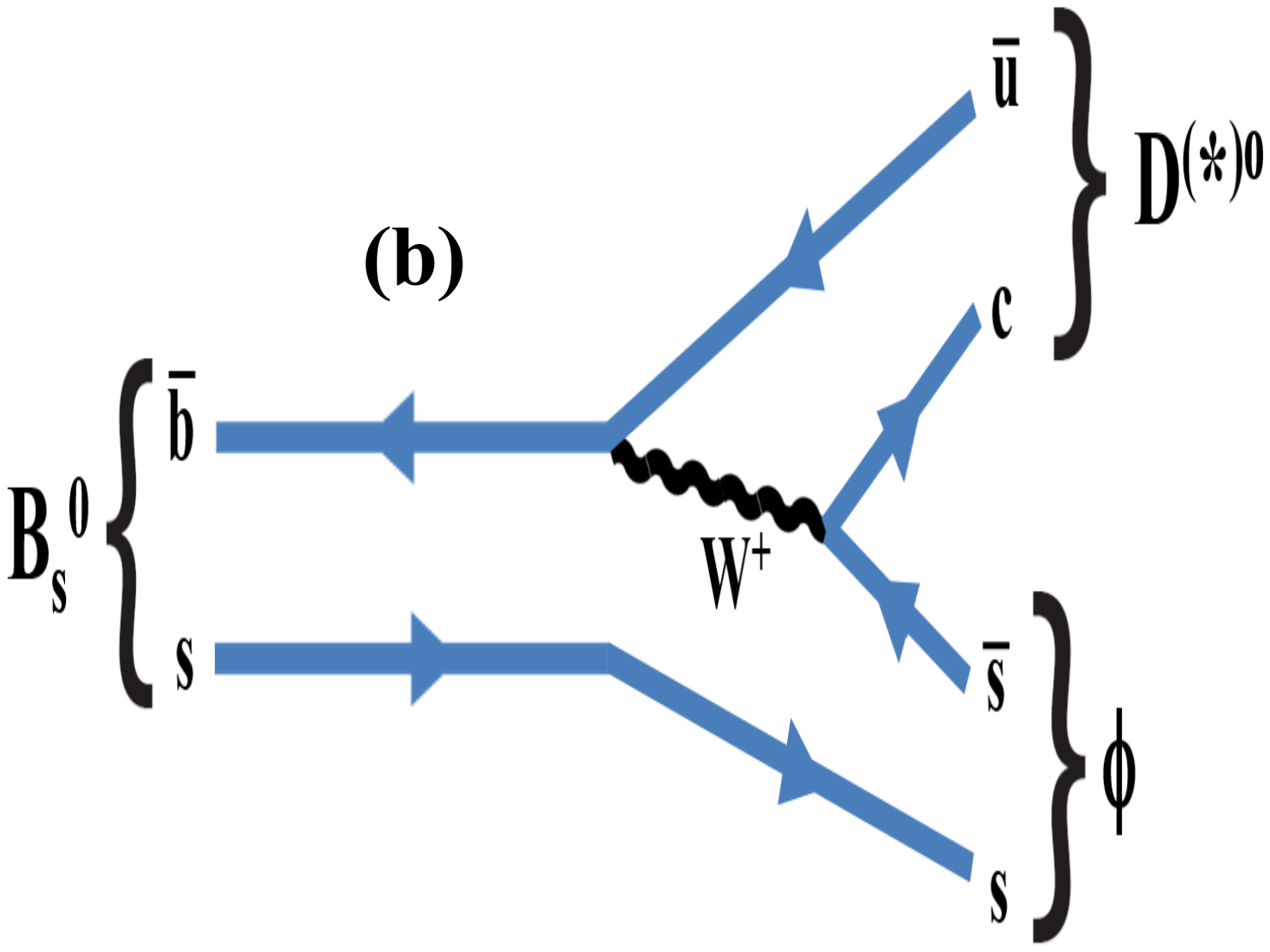}
\caption{\label{fig:Diagrams} Feynman diagrams for (a, left) $\BsDbphi$ and (b, right) $\BsDphi$ decays.}
\end{figure}



\section{Introduction}
\label{sec:intro}

Precision measurements of the CKM~\cite{CKMmatrix} angle $\gamma$~(defined as $\arg[-V_{ud}V^*_{ub}/V_{cd}V^*_{cb}]$) in a variety of $B$-meson decay modes is one of the main goals of flavour physics. Such measurements can be achieved by exploiting the interference of decays that proceed via the $b\rightarrow c\bar{u}s$ and $b\rightarrow u\bar{c}s$ tree-level amplitudes, where the determination of the relative weak phase $\gamma$ is not affected by theoretical uncertainties.

Several methods have been proposed to extract $\gamma$~\cite{Dunietz,GLW,ADS,GGSZ,GLS,gammaHistory}. At LHCb, the best precision is obtained by combining the measurements of many decay modes, which gives $\gamma = (74.0^{+5.0}_{-5.8})^{\circ}$~\cite{LHCb-CONF-2018-002}. This precision dominates the world average on $\gamma$ from tree-level decays. LHCb has presented a new measurement based on the BPGGSZ method~\cite{GGSZ} using the full Run~1 and Run~2 data. The result is $\gamma = (68.7^{+5.2}_{-5.1})^\circ$~\cite{LHCb-CONF-2020-001} and constitutes the single best world measurement on $\gamma$. A $2\sigma$ difference between $\Bp$ and $\Bs$ results was observed since summer 2018. The $\Bs$ measurement is based on a single decay mode only with Run~1 data, {\it i.e.} $\Bs \to D^\mp_s K^\pm$,\footnote{The inclusion of charge-conjugated processes is implied throughout the paper, unless otherwise stated.} and has a large uncertainty~\cite{Aaij:2017lff}. In October 2020, LHCb came with a similar analysis with the decay $\Bs \to D^\mp_s K^\pm \pip\pim$~\cite{BstoDsKpipi} based on Run~1 and Run~2 data for which $\gamma - 2 \beta_s = (42\pm10\pm4\pm5)^{\circ}$. The most
recent LHCb combination is then $\gamma = (67 \pm 4)^{\circ}$~\cite{LHCb-CONF-2020-003} and the new $\Bp$ ($\Bs$) result is $(64^{+4}_{-5})^{\circ}$ $\left( (82^{+17}_{-20})^{\circ} \right)$. Additional $\Bs$ decay modes will help improve the level of measurement precision of $\Bs$ modes and the understanding of a possible discrepancy with respect to the $\Bp$ modes. The two analyses based on $\Bs \to D^\mp_s K^\pm (\pip\pim)$ use time-dependent methods and therefore strongly rely on the $B$-tagging capabilities of the LHCb experiment.  It should be noticed that measurements exist at LHCb for  $\Bz$ mesons~\cite{Aaij:2016Alexis,Aaij:2019Pullen}, they offer also quite good prospects as their present average is $(82^{+8}_{-9})^{\circ}$. They are based on the decay $\Bz \to \Dz K^{*0}$, where the $\Bz$ is self-tagged from the $K^{*0}\rightarrow \Kp\pim$ decay. As opposed to measurements with $\Bs$, those with $\Bz$ and $\Bp$ are also accessible at Belle II~\cite{BelleII}. Prospects on those measurements at LHCb are given in Ref.~\cite{lhcbupgrade}. For the mode $\Bs \to D^\mp_s K^\pm$ one may anticipate a precision on $\gamma$ of the order of $4^\circ$ after the end of LHC Run~3 in 2025, and $1^\circ$ by 2035-2038. When combining with $\Bz$ and $\Bp$ modes the expected sensitivities are $1.5^\circ$ and $0.35^\circ$. The anticipated precision provided by Belle II is $1.5^\circ$.

In this work, $\BsDbphi$ decays, whose observations were published by the LHCb experiment in 2013~\cite{Aaij:2013dda} and 2018~\cite{Aaij:2018jqv}, are used to determine $\gamma$. A novel method presented in Ref.~\cite{Aaij:2018jqv} showed also the feasibility of measuring $\Bs\rightarrow \Dbar^{*0}\phi$ decays with a high purity. It uses a partial reconstruction method for the $\Dbar^{*0}$ meson~\cite{DonalMalcom}.  A time-integrated method~\cite{Gronau:2004gt} is investigated where it was shown that information about \CP violation is preserved in the untagged rate of $\BsDtphi$ (or of $\Bz \rightarrow \Dtz \KS$), and that, if a sufficient number of different $D$-meson final states are included in the analysis, this decay alone can, in principle, be used to measure $\gamma$. The sensitivity to $\gamma$ is expected to be much better with the case of $\BsDtphi$ decay than for $\Bz \rightarrow \Dtz \KS$, as it is proportional to the decay width difference $y=\Delta\Gamma/{2\Gamma}$ defined in Section~\ref{sec:for} and equals to $(6.3\pm0.3)\%$, for $\Bs$ mesons, and $(0.05\pm 0.50)\%$, for $\Bz$~\cite{HFLAV}. Sensitivity to $\gamma$ from $\BsDtphi$ modes comes from the interference between two colour-suppressed diagrams shown in Fig.~\ref{fig:Diagrams}. The relatively large expected value of  the ratio of the $\bar{b}\rightarrow \bar{u}c\bar{s}$ and $\bar{b}\rightarrow \bar{c}u\bar{s}$  tree-level amplitudes ($20-40~\%$, see Sect.~\ref{sec:choiceParams}) is an additional motivation for measuring $\gamma$ in $\BsDtphi$ decays. In this study, five neutral $D$-meson decay modes $K\pi$, $K3\pi$, $K\pi\piz$, $KK$ and $\pi\pi$ are included, whose event yields are estimated using realistic assumptions based on measurements from LHCb~\cite{Aaij:2018jqv,Aaij:2016oso,Aaij:2015jna}. We justify the choice of those decays and also discuss the case of the two decay modes $\tilde{D}^0 \rightarrow \KS\pip\pim$ and $\KS\Kp\Km$ in Section~\ref{sec:yie}.

In Section~\ref{sec:for}, the notations and the choice of $D$-meson decay final states are introduced. In Section~\ref{sec:yie}, the expected signal yields and their uncertainties are presented. In Section~\ref{sec:sen}, the sensitivity which can be achieved using solely these decays is shown, and further improvements are briefly discussed. In Section~\ref{sec:HL-LHC} the future expected precision on $\gamma$ with $\BsDbphi$ at LHCb are discussed for  dataset available after LHC Run~3 by 2025, and after a possible second upgrade of LHCb, by 2038.  Finally, conclusions are made in Section~\ref{sec:con}.


\section{Formalism} \label{sec:for}

Following the formalism introduced in Ref.~\cite{Gronau:2004gt}, we define the amplitudes
\begin{eqnarray}
A(\BsDbphi) & = & A^{(*)}_{B}, \label{EQ__1_} \\
A(\BsDphi) & = & A^{(*)}_{B} \rBst e^{i(\deltaBst +\gamma)},\label{EQ__2_}
\end{eqnarray}
where $A^{(*)}_{B}$ and $r_{B}^{(*)}$  are the magnitude of the $\Bs$ decay amplitude and the amplitude magnitude ratio between the suppressed over the favoured $\Bs$ decay modes, respectively, while $\delta_{B}^{(*)}$ and $\gamma $ are the strong and weak phases, respectively. Neglecting mixing and \CP violation in $D$ decays (see for example Ref.~\cite{GSZ,MartoneZupan}), the amplitudes into the final state $f$ (denoted below as $[f]_{D}$) and its \CP conjugate $\bar{f}$ are defined as
\begin{eqnarray}
A(\Dzb \rightarrow f) = A(\Dz  \rightarrow \bar{f}) & = & A_{f}, \label{EQ__3_}\\
A(\Dz  \rightarrow f) = A(\Dzb \rightarrow \bar{f}) & = & A_{f} r_{D}^{f} e^{i\delta_{D}^{f}}, \label{EQ__4_}
\end{eqnarray}
where $\delta_{D}^{f} $ and $r_{D}^{f}$ are the strong phase difference and relative magnitude, respectively, between the $\Dz \rightarrow f$ and the $\Dzb \rightarrow f$ decay amplitudes.

The amplitudes of the full decay chains are given by
\begin{eqnarray}
A_{Bf} &\equiv& A(\Bs \rightarrow [f]_{D^{(*)} } \phi )\nonumber\\
 & = & A^{(*)}_{B} A^{(*)}_{f} \left[1+\rBst r_{D}^{f} e^{i(\deltaBst + \delta_{D}^{f} + \gamma)} \right], \label{EQ__5_} \\
A_{B\bar{f}} &\equiv& A(\Bs \rightarrow [\bar{f}]_{D^{(*)} } \phi ) \nonumber\\
& = & A^{(*)}_{B} A^{(*)}_{f} \left[\rBst e^{i(\deltaBst +\gamma )} + r_{D}^{f} e^{i\delta_{D}^{f} } \right]. \label{EQ__6_}
\end{eqnarray}
The amplitudes for the \CP-conjugate decays are given by changing the sign of the weak phase $\gamma$
\begin{eqnarray}
\bar{A}_{Bf} &\equiv& A(\Bsb \rightarrow [f]_{D^{(*)}} \phi) \nonumber\\
& = & A^{(*)}_{B} A^{(*)}_{f} \left[\rBst e^{i(\deltaBst - \gamma)} + r_{D}^{f} e^{i\delta_{D}^{f} } \right], \label{EQ__7_} \\
\bar{A}_{B\bar{f}} &\equiv& A(\Bsb \rightarrow [\bar{f}]_{D^{(*)}} \phi ) \nonumber\\
& = & A^{(*)}_{B} A^{(*)}_{f} \left[1+\rBst r_{D}^{f} e^{i(\deltaBst +\delta_{D}^{f} -\gamma)} \right]. \label{EQ__8_}
\end{eqnarray}
Using the standard notations
\begin{eqnarray}
&&\tau =\Gamma_s t,\quad \Gamma_s =\frac{\Gamma_{L} +\Gamma_{H}}{2},\quad \Delta\Gamma_s=\Gamma_{L}-\Gamma_{H},\nonumber\\
&&\quad y=\frac{\Delta \Gamma_s}{2\Gamma_s},\quad \lambda_{f}=\frac{q}{p} . \frac{\bar{A}_{Bf}}{A_{Bf}},\nonumber
\end{eqnarray}
and assuming $|q/p|=1$ ($|q/p|=1.0003\pm 0.0014$ \cite{PDG}), the untagged decay rate for the decay $\Bs/\Bsb \rightarrow [f]_{D^{(*)}} \phi$ is given by (Eq.~(10) of Ref.~\cite{Gronau:2007bh})
\begin{multline} \label{EQ__9_}
\frac{d\Gamma (\Bs(\tau) \rightarrow [f]_{D^{(*)}} \phi )}{d\tau } + \frac{d\Gamma(\Bsb(\tau) \rightarrow [f]_{D^{(*)}} \phi )}{d\tau } \propto \\
e^{-\tau } |A_{Bf}|^{2} \times
\left[(1+\left|\lambda_{f} \right|^{2})\cosh(y\tau)-2\mathrm{Re}(\lambda_{f})\sinh(y\tau)\right].
\end{multline}

\subsection{Time acceptance}
\label{sec:timeAcc}

Experimentally, due to trigger and selection requirements and to inefficiencies in the reconstruction, the decay time distribution is affected by acceptance effects. The acceptance correction has been estimated from pseudoexperiments based on a related publication by the LHCb collaboration~\cite{LHCb:2011aa}. It is described by an empirical acceptance function
\begin{equation} \label{EQ__10_}
\varepsilon_{ta}(\tau )=\frac{(\alpha \tau )^{\beta }}{1+(\alpha \tau )^{\beta }}(1-\xi \tau ),
\end{equation}
with $\alpha =1.5$, $\beta =2.5$ and $\xi =0.01$.

Taking into account this effect, the time-integrated untagged decay rate is
\begin{eqnarray} \label{EQ__11_}
&&\Gamma({{\tilde{B}^0_s}} \rightarrow [f]_{D^{(*)} } \phi ) \nonumber\\
&&=\int_{0}^{\infty}\left[\frac{d\Gamma (\Bs(\tau) \rightarrow [f]_{D^{(*)}} \phi)}{d\tau} \nonumber\right.\\
&&\left.+ \frac{d\Gamma(\Bsb(\tau )\rightarrow [f]_{D^{(*)}} \phi)}{d\tau } \right] \varepsilon_{ta}(\tau ) d\tau.
\end{eqnarray}

Defining the function
\begin{equation} \label{EQ__12_}
g(x)=\int_{0}^{\infty}\frac{e^{-x\tau} (1+\xi \tau (\alpha \tau)^{\beta})}{1+(\alpha \tau )^{\beta }}d\tau,
\end{equation}
and using Eq.~\eqref{EQ__9_}, one gets
\begin{equation} \label{EQ__13_}
\Gamma(\Bs \rightarrow [f]_{D} \phi)\propto \left|A_{Bf} \right|^{2} \left[(1+\left|\lambda_{f} \right|^{2} )\AT-2y\mathrm{Re}(\lambda_{f})\BT\right],
\end{equation}
where  $\AT=1-[g(1-y)+g(1+y)]/2$ and $\BT=1-[g(1-y)-g(1+y)]/2y$. With $y=(0.128\pm 0.009)/2$ for the $\Bs$ meson~\cite{HFLAV}, one gets $\AT=0.488\pm 0.005$ and $\BT=0.773\pm 0.008$. Examples of decay-time acceptance distributions are displayed in Fig.~\ref{fig:timeAcceptcurv}.

\myfigure{\includegraphics[width=0.9\textwidth]{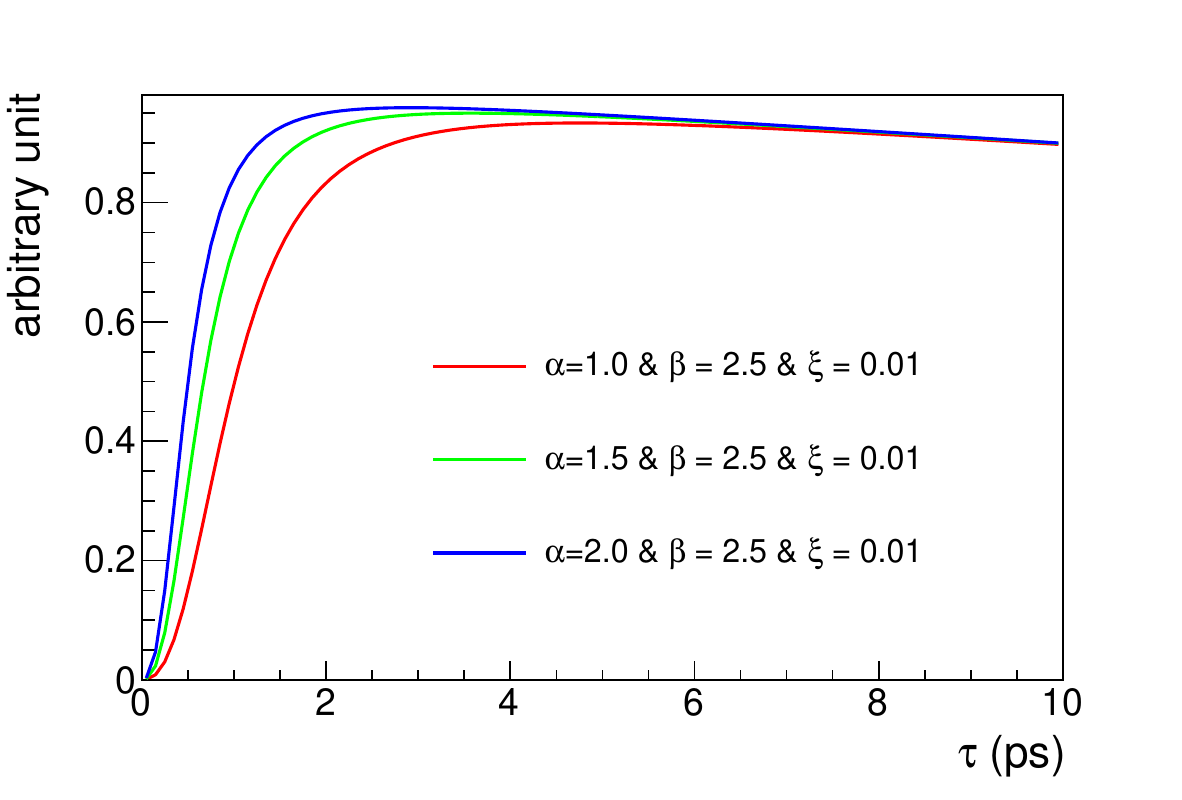}
\figcaption{\label{fig:timeAcceptcurv} Examples of decay-time acceptance distributions for three different sets of parameters $\alpha$, $\beta$,
and  $\xi$ (nominal in green).}}

\subsection{Observables for $\Dz$ decays}
\label{sec:ObsDz}

The $D$-meson decays are reconstructed in quasi flavour-specific modes:
$f^{-}(\equiv f)= \Km \pip$, $\Km 3\pi$, $\Km \pip \pi^{0}$, and their \CP-conjugate modes:
$f^{+}(\equiv \bar{f})= \Kp \pim$, $\Kp 3\pi$, $\Kp \pim \pi^{0}$
as well as \CP-eigenstate modes: $f_{CP}= \Kp \Km$, $\pip \pim$.

In the following, we introduce the weak phase $\beta_{s}$ is defined as $\beta_{s} = \arg \left(-\frac{V_{ts} V_{tb}^{*} }{V_{cs} V_{cb}^{*} } \right)$.
From  Eqs.~\eqref{EQ__5_}, \eqref{EQ__7_}, \eqref{EQ__13_} and with $\lambda_{f}=e^{2i\beta_{s}}\frac{\bar{A}_{Bf}}{A_{Bf}}$,  for a given number of untagged $\Bs$ mesons produced in the $pp$ collisions at the LHCb interaction point, $N(\Bs)$, we can compute the number of $\BstoDbphi$ decays with the $D$ meson decaying into the final state $f^{-}$. For the reference decay mode $f^{-} \equiv \Km \pip$ we obtain
\begin{multline}  \label{EQ__14_}
N\left(\Bs \rightarrow \left[\Km \pip \right]_{D} \left[\Kp \Km \right]_{\phi } \right) = \\
C_{K\pi}\Big[-2\BT y \rB \cos \left(\deltaB + 2\beta_{s} - \gamma \right) \\
+  \AT \left(1+\rB^{2} +4 \rB r_{D}^{K\pi} \cos \deltaB \cos \left(\delta_{D}^{K\pi } + \gamma \right)\right)\Big],
\end{multline}
where, the terms proportional to $(r_{D}^{K\pi})^{2} \ll 1$ and $y r_{D}^{K\pi } \ll 1$ have been neglected ($r_{D}^{K\pi} =5.90^{+0.34}_{-0.25}~\%$~\cite{HFLAV}). The best approximation for the scale factor $C_{K\pi}$ is
\begin{multline} \label{EQ__15_}
C_{K\pi }=N(\Bs)\times \varepsilon(\Bs \rightarrow \left[\Km \pip \right]_{D} \left[\Kp \Km \right]_{\phi})\\
\times Br(\Bs \rightarrow \left[\Km \pip \right]_{D}  \left[\Kp \Km \right]_{\phi}),
\end{multline}
where,   $\varepsilon(\Bs \rightarrow \left[\Km \pip \right]_{D} \left[\Kp \Km \right]_{\phi})$ is the global detection efficiency of this decay mode, and $Br(\Bs \rightarrow \left[\Km \pip \right]_{D} \left[\Kp \Km \right]_{\phi})$ its branching fraction. The value of the scale factor $C_{K\pi}$  is estimated from the LHCb Run 1 data~\cite{Aaij:2018jqv}, the average $f_{s}/f_{d}$ of the $b$-hadron production fraction ratio measured by LHCb~\cite{LHCb:2013lka} and the different branching fractions~\cite{PDG}.

For a better numerical behaviour, we use the Cartesian coordinates parametrisation
\begin{equation}\label{EQ__16_}
x_{\pm}^{(*)} =\rBst \cos (\deltaBst \pm \gamma ) \ \  \textrm{and} \quad y_{\pm}^{(*)}=\rBst \sin(\deltaBst \pm \gamma).
\end{equation}
Then, Eq~\eqref{EQ__14_} becomes
\begin{multline} \label{EQ__18_}
N\left(\Bs \rightarrow \left[\Km \pip \right]_{D} \left[\Kp \Km \right]_{\phi } \right) = \\
C_{K\pi} \Big[ -2 \BT y \left[x_{-} \cos(2\beta_{s}) - y_{-} \sin(2\beta_{s}) \right] + \\
\AT \Big( 1+x_{-}^{2} +y_{-}^{2} +   \\
2r_{D}^{K\pi} \Large[(x_{+} +x_{-}) \cos \delta_{D}^{K\pi}
    - (y_{+} -y_{-}) \sin \delta_{D}^{K\pi} \Large] \Big)\Big].
\end{multline}

For three and four body final states $K3\pi$ and $K\pi \pi^{0} $, there are multiple interfering amplitudes, therefore their amplitudes and phases $\delta_{D}^{f} $ vary across  the decay phase space. However, an analysis which integrates over the phase space can be performed in a very similar way to two body decays with the inclusion of an additional parameter, the so-called coherence factor $R_{D}^{f}$ which has been measured in previous experiments~\cite{Evans:2016tlp}. The strong phase difference $\delta_{D}^{f} $ is then treated as an effective phase averaged over all amplitudes. For these modes, we have an expression similar to \eqref{EQ__18_}
\begin{multline}\label{EQ__19_}
N\left(\Bs \rightarrow \left[f^{-} \right]_{D} \left[\Kp \Km \right]_{\phi} \right) =\\
C_{K\pi} F_{f} \Big[ -2 \BT y\left[x_{-} \cos \left(2\beta_{s} \right) - y_{-} \sin \left(2\beta_{s} \right)\right] + \\
 \AT \Big(1+x_{-}^{2} + y_{-}^{2} \\
  + 2r_{D}^{f} R_{D}^{f} \left[\left(x_{+} +x_{-} \right)\cos \delta_{D}^{f} -\left(y_{+} -y_{-} \right)\sin \delta_{D}^{f} \right]\Big)\Big],
\end{multline}
where  $F_{f}$ is the scale factor of the $f$ decay relative to the $K\pi$ decay and depends on the ratios of detection efficiencies and branching fractions of the corresponding modes
\begin{multline} \label{EQ__20_}
F_{f} =\frac{C_{f}}{C_{K\pi}} = \frac{\varepsilon(D\rightarrow f)}{\varepsilon (D\rightarrow K\pi )} \times \\
\frac{[Br(\Dz \rightarrow f) + Br(\Dzb \rightarrow f)]}{[Br(\Dz \rightarrow \Km \pip)+Br(\Dzb \rightarrow \Km \pip)]}.
\end{multline}
The value of $F_{f}$ for the different modes used in this study is determined from LHCb measurements in $B^{\pm} \rightarrow DK^{\pm}$ and $B^{\pm} \rightarrow D\pi^{\pm }$ modes, with two or four-body $D$ decays~\cite{Aaij:2016oso,Aaij:2015jna}.

The time-integrated untagged decay rate for $\Bs \rightarrow [\bar{f}]_{D} \phi$  is given by Eq.~\eqref{EQ__13_} by substituting  $A_{Bf} \rightarrow \bar{A}_{B\bar{f}}$ and $\lambda_{f} \rightarrow \bar{\lambda}_{\bar{f}} = \lambda_{f}^{-1} = e^{-2i\beta_{s}} (A_{B\bar{f}}/\bar{A}_{B\bar{f}})$ which is equivalent to the change $\beta_{s} \rightarrow -\beta_{s} $  and $\gamma \rightarrow -\gamma $  ({\it i.e.} $x_{\pm} \rightarrow x_{\mp}$ and $y_{\pm} \rightarrow y_{\mp}$). Therefore, the observables are
\begin{multline} \label{EQ__21_}
N\left(\Bs \rightarrow \left[\Kp \pim \right]_{D} \left[\Kp \Km \right]_{\phi} \right) = \\
C_{K\pi} \Big[ -2 \BT y\left[x_{+} \cos \left(2\beta_{s} \right)+y_{+} \sin \left(2\beta_{s} \right)\right] + \\
 \AT \Big(1+x_{+}^{2} +y_{+}^{2} + \\
2r_{D}^{K\pi} \left[\left(x_{+} +x_{-} \right)\cos \delta_{D}^{K\pi} + \left(y_{+} -y_{-} \right)\sin \delta_{D}^{K\pi} \right]\Big)\Big],
\end{multline}
and for the modes $f^{+} \equiv \Kp 3\pi$, $\Kp \pim \pi^{0}$
\begin{multline} \label{EQ__22_}
N\left(\Bs \rightarrow \left[f^{+} \right]_{D} \left[\Kp \Km \right]_{\phi } \right) = \\
C_{K\pi} F_{f} \Big[ -2\BT y\left[x_{+} \cos \left(2\beta_{s} \right)+y_{+} \sin \left(2\beta_{s} \right)\right] + \\
\AT \Big(1+x_{+}^{2} +y_{+}^{2} + \\
2r_{D}^{f} R_{f} \left[\left(x_{+} +x_{-} \right)\cos \delta_{D}^{f} +\left(y_{+} -y_{-} \right)\sin \delta_{D}^{f} \right]\Big)\Big].
\end{multline}
Obviously, any significant asymmetries on the yield of observable corresponding to Eq.~\ref{EQ__18_} with respect to  Eq.~\ref{EQ__21_}, or  Eq.~\ref{EQ__19_} with respect to  Eq.~\ref{EQ__22_}, is a clear signature for \CP violation.

For the \CP-eigenstate modes $D\rightarrow h^{+} h^{-} \, (h\equiv K,\, \pi )$, we have $r_{D}=1$  and $\delta_{D}=0$. Following the same approach than for quasi flavour-specific modes, the observables can be written as
\begin{multline} \label{EQ__23_}
N\left(\Bs \rightarrow \left[h^{+} h^{-} \right]_{D} \left[\Kp \Km \right]_{\phi} \right) = \\
4 C_{K\pi } F_{hh} \Big[ \AT \left(1+x_{+}^{2} +y_{+}^{2} +x_{+} +x_{-} \right) -  \\
    \BT y \Big( \left(1+x_{+} +x_{-} +x_{+} x_{-} +y_{+} y_{-} \right)\cos \left(2\beta_{s} \right)+ \\
    \left(y_{+} -y_{-} +y_{+} x_{-} -x_{+} y_{-} \right)\sin \left(2\beta_{s} \right) \Big) \Big].
\end{multline}
In analogy with $F_{f}$, $F_{hh}$ is defined as
\begin{multline} \label{EQ__24_}
F_{hh} =
\frac{C_{hh} }{C_{K\pi} } =\frac{\varepsilon (D\rightarrow hh)}{\varepsilon(D\rightarrow K\pi )} \times \\
\frac{Br(\Dz \rightarrow hh)}{[Br(\Dz \rightarrow \Km \pip)+ Br(\Dzb \rightarrow \Km \pip )]}
\end{multline}
and their values are determined in the same way than $F_{f}$.

For the modes $\KS\pip\pim$ and $\KS\Kp\Km$ ({\it i.e.} $\KS hh$) one obtains
\begin{multline} \label{EQ__Kshh}
N\left(\Bs \rightarrow \left[ \KS hh\right]_{D} \left[\Kp \Km \right]_{\phi } \right) = 2 C_{K\pi} F_{\KS hh} \times \\
\Big[ -\BT y  \left[(x_{+}+x_{-}) \cos(2\beta_{s}) +(y_{+}-y_{-}) \sin(2\beta_{s})\right] + \\
\AT \Big( 1 +x_{-}^{2} +y_{-}^{2} + 2 (x_{+} +x_{-}) \\
r_{D}^{\KS hh}(m^2_+,m^2_-) \kappa_{D}^{\KS hh}(m^2_{+},m^2_{-}) \cos \delta_{D}^{\KS hh}(m^2_{+},m^2_{-}) \Big) \Big],
\end{multline}
where $F_{\KS hh}$ is defined as for Eq.~\ref{EQ__24_}. The strong parameters $r_{D}^{\KS hh}(m^2_+,m^2_-)$, $\kappa_{D}^{\KS hh}(m^2_+,m^2_-)$, and   $\cos \delta_{D}^{\KS hh}(m^2_+,m^2_-)$ vary over the Dalitz plot $(m^2_+,m^2_-)\equiv (m^2(\KS\pip), \ m^2(\KS\pim))$  and are defined in Sec.~\ref{sec:yie}.

\subsection{Observables for $\Dstz$ decays}
\label{sec:ObsDzst}

For the \Dstz decays, we considered the two modes: $\Dstz\rightarrow \Dz \pi^{0}$  and $\Dstz \rightarrow \Dz \gamma$, where the $\Dz$  mesons are reconstructed, as in the above, in quasi flavour-specific modes: $K\pi$, $K3\pi$, $K\pi \pi^{0}$ and \CP-eigenstate modes: $\pi \pi$ and $KK$ . As shown in Ref.~\cite{Bondar:2004bi}, the formalism for the cascade $\Bs \rightarrow \Dstzb \phi ,\; \Dstzb \rightarrow \Dzb \pi^{0}$  is similar to the $\Bs \rightarrow \Dzb \phi$. Therefore, the relevant observables can be written similarly to Eqs.~\eqref{EQ__18_}, \eqref{EQ__19_}, \eqref{EQ__21_}, \eqref{EQ__22_} and \eqref{EQ__23_}, by substituting $C_{K\pi} \rightarrow C_{K\pi,D \piz}$, $\rB \rightarrow \rstB$  and $\deltaB \rightarrow \deltastB$ ({\it i.e.} $x_{\pm} \rightarrow x_{\pm}^{*}$ and $y_{\pm} \rightarrow y_{\pm}^{*}$)
\begin{multline} \label{EQ__25_}
N\left(\Bs \rightarrow \left[\left[\Km \pip \right]_{D} \pi^{0} \right]_{D^{*}} \left[\Kp \Km \right]_{\phi} \right) = \\
C_{K\pi,D\piz} \Big[ -2 \BT y\left[x_{-}^{*} \cos \left(2\beta_{s} \right)-y_{-}^{*} \sin \left(2\beta_{s} \right)\right] + \\
\AT \Big(1+x_{-}^{*2} +y_{-}^{*2} + \\
2r_{D}^{K\pi} \left[\left(x_{+}^{*} +x_{-}^{*} \right)\cos \delta_{D}^{K\pi} - \left(y_{+}^{*} - y_{-}^{*} \right) \sin \delta_{D}^{K\pi} \right]\Big)\Big],
\end{multline}
\begin{multline} \label{EQ__26_}
N\left(\Bs \rightarrow \left[\left[\Kp \pim \right]_{D} \pi^{0} \right]_{D^{*}} \left[\Kp \Km \right]_{\phi } \right) = \\
C_{K\pi,D\piz} \Big[ -2 \BT y\left[x_{+}^{*} \cos \left(2\beta_{s} \right)+y_{+}^{*} \sin \left(2\beta_{s} \right)\right] + \\
\AT \Big(1+x_{+}^{*2} + y_{+}^{*2} + \\
2r_{D}^{K\pi} \left[ \left(x_{+}^{*} + x_{-}^{*} \right)\cos \delta_{D}^{K\pi} + \left(y_{+}^{*} - y_{-}^{*} \right) \sin \delta_{D}^{K\pi} \right]\Big)\Big],
\end{multline}
\begin{multline} \label{EQ__27_}
N\left(\Bs \rightarrow \left[\left[f^{-} \right]_{D} \pi^{0} \right]_{D^{*} } \left[\Kp \Km \right]_{\phi} \right) = \\
C_{K\pi,D\piz} F_{f} \Big[ - 2 \BT y\left[x_{-}^{*} \cos \left(2\beta_{s} \right)-y_{-}^{*} \sin \left(2\beta_{s} \right)\right] + \\
\AT \Big(1+x_{-}^{*2} +y_{-}^{*2} + \\
    2r_{D}^{f} R_{f} \left[\left(x_{+}^{*} +x_{-}^{*} \right)\cos \delta_{D}^{f} -\left(y_{+}^{*} -y_{-}^{*} \right)\sin \delta_{D}^{f} \right]\Big)\Big],
\end{multline}
\begin{multline} \label{EQ__28_}
N\left(\Bs \rightarrow \left[\left[f^{+} \right]_{D} \pi^{0} \right]_{D^{*}} \left[\Kp \Km \right]_{\phi} \right) = \\
C_{K\pi,D\piz} F_{f} \Big[ -2\BT y\left[x_{+}^{*} \cos \left(2\beta_{s} \right)+y_{+}^{*} \sin \left(2\beta_{s} \right)\right] + \\
\AT \Big(1+x_{+}^{*2} +y_{+}^{*2} + \\
2r_{D}^{f} R_{f} \left[\left(x_{+}^{*} +x_{-}^{*} \right)\cos \delta_{D}^{f} +\left(y_{+}^{*} -y_{-}^{*} \right)\sin \delta_{D}^{f} \right]\Big)\Big],
\end{multline}
\begin{multline} \label{EQ__29_}
N\left(\Bs \rightarrow \left[\left[h^{+} h^{-} \right]_{D} \pi^{0} \right]_{D^{*}} \left[\Kp \Km \right]_{\phi} \right) = \\
4 C_{K\pi,D\piz} F_{hh} \Big[ \AT \left(1+x_{+}^{*2} +y_{+}^{*2} +x_{+}^{*} +x_{-}^{*} \right)  - \\
\BT y \Big( \left(1+x_{+}^{*} +x_{-}^{*} +x_{+}^{*} x_{-}^{*} +y_{+}^{*} y_{-}^{*} \right)\cos \left( 2\beta_{s} \right) \\
+\left(y_{+}^{*} -y_{-}^{*} +y_{+}^{*} x_{-}^{*} -x_{+}^{*} y_{-}^{*} \right)\sin \left(2\beta_{s} \right) \Big)\Big].
\end{multline}

In the case $\Dstz \rightarrow \Dz \gamma$, the formalism is very similar, except that there is an effective strong phase shift of $\pi$ with respect to the $\Dstz \rightarrow \Dz \pi^{0}$~\cite{Bondar:2004bi}. The observables can be derived from the previous ones substituting $C_{K\pi,D\piz} \rightarrow C_{K\pi,D\gamma}$  and $\delta_{B}^{*} \rightarrow \delta_{B}^{*} +\pi $  ({\it i.e.} $x_{\pm}^{*} \rightarrow -x_{\pm}^{*}$ and $y_{\pm}^{*} \rightarrow -y_{\pm}^{*} $)
\begin{multline} \label{EQ__30_}
N\left(\Bs \rightarrow \left[\left[\Km \pip \right]_{D} \gamma \right]_{D^{*}} \left[\Kp \Km \right]_{\phi} \right) = \\
C_{K\pi,D\gamma } \Big[ 2 \BT y\left[x_{-}^{*} \cos \left(2\beta_{s} \right)-y_{-}^{*} \sin \left(2\beta_{s} \right)\right]  + \\
\AT \Big(1+x_{-}^{*2} +y_{-}^{*2} + \\
2r_{D}^{K\pi} \left[-\left(x_{+}^{*} +x_{-}^{*} \right)\cos \delta_{D}^{K\pi} +\left(y_{+}^{*} -y_{-}^{*} \Big)\sin \delta_{D}^{K\pi } \right]\right)\Big],
\end{multline}
\begin{multline} \label{EQ__31_}
N\left(\Bs \rightarrow \left[\left[\Kp \pim \right]_{D} \gamma \right]_{D^{*}} \left[\Kp \Km \right]_{\phi} \right) = \\
C_{K\pi,D\gamma} \Big[2 \BT y\left[x_{+}^{*} \cos \left(2\beta_{s} \right)+y_{+}^{*} \sin \left(2\beta_{s} \right)\right]   + \\
\AT \Big(1+x_{+}^{*2} +y_{+}^{*2} + \\
2r_{D}^{K\pi} \left[-\left(x_{+}^{*} +x_{-}^{*} \right)\cos \delta_{D}^{K\pi } -\left(y_{+}^{*} -y_{-}^{*} \Big)\sin \delta_{D}^{K\pi } \right]\right)\Big],
\end{multline}


\begin{table}[htbp]
\centering
\caption{\label{tab:lum} Integrated luminosities and cross-sections of LHCb Run 1 and Run 2 data. The integrated luminosities come from Ref.~\cite{LHCbPage} and cross-sections from Refs.~\cite{Aaij:2013noa,Aaij:2016avz}  }
\small
\begin{tabular}{lcccc}
 \hline \hline
Years/Run & $\sqrt{s}$ (TeV) & int. lum.(\invfb) & cross section & equiv. 7 TeV data \\
\hline
2011    & 7     & 1.1       & $\sigma_{2011}=38.9~\upmu\mathrm{b}$ & 1.1 \\
2012    & 8     & 2.1       & $1.17 \times \sigma_{2011}$     & 2.4 \\
\hline
Run 1   &   \textendash    &  3.2      &    \textendash                          & 3.5 \\
\hline
2015-2018 (Run 2) & 13     & 5.9  &  $2.00 \times \sigma_{2011}$       & 11.8 \\
\hline
Total     &    \textendash   & 9.1     &             \textendash              & 15.3 \\
\hline
\end{tabular}

\end{table}


%
\begin{multline} \label{EQ__32_}
N\left(\Bs \rightarrow \left[\left[f^{-} \right]_{D} \gamma \right]_{D^{*}} \left[\Kp \Km \right]_{\phi} \right) = \\
C_{K\pi,D\gamma} F_{f} \Big[2 \BT y\left[x_{-}^{*} \cos \left(2\beta_{s} \right)-y_{-}^{*} \sin \left(2\beta_{s} \right)\right] + \\
\AT \Big(1+x_{-}^{*2} +y_{-}^{*2} + \\
2r_{D}^{f} R_{f} \left[-\left(x_{+}^{*} +x_{-}^{*} \right)\cos \delta_{D}^{f} +\left(y_{+}^{*} -y_{-}^{*} \Big)\sin \delta_{D}^{f} \right]\right)\Big],
\end{multline}

\begin{multline} \label{EQ__33_}
N\left(\Bs \rightarrow \left[\left[f^{+} \right]_{D} \gamma \right]_{D^{*}} \left[\Kp \Km \right]_{\phi} \right) = \\
C_{K\pi,D\gamma} F_{f} \Big[2 \BT y\left[x_{+}^{*} \cos \left(2\beta_{s} \right)+y_{+}^{*} \sin \left(2\beta_{s} \right)\right] + \\
 \AT \Big(1+x_{+}^{*2} +y_{+}^{*2} + \\
 2r_{D}^{f} R_{f} \left[-\left(x_{+}^{*} +x_{-}^{*} \right) \cos \delta_{D}^{f} - \left(y_{+}^{*} -y_{-}^{*} \right)\sin \delta_{D}^{f} \right]\Big)\Big],
\end{multline}
\begin{multline} \label{EQ__34_}
N\left(\Bs \rightarrow \left[\left[h^{+} h^{-} \right]_{D} \gamma \right]_{D^{*} } \left[\Kp \Km \right]_{\phi} \right) = \\
4C_{K\pi,D\gamma} F_{hh} \Big[ \AT \left(1+x_{+}^{*2} +y_{+}^{*2} -x_{+}^{*} -x_{-}^{*} \right)  - \\
 \BT y \Big( \left(1-x_{+}^{*} -x_{-}^{*} +x_{+}^{*} x_{-}^{*} +y_{+}^{*} y_{-}^{*} \right) \cos \left(2\beta_{s} \right) \\
  + \left(-y_{+}^{*} + y_{-}^{*} + y_{+}^{*} x_{-}^{*} -x_{+}^{*} y_{-}^{*} \right)\sin \left(2\beta_{s} \right) \Big) \Big].
\end{multline}
$C_{K\pi,D\pi^{0}}$  and $C_{K\pi,D\gamma}$ are determined in the same way $C_{K\pi }$, {\it i.e.} from the LHCb Run 1 data~\cite{Aaij:2018jqv} and taking into account the fraction of longitudinal polarization in the decay $\Bs \rightarrow \Dstz \phi$: $f_{L} =(73\pm 15\pm 4)\%$~\cite{Aaij:2018jqv} and the branching fractions $Br(\Dstzb \rightarrow \Dzb \pi^{0})$  and  $Br(\Dstzb \rightarrow \Dzb \gamma)$~\cite{PDG}.


\section{Expected yields}
\label{sec:yie}

The LHCb collaboration has measured the yields of $\BsDtphi$, $\tilde{D}^0\rightarrow K\pi$ modes using Run 1 data, corresponding to an integrated luminosity of 3~$\invfb$ (Ref.~\cite{Aaij:2018jqv}). Taking into account cross-section differences among different centre-of-mass energies, the equivalent integrated luminosities in different data taking years at LHCb are summarized in Table~\ref{tab:lum}. The corresponding expected yields of $\tilde{D}^0$ meson decaying into other modes are also estimated according to Ref.~\cite{Aaij:2016oso}, \cite{Aaij:2015jna}, and \cite{2014GGSZ}, the scaled results are listed in Table~\ref{tab:yields}, where the longitudinal polarisation fraction $f_L=(73\pm 15\pm 4)~\%$~\cite{Aaij:2018jqv} of $\Bs\rightarrow \tilde{D}^{*0}\phi$ is considered so that the \CP eigenvalue of the final state is well defined and similar to that of the $\Bs\rightarrow \tilde{D}^{0}\phi$  mode.

There are some extra parameters used in the sensitivity study, as shown in Table~\ref{tab:para}. Most of which come from $D$ decays, and the scale factors $F$ are calculated by using the data from Ref.~\cite{Aaij:2016oso} and \cite{Aaij:2015jna}, and branching fractions from PDG~\cite{PDG}.

\begin{center}
\tabcaption{\label{tab:yields} Expected yield of each mode for $9.1 \invfb$~(Run 1 and Run 2 data). The expected yields for the $\Bs\rightarrow \tilde{D}^{*0}\phi$ sub-modes are scaled by the longitudinal fraction of polarisation $f_L=(73\pm15)~\%$. To be scaled by 6.3 (90) for prospects after 2025 (2038)(see Section~\ref{sec:HL-LHC}).}
\begin{tabular}{lcrr}
\hline \hline
  & &  \multicolumn{2}{c}{Expect. yield (Run 1 only)} \\
\hline
$\Bs\rightarrow \tilde{D}^0(K\pi)\phi$    &    & \multicolumn{2}{c}{$577$ ($132\pm 13$~\cite{Aaij:2018jqv})} \\
$\Bs\rightarrow \tilde{D}^0(K3\pi)\phi$    &   & \multicolumn{2}{c}{$218$} \\
$\Bs\rightarrow \tilde{D}^0(K\pi\piz)\phi$ &  & \multicolumn{2}{c}{$58$} \\
$\Bs\rightarrow \tilde{D}^0(KK)\phi$        &  & \multicolumn{2}{c}{$82$} \\
$\Bs\rightarrow \tilde{D}^0(\pi\pi)\phi$    &  & \multicolumn{2}{c}{$24$} \\
$\Bs\rightarrow \tilde{D}^0(\KS \pi\pi)\phi$        &  & \multicolumn{2}{c}{$54$} \\
$\Bs\rightarrow \tilde{D}^0(\KS KK)\phi$    &  & \multicolumn{2}{c}{$8$} \\
\hline
$\Bs\rightarrow \tilde{D}^{*0}\phi$ mode & & $D^0\piz$  & $D^0\gamma$ \\
\hline
$\Bs\rightarrow \tilde{D}^{*0}(K\pi)\phi$ &     & $337$ & $184$ \\
                                  &  & \multicolumn{2}{c}{(119~\cite{Aaij:2018jqv})} \\
$\Bs\rightarrow \tilde{D}^{*0}(K3\pi)\phi$ &      & $127$ & $69$ \\
$\Bs\rightarrow \tilde{D}^{*0}(K\pi\piz)\phi$ &  & $34$ & $18$ \\
$\Bs\rightarrow \tilde{D}^{*0}(KK)\phi$       &   & $48$ & $26$ \\
$\Bs\rightarrow \tilde{D}^{*0}(\pi\pi)\phi$    &  & $14$ & $8$ \\
\hline
\end{tabular}
\end{center}

The expected numbers of signal events are also calculated from the full expressions given in Sections~\ref{sec:ObsDz} and~\ref{sec:ObsDzst}, by using detailed branching fraction derivations explained in Ref.~\cite{Aaij:2018jqv} and scaling by the LHCb Run 1 and Run 2 integrated luminosities as listed in Table~\ref{tab:lum}. The obtained normalisation factors $C_{K\pi}$, $C_{K\pi,D\piz}$, and $C_{K\pi,D\gamma}$ are respectively $608\pm67$, $347\pm56$, and $189\pm31$. To compute the uncertainty on the normalisation factors, we made the assumption that it is possible to improve by a factor 2 the global uncertainty on the measurement of the branching fraction of the decay modes $\BsDbphi$, and of the polarisation of the mode $\BsDstbphi$, when adding LHCb data from Run 2~\cite{Aaij:2018jqv}.  The values of the three normalisation factors are in good agreement with the yields listed in Table~\ref{tab:yields}.

\begin{center}
\tabcaption{\label{tab:para} Other external parameters used in the sensitivity study. The scale factors $F$ are also listed.}
\centering
\begin{tabular}{lr}
\hline
\hline
Parameter      & Value \\
\hline
-2$\beta_S$ [mrad]     & $-36.86\pm0.82$~\cite{CKMfitterSUMMER19} \\
$y=\Delta\Gamma_s/2\Gamma_s$  (\%)   & $6.40\pm0.45$~\cite{HFLAV}  \\
$r^{K\pi}_D$ (\%)    & $5.90^{+0.34}_{-0.25}$~\cite{HFLAV} \\
$\delta^{K\pi}_D$ [deg]   & $188.9^{+8.2}_{-8.9}$~\cite{HFLAV} \\
$r^{K3\pi}_D$ (\%)    & $5.49\pm0.06$~\cite{Evans:2016tlp} \\
$R^{K3\pi}_D$ (\%)    & $43^{+17}_{-13}$~\cite{Evans:2016tlp} \\
$\delta^{K3\pi}_D$  [deg]   & $128^{+28}_{-17}$~\cite{Evans:2016tlp} \\
$r^{K\pi\piz}_D$ (\%)    & $4.47\pm0.12$~\cite{Evans:2016tlp} \\
$R^{K\pi\piz}_D$ (\%)    & $81\pm6$~\cite{Evans:2016tlp} \\
$\delta^{K\pi\piz}_D$ [deg]   & $198^{+14}_{-15}$~\cite{Evans:2016tlp} \\
\hline
Scale factor (wrt $K\pi$) & (stat. uncertainty only) \\
\hline
$F_{K3\pi}$ (\%)    & $37.8\pm0.1$~\cite{Aaij:2016oso} \\
$F_{K\pi\piz}$ (\%)   & $10.0\pm0.1$~\cite{Aaij:2015jna} \\
$F_{KK}$ (\%)   & $14.2\pm 0.1$~\cite{Aaij:2016oso} \\
$F_{\pi\pi}$ (\%)    & $4.2\pm0.1$~\cite{Aaij:2016oso} \\
\hline
\end{tabular}
\end{center}

The number of expected event yields and the value of the coherence factor, $R_D$,  listed in Table~\ref{tab:yields} and~\ref{tab:para} justify a posteriori our choice of performing the sensitivity study on $\gamma$ with the $D$-meson decay modes $K\pi$, $K3\pi$, $K\pi\piz$, $KK$ and $\pi\pi$. By definition the
value of $R_D$ is one for two-body decays and $r_D=1$ for \CP-eigenstates, while for $K3\pi$ $R_D$ is about $43\%$ and larger, $81\%$, for $K\pi\piz$. The larger is $R_D$ the strongest is the sensitivity to $\gamma$. As from Eqs.~\ref{EQ__21_}, \ref{EQ__22_}, and \ref{EQ__23_} it is clear that the largest sensitivity to $\gamma$ is expected to be originated from the ordered $\Dz$ decay modes: $KK$, $\pi\pi$, $K\pi$, $K\pi\piz$, and $K3\pi$, for the same number of selected events. Therefore, even with lower yields the modes $K\pi\piz$ and $\pi\pi$ should be of interest; it is discussed in Section~\ref{sec:effectOfDpipi_Kpipi0}.

Coming back to the modes $\Dz\rightarrow \KS\pi\pi$ and $\KS KK$,  the scale factors are $F_{\KS\pi\pi} = (9.3\pm0.1)\%$ and $F_{\KS KK}=(1.4\pm0.1)\%$~\cite{2014GGSZ}. The strong parameters $r_{D}^{\KS hh}(m^2_+,m^2_-)$, $\kappa_{D}^{\KS hh}(m^2_+,m^2_-)$, and   $\cos \delta_{D}^{\KS hh}(m^2_+,m^2_-)$ can be defined following the effective method presented in Ref.~\cite{BondarPoluektov},  while using quantum-correlated $\Dtz$ decays and where the phase space  $(m^2_+, \ m^2_-)$ is split in $\cal N$ tailored regions or ``bins"~\cite{CLEOcstrong}, such that in bin of index $i$
\begin{align}
\sqrt{K_i/K_{-i}}  &= r_{D,i}^{\KS hh}, \nonumber \\
c_i  & = \kappa_{D,i}^{\KS hh} \cos \delta_{D,i}^{\KS hh}, \ \nonumber  \\
{\rm and} \ s_i  & = \kappa_{D,i}^{\KS hh} \sin \delta_{D,i}^{\KS hh}, \nonumber
\end{align}
where, $\delta_{D,i}^{\KS hh}$ is the strong phase difference and $\kappa_{D,i}^{\KS hh}$ is the coherence factor. There is a recent publication by the BES-III collaboration~\cite{BESIIIstrong} that combines its data together with the results of CLEO-c~\cite{CLEOcstrong}, while applying the same technique to obtain the value of the $c_i$, $s_i$, and $K_{\pm i}$ parameters varying of the phase space. The binning schemes are symmetric with respect to the diagonal in the Dalitz plot $(m^2_+, \ m^2_-)$ ({\it i.e.} $\pm i$). Those results are also compared to an amplitude model from the B-factories BaBar and Belle~\cite{BaBarBellestrong}. When porting result between the BES-III/CLEO-c combination, obtained with quantum correlated $\Dtz$ decays, and LHCb for $\BsDtphi$ measurements, one needs to be careful about the  bin conventions so there might be a minus sign in phase (which only affects $s_i$).
The expected yield listed in Table~\ref{tab:yields} for the mode $\Dz\rightarrow \KS\pi\pi$  is 54 events. It is 8 events for the decay $\Dz\rightarrow \KS KK$. Though the binning scheme that latter case is only $2\times2$, it  has definitely a too small expected yield to be further considered. For $\Dz\rightarrow \KS\pi\pi$,  the binned method of Refs.~\cite{BESIIIstrong,CLEOcstrong} supposes to split the selected $\Dtz$ events over $2\times8$ bins such that with Run~1 and Run~2 only about 3 events only may populate each bin. That is the reason why, though the related observable is presented in Eq.~\ref{EQ__Kshh}, we decided to not include that mode in the sensitivity study. This choice could eventually be revisited after Run~3, when about 340 $\Bs \rightarrow \Dz(\KS\pi\pi)\phi$ events should be available, and then about 20 events may populate each bin.


\section{Sensitivity study for Run 1 \& 2 LHCb dataset}
\label{sec:sen}

The sensitivity study consists in testing and measuring the value of the unfolded $\gamma$, $\rBst$, and $\deltaBst$ parameters and their expected resolution, after having computed the values of the observables according to various initial configurations and given external inputs for the other involved physics parameters or associated experimental observables. To do this, a procedure involving global $\chi^2$ fit based on the CKMfitter package~\cite{CKMfitter} has been established to generate pseudoexperiments and fit samples of $\BsDtphi$ events.

This Section is organized as follow. In Subsection~\ref{sec:choiceParams} we explain the various configurations that we tested for the nuisance strong parameters $\rBst$, and $\deltaBst$, as well as the value of $\gamma$. Then, in Subsection~\ref{sec:pseudoExp} we explain how the pseudoexperiments have been generated. In Subsection~\ref{sec:p-values} the first step of the method is illustrated with one- and two-dimension $p$-value profiles for the  $\gamma$, $\rBst$, and $\deltaBst$ parameters. Before showing how the $\gamma$, $\rBst$, and $\deltaBst$ parameters are unfolded from the generated pseudoexperiments in Subsection~\ref{sec:FittingpseudoExp}, we discuss the stability of the former one-dimension $p$-value profile for $\gamma$ against changing the time acceptance parameters (Subsection~\ref{sec:ta}) and for a newly available binning scheme for the $D \rightarrow K3\pi$ decay
(Subsection~\ref{sec:newbinschem}). Then, unfolded values for $\gamma$ and precisions, {\it i.e.} sensitivity for Run 1 \& 2 LHCb dataset, for the various generated configurations of  $\deltaBst$ and $\rBst$ are presented in Subsection~\ref{sec:varyrBanddeltaB}. We finally conclude this Section with Subsections~\ref{sec:74deg} and~\ref{sec:effectOfDpipi_Kpipi0}, in which we study the intriguing case where $\gamma=74^\circ$ (see LHCb 2018 combination~\cite{LHCb-CONF-2018-002}, recently superseded by~\cite{LHCb-CONF-2020-003}) and we test the effect of dropping or not the least abundant expected decays modes $\Bs\rightarrow \tilde{D}^{(*)0}(\pi\pi)\phi$ and $\Bs\rightarrow \tilde{D}^{(*)0}(K\pi\piz)\phi$  in Run 1 \& 2 LHCb dataset.

\subsection{The various configurations of the $\gamma$, $\rBst$, and $\deltaBst$ parameters}
\label{sec:choiceParams}

The sensitivity study was performed with the CKM angle $\gamma$ true value set to be $(65.66^{+0.90}_{-2.65})^{\circ}$ ({\it i.e.} 1.146 rad) as obtained by the CKMfitter group, while excluding any measured values of $\gamma$ in its global fit~\cite{CKMfitterSUMMER19}. As a reminder, the average of the LHCb measurements is  $\gamma = (74.0^{+5.0}_{-5.8})^{\circ}$~\cite{LHCb-CONF-2018-002}, therefore, the value $\gamma =74^{\circ}$, is also tested (see Sec.~\ref{sec:74deg}).

The value of the strong phases $\deltaBst$ is a nuisance parameter  that cannot be predicted or guessed by any argument, and therefore, six different values are assigned to it:  0, 1, 2, 3, 4, 5 rad ($0^{\circ}$, $57.3^{\circ}$,  $114.6^{\circ}$, $171.9^{\circ}$, $229.2^{\circ}$, $286.5^{\circ}$). This corresponds to 36 tested configurations ({\it i.e.} $6\times6$).

Since both interfering diagrams displayed in  Fig.~\ref{fig:Diagrams} are colour-suppressed, the value of the ratio of the $\bar{b}\rightarrow \bar{u}c\bar{s}$ and $\bar{b}\rightarrow \bar{c}u\bar{s}$  tree-level amplitudes, $\rBst$, is expected to be $|V_{ub}V_{cs}|/|V_{cb}V_{us}|\sim0.4$. This assumption is well supported by the study performed with $\Bs \rightarrow D_s^{\mp} K^{\pm}$ decays by the LHCb collaboration , for which a value $\rB=0.37^{+0.10}_{-0.09}$ has been measured~\cite{Aaij:2017lff}. But, as the decay  $\Bs \rightarrow D_s^{\mp} K^{\pm}$ is colour-favoured, it is important to test other values originated from  already measured colour-suppressed $B$-meson decays, as non factorizing final state interactions can modify the decay dynamics~\cite{BaBarColSup}. Among them, the decay $\Bz \rightarrow D K^{*0}$ plays such a role for which LHCb obtains $\rB=0.22^{+0.17}_{-0.27}$~\cite{CKMfitterSUMMER19}, confirmed by a more recent and accurate computation: $\rB=0.265\pm0.023$~\cite{Aaij:2019Pullen}. The value of $\rB$ is known to strongly impact the precision on $\gamma$ measurements as $1/\rB$~\cite{tisserand}. Therefore, the two extreme values $0.22$ and $0.40$ for $\rBst$ have been tested for the sensitivity study, while the values for $\rB$ and $\rstB$ are expected to be similar.

This leads to a total of 72 tested configurations for the $\rBst$, $\delta_B$, and $\delta^*_{B}$ parameters ({\it i.e.} $2\times6 \times 6$).

\subsection{Generating pseudoexperiments for various configurations of parameters}
\label{sec:pseudoExp}

At a first step, different configurations for observables corresponding to Sec.~\ref{sec:ObsDz} and~\ref{sec:ObsDzst} are computed. The observables are obtained with the value of the angle $\gamma$ and of the four nuisance parameters $\rBst$ and $\deltaBst$ fixed to various sets of initial true values (see Sec.~\ref{sec:choiceParams}), while the external parameters listed in Table~\ref{tab:para} and the normalisation factors $C_{K\pi}$, $C_{K\pi,D\piz}$, and $C_{K\pi,D\gamma}$ have been left free to vary within their uncertainties. In a second step, for the obtained observables, including their uncertainties that we assume to be their square root, and all the other parameters, except $\gamma$, $\rBst$, and $\deltaBst$, a global $\chi^2$ fit is performed to compute the resulting $p$-value distributions of the $\gamma$, $\rBst$, and $\deltaBst$ parameters. Then, at a third step, for the obtained observables, including their uncertainties, and all the other parameters, except $\gamma$, $\rBst$, and $\deltaBst$, 4000 pseudoexperiments are generated according to Eqs.~\eqref{EQ__14_}-\eqref{EQ__34_}, for the various above tested configurations. And in a fourth step, for each of the generated pseudoexperiment, all the quantities are varied within their uncertainties. Then a global $\chi^2$ fit is performed to unfold the value of the parameters $\gamma$, $r^{(*)}_{\Bs}$, and $\delta^{(*)}_{\Bs}$, for each of the 4000 generated pseudoexperiments. In a fifth step, for the distribution of the 4000 values of  fitted $\gamma$, $\rBst$, and $\deltaBst$, an extended unbinned maximum likelihood fit  is performed to compute the most probable value for each of the former five parameters, together with their dispersion.  The resulting values are compared to their injected initial true values. The sensitivity to $\gamma$, $\rBst$, and $\deltaBst$ is finally deduced and any bias correlation is eventually highlighted and studied.

\begin{figure}[h]
\centering
\includegraphics[width=0.425\textwidth,height=0.125\textheight]{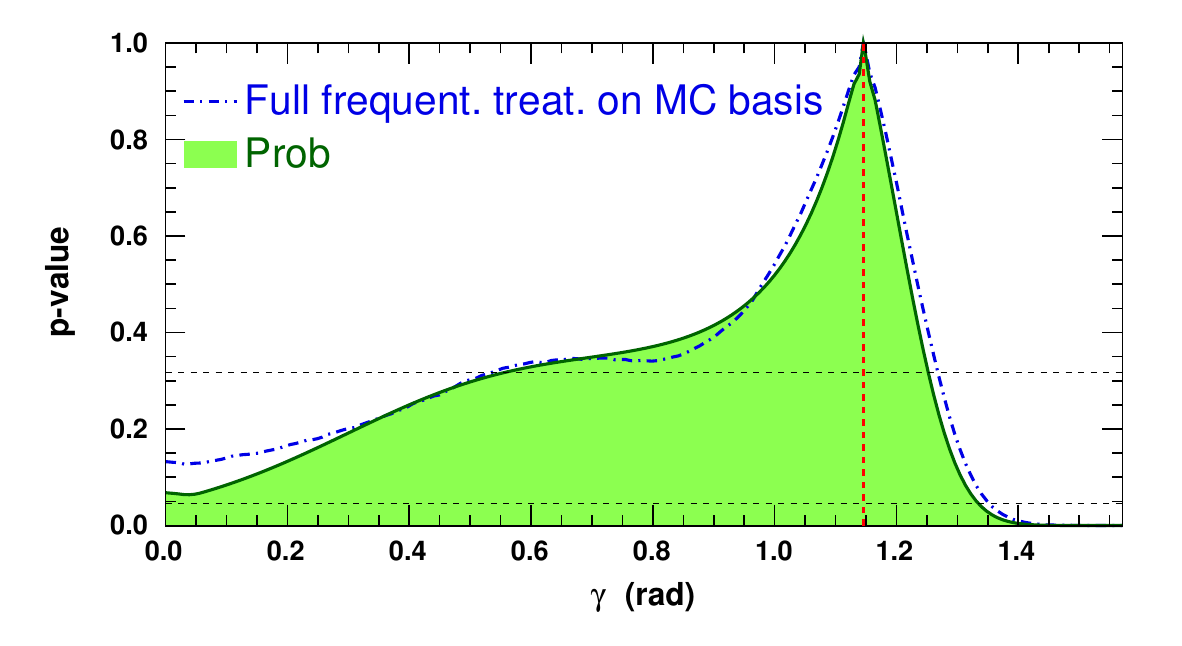}
\includegraphics[width=0.425\textwidth,height=0.125\textheight]{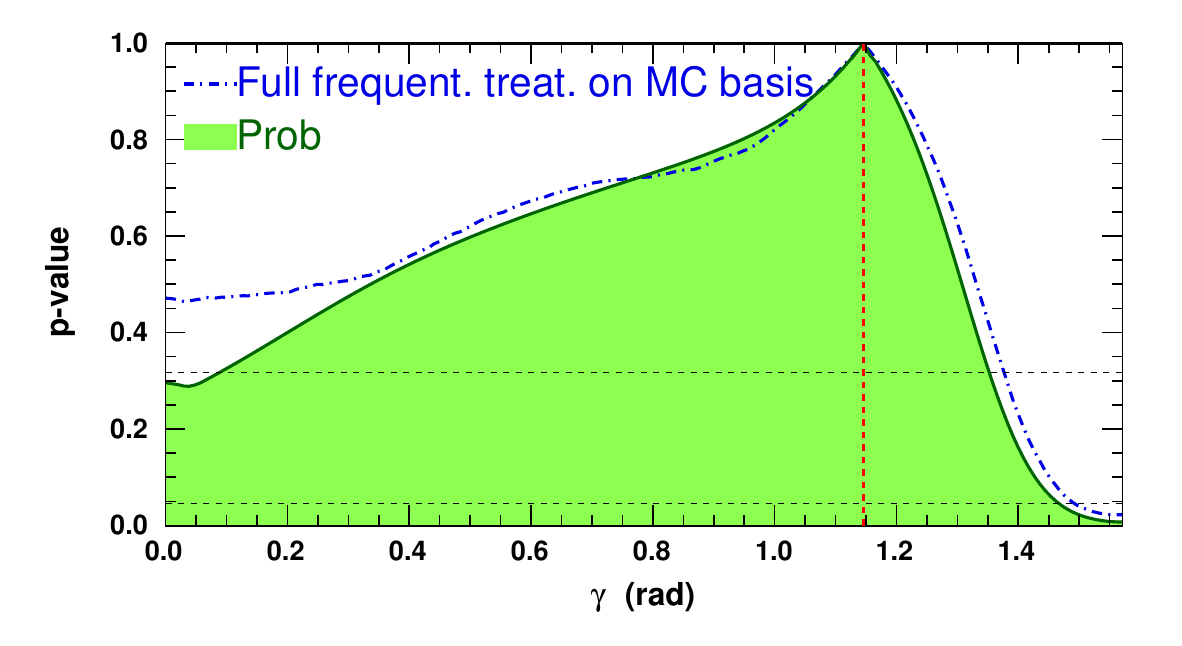} \\
\includegraphics[width=0.425\textwidth,height=0.125\textheight]{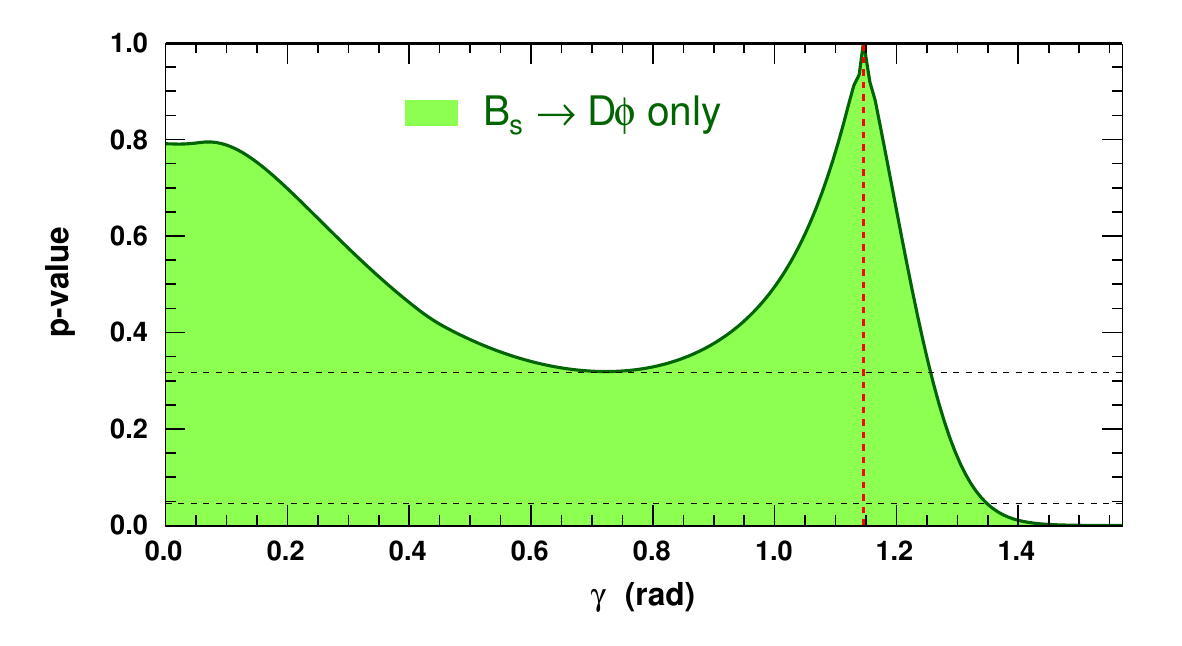}
\includegraphics[width=0.425\textwidth,height=0.125\textheight]{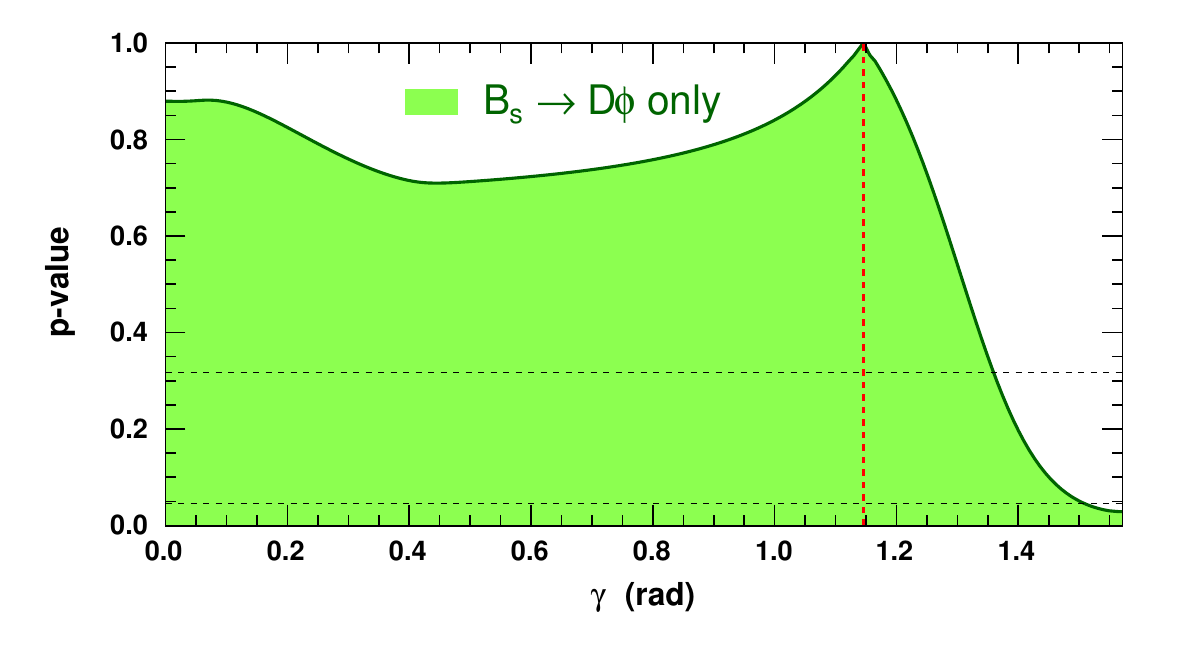}
\caption{\label{fig:gamma_1D}  Profile of the $p$-value of the global $\chi^2$ fit to $\gamma$ after the observables are computed (top left), for the set of true initial parameters: $\gamma=1.146$ rad, $\rBst=0.4$, $\deltaB=3.0$ rad, and $\deltastB=2.0$ rad (the corresponding distribution obtained from a full frequentist treatment on Monte-Carlo simulation basis~\cite{plugin} is superimposed to the same distribution). The related $p$-value profile for $\rBst=0.22$ is also presented (top right). The integrated luminosity assumed here is that of LHCb data collected in Run~1 \& 2.  Profile of the $p$-value of the global $\chi^2$ fit to $\gamma$ after the observables are computed, where only the decay mode $\Bs\rightarrow \tilde{D}^0\phi$ is used and for the set of true initial parameters: $\gamma=1.146$ rad, $\rBst=0.4$ (bottom left) and $0.22$ (bottom right), $\deltaB=3.0$ rad, and $\deltastB=2.0$ rad. On each figure the vertical dashed red line indicates the initial $\gamma$ true value, and the two horizontal dashed black lines, refer to 68.3 and 95.4~\% CL.}
\includegraphics[width=0.425\textwidth,height=0.125\textheight]{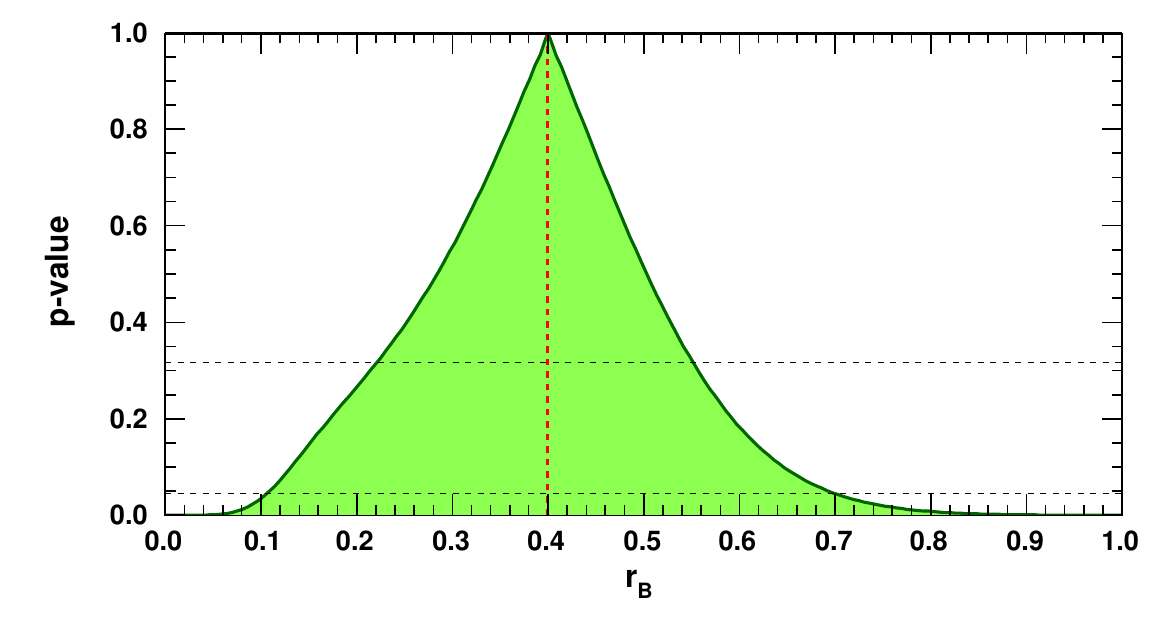}
\includegraphics[width=0.425\textwidth,height=0.125\textheight]{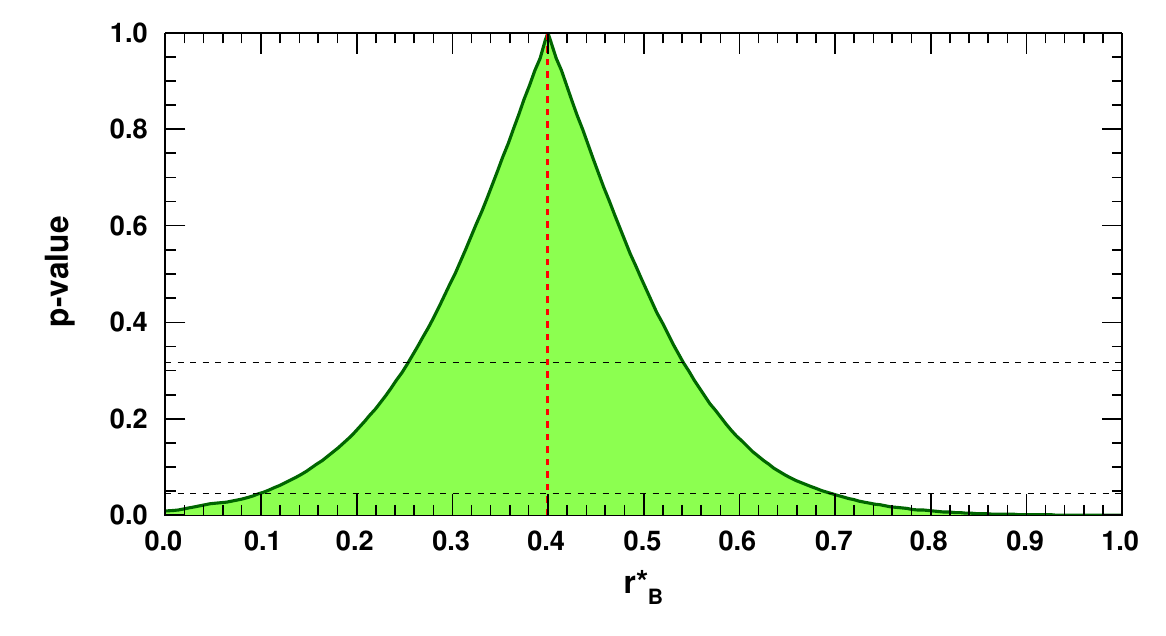} \\
\includegraphics[width=0.425\textwidth,height=0.125\textheight]{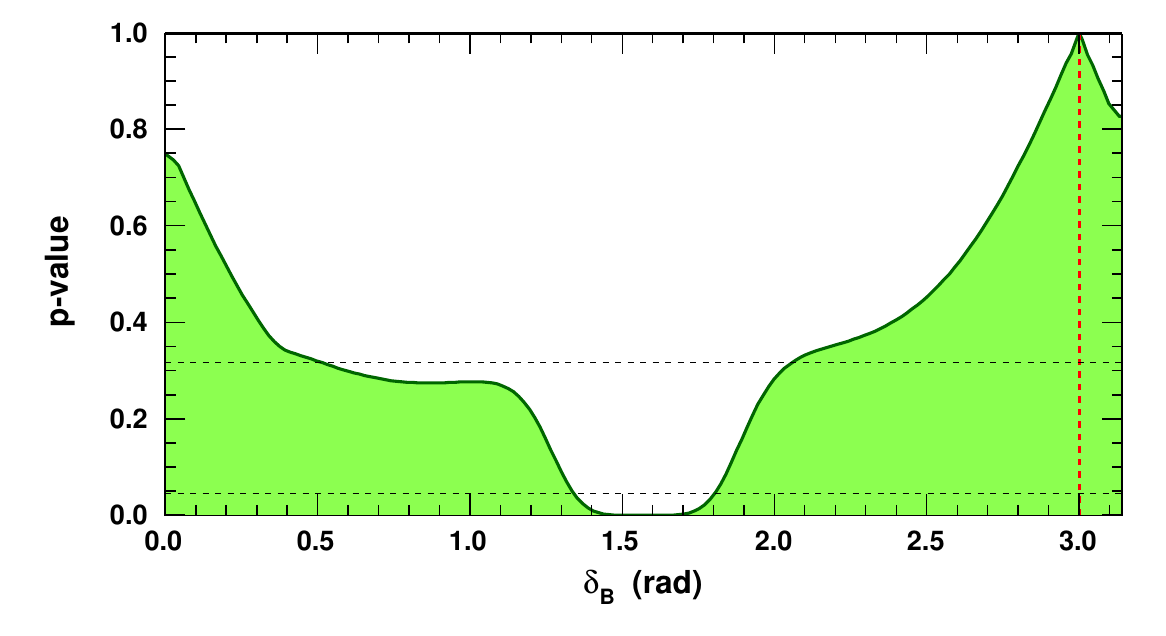}
\includegraphics[width=0.425\textwidth,height=0.125\textheight]{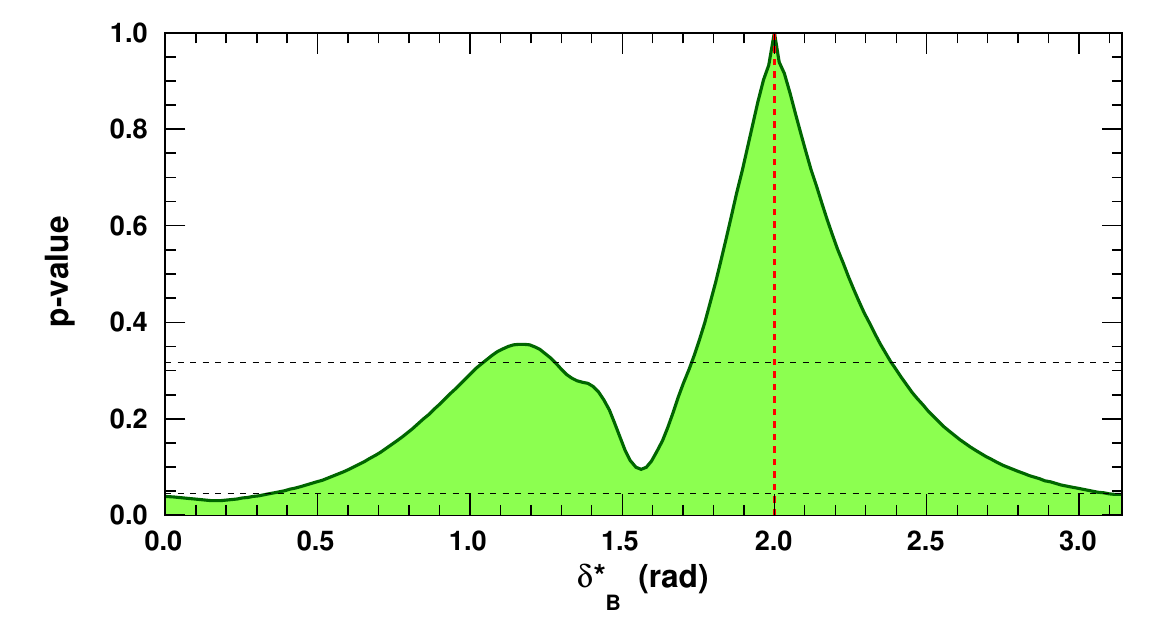}
\caption{\label{fig:others_1D} Profile of the $p$-value distribution of the global $\chi^2$ fit to $\rBst$ (top left (right)) and $\deltaBst$ (bottom left (right)), after that the observables are computed, for the set of initial true parameters: $\gamma=65.66^{\circ}$ (1.146 rad), $\rBst=0.4$, $\deltaB=3.0$ rad, and $\deltastB=2.0$ rad. On each figure the vertical dashed red line indicates the initial  $\rBst$ and $\deltaBst$ true values, and the two horizontal dashed black lines, refer to 68.3 and 95.4~\% CL.}
\end{figure}

\subsection{One- and two-dimension $p$-value profiles for the  $\gamma$, $\rBst$, and $\deltaBst$ parameters}
\label{sec:p-values}

Figure~\ref{fig:gamma_1D} displays the one-dimension $p$-value  profile of $\gamma$, at the step two of the procedure described in Sec.~\ref{sec:pseudoExp}. The Figure is obtained for an example set of initial parameters: $\gamma=65.66^{\circ}$ (1.146 rad), $\rBst=0.4$, $\deltaB=3.0$ rad, and $\deltastB=2.0$ rad. The integrated luminosity assumed here is that of LHCb data collected in Run 1 \& 2. The corresponding fitted value is $\gamma=\left(65.7^{+6.3}_{-33.8}\right)^\circ$, thus in excellent agreement with the initial tested true value. The Fig.~\ref{fig:gamma_1D} also shows the  corresponding distribution obtained from a full frequentist treatment on Monte-Carlo simulation basis~\cite{plugin}, where $\gamma=\left(65.7^{+6.9}_{-34.9}\right)^\circ$. This has to be considered as a demonstration that the two estimates on $\gamma$ are in quite fair agreement at least at the $68.3~\%$ confidence level (CL), such that no obvious under-coverage is experienced with the nominal method, based on the {\ttfamily ROOT} function {\ttfamily  TMath::Prob}~\cite{PDGstat}. On the upper part of the distribution the relative under-coverage of the ``{\ttfamily  Prob}'' method is about $6.3/6.9\simeq91~\%$. As opposed to the full frequentist treatment on Monte-Carlo simulation basis, the nominal retained method allows performing computations of very large number of pseudoexperiments within a reasonable amount of time and for non-prohibitive CPU resources. For the LHCb Run 1 \& 2 dataset, 72 configurations of 4000 pseudoexperiments were generated ({\it i.e.} 288 000 pseudoexperiments in total). The whole study was repeated another two times for prospective studies with future anticipated LHCb data, such that more than about 864 000 pseudoexperiments were generated for this publication (see Sec.~\ref{sec:HL-LHC}). In the same Figure, one can also see the effect of modifying the value of $\rBst$ from $0.4$ to $0.22$, for which  $\gamma = \left(65.7^{+12.0}_{-60.7}\right)^\circ$, where the upper uncertainty scales roughly as expected as $1/\rBst$ ({\it i.e.} $6.3\times 0.4/0.22=12.6$).
Compared to the full frequentist treatment on Monte-Carlo simulation, where $\gamma = \left(65.7^{+13.2}_{-\infty}\right)^\circ$, the relative under-coverage of the ``{\ttfamily  Prob}'' method is about $12.0/13.2\simeq 91~\%$.   Finally, the $p$-value  profile of $\gamma$ is also displayed when dropping the information provided by the $\Bs\rightarrow \Dtstz \phi$ mode, and thus keeping only that of the $\Bs\rightarrow \Dtz\phi$ mode. In that case, $\gamma$ is equal to  $\left(65.7^{+6.3(12.0)}_{-\infty}\right)^\circ$, $\rBst=0.4$ ($0.22$),  such that the CL interval is noticeably enlarged on the lower side of the $\gamma$ angle distribution (more details can be found in Sec.~\ref{sec:effectOfDstarphi}).

For the same set of initial parameters ({\it i.e.} $\gamma=65.66^{\circ}$ (1.146 rad), $\rBst=0.4$, $\deltaB=3.0$ rad, and $\deltastB=2.0$ rad) and the same projected integrated luminosity,  Fig.~\ref{fig:others_1D} displays the one-dimension $p$-value  profile of the nuisance parameters $\rBst$ and $\deltaBst$. It can be seen that the $p$-value is maximum at the initial tested value, as expected.

Two-dimension $p$-value  profile of the nuisance parameters $\rBst$ and $\deltaBst$ as a function of $\gamma$ are provided in Fig.~\ref{fig:2D_2}-\ref{fig:2D_1}. Figures~\ref{fig:2D_2} and~\ref{fig:2D_3} correspond to two other example configurations $\gamma=1.146$ rad, $\deltaB=1.0$ rad, and $\deltastB=5.0$ rad, and  $\rBst=0.4$ and $\rBst=0.22$, respectively. Figures~\ref{fig:2D_0} and~\ref{fig:2D_1} stand for the configurations $\gamma=1.146$ rad, $\deltaB=3.0$ rad, $\deltastB=2.0$ rad, and  $\rBst=0.4$ and $\rBst=0.22$, respectively.   Those two-dimension views allows to see the correlation between the different parameters. In general large correlations between $\delta^{(*)}_{B}$ and $\gamma$ are observed. In case of configurations where $\rBst=0.4$, large fraction of the $\deltaBst$ {\it vs.}  $\gamma$ plane can be excluded at $95~\%$ CL, while the fraction is significantly reduced for the corresponding $\rBst=0.22$ configurations. For the $\deltastB$ {\it vs.} $\gamma$ plane, one can easily see the advantage of our Cartesian coordinates approach (see Sec.~\ref{sec:ObsDz} and~\ref{sec:ObsDzst}) together with the fact that in the case of the mode $\Dstz \rightarrow \Dz \gamma$ there is an effective strong phase shift of $\pi$ with respect to the $\Dstz \rightarrow \Dz \pi^{0}$~\cite{Bondar:2004bi}, such that additional constraints allow to remove fold-ambiguities with respect to the associated $\deltaB$ {\it vs.} $\gamma$ plane.

\subsection{Effect of the time acceptance parameters}\label{sec:ta}

\begin{figure}[h]
\centering
\includegraphics[width=0.425\textwidth]{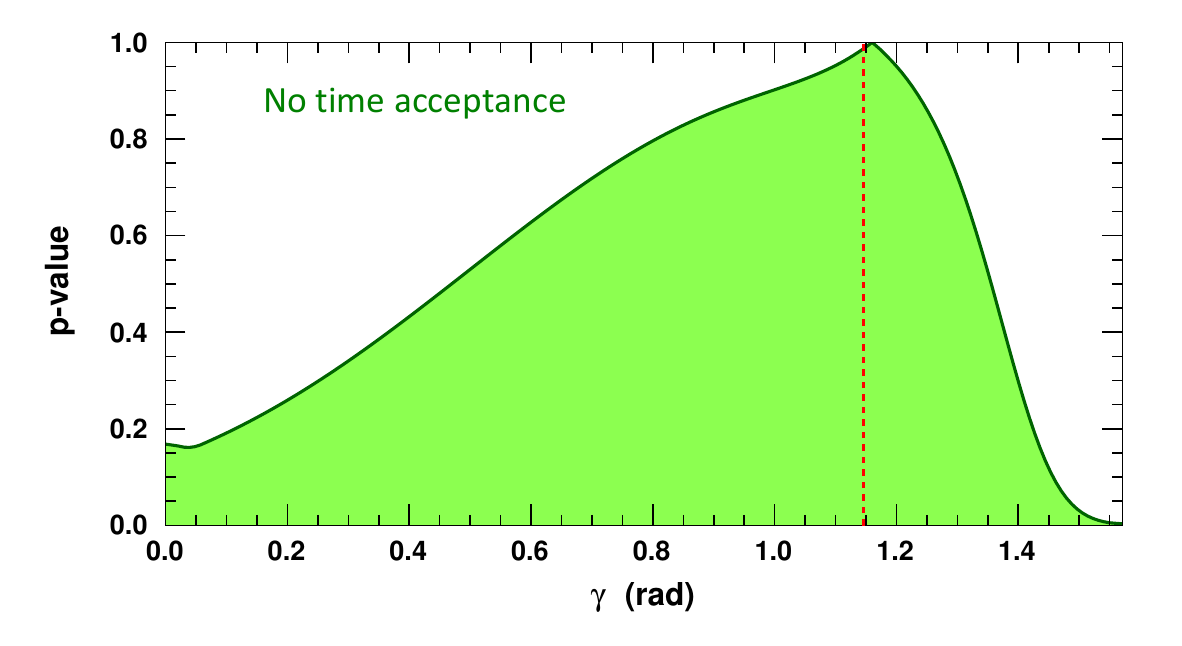}
\includegraphics[width=0.425\textwidth]{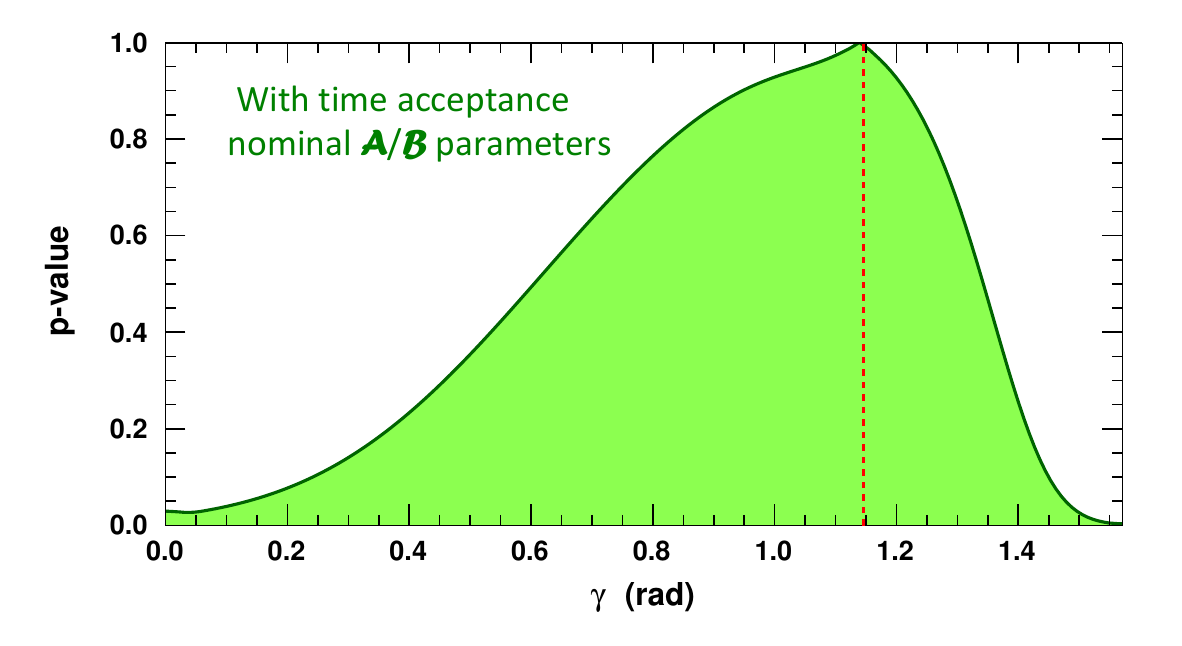}
\caption{\label{fig:gamma_timeAccept}
Profile of the $p$-value distribution  of the global $\chi^2$ fit to $\gamma$, for the set of true initial parameters: $\gamma=1.146$ rad, $\rBst=0.4$, and $\deltaBst=1.0$ rad. The integrated luminosity assumed here is that of LHCb data collected in Run 1 \& 2, when the time acceptance $\AT$ and $\BT$ are set to 1 in Eqs.~\eqref{EQ__14_}-\eqref{EQ__34_}: \emph{no time acceptance}  (top left) or to their nominal values  $\AT=0.488\pm 0.005$ and $\BT=0.773\pm 0.008$ (top right), as computed in Sec.~\ref{sec:timeAcc}. The dashed red line shows the initial $\gamma$ true value $\gamma=65.66^\circ$ (1.146 rad).}
\end{figure}

Figure~\ref{fig:gamma_timeAccept} shows for a tested configuration $\gamma=1.146$ rad, $\rBst=0.4$, and $\deltaBst=1.0$ rad, that the impact of the time acceptance parameters $\AT$ and $\BT$ can eventually be non negligible and has an impact on the  profile distribution of the $p$-value of the global $\chi^2$ fit to $\gamma$. For the given example the fitted value of $\gamma$ is either  $\left(65.3^{+14.3}_{-38.4}\right)^\circ$ or  $\left(66.5^{+13.8}_{-51.0}\right)^\circ$, when the time acceptance is either or not accounted for. The reason why the precision improves when the time acceptance is taken into account may be not intuitive.
 This is because for $\BT/\AT\simeq1.6$, as opposed to the case $\BT/\AT\simeq1.0$, the impact of the first term in Eq.~\ref{EQ__14_}, which is directly proportional to $\cos \left(\deltaB + 2\beta_{s} - \gamma \right)$, is amplified with respect to the second term, for which the sensitivity to $\gamma$ is more diluted.

\begin{center}
\tabcaption{\label{tab:diffTA} Expected value of $\gamma$, as a function of different time acceptance parameters. The second line corresponds to the nominal values. The nominal set of parameters $\AT$ and $\BT$ is written in bold style.}
\footnotesize
\begin{tabular*}{80mm}{rrcccc|c}
\hline \hline
$\alpha$  & $\beta$ & $\xi$ & $\AT$   & $\BT$ & $\BT/\AT$  & fitted $\gamma$  ($^{\circ}$)  \\
\hline
1.0 & 2.5 & 0.01 & 0.367 & 0.671 & 1.828 & $66.5^{+13.8}_{-40.1}$ \\
{\bf 1.5} & {\bf 2.5} & {\bf 0.01} & {\bf 0.488} & {\bf 0.773} & {\bf 1.584} & {\boldm ${ 65.3^{+14.3}_{-38.4}}$} \\
2.0 & 2.5 & 0.01 & 0.570 & 0.851 & 1.493 & $65.3^{+13.2}_{-37.8}$ \\
1.5 & 2.0 & 0.01 & 0.484 & 0.751 & 1.552 & $65.9^{+13.2}_{-39.0}$ \\
1.5 & 3.0 & 0.01 & 0.491 & 0.789 & 1.607 & $66.5^{+13.2}_{-38.4}$ \\
1.5 & 2.5 & 0.02 & 0.480 & 0.755 & 1.573 & $66.5^{+13.8}_{-39.5}$ \\
1.5 & 2.5 & 0.005 & 0.492 & 0.783 & 1.591 & $65.3^{+13.8}_{-36.7}$ \\

\end{tabular*}
\end{center}

Even if  the parameters $\AT$ and $\BT$ are computed to a precision at the percent level (Sec.~\ref{sec:timeAcc}), we investigated further the impact of changing their values. Note that for this study, the overall efficiency is kept constant, while the shape of the acceptance function is varied.   The values  $\alpha$, $\beta$ and $\xi$ were changed in Eq.~\ref{EQ__10_}, and the results of those changes are listed in Table~\ref{tab:diffTA}. When $\alpha$ increases, both $\AT$ and $\BT$ turn larger, but the value of the ratio $\BT/\AT$ decreases. When $\beta$ or $\xi$ decreases, the 3 values of $\AT$, $\BT$, and $\BT/\AT$ increase. The effect of changing  $\beta$ or $\xi$ alone is small. A modification of $\alpha$ has a much larger impact on $\AT$ and $\BT$. However, all these changes have a weak impact on the precision of the fitted $\gamma$ value.  This is good news as this means that the relative  efficiency loss caused by time acceptance effects will not cause much change in the sensitivity to the CKM $\gamma$ angle. As a result, time acceptance requirements can be varied without much worry to improve the signal purity and statistical significance, when analysing the $\BsDtphi$ decays with LHCb data.

\subsection{Effect of a new binning scheme for the $D \rightarrow K3\pi$ decay}
\label{sec:newbinschem}
According to Ref.~\cite{Evans:2019wza}, averaged values of the $K3\pi$ input parameters over the phase space defined as
\begin{eqnarray}
R^{K3\pi}_D e^{-i\delta^{K3\pi}_{D}}=\frac{\int A^*_{\Dzb \rightarrow K3\pi}(x)A_{\Dz \rightarrow K3\pi}(x)dx}{A_{\Dzb \rightarrow K3\pi}A_{\Dz \rightarrow K3\pi}},
\end{eqnarray}

\begin{figure}[h]
\centering
\includegraphics[width=0.425\textwidth,height=0.15\textheight]{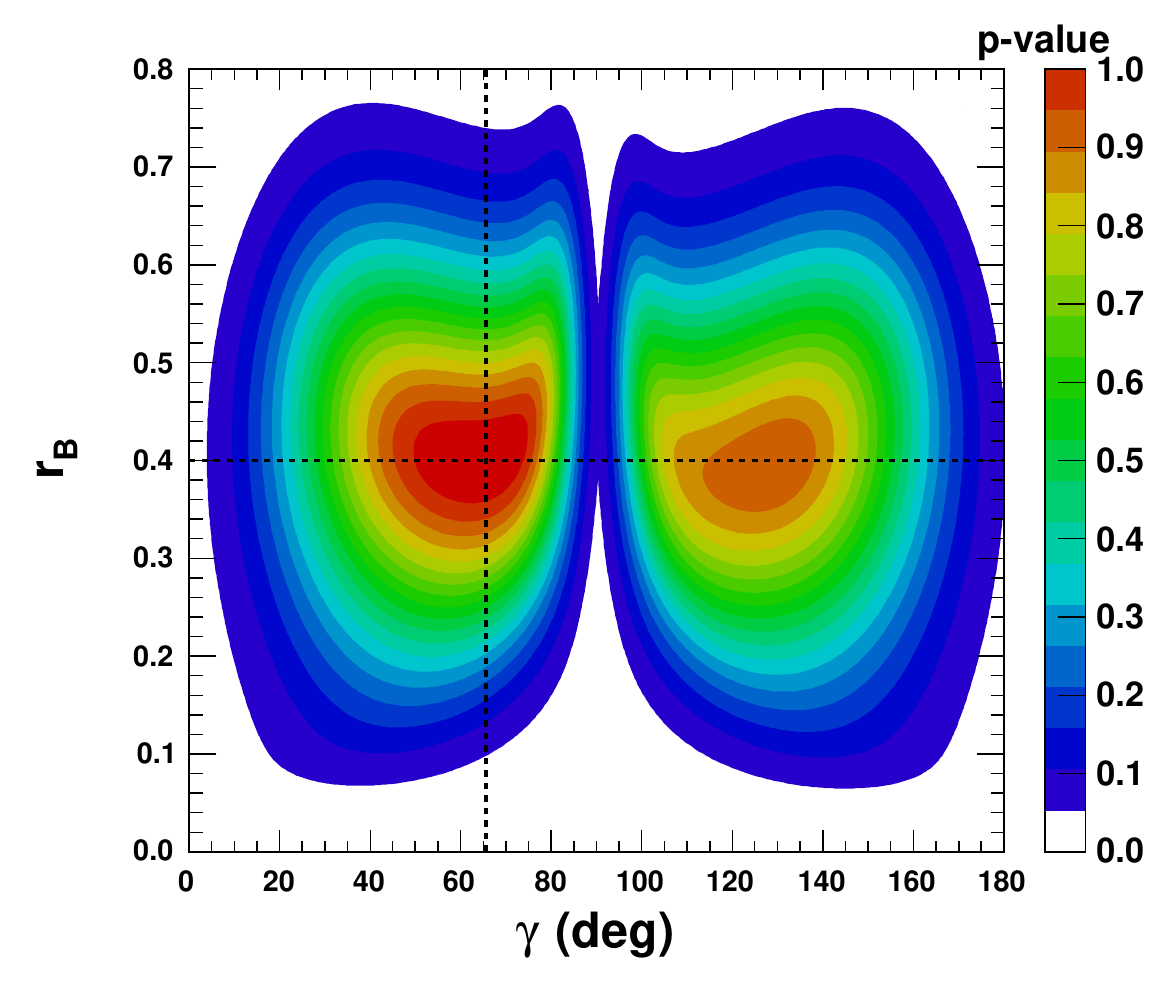}
\includegraphics[width=0.425\textwidth,height=0.15\textheight]{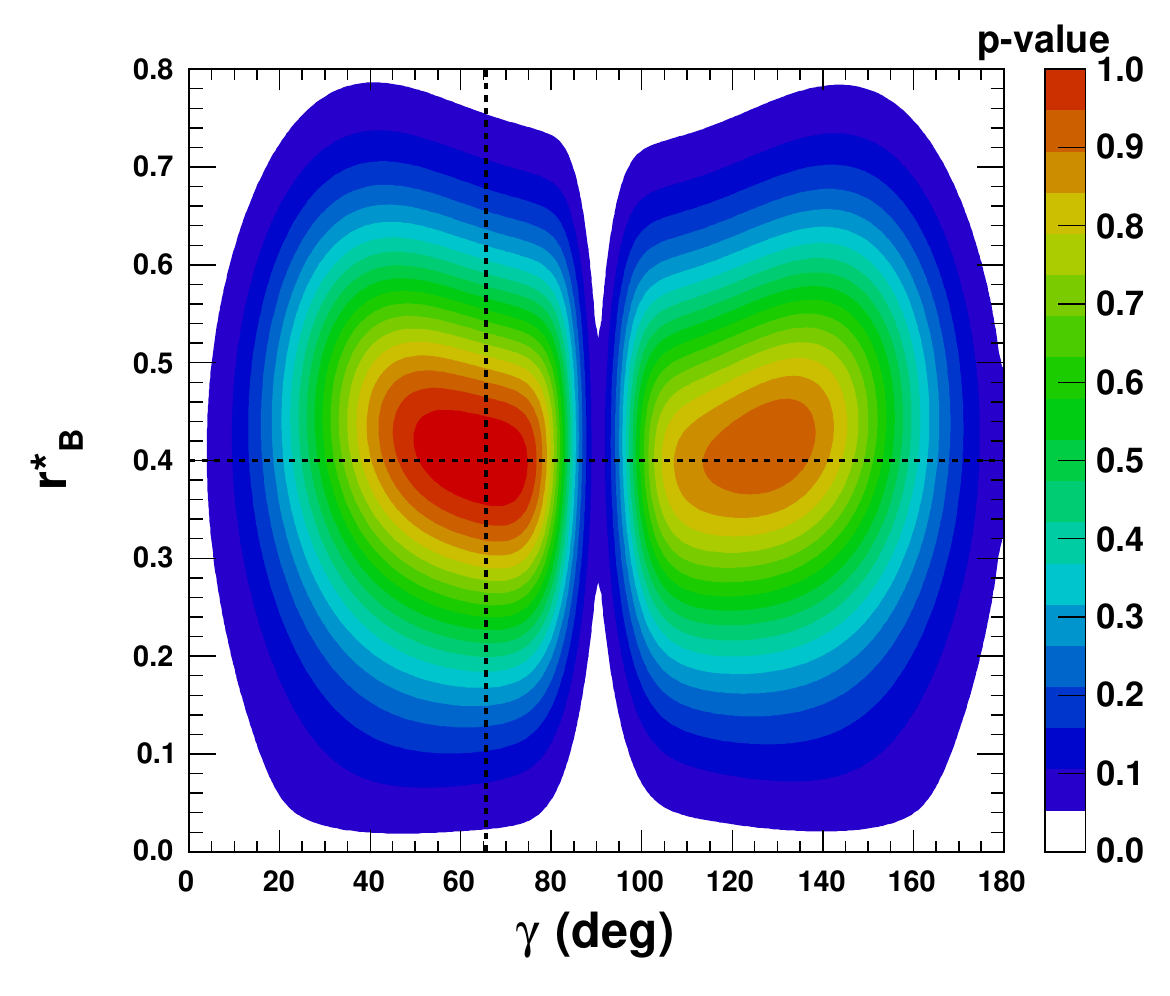} \\
\includegraphics[width=0.425\textwidth,height=0.15\textheight]{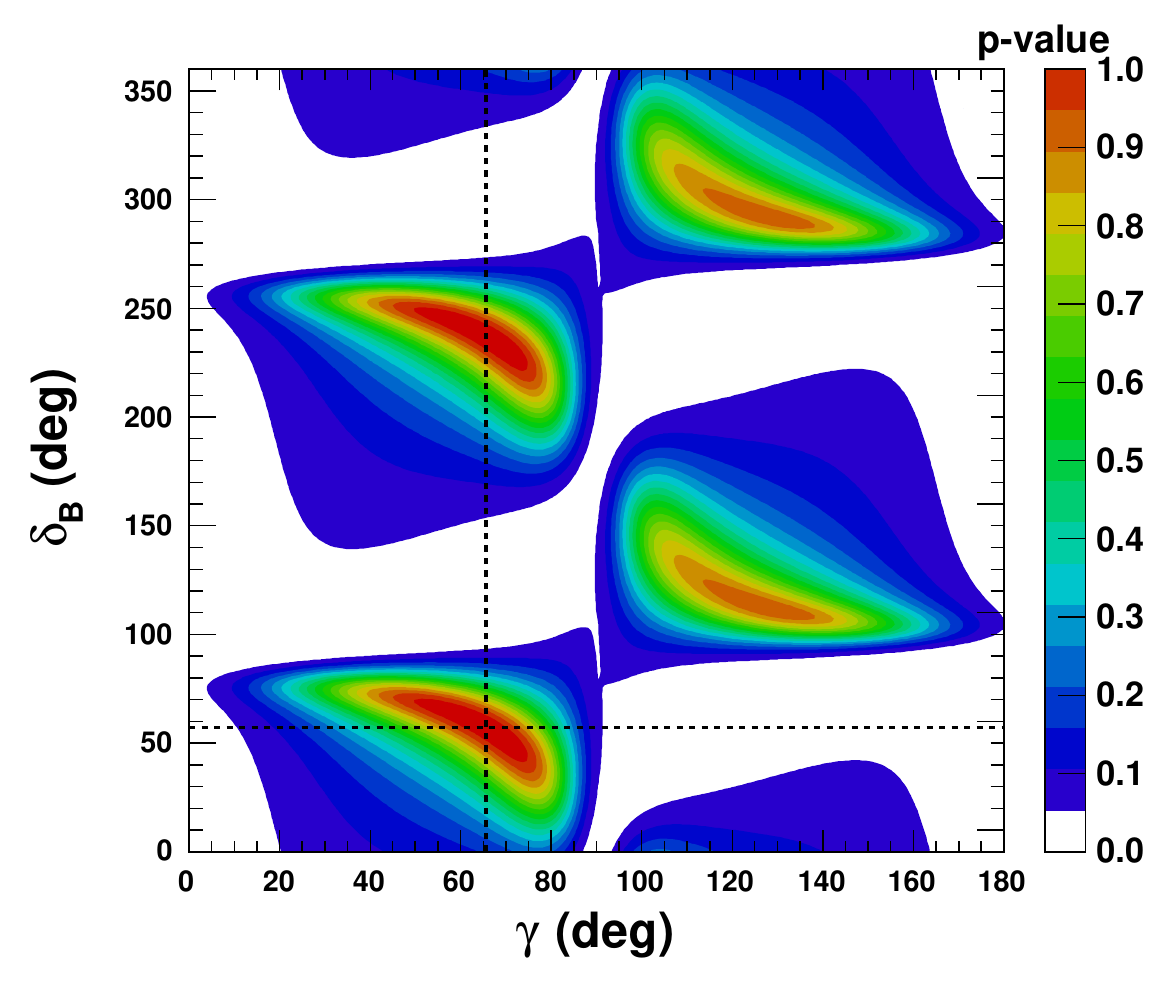}
\includegraphics[width=0.425\textwidth,height=0.15\textheight]{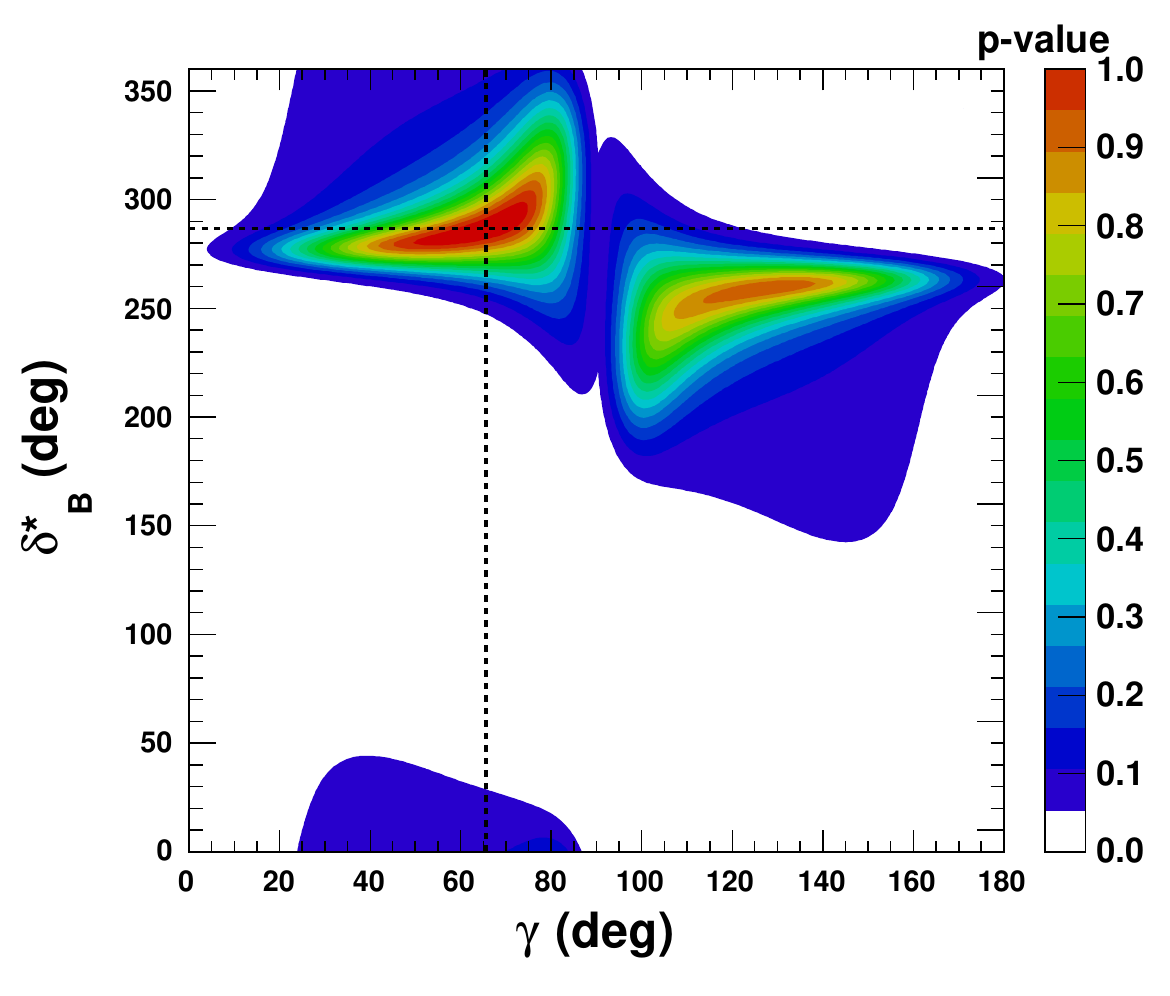}
\caption{\label{fig:2D_2} Two-dimension $p$-value  profile distribution of the nuisance parameters $\rBst$ and $\deltaBst$ as a function of $\gamma$. On each figure the dashed black  lines indicate the initial true values: $\gamma=65.66^\circ$ (1.146 rad), $\deltaB=57.3^\circ$ (1.0 rad), and $\deltastB=286.5^\circ$ (5.0 rad), and  $\rBst=0.4$.}
\includegraphics[width=0.425\textwidth,height=0.15\textheight]{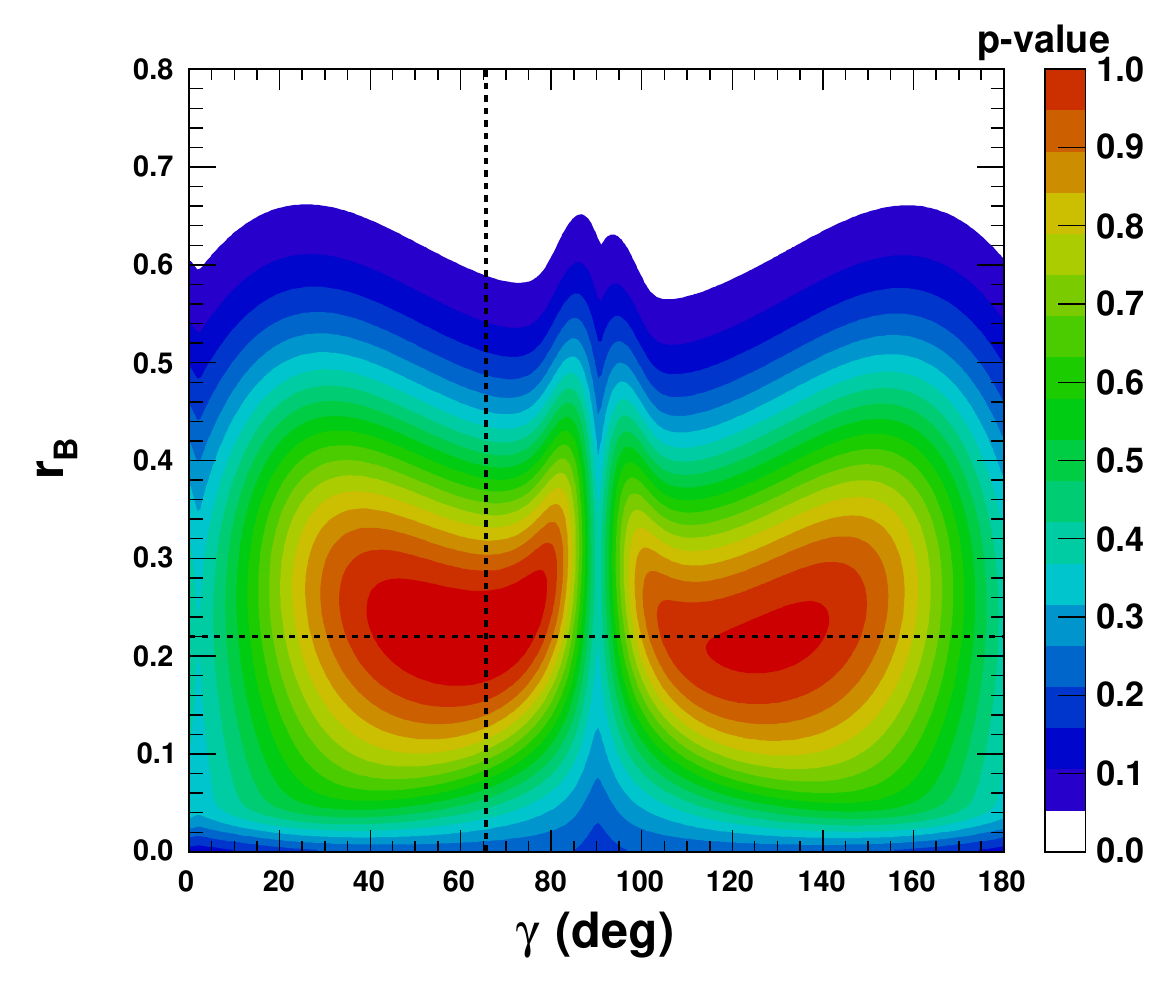}
\includegraphics[width=0.425\textwidth,height=0.15\textheight]{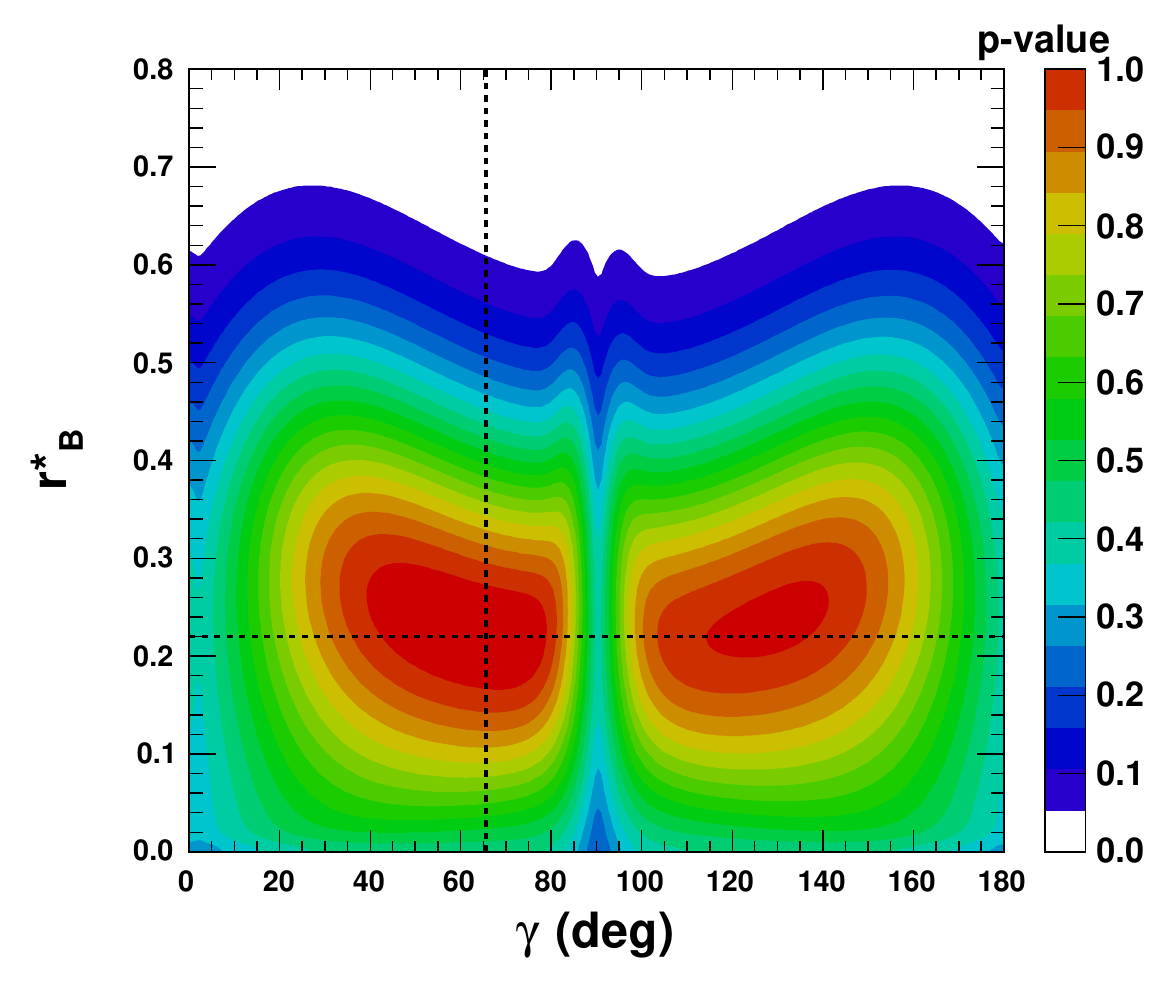} \\
\includegraphics[width=0.425\textwidth,height=0.15\textheight]{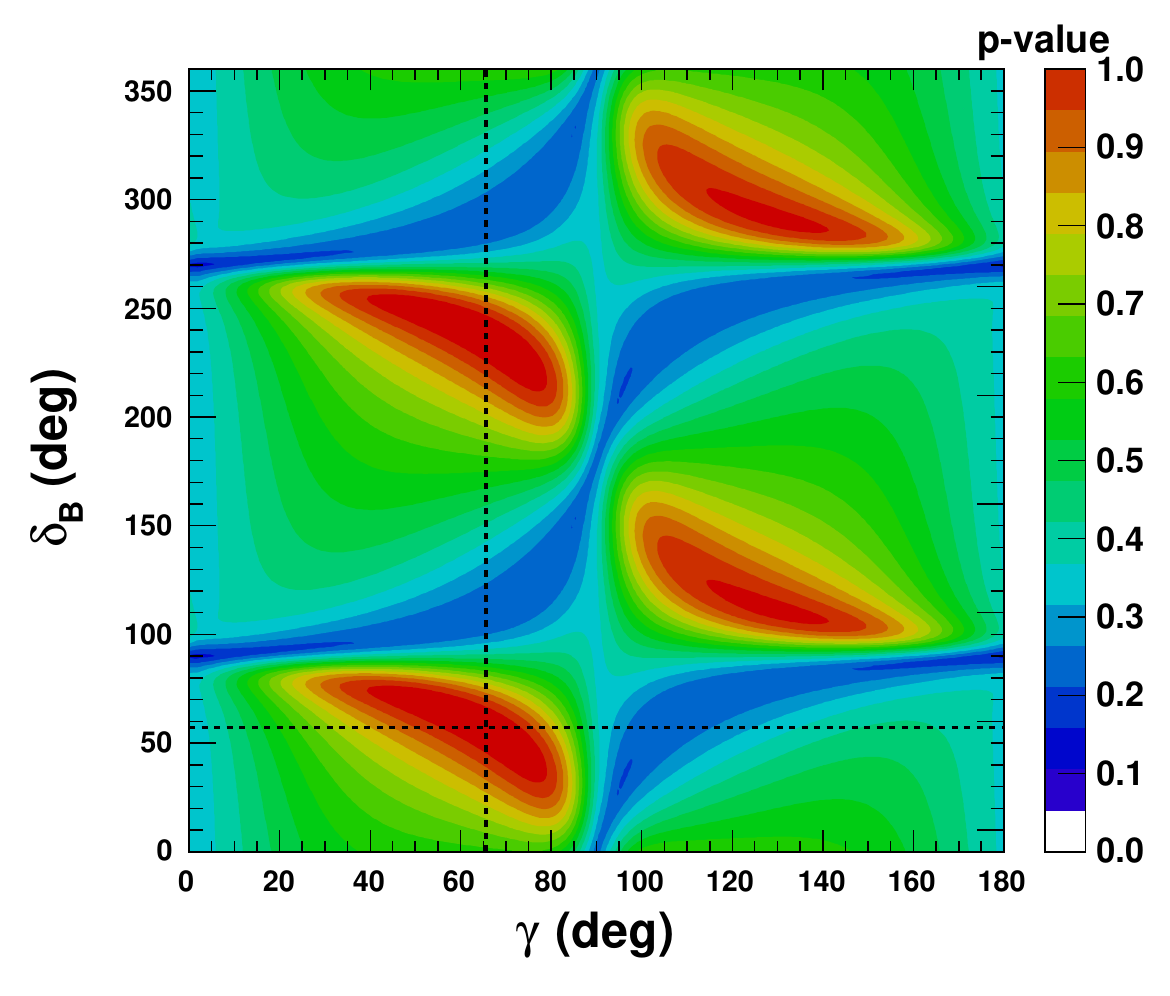}
\includegraphics[width=0.425\textwidth,height=0.15\textheight]{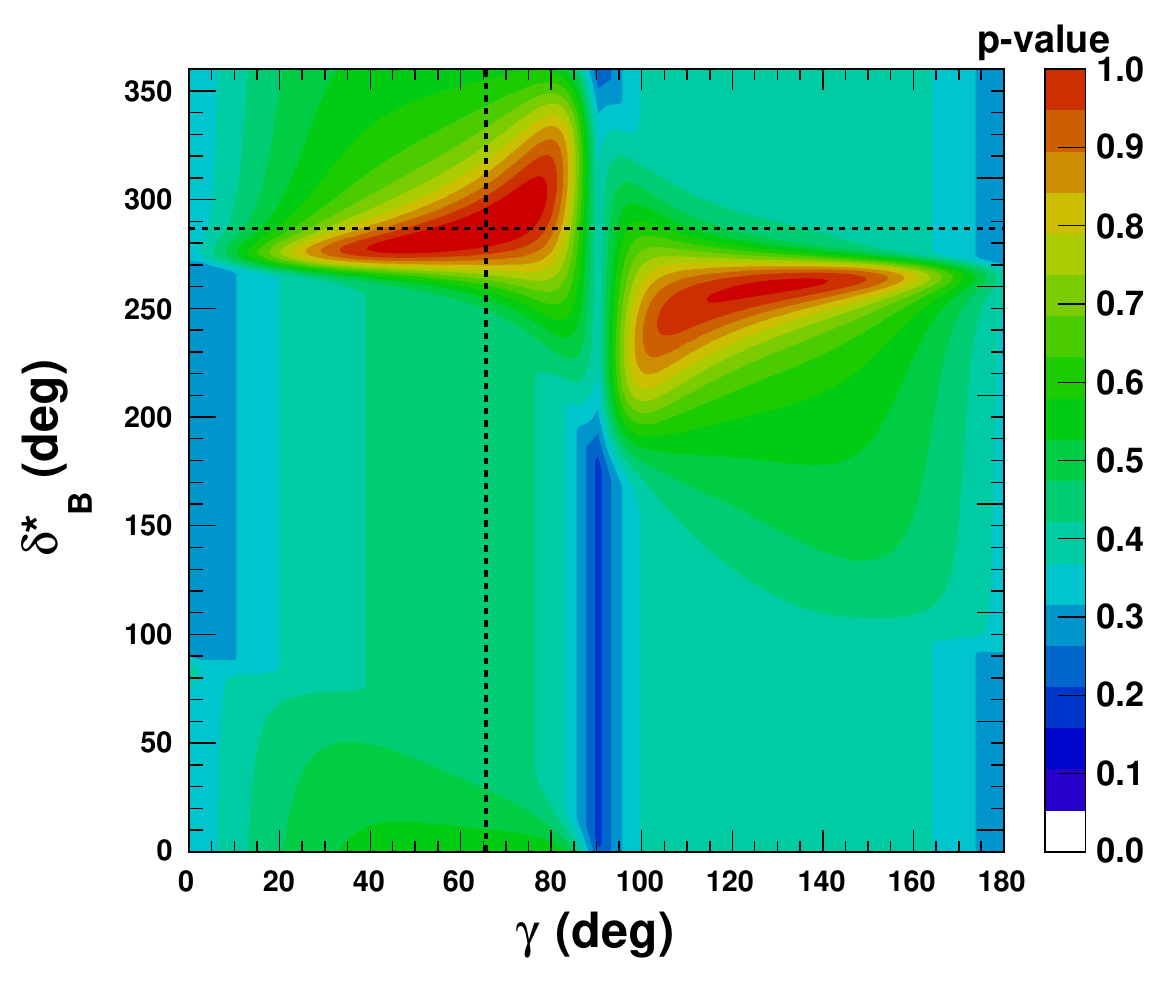}
\caption{\label{fig:2D_3} Two-dimension $p$-value  profile distribution of the nuisance parameters $\rBst$ and $\deltaBst$ as a function of $\gamma$. On each figure the dashed black  lines indicate the initial true values: $\gamma=65.66^\circ$ (1.146 rad), $\deltaB=57.3^\circ$ (1.0 rad), and $\deltastB=286.5^\circ$ (5.0 rad), and  $\rBst=0.22$.}
\end{figure}

\begin{figure}[h]
\centering
\includegraphics[width=0.4\textwidth,height=0.15\textheight]{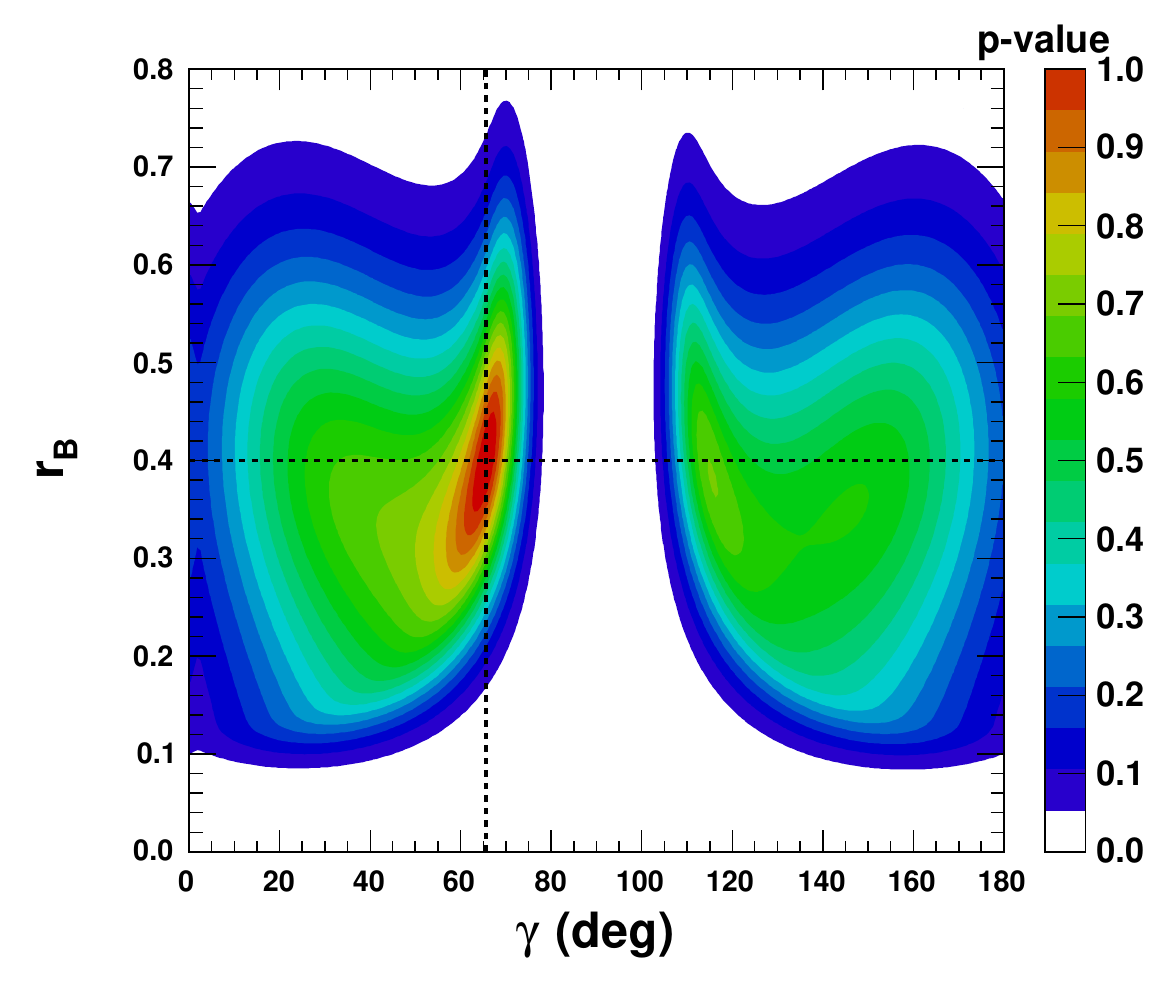}
\includegraphics[width=0.4\textwidth,height=0.15\textheight]{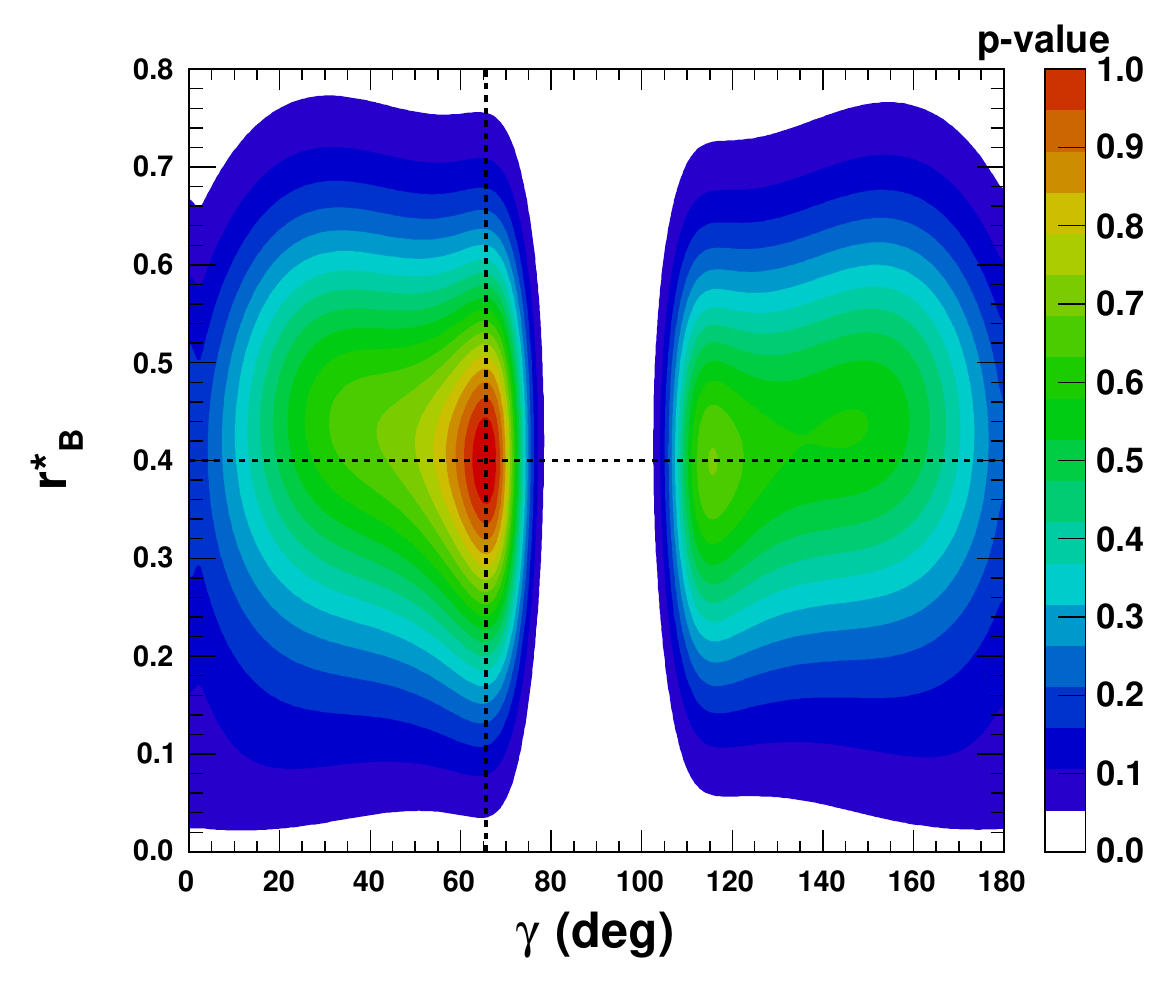} \\
\includegraphics[width=0.4\textwidth,height=0.15\textheight]{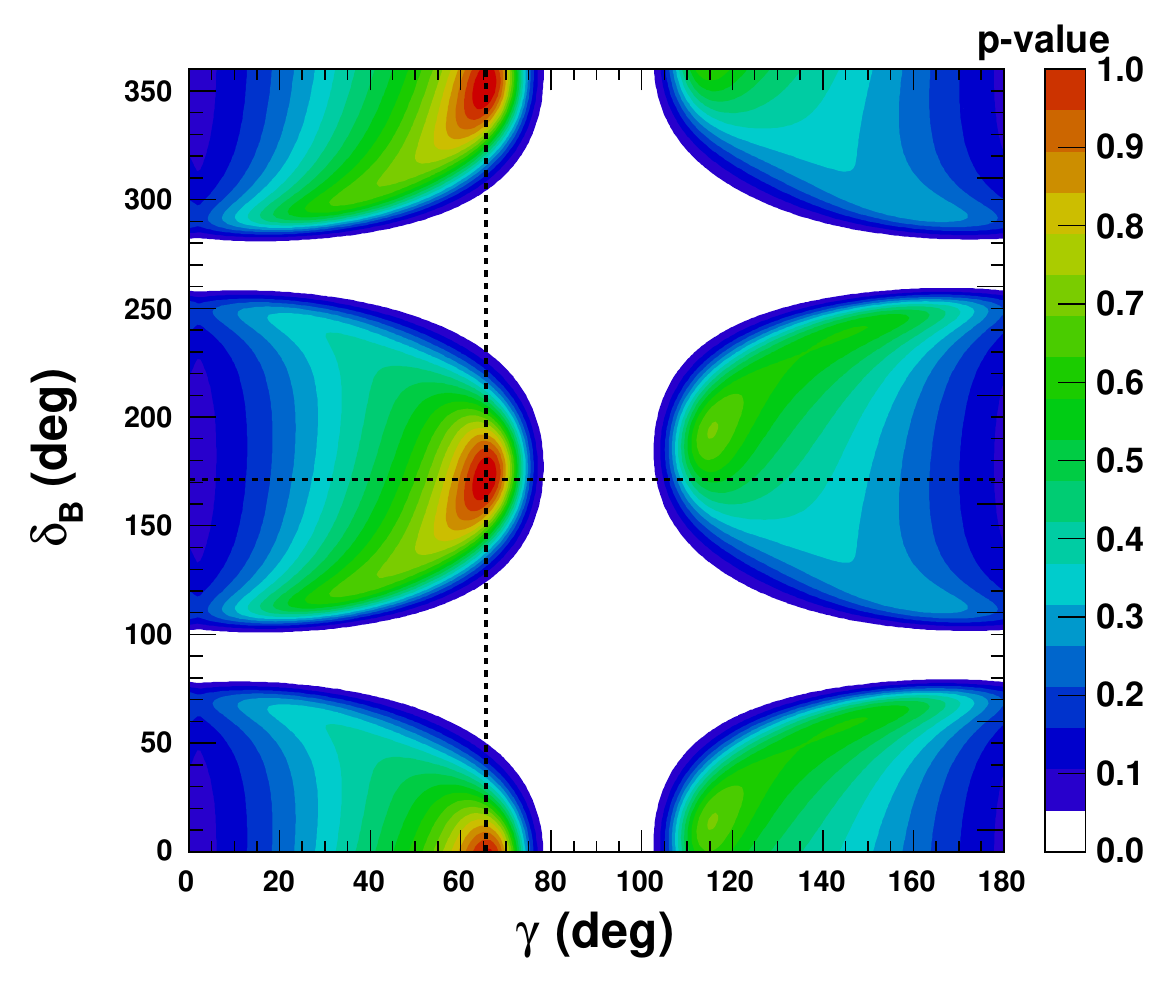}
\includegraphics[width=0.4\textwidth,height=0.15\textheight]{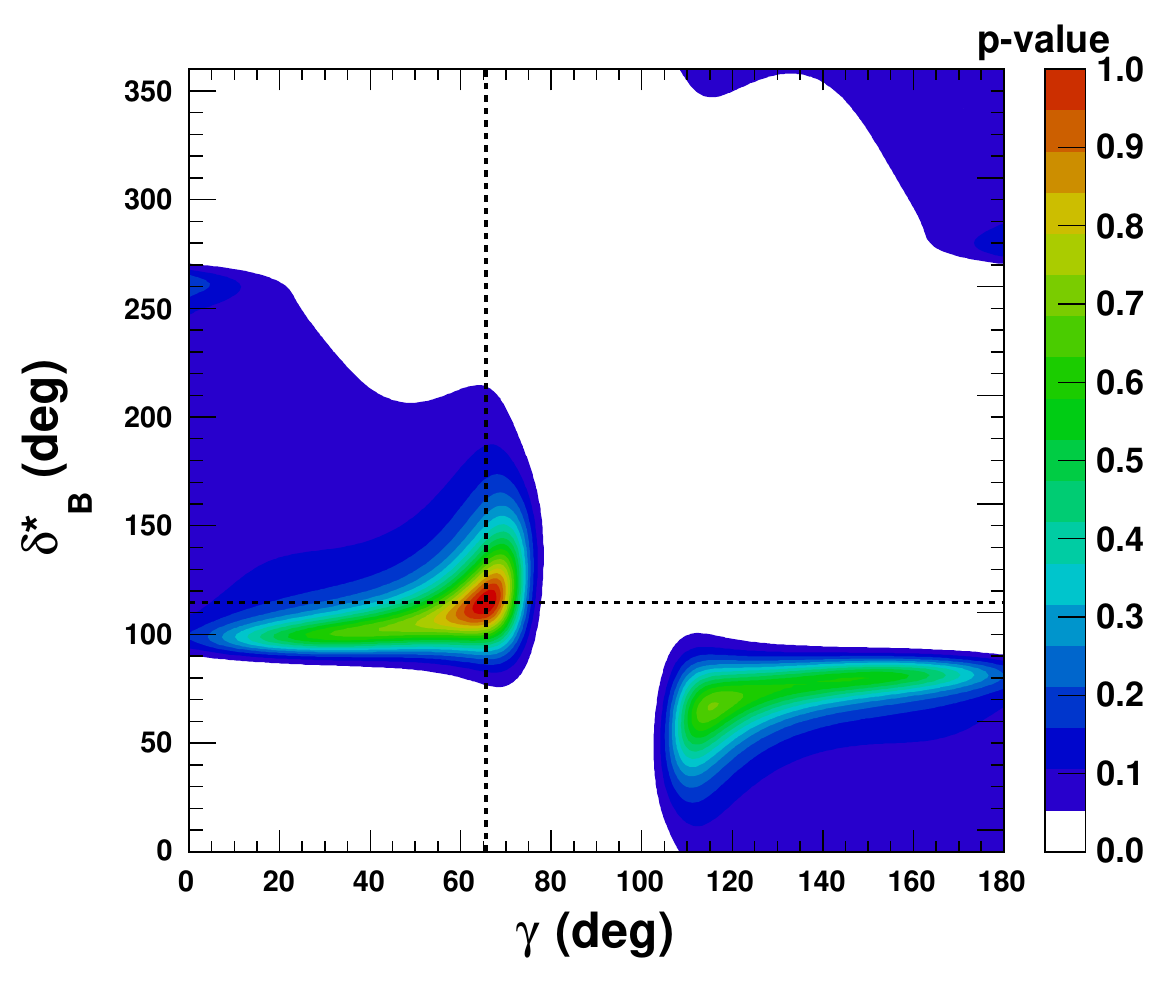}
\caption{\label{fig:2D_0} Two-dimension $p$-value  profile distribution of the nuisance parameters $\rBst$ and $\deltaBst$ as a function of $\gamma$. On each figure the dashed black  lines indicate the initial  true values: $\gamma=65.66^\circ$ (1.146 rad), $\deltaB=171.9^\circ$ (3.0 rad), and $\deltastB=114.6^\circ$ (2.0 rad), and  $\rBst=0.4$.}
\includegraphics[width=0.425\textwidth,height=0.15\textheight]{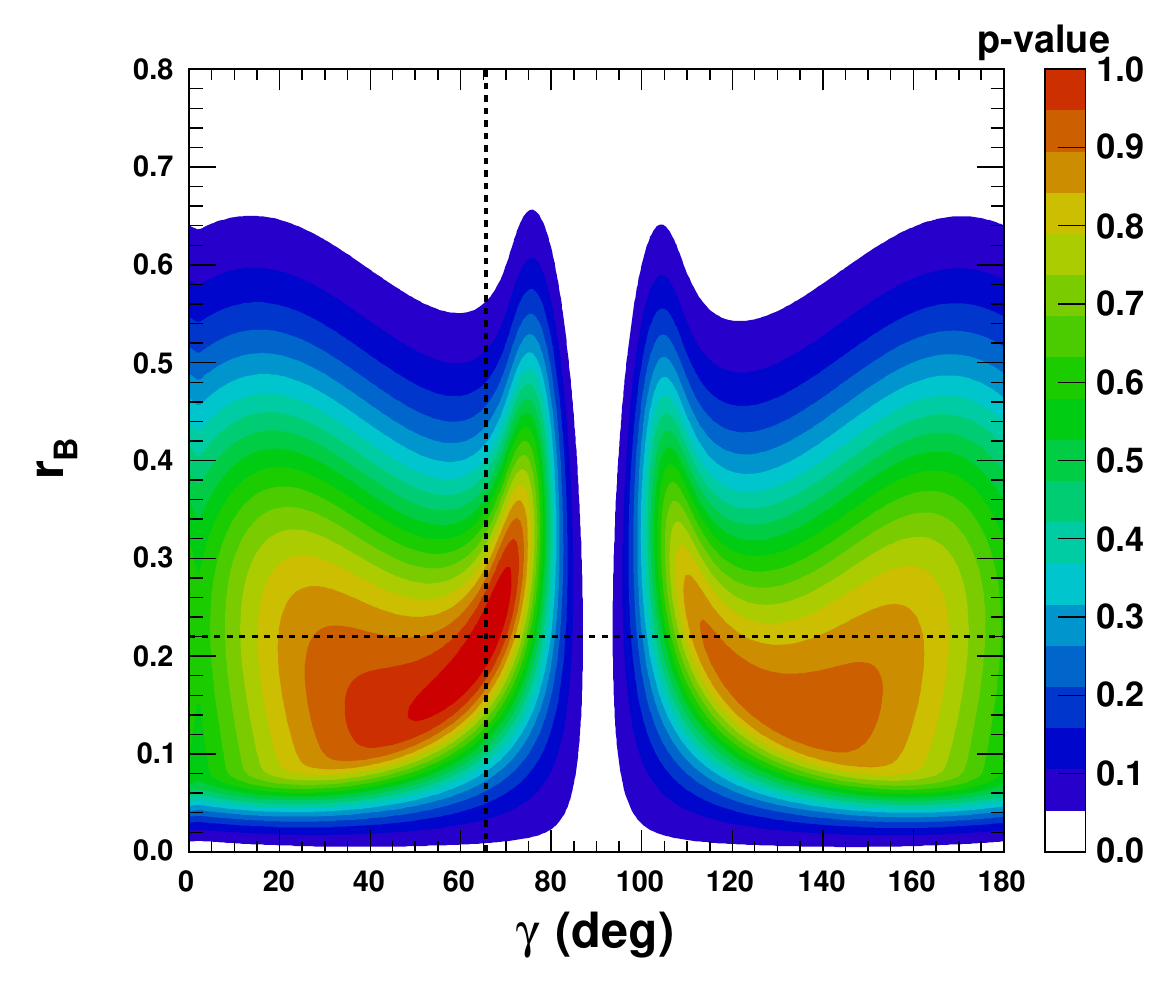}
\includegraphics[width=0.425\textwidth,height=0.15\textheight]{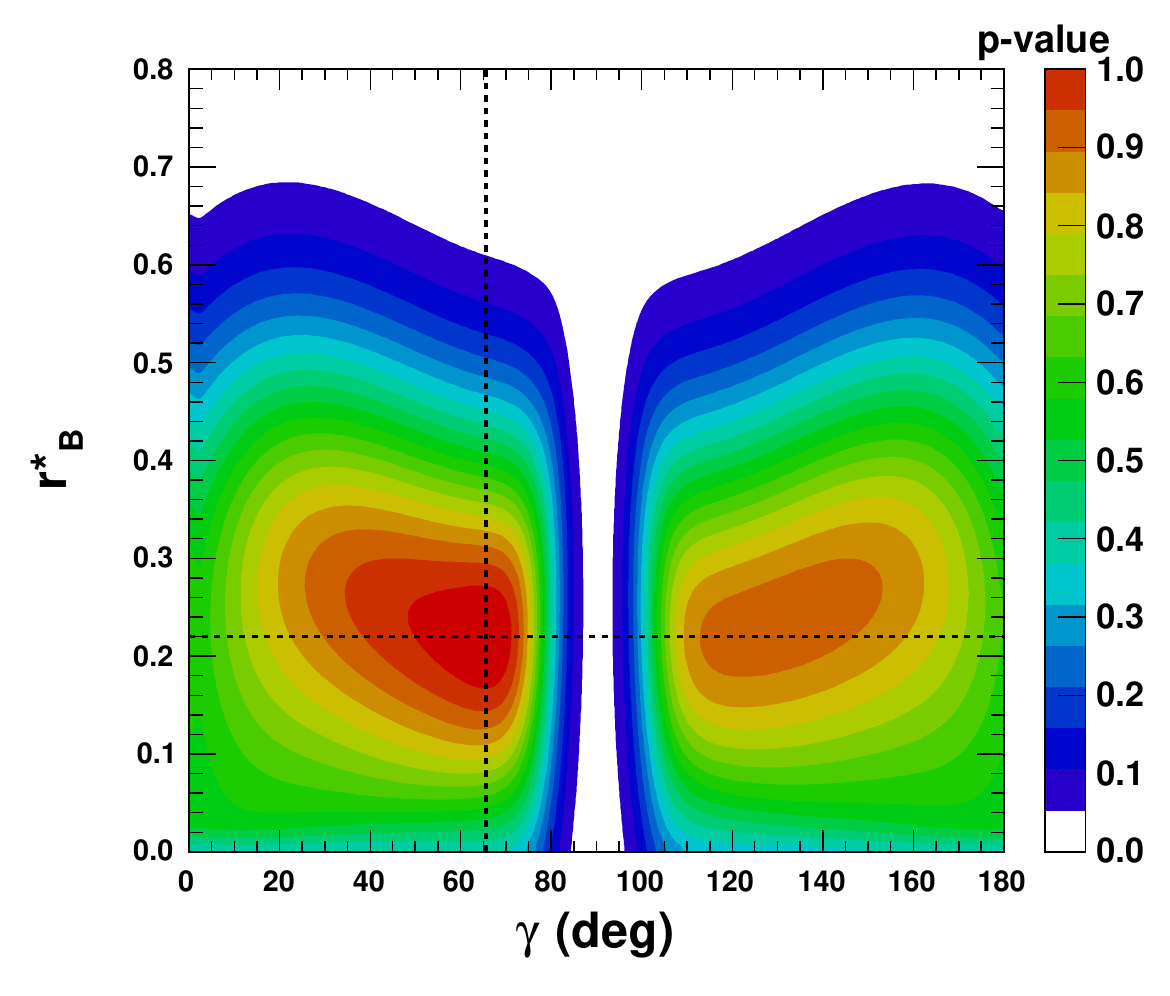} \\
\includegraphics[width=0.425\textwidth,height=0.15\textheight]{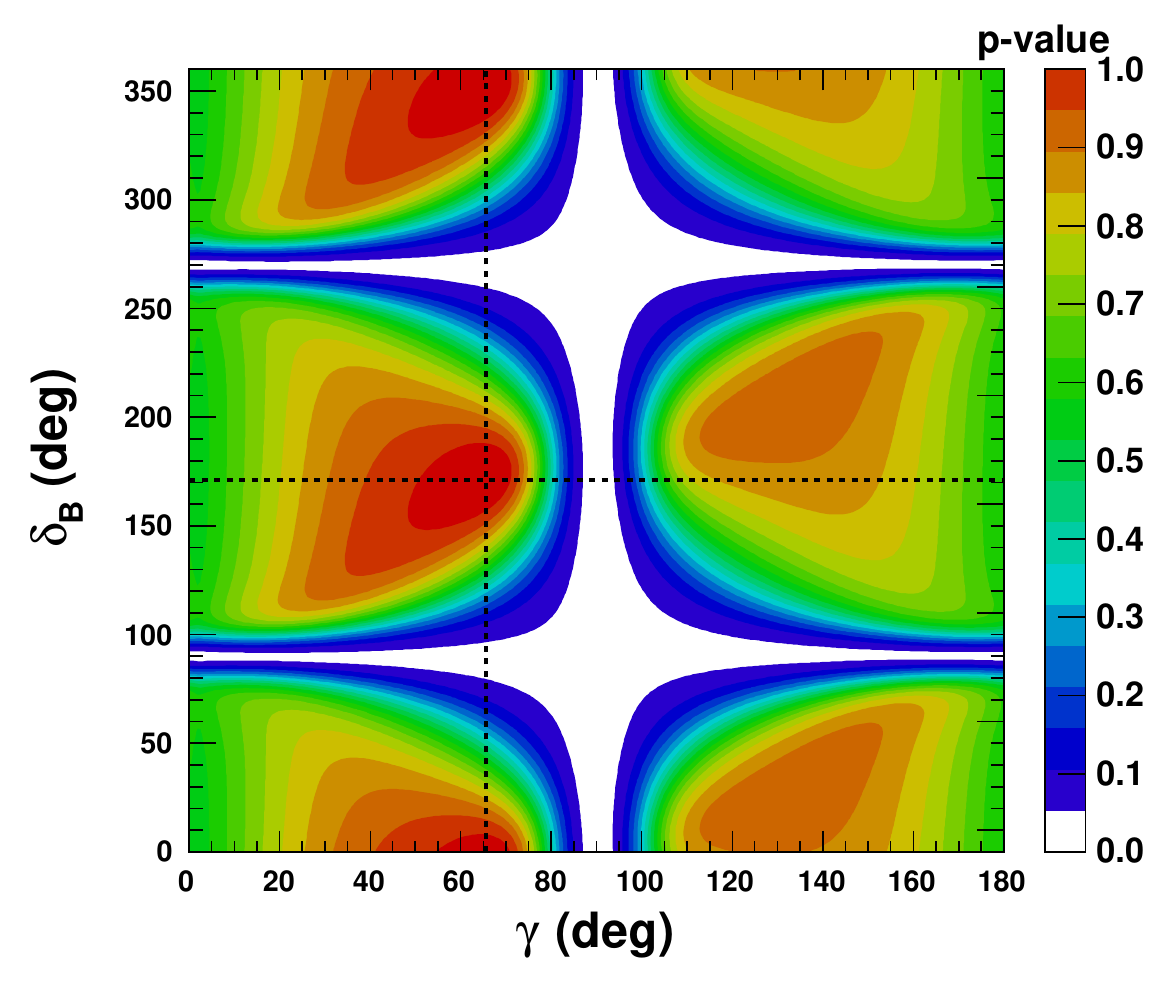}
\includegraphics[width=0.425\textwidth,height=0.15\textheight]{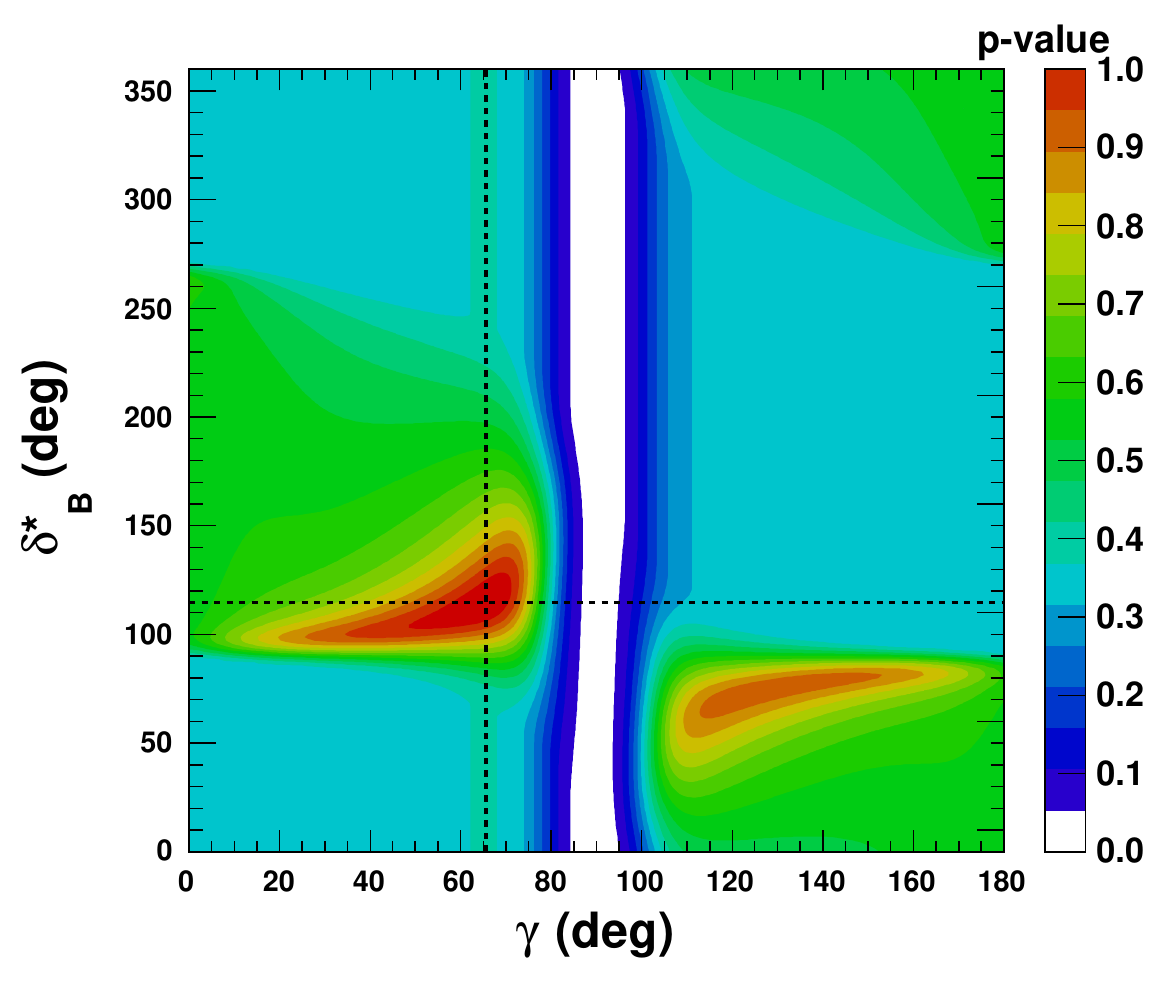}
\caption{\label{fig:2D_1} Two-dimension $p$-value  profile distribution of the nuisance parameters $\rBst$ and $\deltaBst$ as a function of $\gamma$. On each figure the dashed black  lines indicate the true values: $\gamma=65.66^\circ$ (1.146 rad), $\deltaB=171.9^\circ$ (3.0 rad), and $\deltastB=114.6^\circ$ (2.0 rad), and  $\rBst=0.22$.}
\end{figure}

\clearpage

are used here and corresponds to relatively limited value for the coherence factor: $R^{K3\pi}_D= (43^{+17}_{-13})\%$~\cite{Evans:2016tlp}. A more attractive approach could be to perform the analysis in disjoint bins of the phase space. In this case, the parameters are re-defined within each bin.
New values for $R^{K3\pi}_D$ and $\delta^{K3\pi}_{D}$ in each bins from Ref.~\cite{Evans:2019wza} have alternatively been employed. No noticeable change on $\gamma$ and $\rBst$ fitted $p$-value profiles were seen, but it is possible that some fold-effects  on $\deltaBst$, as seen {\it e.g.} in Figs.~\ref{fig:2D_0}-\ref{fig:2D_2},  become less probable. The lack of significant improvement is expected, as the $\Dtz \rightarrow K3\pi$ mode is not the dominant decay and also because the new measurements of $R^{K3\pi}_{D}$ and $\delta^{K3\pi}_{D}$ in each bin still have large uncertainties.

\subsection{Unfolding the $\gamma$, $\rBst$, and $\deltaBst$ parameters from the generated pseudoexperiments}
\label{sec:FittingpseudoExp}

\begin{figure}[t]
\centering
\includegraphics[width=0.425\textwidth]{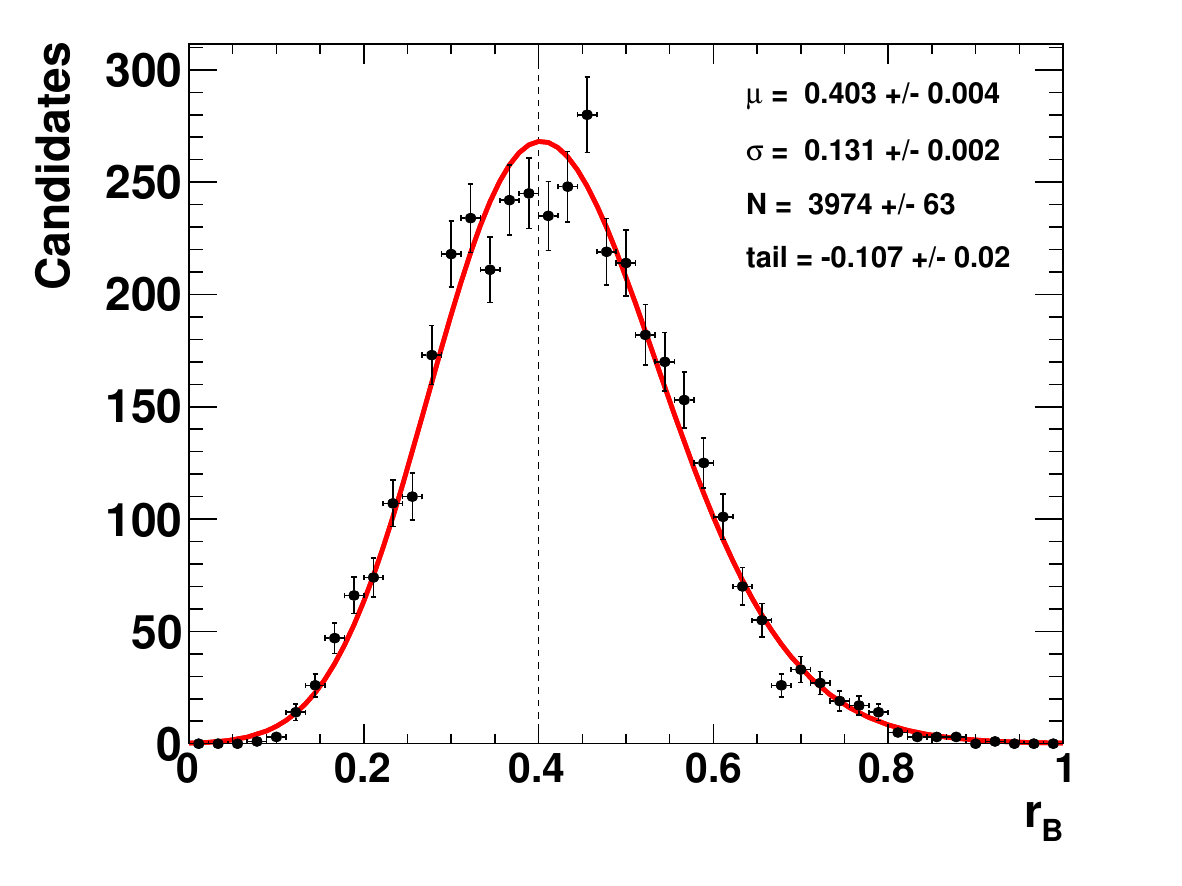}
\includegraphics[width=0.425\textwidth]{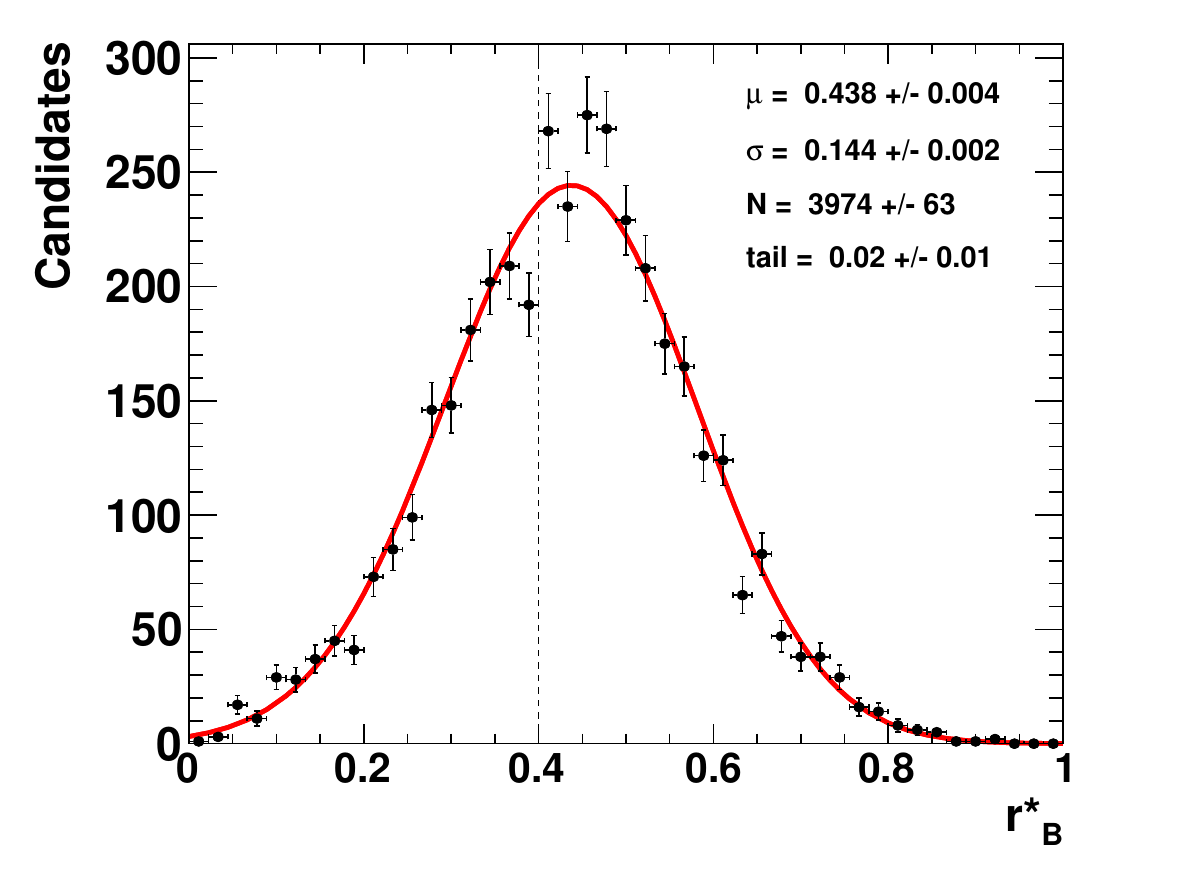} \\
\includegraphics[width=0.425\textwidth]{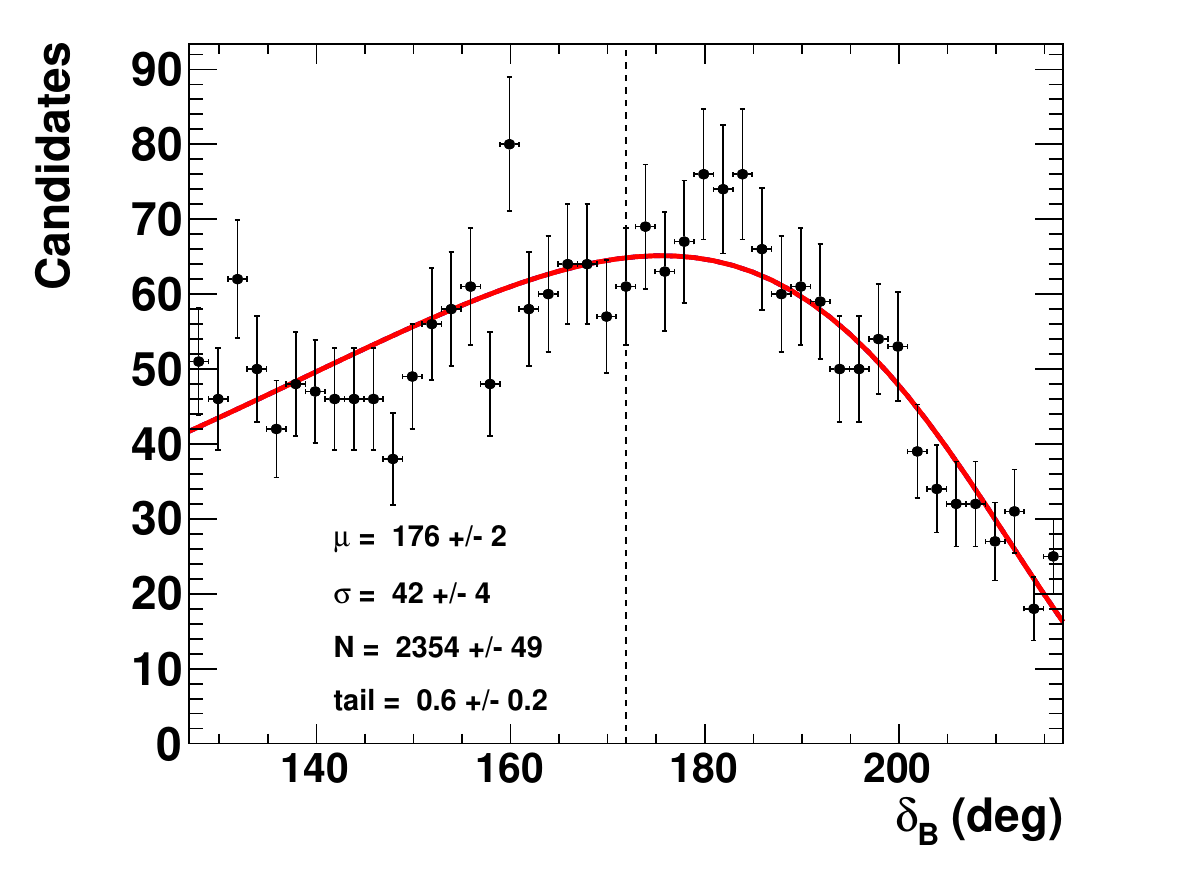}
\includegraphics[width=0.425\textwidth]{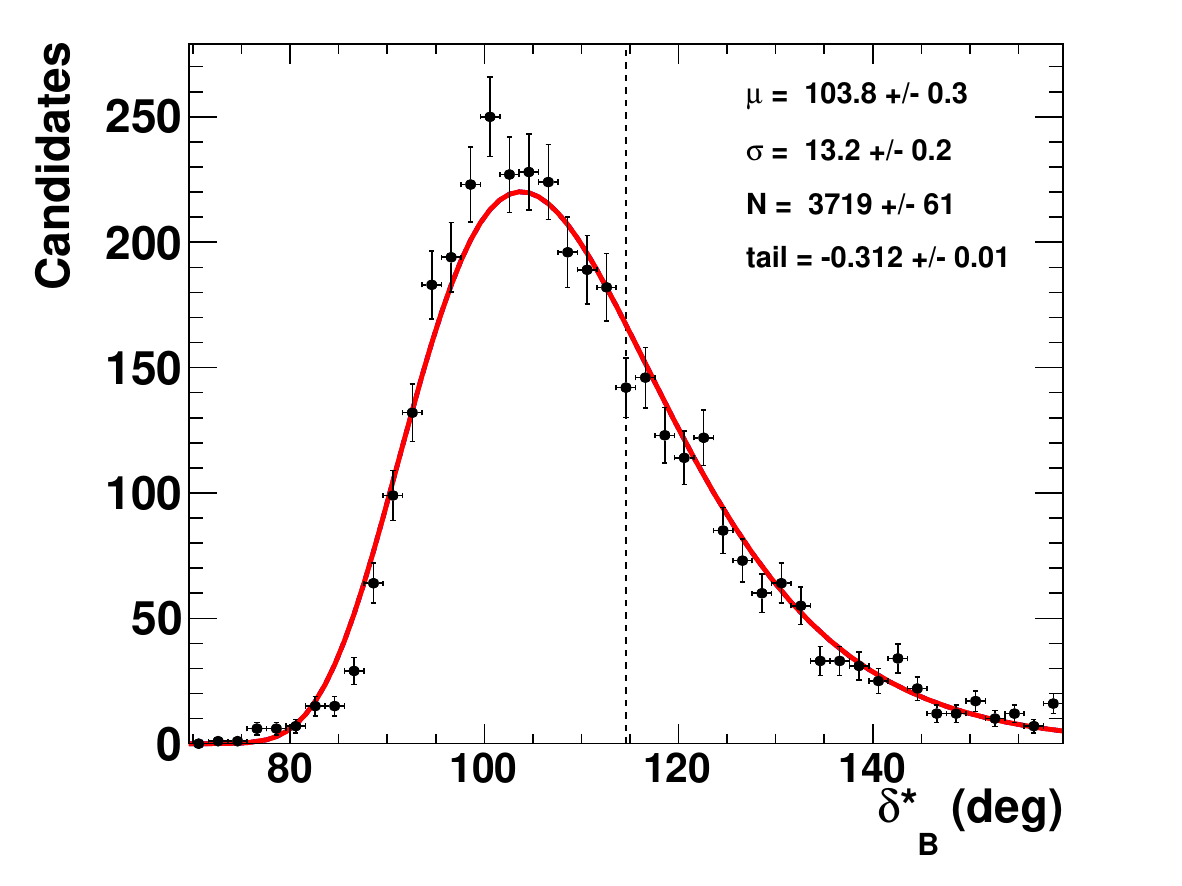}
\caption{\label{fig:Fit_dBstAndrBst} Fit to the distributions of the  nuisance parameters  $\rBst$ (top left (right)) and $\deltaBst$ (bottom left (right)) obtained from 4000 pseudoexperiments. The initial configuration is $\gamma=65.66^{\circ}$ (1.146 rad), $\rBst=0.4$, $\deltaB=171.9^\circ$ (3 rad), and $\deltastB=114.6^\circ$ (2 rad). The distributions of $\deltaBst$ are plotted and fitted within $\pm 45^\circ$ their initial true value. In the distributions, only the candidates with a value  $\gamma\  \in \ [0^\circ, \  90^\circ]$ are considered. }
\end{figure}

As explained in Sec.~\ref{sec:pseudoExp}, for each of the tested $\gamma$, $\rBst$, and $\deltaBst$ configurations,  4000 pseudoexperiments are generated, which values of $\gamma$, $\rBst$, and $\deltaBst$ are unfolded from global $\chi^2$ fits (See Sec.~\ref{sec:p-values} for illustrations). Figure~\ref{fig:Fit_dBstAndrBst} displays the extended unbinned maximum likelihood fits to the nuisance parameters  $\rBst$ and $\deltaBst$. The initial configuration is $\gamma=65.66^{\circ}$ (1.146 rad), $\rBst=0.4$, $\deltaB=171.9^\circ$ (3 rad), and $\deltastB=114.6^\circ$ (2 rad) and an integrated  luminosity equivalent of LHCb Run 1 \& 2 data. It can be compared with Fig.~\ref{fig:others_1D}. All the distributions are fitted with the \emph{Novosibirsk} empirical function, whose description contains a Gaussian core part and a left or right tail, depending on the sign of the tail parameter~\cite{Novosibirsk}.  The fitted values of $\rBst$ are centered at their initial tested values 0.4, with a resolution of 0.14, and no bias is observed. For $\deltaBst$, the fitted value is $(176\pm42)^\circ$ ($(104\pm13)^\circ$) for an initial true value equal to $171.9^\circ$ ($114.6^\circ$). The fitted value for $\deltastB$ is slightly shifted by about $2/3$ of a standard deviation, but its measurement is much more precise than that of $\deltaB$, as it is measured both from the  $\Dstz \rightarrow \Dz \gamma$ and the $\Dstz \rightarrow \Dz \pi^{0}$ observables.

Figure~\ref{fig:Fit_gamma} shows the corresponding fit to the CKM angle $\gamma$, where the value $\rBst=0.22$ is also tested. This Figure can be compared to the initial $p$-value profiles shown in Fig.~\ref{fig:gamma_1D}. As shown in Figs.~\ref{fig:2D_0} and~\ref{fig:2D_1}, $\gamma$ is correlated with the nuisance parameters $\rBst$ and $\deltaBst$, such correlations may generate long tails in its distribution as obtained from 4000 pseudoexperiments. To account for those tails, extended unbinned maximum likelihood fits, constituted of two \emph{Novosibirsk} functions, with opposite-side tails, are performed to the $\gamma$ distributions. With an initial value of $65.66^\circ$, the fitted value for $\gamma$ returns a central value equal to  $\mu_{\gamma}=(65.9\pm0.3)^\circ$, with a resolution of $\sigma_{\gamma}=(8.8\pm0.2)^\circ$, when $\rBst=0.4$ and respectively, $\mu_{\gamma}=(66.6\pm0.7)^\circ$, with a resolution of $\sigma_{\gamma}=(14.4\pm0.5)^\circ$, when $\rBst=0.22$. The worse resolution obtained with $\rBst=0.22$ follows the empirical behaviour $1/\rBst$ ({\it i.e.} $8.8\times0.4/0.22\simeq 16.0$). There again, no bias is observed.

Finally, Fig.~\ref{fig:2Dtoy} displays the two-dimension distributions of the nuisance parameters $\rBst$ and $\deltaBst$ as a function of $\gamma$ obtained from 4000 pseudoexperiments. It can be compared with the corresponding $p$-value profiles shown in Fig.~\ref{fig:2D_0}.


\subsection{Varying $\deltaBst$ and $\rBst$}
\label{sec:varyrBanddeltaB}

According to Sec.~\ref{sec:choiceParams}, 72 configurations of nuisance parameters $\deltaBst$ and $\rBst$ have been tested for $\gamma=65.66^\circ$ (1.146 rad) and 4000 pseudoexperiments have been generated for each set, according to the procedure described in Sec.~\ref{sec:pseudoExp} and illustrated in Sec.~\ref{sec:FittingpseudoExp}. The integrated luminosity assumed in this Section is that of LHCb data collected in Run~1~\&~2.

The fitted mean value of $\gamma$ ($\mu_{\gamma}$), for $\rBst=0.4$ and $0.22$, as a function of $\deltaBst$, for an initial true value of  $65.66^\circ$ (1.146 rad) are given in Table~\ref{tab:dbeffect_rBst04_022_mean}, while the corresponding resolutions  ($\sigma_{\gamma}$) are listed in Table~\ref{tab:dbeffect_rBst04_022_resol}. The fitted means are in general compatible with the true $\gamma$ value within less than one standard deviation. For $\rBst=0.4$, the resolution varies from $\sigma_{\gamma}=8.3^\circ$ to $12.9^\circ$.  For $\rBst=0.22$, the resolution is worse, as expected, it  varies from $\sigma_{\gamma}=13.9^\circ$ to $18.7^\circ$. For $\rBst=0.22$, the distribution of $\gamma$ of the 4000 pseudoexperiments has its maximum above $90^\circ$ for  $\deltaBst=286.5^\circ$ and is therefore not considered.

\begin{figure}[tbp]
\centering
\includegraphics[width=0.45\textwidth]{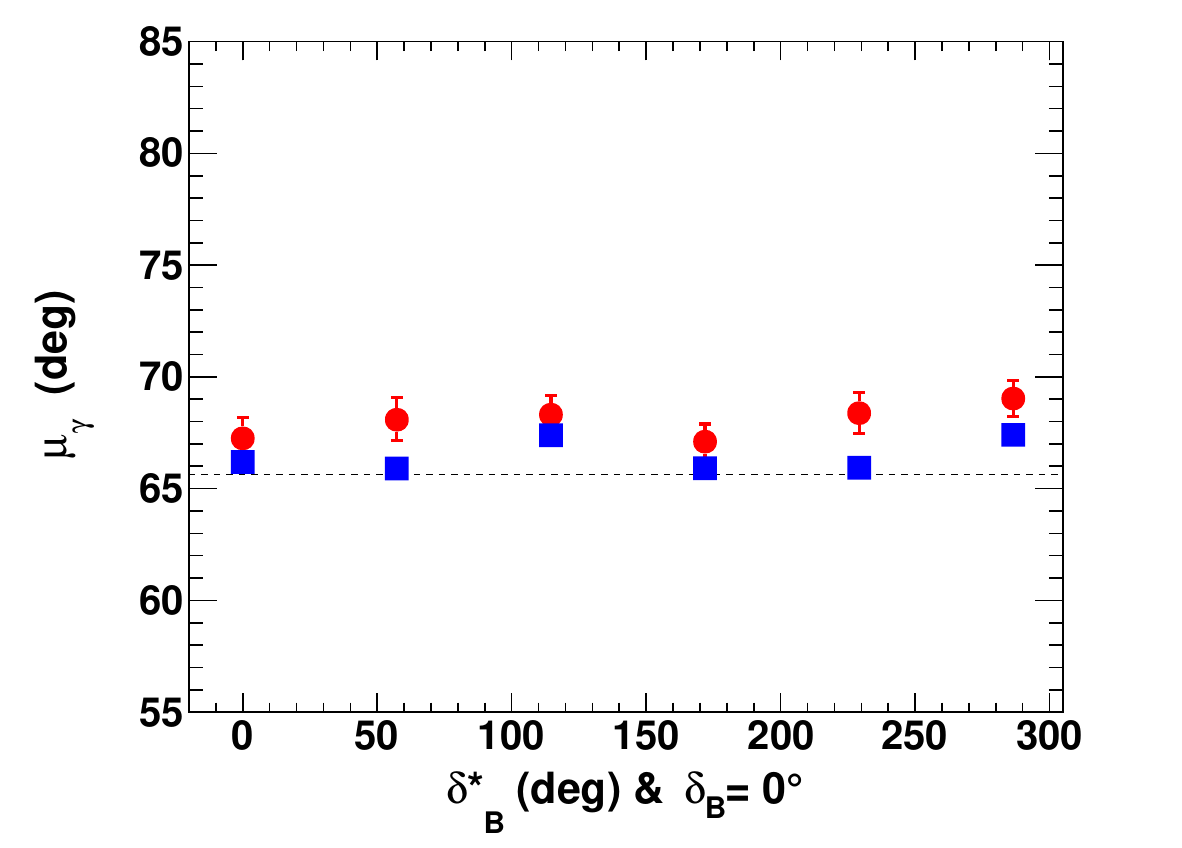}
\includegraphics[width=0.45\textwidth]{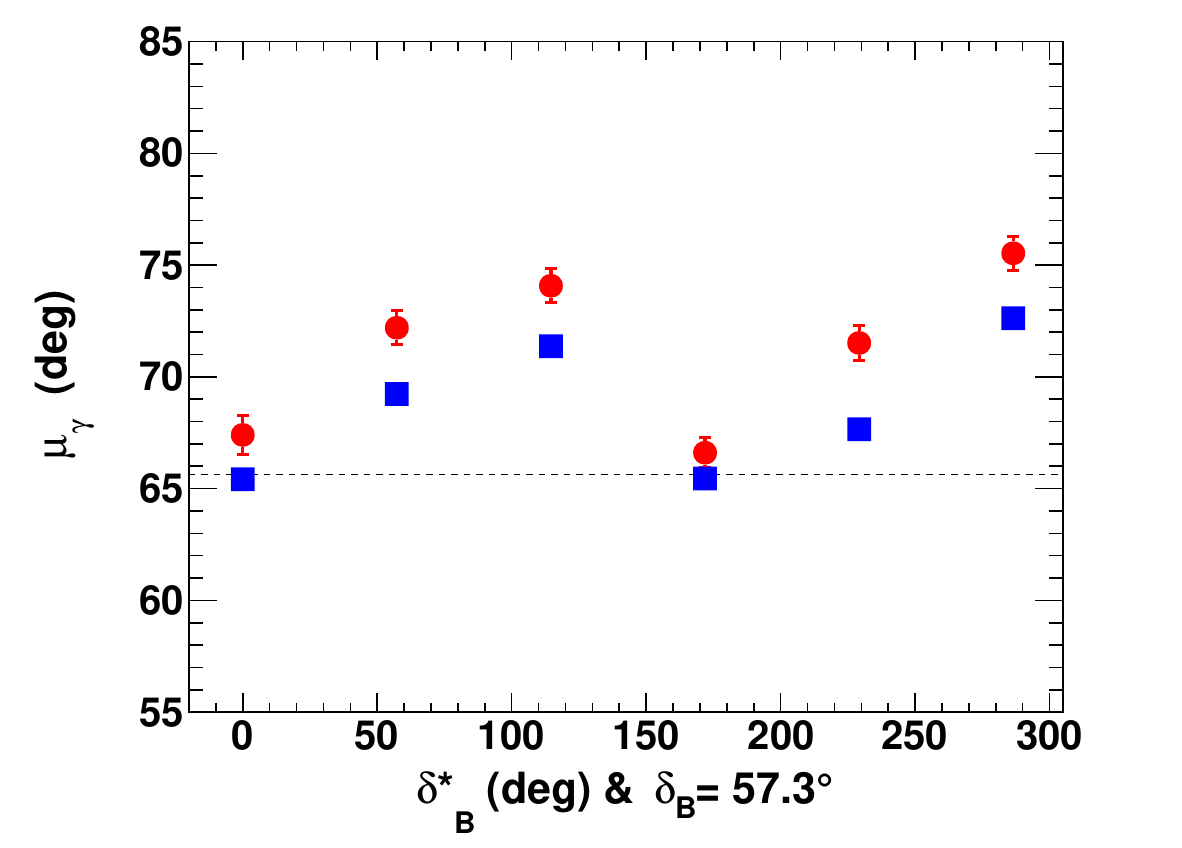} \\
\includegraphics[width=0.45\textwidth]{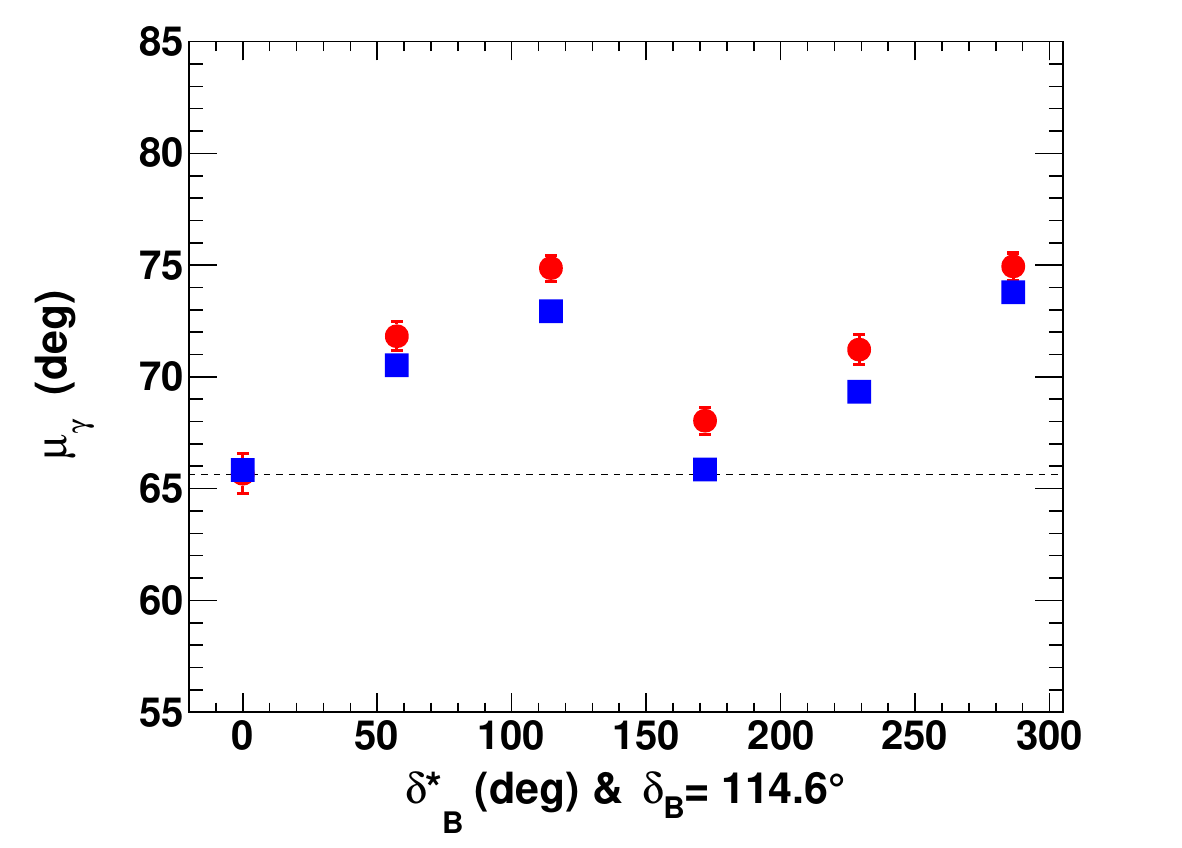}
\includegraphics[width=0.45\textwidth]{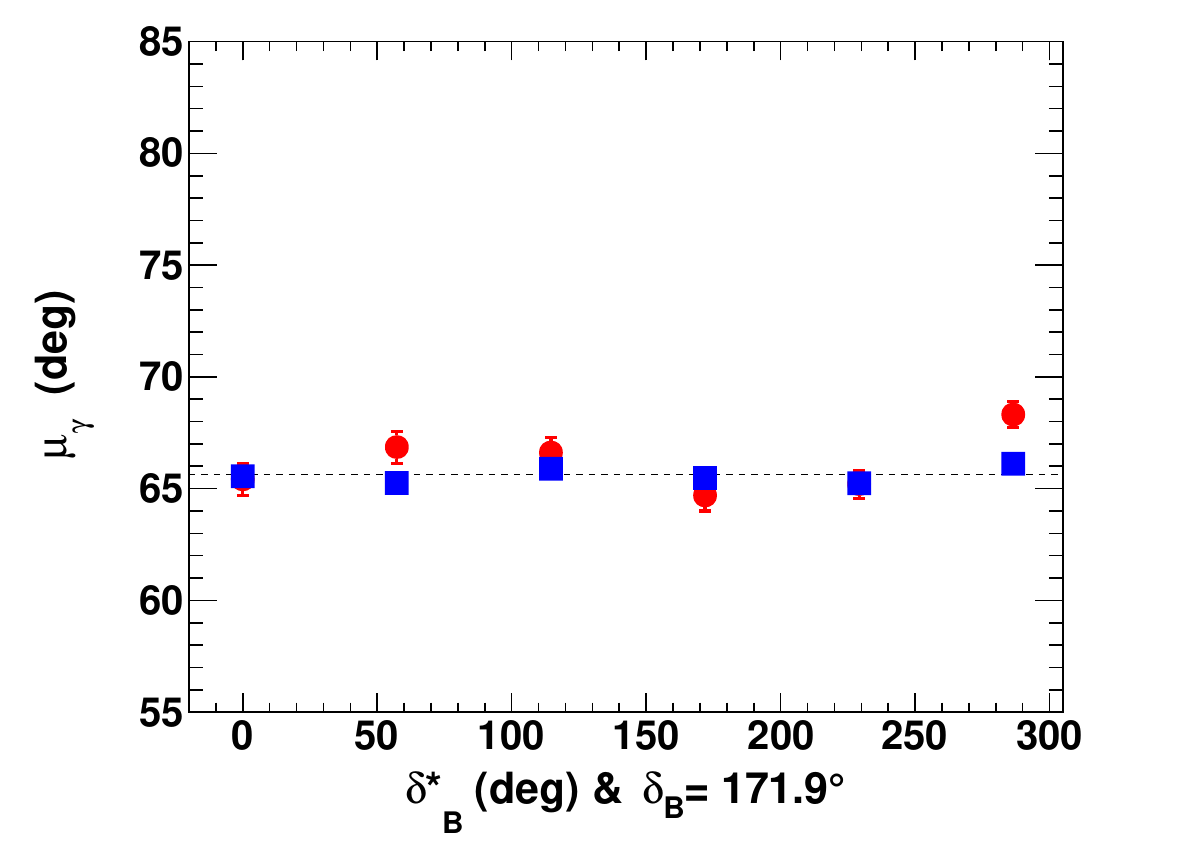} \\
\includegraphics[width=0.45\textwidth]{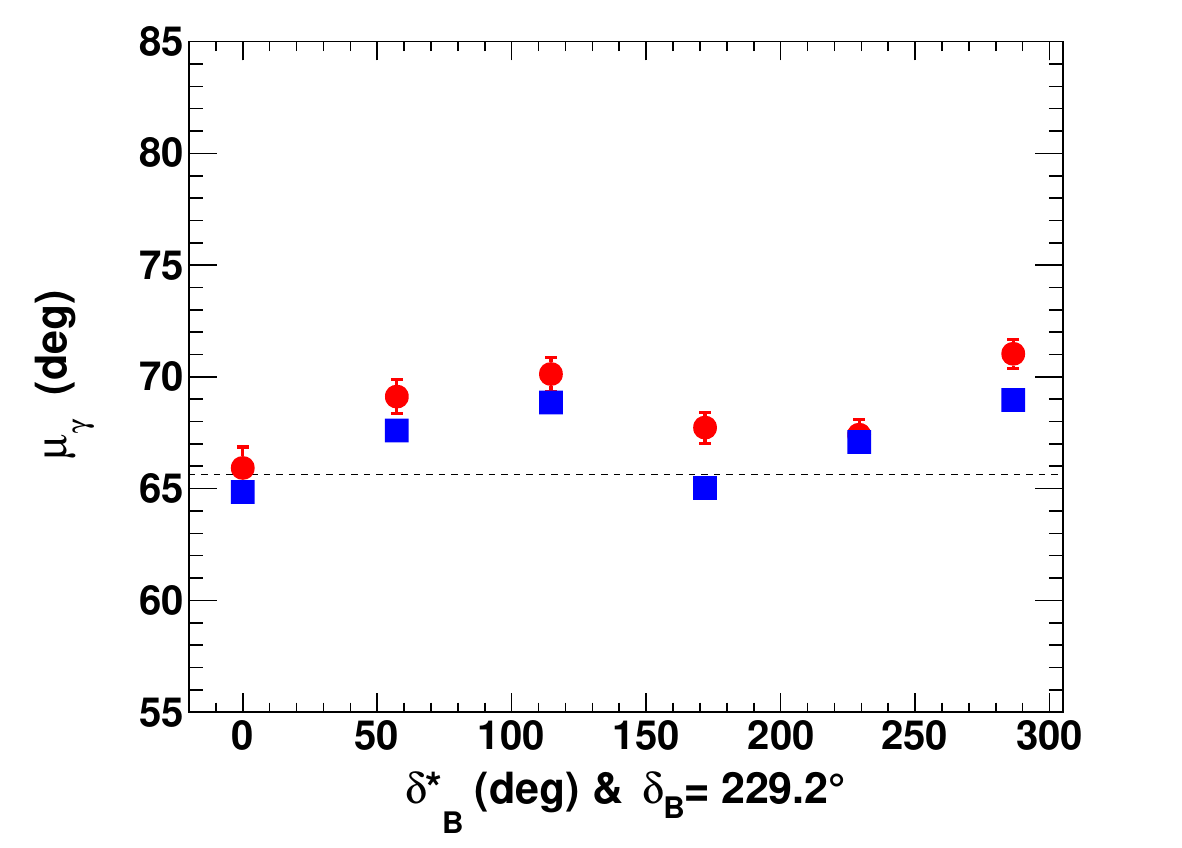}
\includegraphics[width=0.45\textwidth]{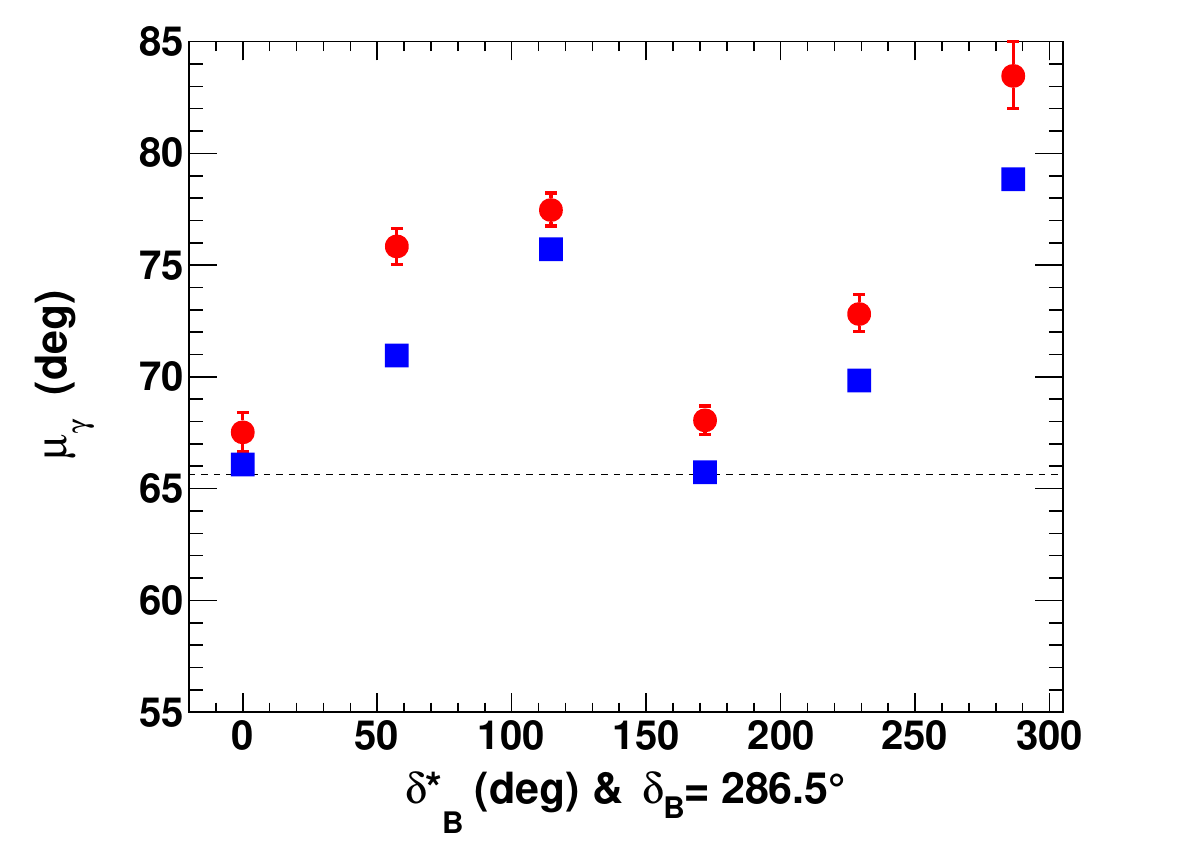}
\caption{\label{fig:dbeffect_rBst022_04_mean}Fitted mean value of $\gamma$ ($\mu_{\gamma}$), for $\rBst=0.22$ (red circles) and 0.4 (blue squares), as a function of $\deltaBst$, for an initial true value of  $65.66^\circ$ (1.146 rad). All the listed values are in [deg]. On each figure, the horizontal dashed black  line indicates the initial  $\gamma$ true value. All the plotted uncertainties are statistical only. }
\end{figure}

\begin{figure}[tbp]
\centering
\includegraphics[width=0.45\textwidth]{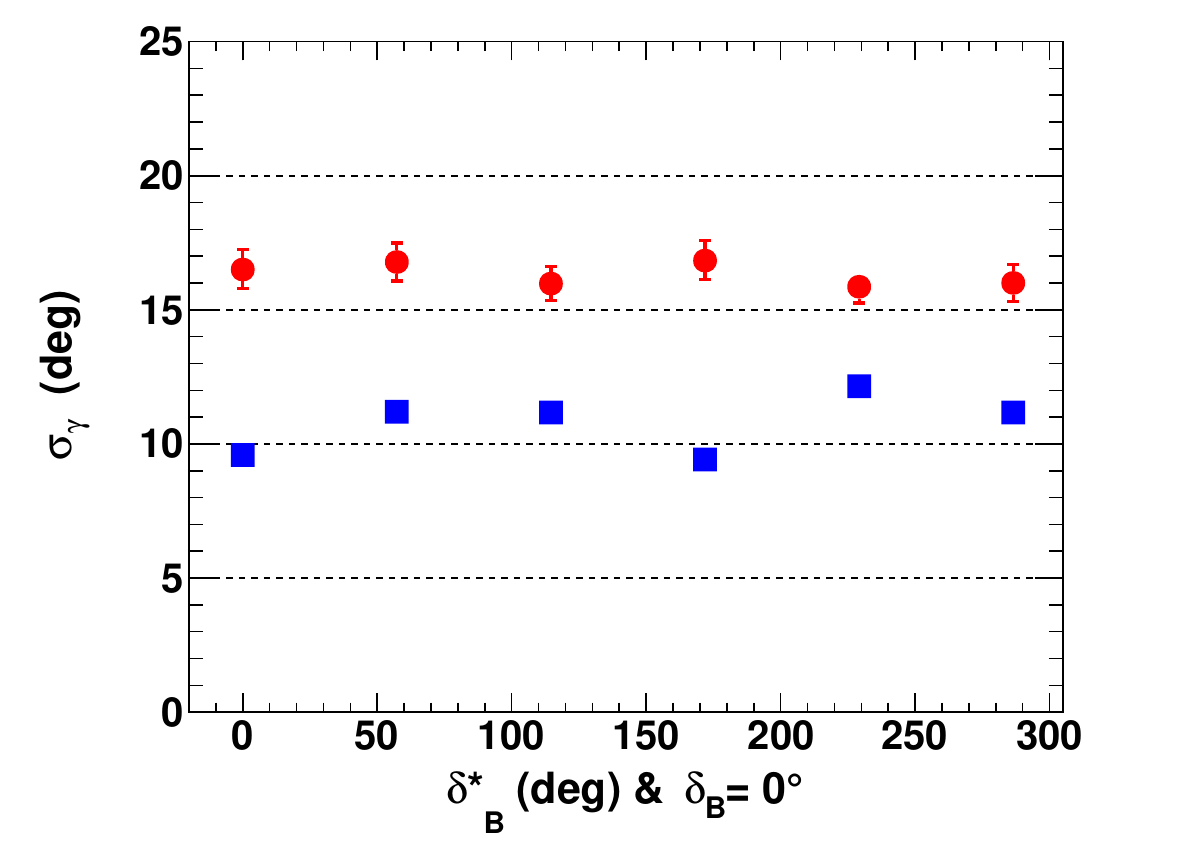}
\includegraphics[width=0.45\textwidth]{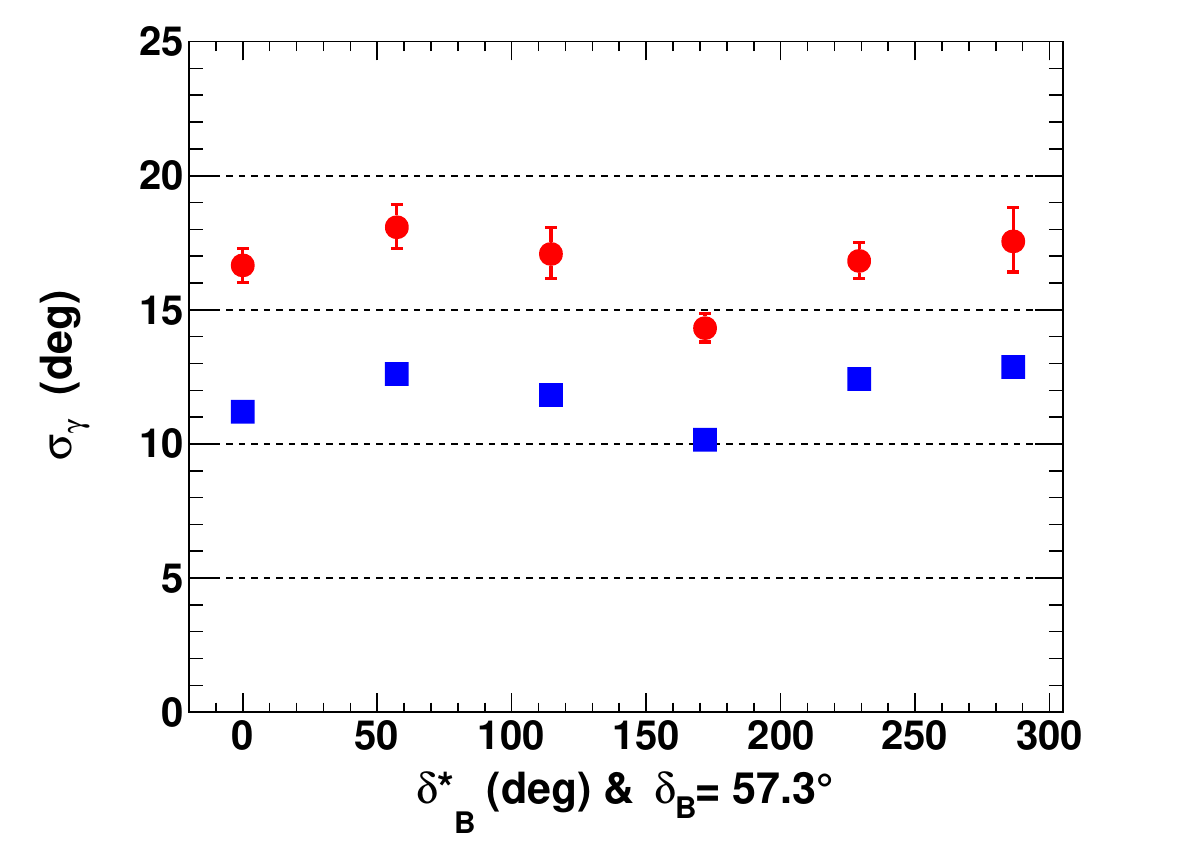} \\
\includegraphics[width=0.45\textwidth]{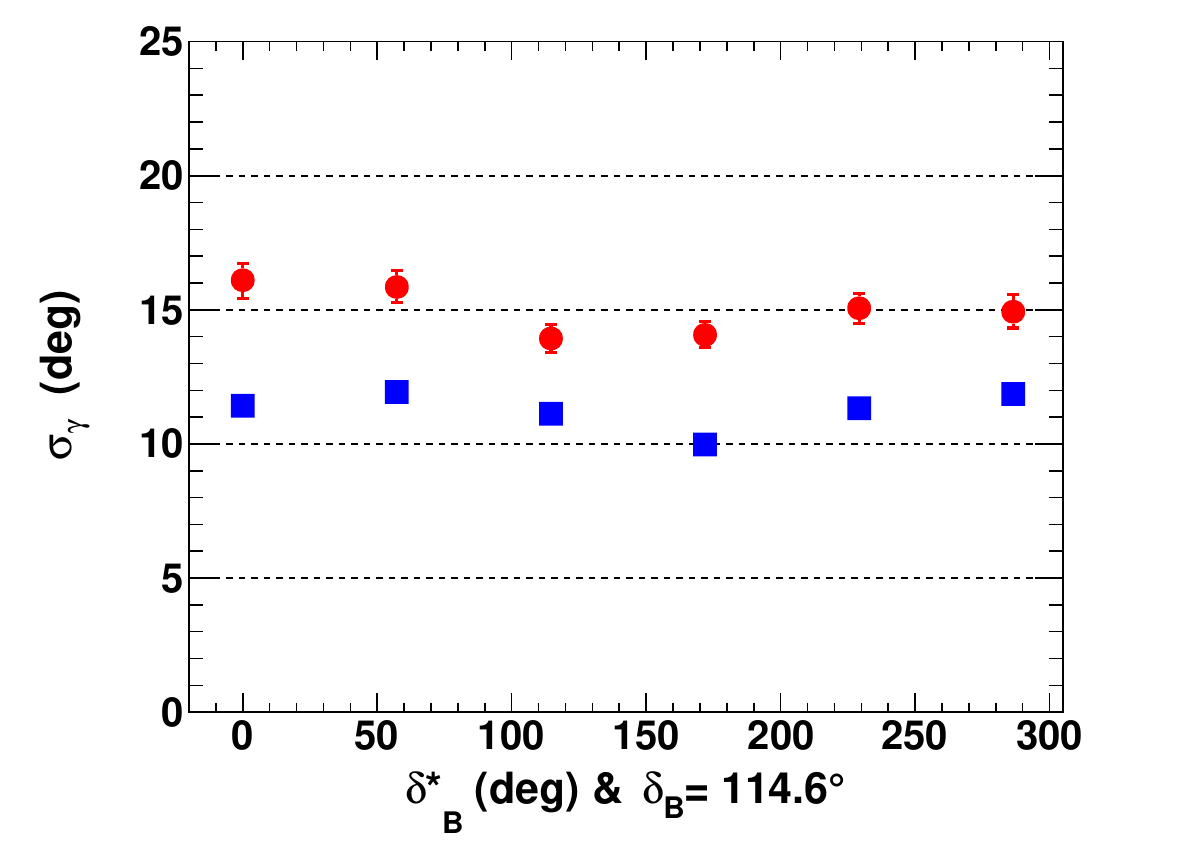}
\includegraphics[width=0.45\textwidth]{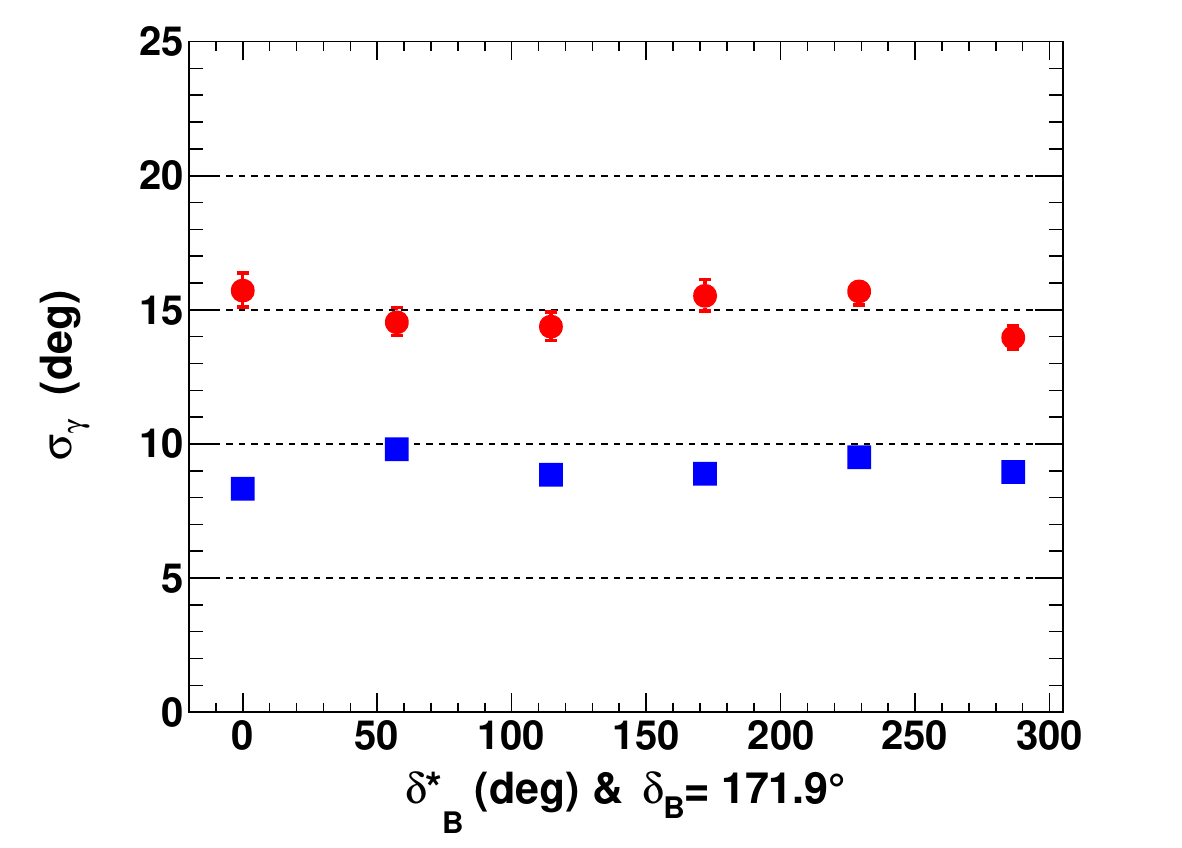}  \\
\includegraphics[width=0.45\textwidth]{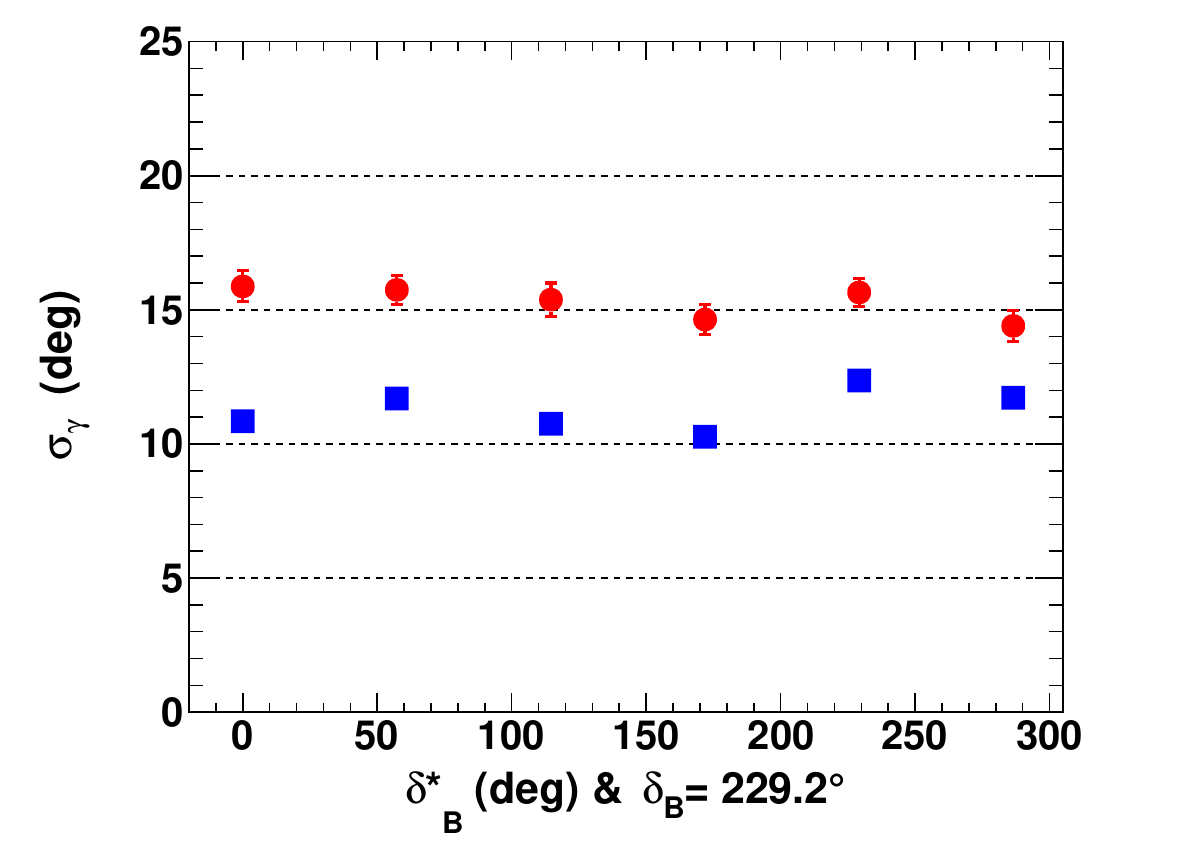}
\includegraphics[width=0.45\textwidth]{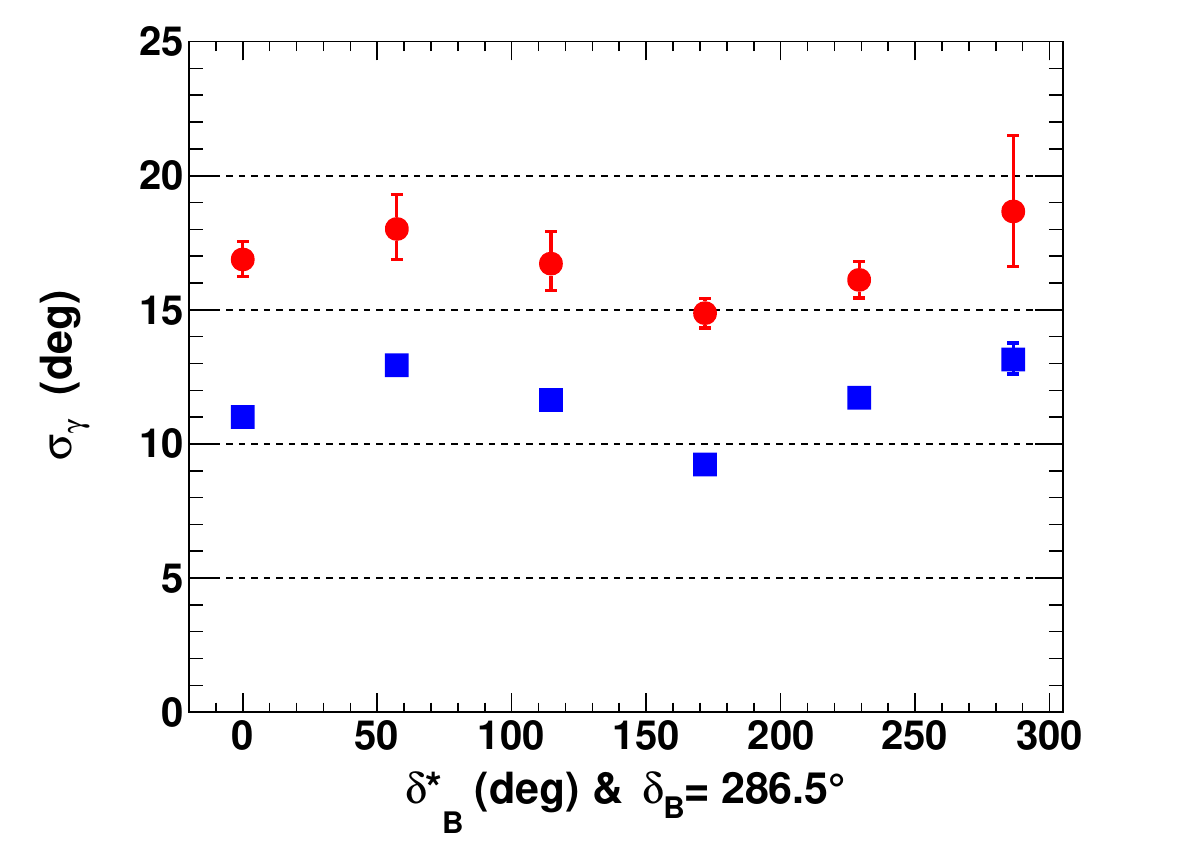}
\caption{\label{fig:dbeffect_rBst022_04_resol}Fitted resolution of $\gamma$ ($\sigma_{\gamma}$), for $\rBst=0.22$ (red circles) and 0.4 (blue squares), as a function of $\deltaBst$, for an initial true value of  $65.66^\circ$ (1.146 rad). All the listed values are in [deg]. On each figure, the horizontal dashed black  lines are guide for the eye at $\sigma_{\gamma}= 5^\circ$, $10^\circ$, $15^\circ$, and $20^\circ$.  All the plotted uncertainties are statistical only. }
\end{figure}

The obtained values for $\mu_{\gamma}$ and $\sigma_{\gamma}$ are also displayed in Figs.~\ref{fig:dbeffect_rBst022_04_mean} and~\ref{fig:dbeffect_rBst022_04_resol}. It is clear that the resolution on $\gamma$ depends to first order on $\rBst$, then to the second order on $\deltaBst$. The best agreement with respect to the tested initial true value of $\gamma$ is obtained when $\deltaBst=0^\circ$ (0 rad) or $180^\circ$ ($\pi$ rad), and,  there also, the best resolutions are obtained ({\it i.e.} the lowest values of $\sigma_{\gamma}$). The largest \CP violation effects and the best sensitivity to $\gamma$ are there.  At the opposite, the worst sensitivity is obtained when $\deltaBst=90^\circ$ ($\pi/2$ rad) or $270^\circ$ ($3\pi/2$ rad). The other best and worst positions for $\deltaBst$, can easily be deduced from Eq.~\ref{EQ__14_}. In most of the cases, for $\rBst=0.4$ (0.22), the value of the resolution is $\sigma_{\gamma}\sim10^\circ$ ($15^\circ$) and the fitted mean value $\mu_{\gamma} \sim 65.66^\circ$, or slightly larger.

For completeness, the fitted means and resolutions for the nuisances parameters $\rBst$ and $\deltaBst$ are presented on  Figs.~\ref{fig:dbeffect_rBst022_04_rB}-\ref{fig:dbeffect_rBst022_04_dBst} in appendix~\ref{sec:appendA}. It can be seen that the fitted  mean values of $\rBst$ and $\deltaBst$ are in good agreement with their initial tested true values, within one standard deviation of their fitted resolutions.

\subsection{The case $\gamma$ equals $74^\circ$}
\label{sec:74deg}

\begin{figure}[tbp]
\centering
\includegraphics[width=0.425\textwidth,height=0.2\textheight]{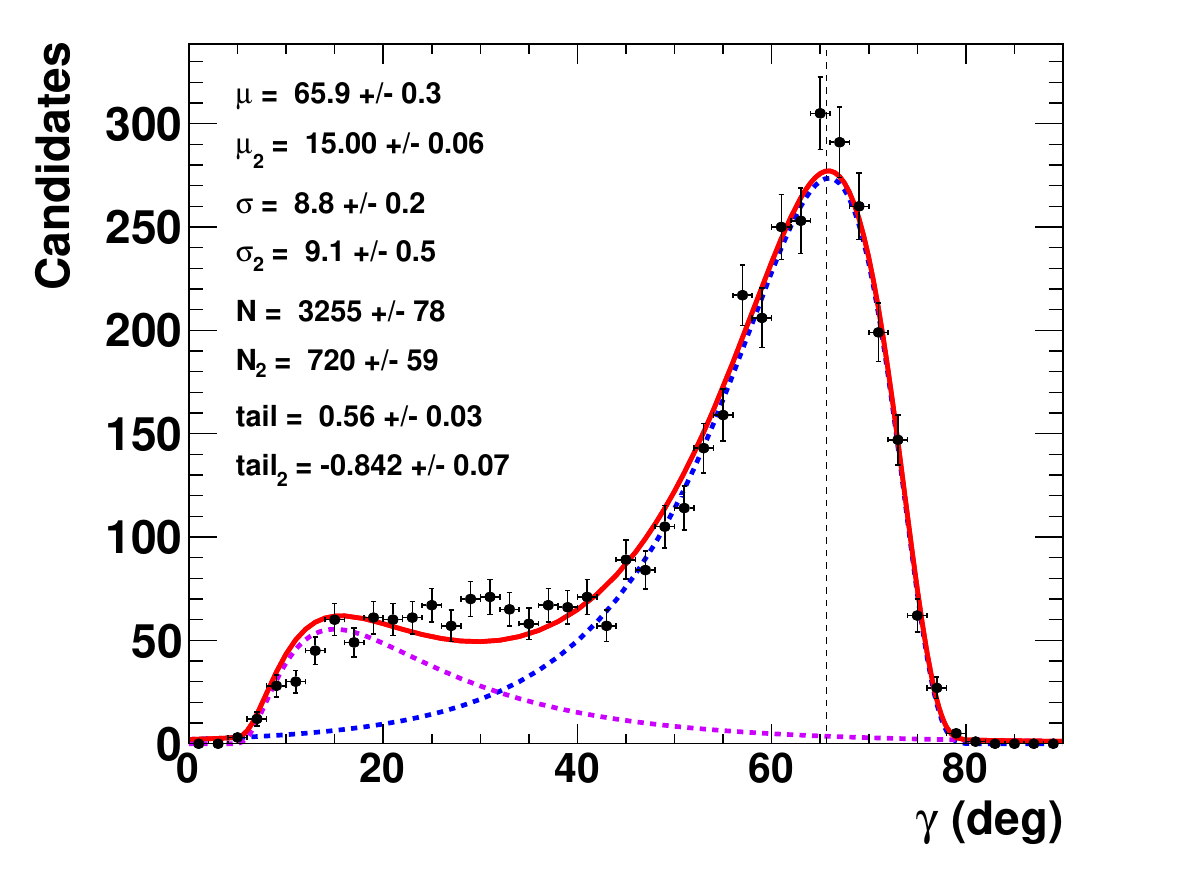}
\includegraphics[width=0.425\textwidth,height=0.2\textheight]{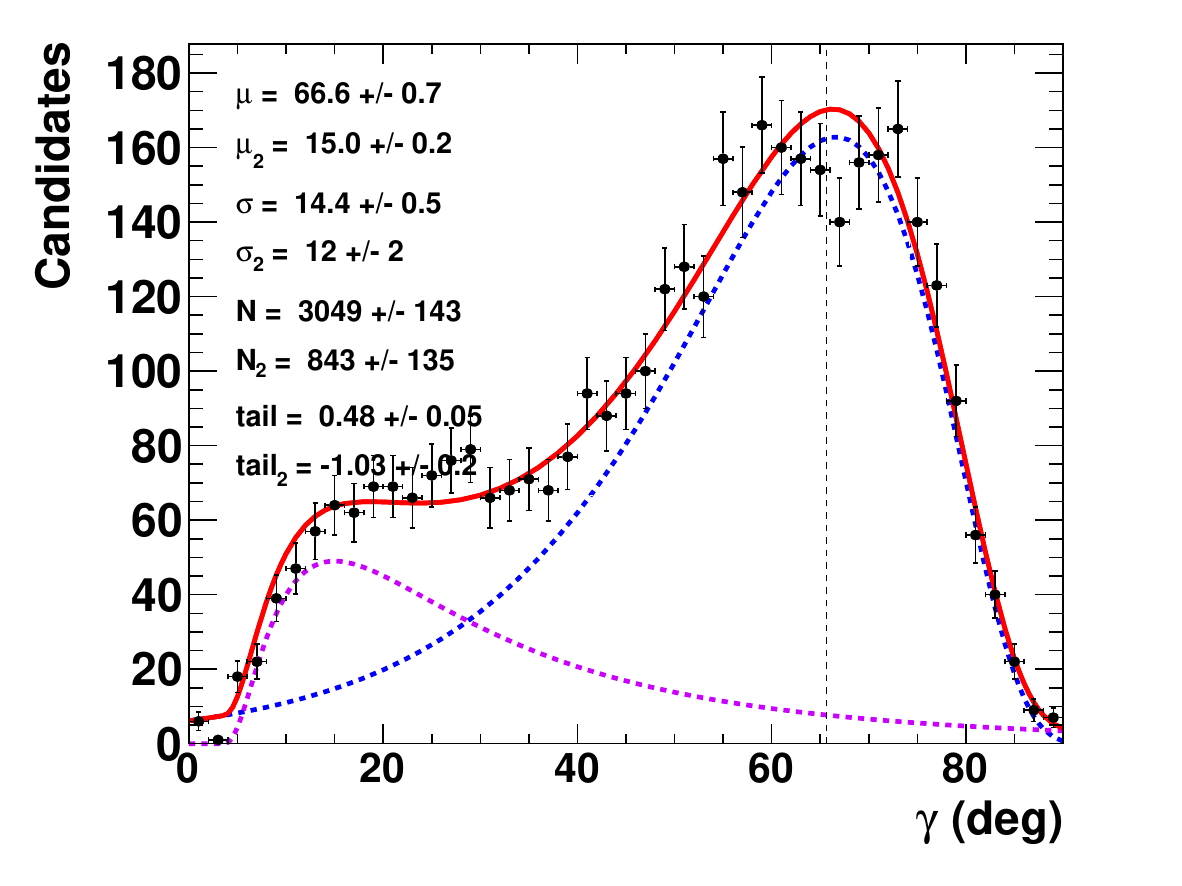} \\
\caption{\label{fig:Fit_gamma}  Fit to the distributions of the  nuisance parameters  $\gamma$ obtained from 4000 pseudoexperiments. The initial configuration is $\gamma=65.66^{\circ}$ (1.146 rad), $\rBst=0.4$ (left) and 0.22 (right), $\deltaB=171.9^\circ$ (3 rad), and $\deltastB=114.6^\circ$ (2 rad). In the distributions, only the candidates with a value  $\gamma\  \in \ [0^\circ, \  90^\circ]$ are considered. The purple dashed curve accounts for tails generated by the correlations with the nuisance parameters $\rBst$ and $\deltaBst$, while the blue dashed curve is the core part of the distribution, the plain red line is the sum of the two components of the fit.}
\includegraphics[width=0.425\textwidth]{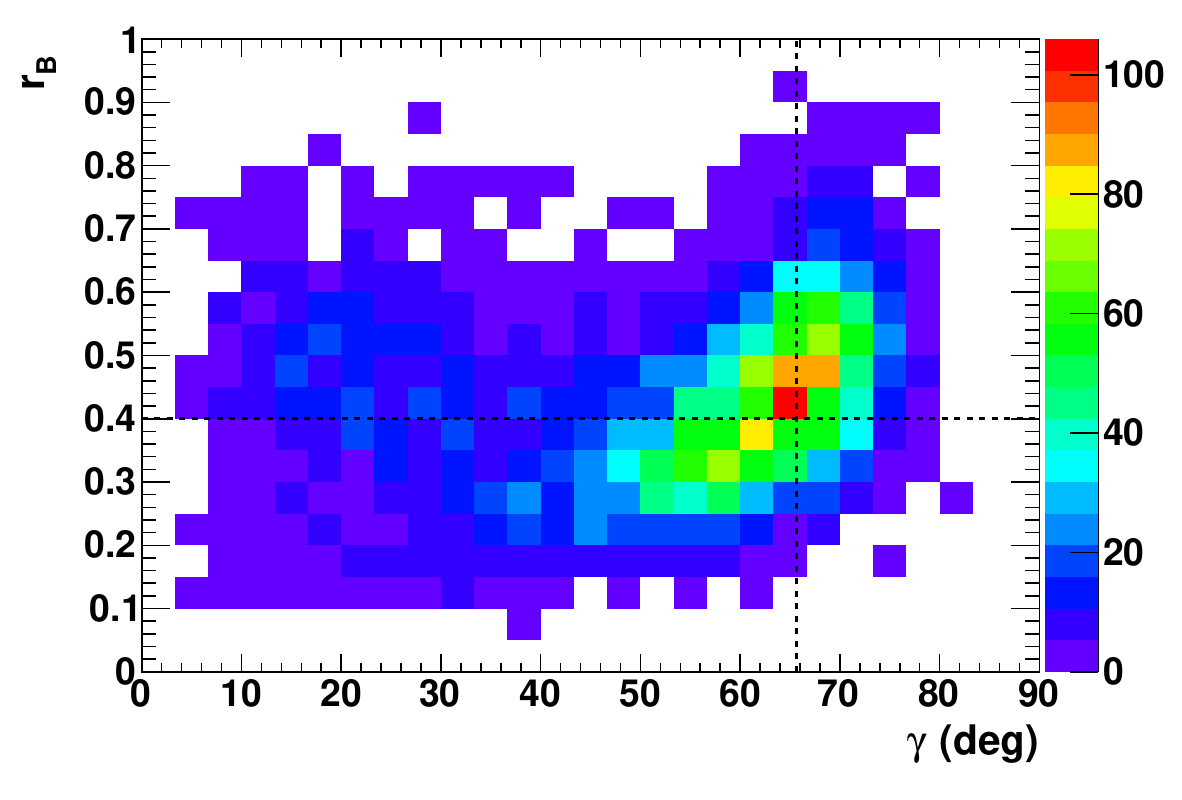}
\includegraphics[width=0.425\textwidth]{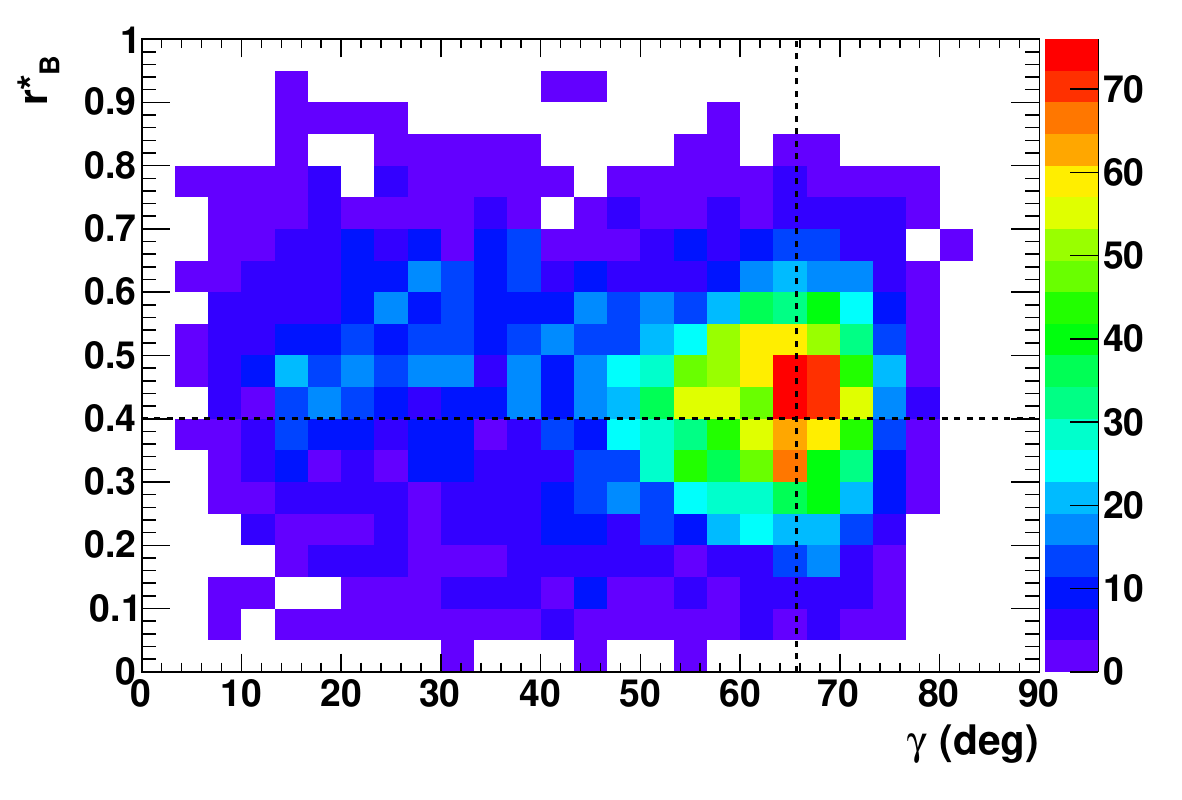} \\
\includegraphics[width=0.425\textwidth]{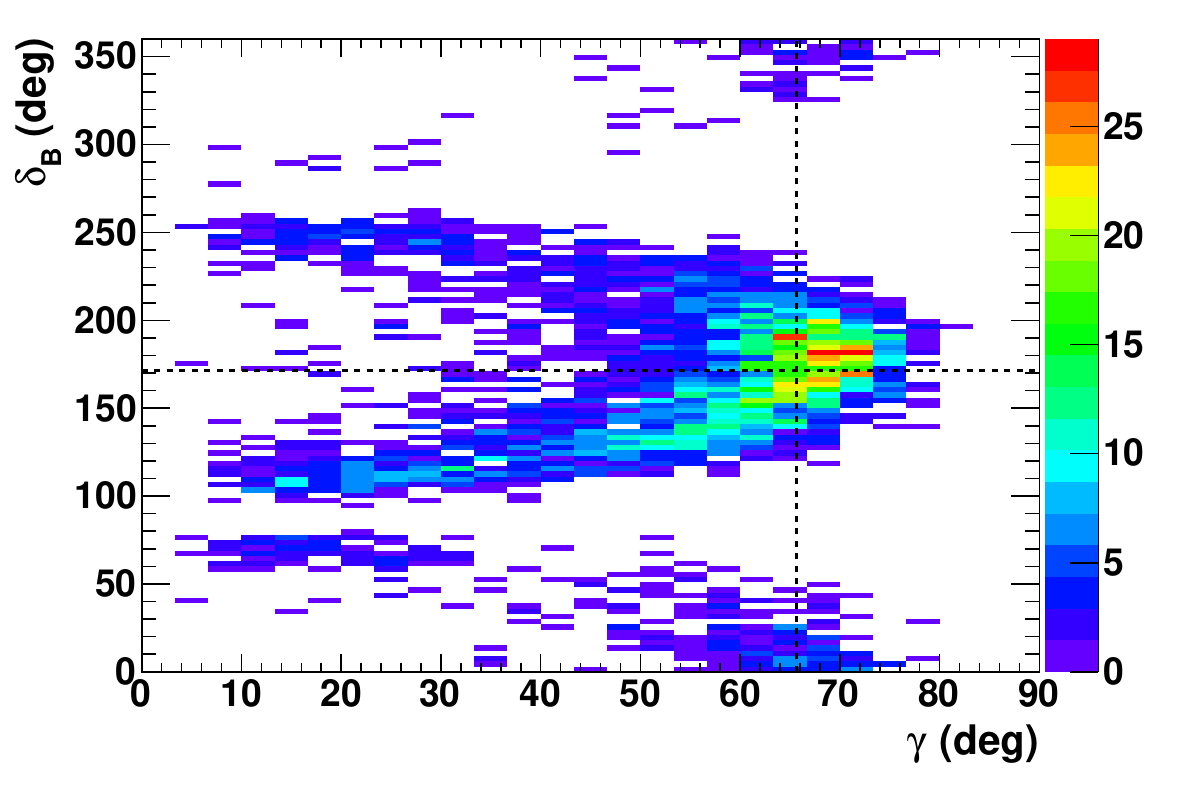}
\includegraphics[width=0.425\textwidth]{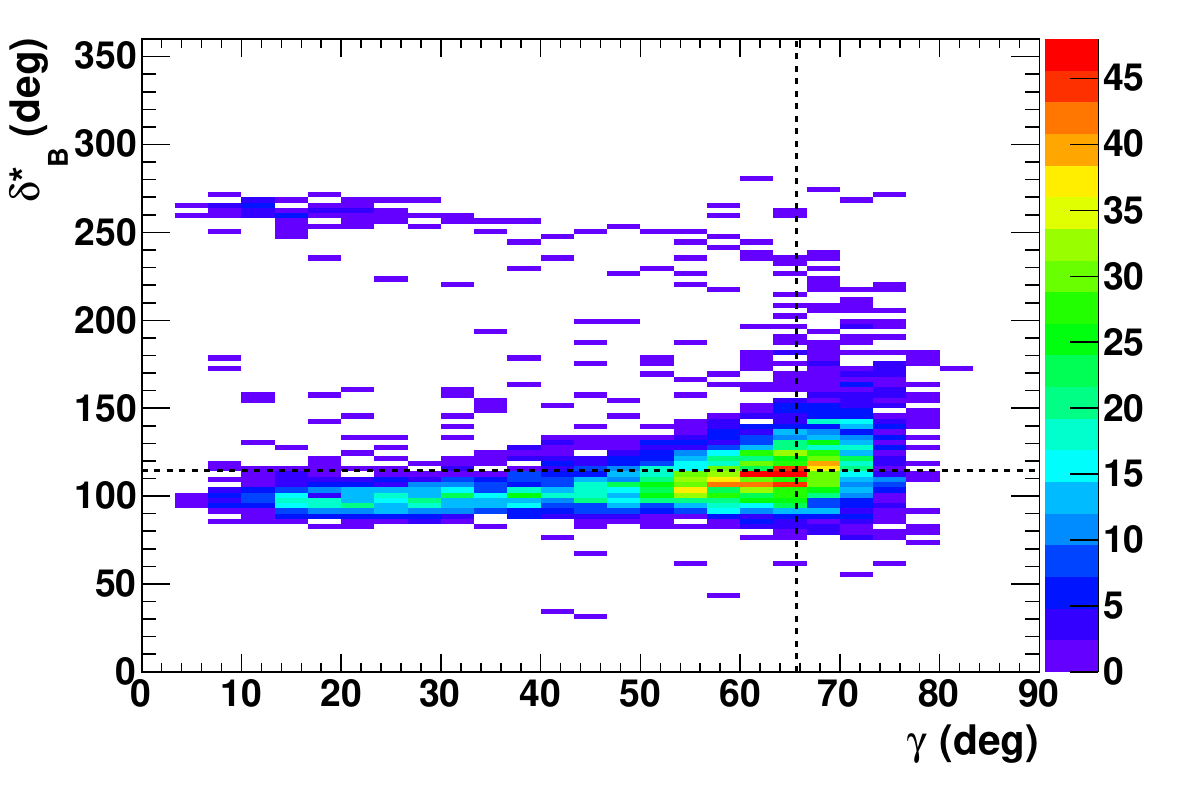}
\caption{\label{fig:2Dtoy} Two-dimension distributions of the nuisance parameters $\rBst$ and $\deltaBst$ as a function of $\gamma$ obtained from 4000 pseudoexperiments. On each figure the horizontal dashed black  lines indicate the initial true values: $\gamma=65.66^\circ$ (1.146 rad), $\deltaB=171.9^\circ$ (3.0 rad), and $\deltastB=114.6^\circ$ (2.0 rad), and  $\rBst=0.4$.}
\end{figure}

Configurations where  $\gamma = 74^{\circ}$ (see Ref.~\cite{LHCb-CONF-2018-002}) have also been tested. The potential problem in that case is that, as the true value of $\gamma$ is closer to the $90^\circ$ boundary, the unfolding of this parameter may become more difficult for many configurations of the nuisance parameters $\rBst$  and $\deltaBst$. It is clear from Eq.~\ref{EQ__14_} that the sensitivity to $\gamma$ is null at $90^\circ$. This is illustrated in Fig.~\ref{fig:74deg_dB3dBst2}, in appendix~\ref{sec:appendB},  and can be compared with Fig.~\ref{fig:Fit_gamma}. In this case, the initial tested configuration is $\gamma=74^{\circ}$, $\rBst=0.4$ and 0.22, $\deltaB=171.9^\circ$ (3 rad), and $\deltastB=114.6^\circ$ (2 rad). For those configurations:  $\mu_{\gamma}=\left( 73.7 \pm 0.3 \right)^\circ$ ($\left( 74.2 \pm 0.7)^\circ\right)$ and $\sigma_{\gamma}=\left( 7.7 \pm 0.3 \right)^\circ$ ($\left( 14.7 \pm 0.6)^\circ \right)$, for $\rBst=0.4$ (0.22). There is limited degradation of the resolution compared to the corresponding configuration, when the true value of $\gamma$ is $65.66^\circ$. In Fig.~\ref{fig:74deg_dB1dBst5}, in appendix~\ref{sec:appendB},  the fit to $\gamma$ for the pseudoexperiments corresponding to the configuration:  $\gamma=74^{\circ}$, $\rBst=0.4$ and 0.22, $\deltaB=57.3^\circ$ (1 rad), and $\deltastB=286.5^\circ$ (5 rad), is presented. For $\rBst=0.22$, on can clearly see that the fitted $\gamma$ value is approaching the boundary limit $90^\circ$ and the corresponding resolution is about $18^\circ$. Such a behavior can clearly be understood from the 2-D distribution shown in Fig.~\ref{fig:2D_3}.  This is comparable to the case listed in Tables~\ref{tab:dbeffect_rBst04_022_mean} and~\ref{tab:dbeffect_rBst04_022_resol}, when $\deltaBst=286.5^\circ$ ({\it i.e.} near $3\pi/2$ rad) and $\rBst=0.22$.

\clearpage
\begin{table}[t]
\centering
\caption{\label{tab:dbeffect_rBst04_022_mean} Fitted mean value of $\gamma$ ($\mu_{\gamma}$) (in [deg]), for $\rBst=0.4$ (top) and $0.22$ (bottom), as a function of $\deltaBst$, for an initial true value of  $65.66^\circ$.}
\begin{tabular}{c|cccccc}
\hline \hline
\diagbox{$\deltaB$}{$\deltastB$}  & 0 & 57.3 & 114.6 & 171.9 & 229.2 & 286.5  \\
\hline
$\rBst=0.4$ & & & & & & \\
\hline
0.0 & $66.2 \pm0.4$ & $65.9 \pm0.4$ & $67.4 \pm0.4$ & $65.9 \pm0.4$ & $65.9 \pm0.4$ & $67.4 \pm0.4$ \\
57.3 & $65.4 \pm0.4$ & $69.2 \pm0.4$ & $71.4 \pm0.4$ & $65.5 \pm0.3$ & $67.7 \pm0.4$ & $72.6 \pm0.4$ \\
114.6 & $65.8 \pm0.4$ & $70.5 \pm0.4$ & $72.9 \pm0.4$ & $65.9 \pm0.3$ & $69.3 \pm0.4$ & $73.8 \pm0.4$ \\
171.9 & $65.6_{-0.4}^{+0.5}$ & $65.2 \pm0.3$ & $65.9 \pm0.3$ & $65.5 \pm0.3$ & $65.2 \pm0.4$ & $66.1 \pm0.3$ \\
229.2 & $64.9 \pm0.4$ & $67.6 \pm0.4$ & $68.9 \pm0.4$ & $65.0 \pm0.4$ & $67.1 \pm0.4$ & $69.0 \pm0.4$ \\
286.5 & $66.1 \pm0.4$ & $71.0 \pm0.4$ & $75.7 \pm0.4$ & $65.7 \pm0.3$ & $69.8 \pm0.4$ & $78.8 \pm0.4$ \\
\hline
$\rBst=0.22$ & & & & & & \\
\hline
0.0 & $67.2_{-1.0}^{+0.9}$ & $68.1_{-0.9}^{+1.0}$ & $68.3_{-0.8}^{+0.9}$ & $67.1 \pm0.8$ & $68.4 \pm0.9$ & $69.0 \pm0.8$ \\
57.3  & $67.4 \pm0.9$ & $72.2 \pm0.8$ & $74.1_{-0.7}^{+0.8}$ & $66.6 \pm0.7$ & $71.5 \pm0.8$ & $75.5 \pm0.8$ \\
114.6 & $65.7 \pm0.9$ & $71.8 \pm0.6$ & $74.9 \pm0.6$ & $68.0 \pm0.6$ & $71.2 \pm0.7$ & $74.9 \pm0.6$ \\
117.9 & $65.4 \pm0.7$ & $66.9 \pm0.7$ & $66.6 \pm0.7$ & $64.7 \pm0.7$ & $65.2 \pm0.6$ & $68.3 \pm0.6$ \\
229.2 & $65.9_{-1.0}^{+0.9}$ & $69.1 \pm0.8$ & $70.1_{-0.8}^{+0.7}$ & $67.7 \pm0.7$ & $67.4 \pm0.7$ & $71.0_{-0.7}^{+0.6}$ \\
286.5 & $67.5 \pm0.9$ & $75.8 \pm0.8$ & $77.5_{-0.7}^{+0.8}$ & $68.1 \pm0.6$ & $72.8_{-0.8}^{+0.9}$ & $83.5_{-1.5}^{+2.4}$ \\
\end{tabular}
\end{table}

\begin{table}[b]
\centering
\caption{\label{tab:dbeffect_rBst04_022_resol} Fitted resolution of $\gamma$ ($\sigma_{\gamma}$) (in [deg]), for $\rBst=0.4$ (top) and $0.22$ (bottom), as a function of $\deltaBst$.  }
\begin{tabular}{c|rrrrrr}

\hline   \hline
\diagbox{$\deltaB$}{$\deltastB$}  & 0 & 57.3 & 114.6 & 171.9 & 229.2 & 286.5  \\
\hline
$\rBst=0.4$ & & & & & & \\
\hline
0.0 & $9.6 \pm0.4$ & $11.2 \pm0.3$ & $11.2 \pm0.3$ & $9.4 \pm0.4$ & $12.2 \pm0.4$ & $11.2 \pm0.3$ \\
57.3 & $11.2 \pm0.3$ & $12.6 \pm0.3$ & $11.8 \pm0.3$ & $10.2 \pm0.3$ & $12.4 \pm0.3$ & $12.9 \pm0.3$ \\
114.6 & $11.4 \pm0.3$ & $11.9 \pm0.3$ & $11.1 \pm0.3$ & $10.0 \pm0.3$ & $11.3 \pm0.3$ & $11.9 \pm0.3$ \\
171.9 & $8.3 \pm0.4$ & $9.8 \pm0.3$ & $8.8 \pm0.2$ & $8.9 \pm0.3$ & $9.5 \pm0.3$ & $9.0 \pm0.2$ \\
229.2 & $10.8 \pm0.3$ & $11.7 \pm0.3$ & $10.8 \pm0.3$ & $10.3 \pm0.3$ & $12.4 \pm0.3$ & $11.7 \pm0.3$ \\
286.5 & $11.0 \pm0.3$ & $12.9 \pm0.3$ & $11.6 \pm0.3$ & $9.2 \pm0.2$ & $11.7 \pm0.3$ & $13.2_{-0.5}^{+0.6}$ \\
\hline
$\rBst=0.22$ & & & & & & \\
\hline
0.0 & $16.5 \pm0.7$ & $16.8 \pm0.7$ & $16.0 \pm0.6$ & $16.8_{-0.7}^{+0.8}$ & $15.9_{-0.6}^{+0.0}$ & $16.0 \pm0.7$ \\
57.3 & $16.7 \pm0.6$ & $18.1_{-0.8}^{+0.9}$ & $17.1_{-0.9}^{+1.0}$ & $14.3 \pm0.5$ & $16.8 \pm0.7$ & $17.6_{-1.1}^{+1.3}$ \\
114.6  & $16.1_{-0.7}^{+0.6}$ & $15.9 \pm0.6$ & $13.9 \pm0.5$ & $14.1 \pm0.5$ & $15.1 \pm0.6$ & $14.9 \pm0.6$ \\
171.9 & $15.7_{-0.6}^{+0.7}$ & $14.5 \pm0.5$ & $14.4 \pm0.5$ & $15.5 \pm0.6$ & $15.7_{-0.5}^{+0.0}$ & $14.0 \pm0.5$ \\
229.2 & $15.9 \pm0.6$ & $15.7_{-0.6}^{+0.5}$ & $15.4 \pm0.6$ & $14.6_{-0.5}^{+0.6}$ & $15.6 \pm0.5$ & $14.4 \pm0.6$ \\
286.5 & $16.9_{-0.6}^{+0.7}$ & $18.0_{-1.1}^{+1.3}$ & $16.7_{-1.0}^{+1.2}$ & $14.9_{-0.6}^{+0.5}$ & $16.1 \pm0.7$ & $18.7_{-2.1}^{+2.8}$ \\
\end{tabular}
\end{table}

\clearpage

\clearpage


\subsection{Effect of using or not the $\Bs\rightarrow \tilde{D}^{(*)0}(\pi\pi)\phi$ and $\Bs\rightarrow \tilde{D}^{(*)0}(K\pi\piz)\phi$  decays}
\label{sec:effectOfDpipi_Kpipi0}

As listed in Table~\ref{tab:yields} the expected yields for the $D$-meson decays to $\pi\pi$ and $K\pi\piz$ are somewhat lower than for the other modes, down to few tens of events. Again those yields have been computed from LHCb studies on $B^\pm \rightarrow \tilde{D}^{0}(\pi/K)^\pm$ reported in Refs.~\cite{Aaij:2016oso} and~\cite{Aaij:2015jna} and normalised to Ref.~\cite{Aaij:2018jqv}, with respect to the mode $\Bs\rightarrow \tilde{D}^{(*)0}(K\pi)\phi$. Therefore the selections are not necessarily against the signals $\Bs\rightarrow \tilde{D}^{(*)0}(\pi\pi)\phi$ and $\Bs\rightarrow \tilde{D}^{(*)0}(K\pi\piz)\phi$ and the expected yields may be underestimated as well as all the sub-decays listed in Table~\ref{tab:yields}. It should also be noticed that the mode $\pi\pi$ is a $CP$-eigenstate, while the $K\pi\piz$ 3-body decay has also a large coherence factor value $R^{K\pi\piz}_D=(81\pm6)\%$~\cite{Evans:2016tlp}.   Nevertheless the effect of using or not the $\Bs\rightarrow \tilde{D}^{(*)0}(\pi\pi)\phi$ and $\Bs\rightarrow \tilde{D}^{(*)0}(K\pi\piz)\phi$  decays has been studied and is reported here, while in Sec.~\ref{sec:effectOfDstarphi} the effect of including or not the decays $\Bs \rightarrow \Dtstz\phi$  is discussed also for future more abundant datasets.

According to Fig.~\ref{fig:1-2_nopipi}, in appendix~\ref{sec:appendC}, there is a relative loss on precision to the unfolded value of $\gamma$ of about $3$ to $15\%$, when the  $\Bs\rightarrow \tilde{D}^{(*)0}(\pi\pi)\phi$ decays are not used. Figure~\ref{fig:1-2_noKpipi0}, in appendix~\ref{sec:appendC}, shows that a relative loss in precision of about $3$ to $22\%$ is seen, when the  $\Bs\rightarrow \tilde{D}^{(*)0}(K\pi\piz)\phi$ decays are not used.

\section{Prospective on the sensitivity to $\gamma$ for Run $1-3$ and for the full High-Luminosity LHC (HL-LHC) LHCb datasets}
\label{sec:HL-LHC}

The prospective on the sensitivity to the CKM angle $\gamma$ with $\BsDtphi$ decays have also been studied for the foreseen LHCb integrated luminosities at the end of the LHC Run 3 and for the possible full HL-LHC future LHCb program. According to Ref.~\cite{lhcbupgrade}, the LHCb trigger efficiency will be improved by a factor of 2, at the beginning of LHC Run  3. The full expected LHCb dataset of $pp$ collisions at $\sqrt{s}=13$ TeV, corresponding to the sum of  Run 1, 2, and 3 LHCb dataset should be equal to 23\invfb by 2025, while, it is expected to be 300\invfb by the second half of the 2030 decade. The final integrated LHCb luminosity accounts for a LHCb detector upgrade phase II. In the following, the projected event yields as listed in Table~\ref{tab:yields}, after 2025 and after 2038 have been scaled by a factor $F_{later}=6.3$ and $90$, respectively and with uncertainties on observables as $1/\sqrt{F_{later}}$.

\subsection{Projected precision on $\gamma$ determination with $\BsDtphi$ decays}
\label{sec:projectedPrecision}

For this prospective sensitivity study we have made the safe assumption that the precision on the strong parameters of $D$-meson decays to $K\pi$, $K3\pi$, $K\pi\piz$ listed in Table~\ref{tab:para} should be improved by a factor two at the end of the LHCb program (see the BES-III experiment prospectives~\cite{Asner:2008nq}). The procedure described for LHCb Run~1~\&~2 data in Sec.~\ref{sec:sen} has been repeated. The values of the normalisation factors $C_{K\pi}$, $C_{K\pi,D\piz}$, and $C_{K\pi,D\gamma}$ obtained for Run~1~\&~2 (see Sec.\ref{sec:yie}) have been scaled to their expected equivalent rate for Run 1 to 3 and full HL-LHC LHCb datasets. The statistical uncertainties of the computed observables (see Sect.~\ref{sec:p-values}) obtained for Run~1~\&~2 LHCb data have been scaled by the square root of a factor two times (trigger improvement) the relative increase of the anticipated collected $\Bs$-meson yield: 2.2 (8.8) for Run 1 to 3 (full HL-LHC) LHCb dataset. Then as for Run~1~\&~2 sensitivity studies, the same $2 \times 6 \times 6$ configurations of the $\rBst$, and $\deltaBst$ nuisance parameters have been tested ($\rBst=0.22$ or 0.4 and $\deltaBst=0$, 1, 2, 3, 4, 5 rad, and $\gamma=65.66^{\circ}$ (1.146 rad)).

\begin{figure}[h]
\centering
\includegraphics[width=0.425\textwidth,height=0.15\textheight]{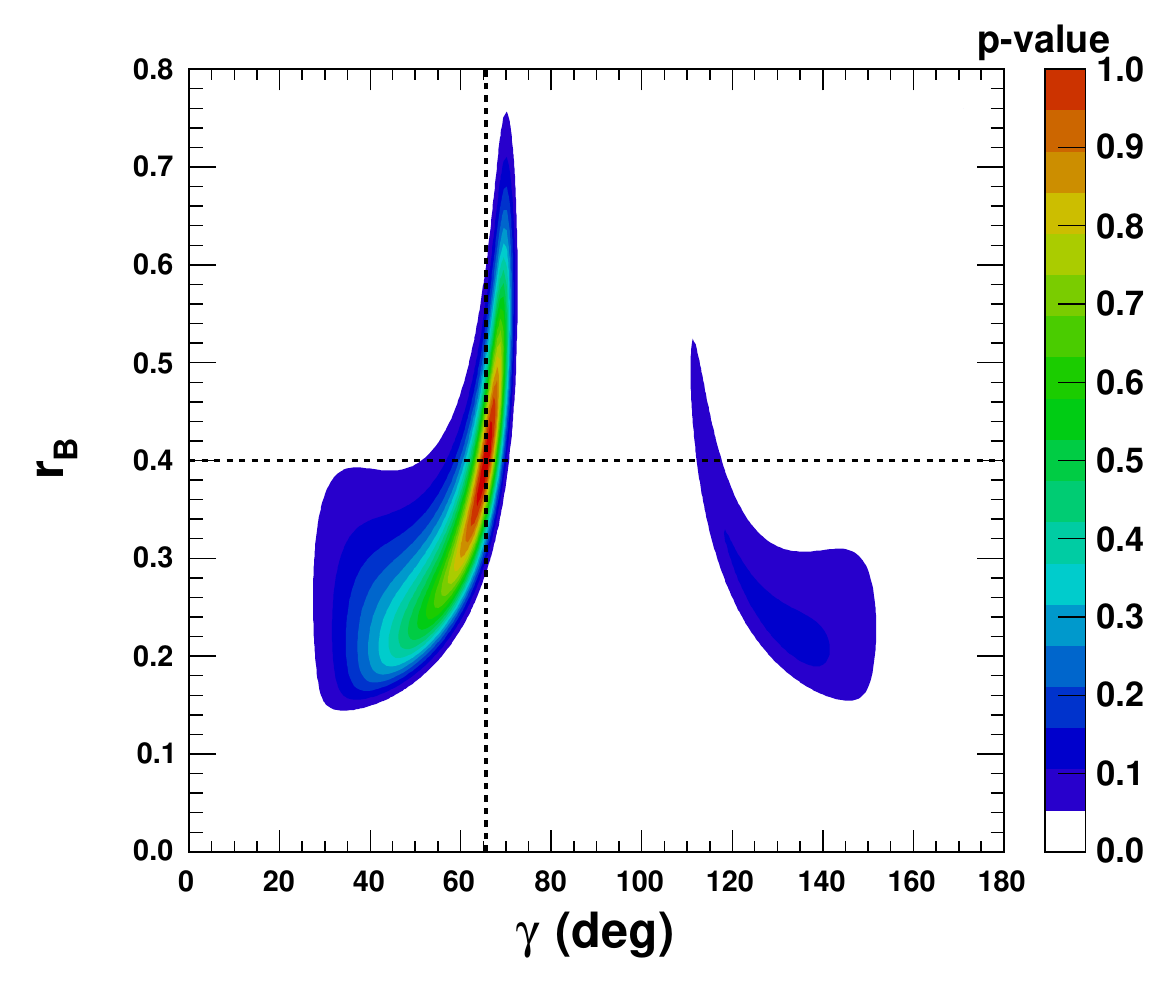}
\includegraphics[width=0.425\textwidth,height=0.15\textheight]{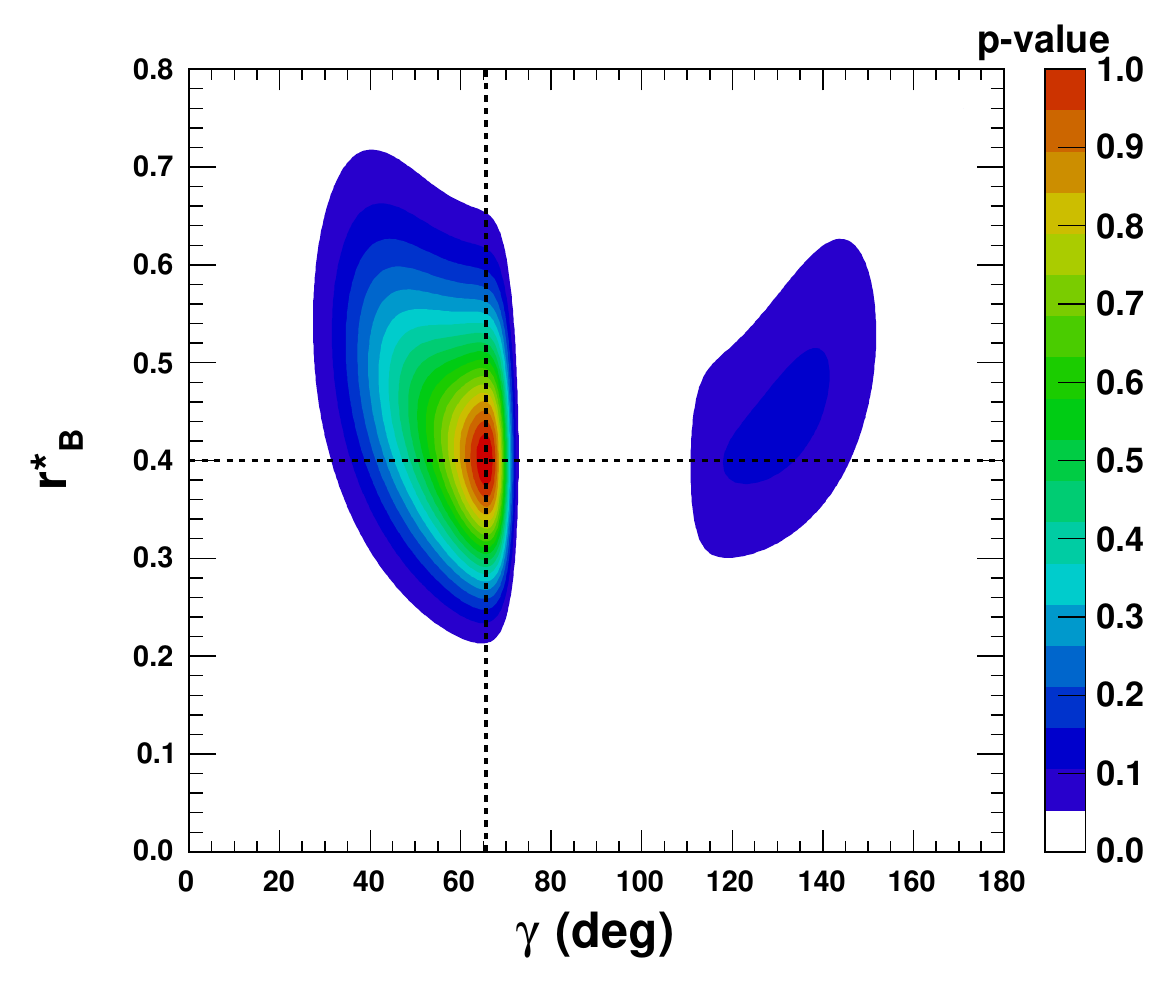} \\
\includegraphics[width=0.425\textwidth,height=0.15\textheight]{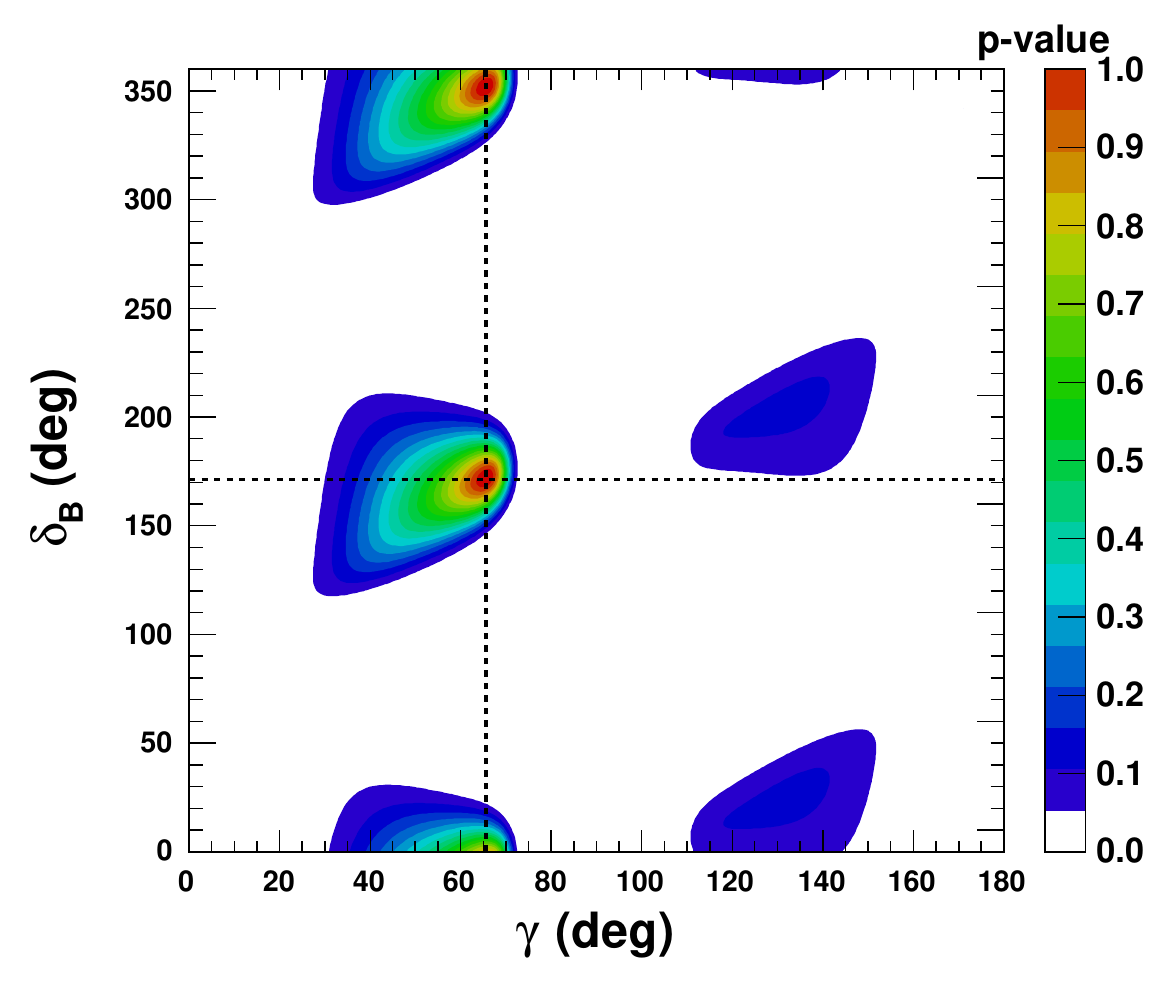}
\includegraphics[width=0.425\textwidth,height=0.15\textheight]{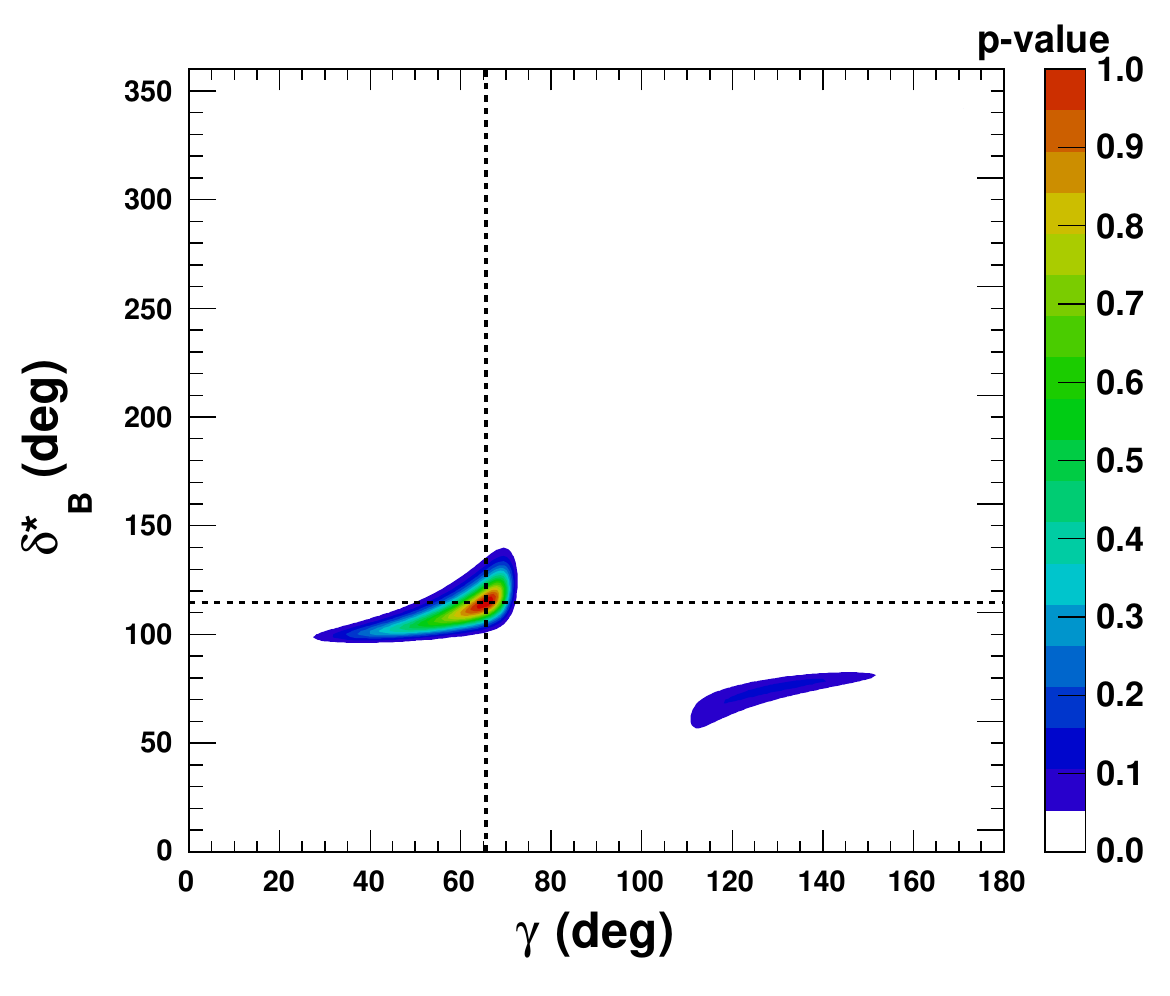}
\caption{\label{fig:2D-Run1-3_rBst04} Two-dimension $p$-value  profile of the nuisance parameters $\rBst$ and $\deltaBst$ (Run $1-3$),  as a function of $\gamma$. On each figure the dashed black  lines indicate the initial true values: $\gamma=65.66^\circ$ (1.146 rad), $\deltaB=171.9^\circ$ (3.0 rad), and $\deltastB=114.6^\circ$ (2.0 rad), and  $\rBst=0.4$.}
\centering
\includegraphics[width=0.425\textwidth,height=0.15\textheight]{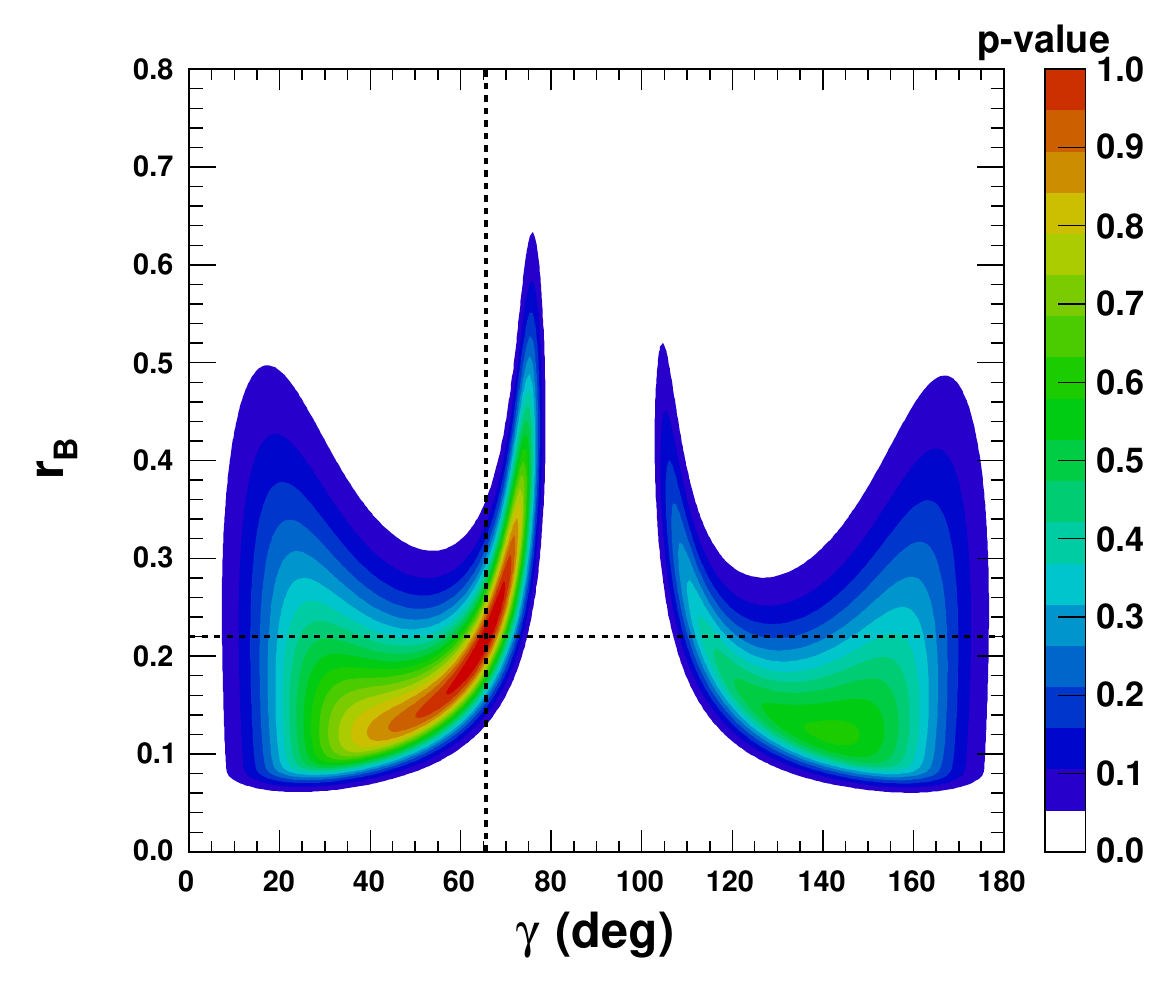}
\includegraphics[width=0.425\textwidth,height=0.15\textheight]{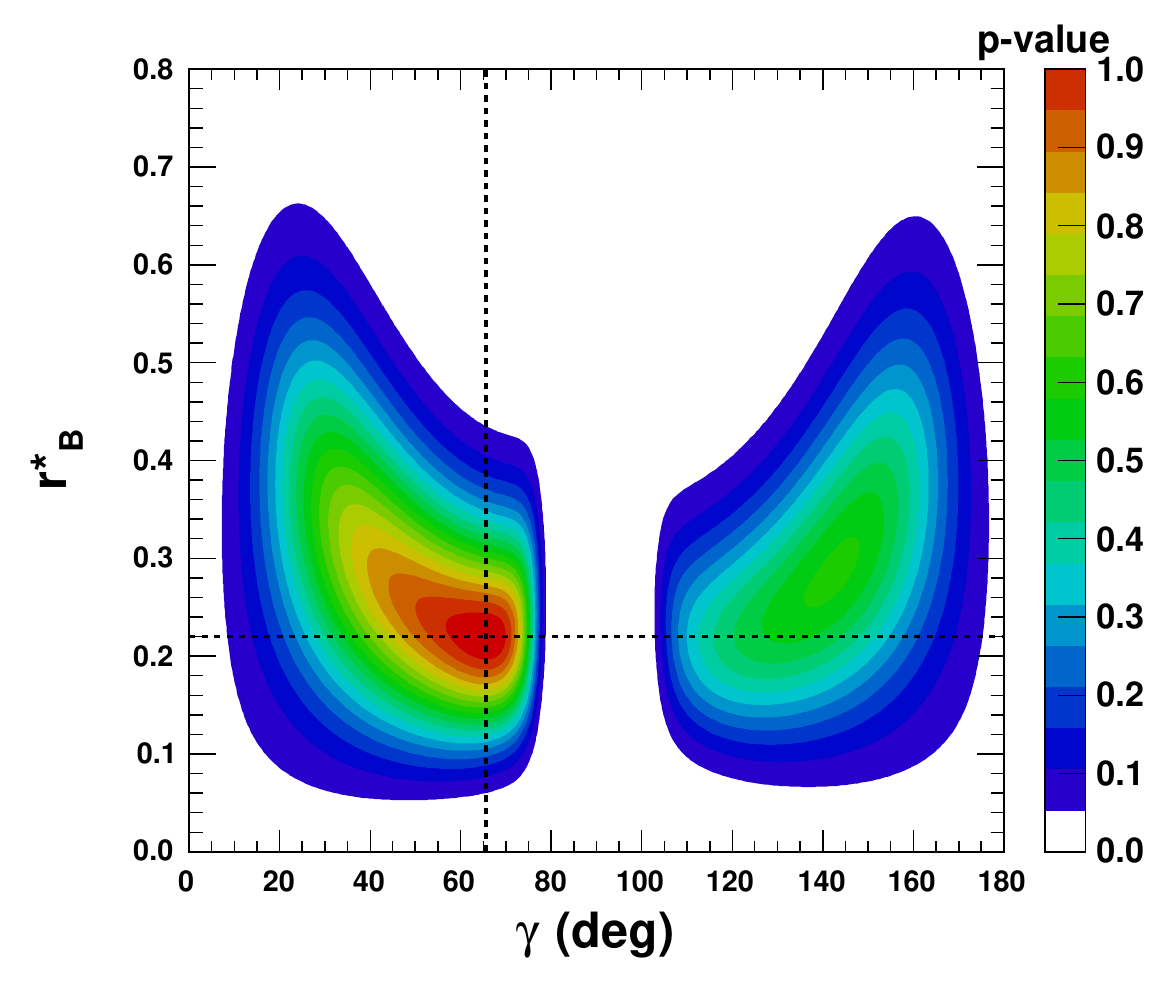} \\
\includegraphics[width=0.425\textwidth,height=0.15\textheight]{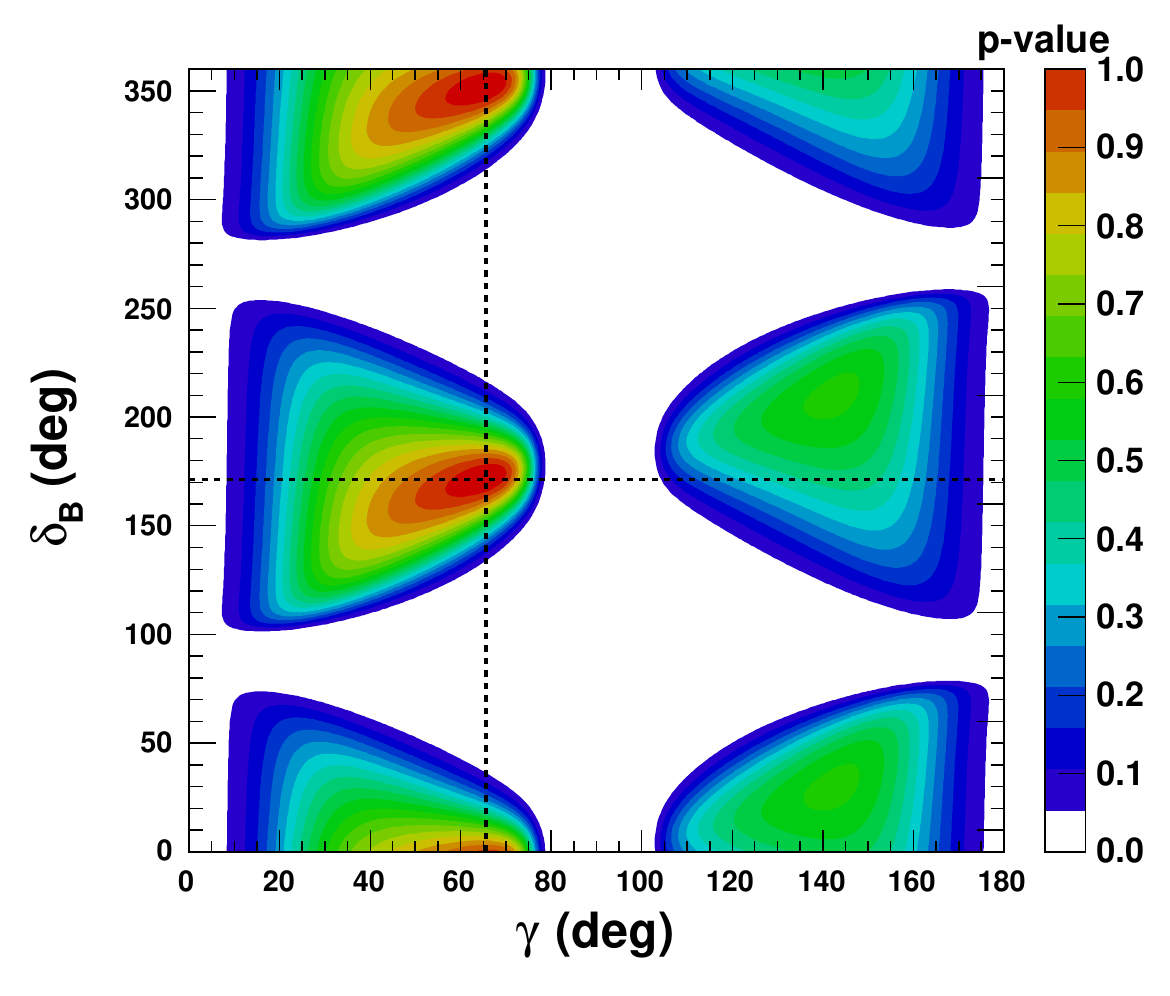}
\includegraphics[width=0.425\textwidth,height=0.15\textheight]{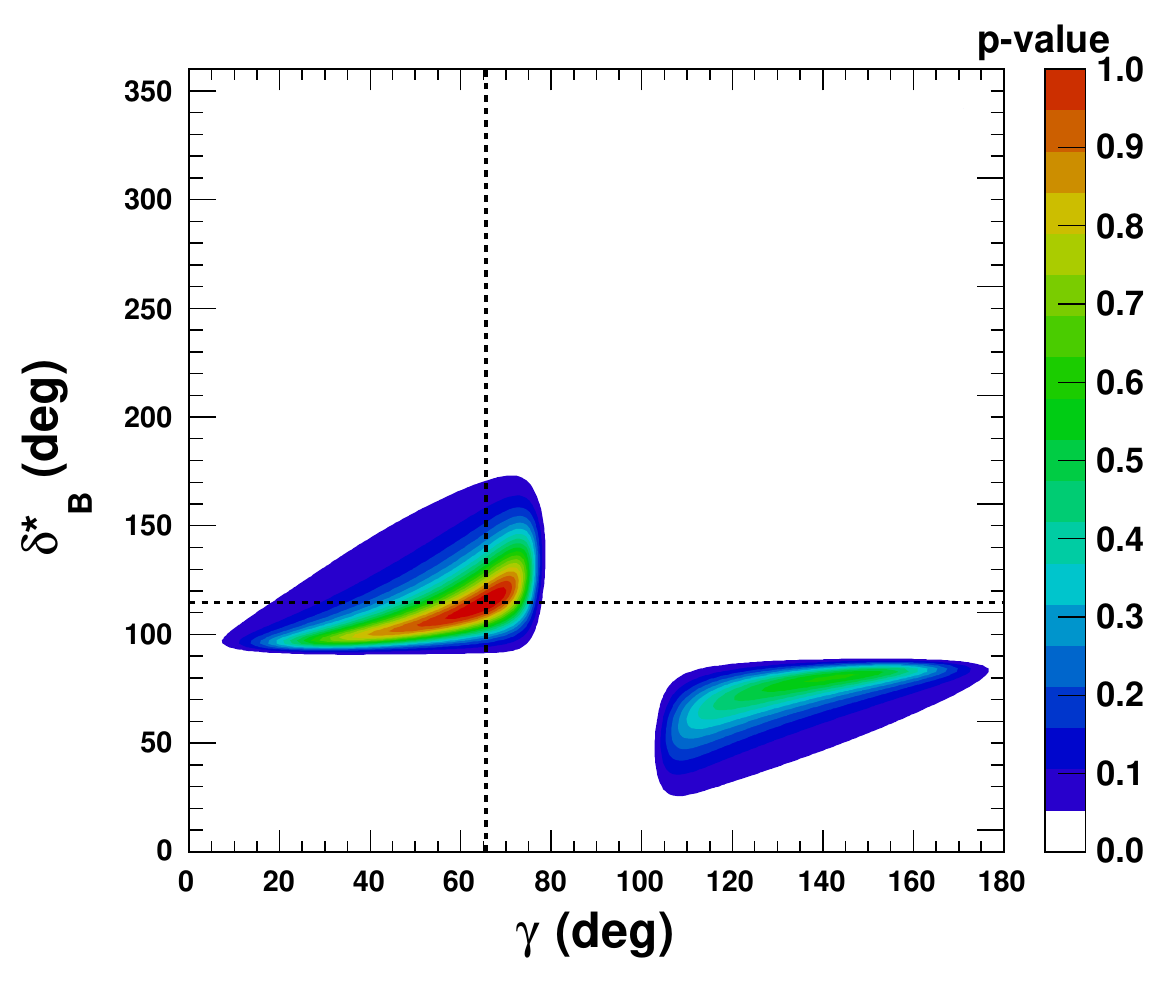}
\caption{\label{fig:2D-Run1-3_rBst022} Two-dimension $p$-value  profile of the nuisance parameters $\rBst$ and $\deltaBst$, for Run $1-3$ LHCb dataset,  as a function of $\gamma$. On each figure the dashed black  lines indicate the initial  true values: $\gamma=65.66^\circ$ (1.146 rad), $\deltaB=171.9^\circ$ (3.0 rad), and $\deltastB=114.6^\circ$ (2.0 rad), and  $\rBst=0.22$.}
\end{figure}

\begin{figure}[h]
\centering
\includegraphics[width=0.425\textwidth,height=0.15\textheight]{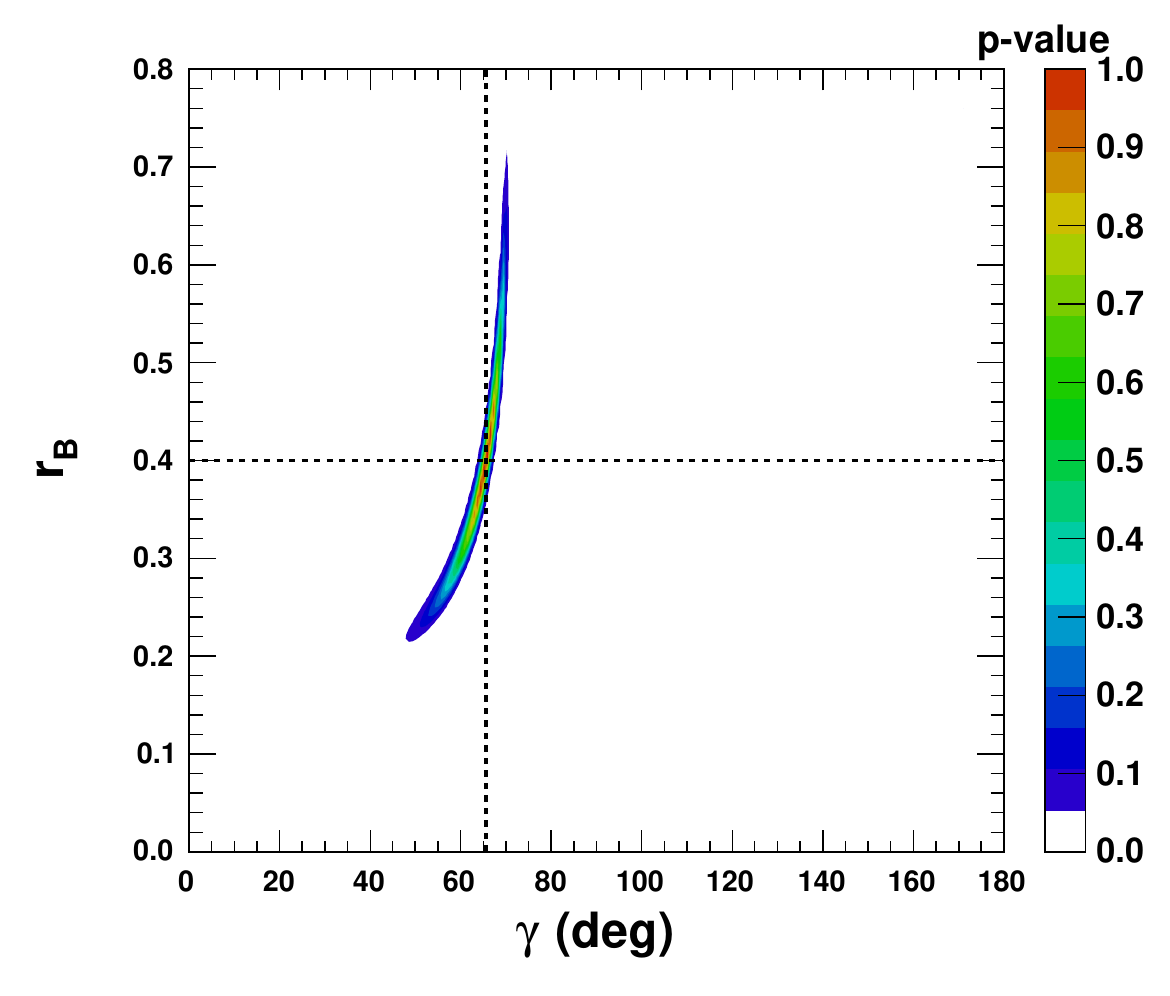}
\includegraphics[width=0.425\textwidth,height=0.15\textheight]{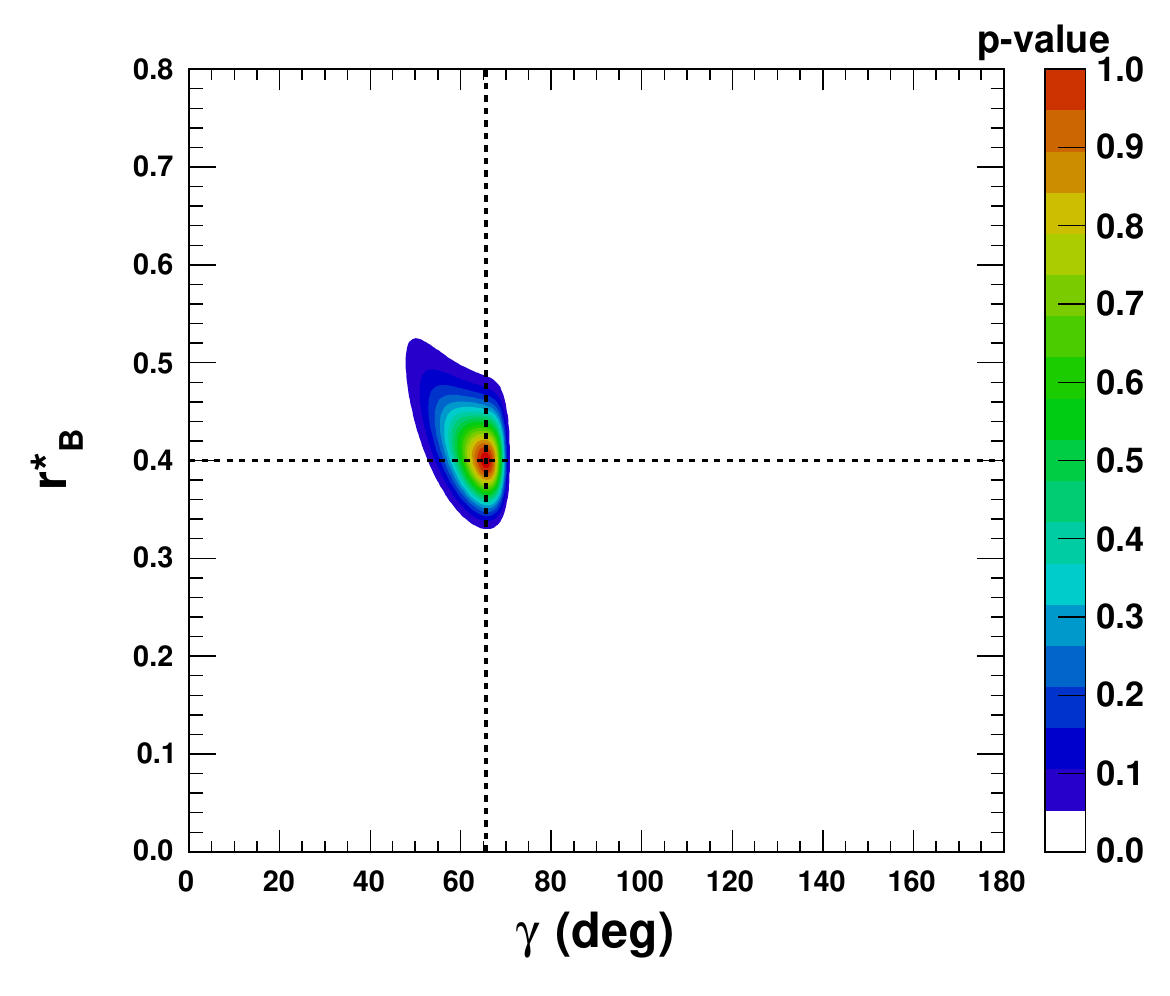} \\
\includegraphics[width=0.425\textwidth,height=0.15\textheight]{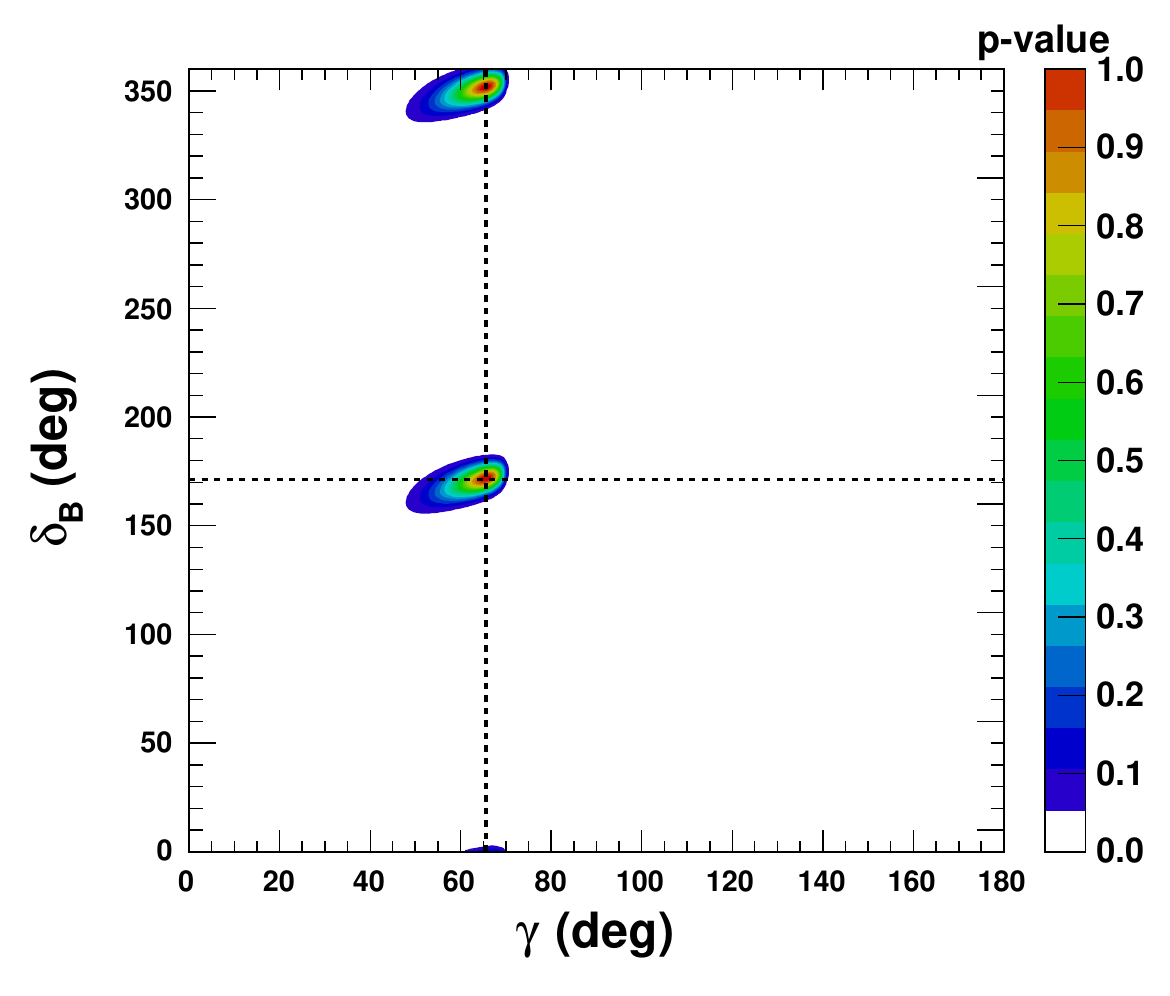}
\includegraphics[width=0.425\textwidth,height=0.15\textheight]{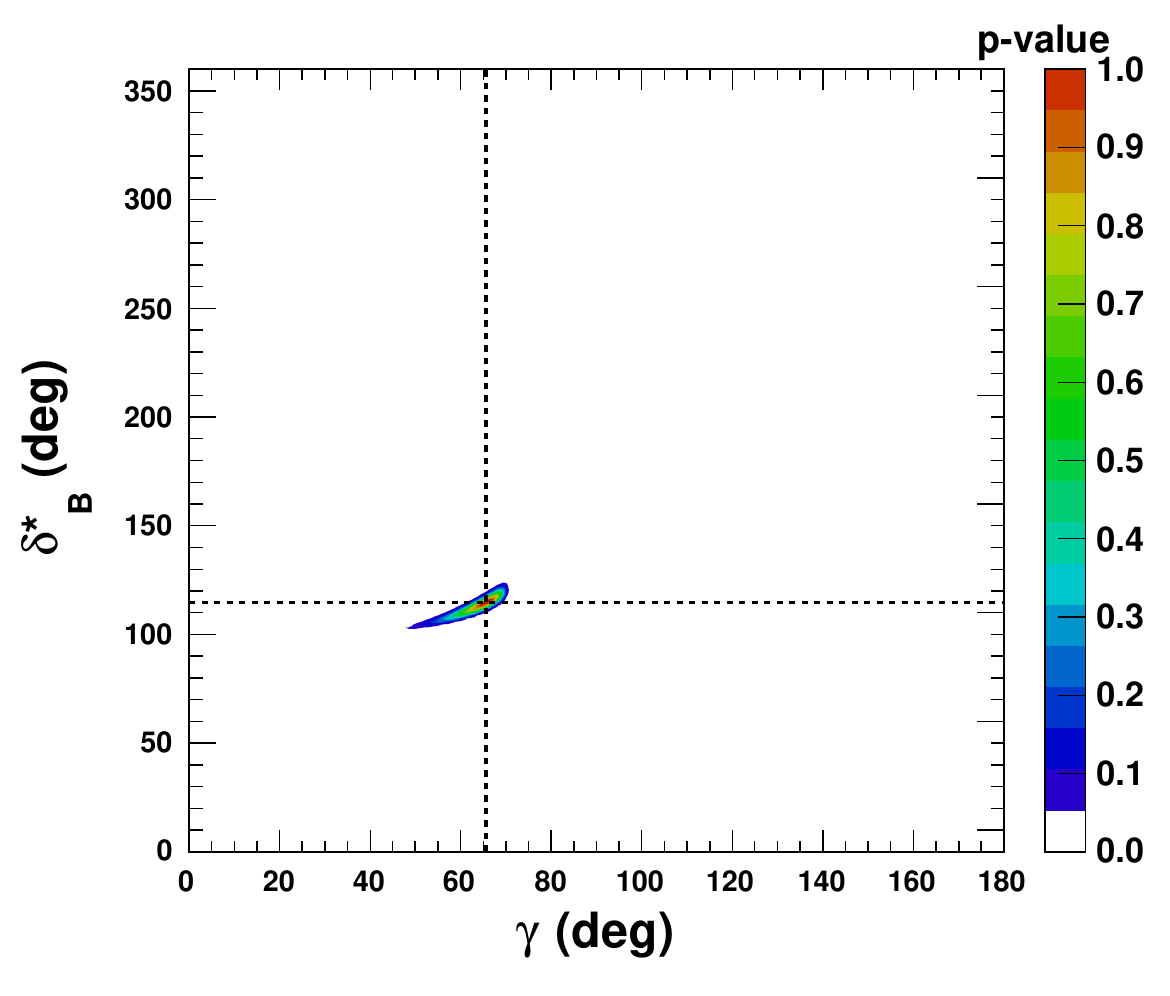}
\caption{\label{fig:2D-RunHL-LHC_rBst04} Two-dimension $p$-value  profile of the nuisance parameters $\rBst$ and $\deltaBst$, for the full HL-LHC LHCb dataset,  as a function of $\gamma$. On each figure the dashed black  lines indicate the initial  true values: $\gamma=65.66^\circ$ (1.146 rad), $\deltaB=171.9^\circ$ (3.0 rad), and $\deltastB=114.6^\circ$ (2.0 rad), and  $\rBst=0.4$.}
\centering
\includegraphics[width=0.425\textwidth,height=0.15\textheight]{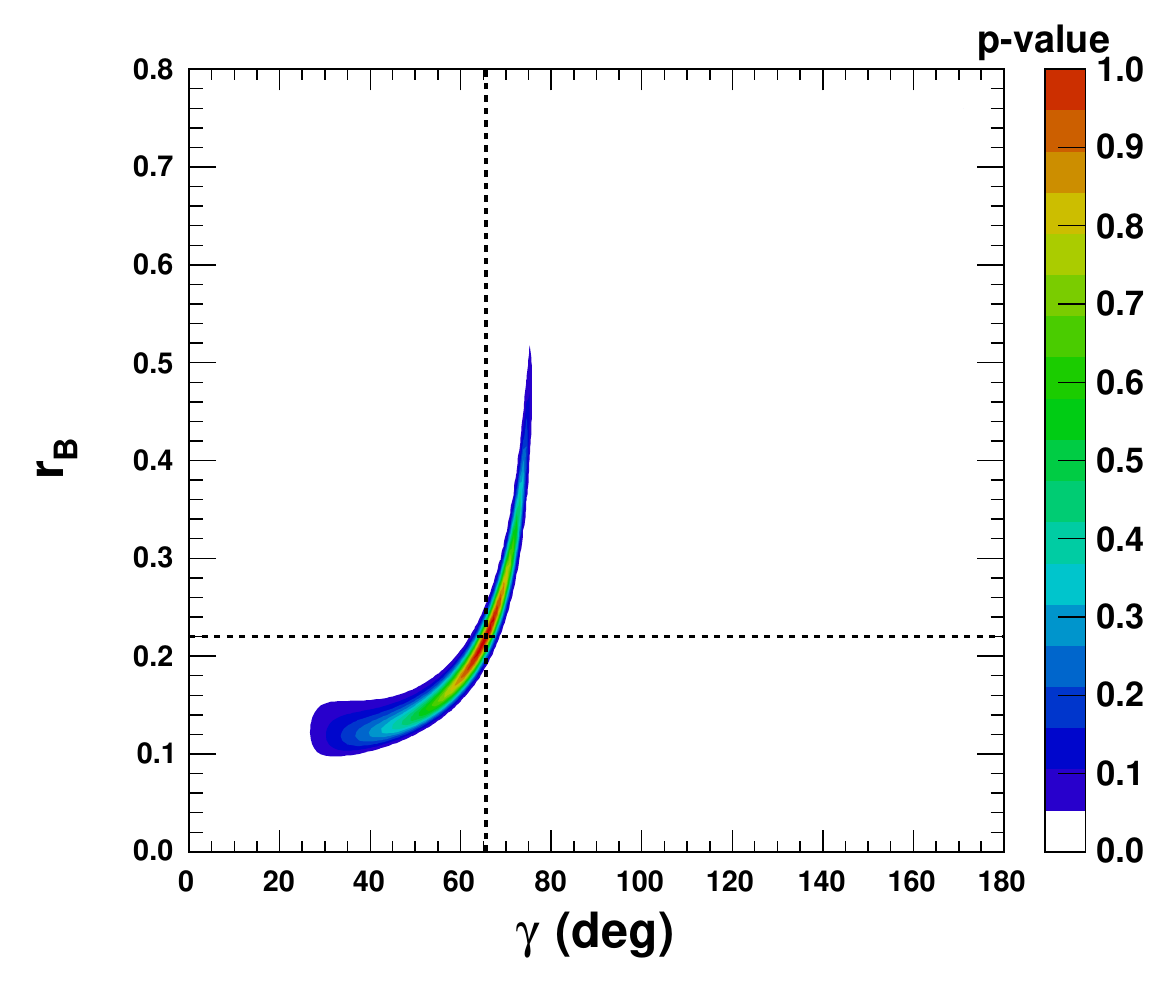}
\includegraphics[width=0.425\textwidth,height=0.15\textheight]{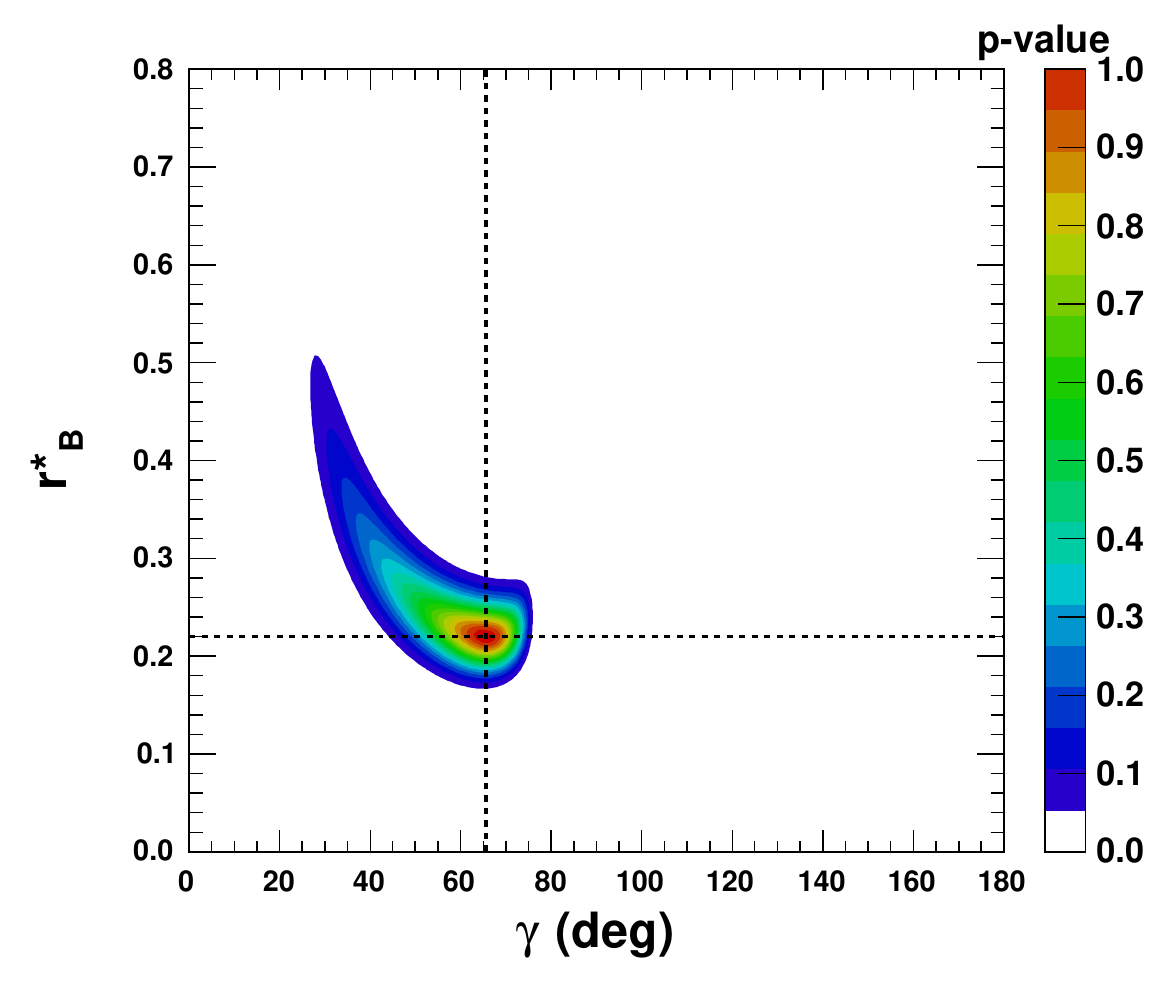} \\
\includegraphics[width=0.425\textwidth,height=0.15\textheight]{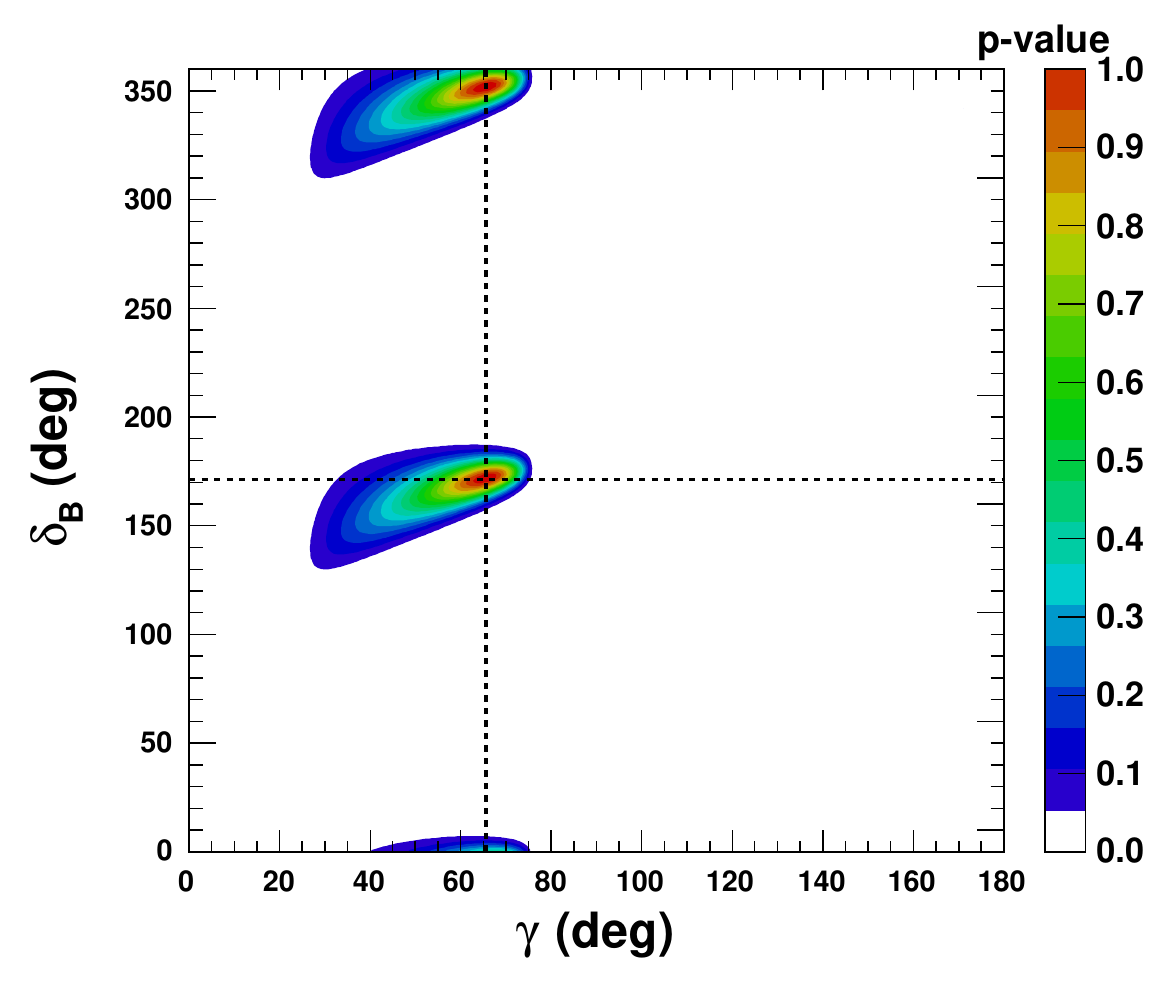}
\includegraphics[width=0.425\textwidth,height=0.15\textheight]{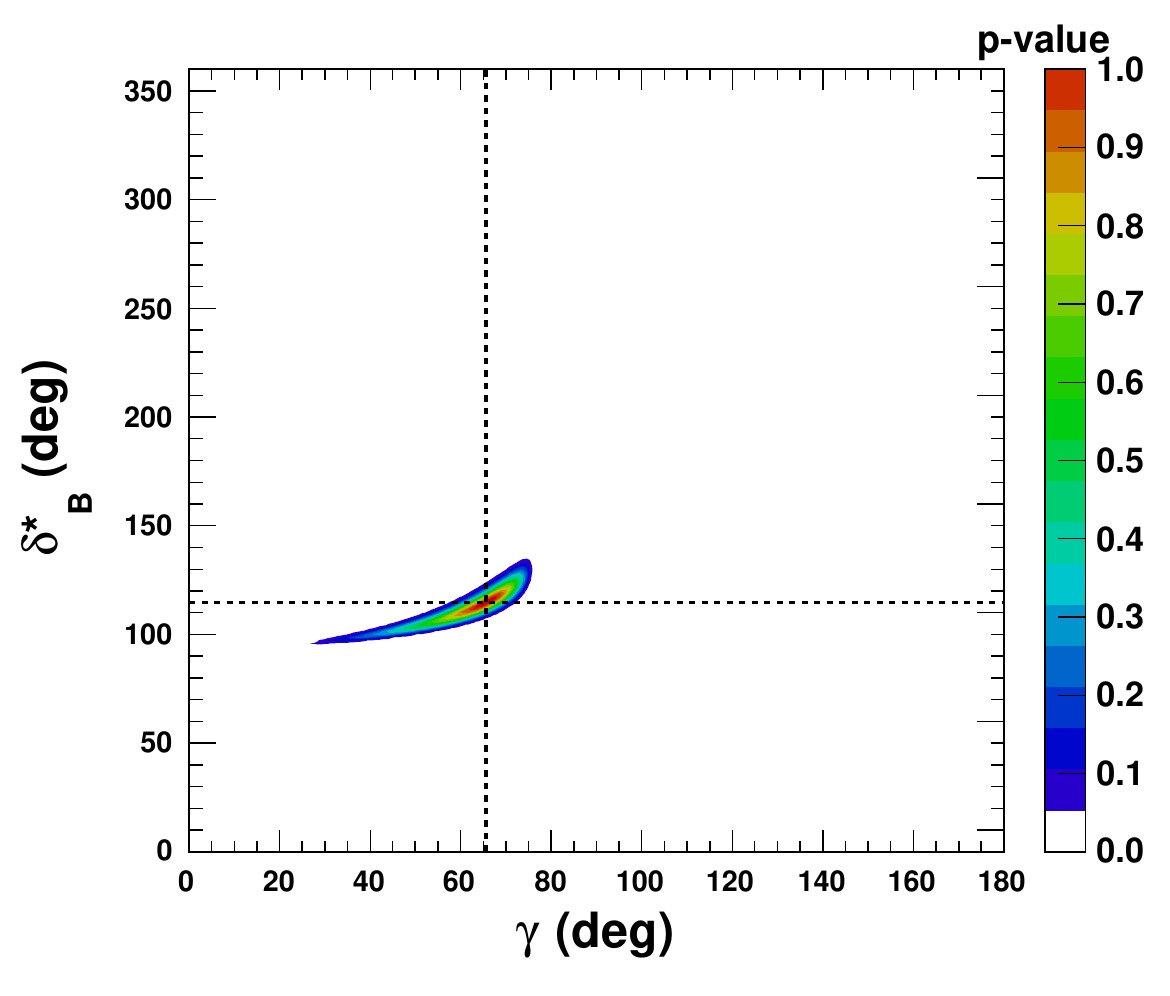}
\caption{\label{fig:2D-RunHL-LHC_rBst022} Two-dimension $p$-value  profile of the nuisance parameters $\rBst$ and $\deltaBst$, for the full HL-LHC LHCb dataset,  as a function of $\gamma$. On each figure the dashed black  lines indicate the initial  true values: $\gamma=65.66^\circ$ (1.146 rad), $\deltaB=171.9^\circ$ (3.0 rad), and $\deltastB=114.6^\circ$ (2.0 rad), and  $\rBst=0.22$.}
\end{figure}

Two-dimension $p$-value  distribution profiles of the nuisance parameters $\rBst$ and $\deltaBst$ as a function of $\gamma$ are provided in Figs.~\ref{fig:2D-Run1-3_rBst04} and~\ref{fig:2D-Run1-3_rBst022}, for the expected Run $1-3$ LHCb dataset, and in Figs.~\ref{fig:2D-RunHL-LHC_rBst04} and~\ref{fig:2D-RunHL-LHC_rBst022}, for the full HL-LHC LHCb dataset. For the purpose of those illustrations the initial configuration of true values is : $\gamma=65.66^\circ$ (1.146 rad), $\deltaB=171.9^\circ$ (3.0 rad), and $\deltastB=114.6^\circ$ (2.0 rad), and  $\rBst=0.4$ (0.22). The distributions can therefore directly be compared to those shown in Figs.~\ref{fig:2D_0} and~\ref{fig:2D_1}. The surface of the excluded regions at 95.4~\% CL in the $\rBst \ vs. \ \gamma$ and  $\deltaBst \ vs. \ \gamma$, clearly increase with the additional data, but even in the semi-asymptotic regime, for the full expected HL-LHC LHCb dataset, one can clearly see possible strong correlations between $\gamma$ and the nuisance parameters $\rBst$ and $\deltaBst$. This is also visible in Figs.~\ref{fig:2D-RunHL-LHC_rBst04_15} and~\ref{fig:2D-RunHL-LHC_rBst022_15} in appendix~\ref{sec:appendD}, which are the equivalent version for the full expected HL-LHC LHCb dataset of Run~1~\&~2 LHCb dataset presented in Figs.~\ref{fig:2D_2} and~\ref{fig:2D_3}, for the configurations:  $\gamma=65.66^\circ$ (1.146 rad), $\deltaB=57.3^\circ$ (1.0 rad), and $\deltastB=286.5^\circ$ (5.0 rad), and  $\rBst=0.4$ (0.22).

\begin{figure}[h]
\centering
\includegraphics[width=0.425\textwidth]{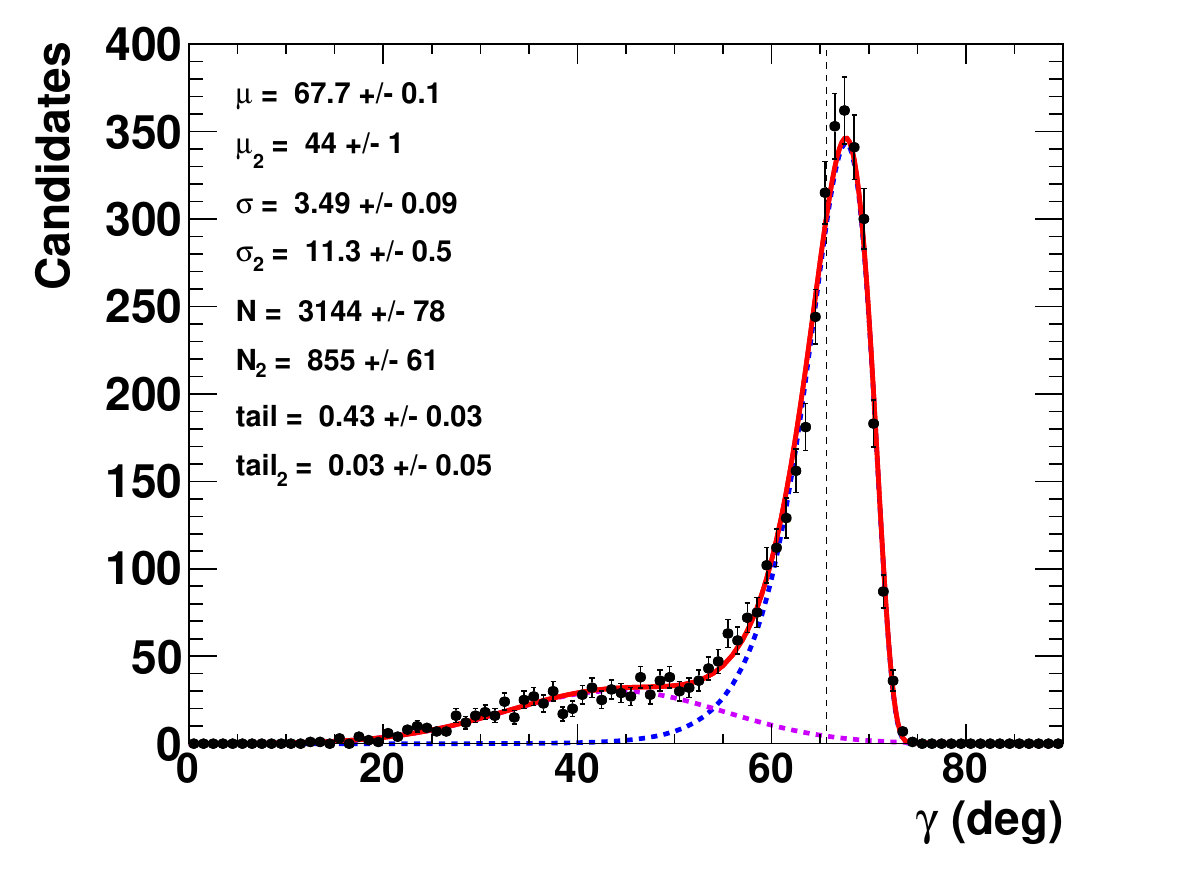}
\includegraphics[width=0.425\textwidth]{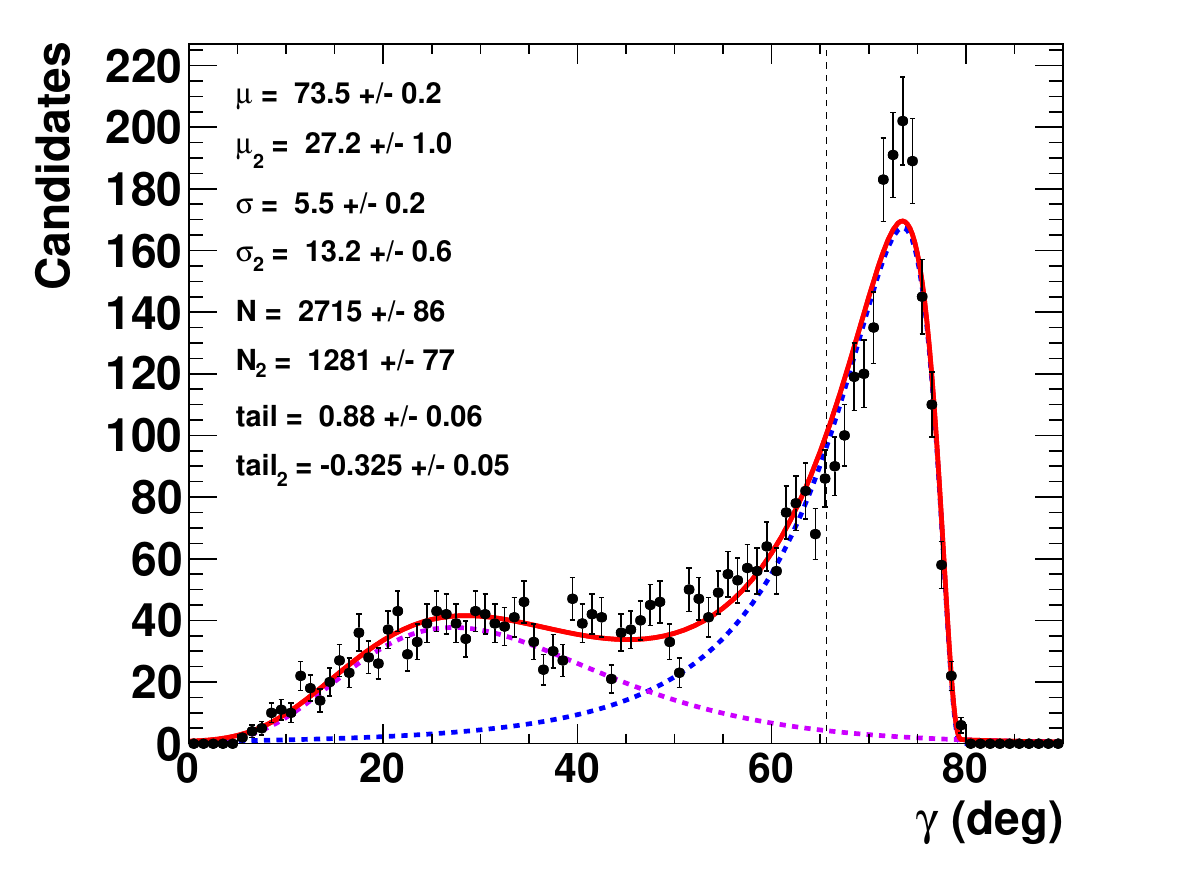}
\caption{\label{fig:1-3_dB3dBst2} Fit to the distributions of  $\gamma$ obtained from 4000 pseudoexperiments, for an the expected Run $1-3$ LHCb dataset. The initial configuration is $\gamma=65.66^{\circ}$, $\rBst=0.4$ (left) and 0.22 (right), $\deltaB=171.9^\circ$ (3 rad), and $\deltastB=114.6^\circ$ (2 rad). The purple dashed curve accounts for tails generated by the correlations with the nuisance parameters $\rBst$ and $\deltaBst$, while the blue dashed curve is the core part of the distribution, the plain red line is the sum of the two components of the fit.}
\includegraphics[width=0.425\textwidth]{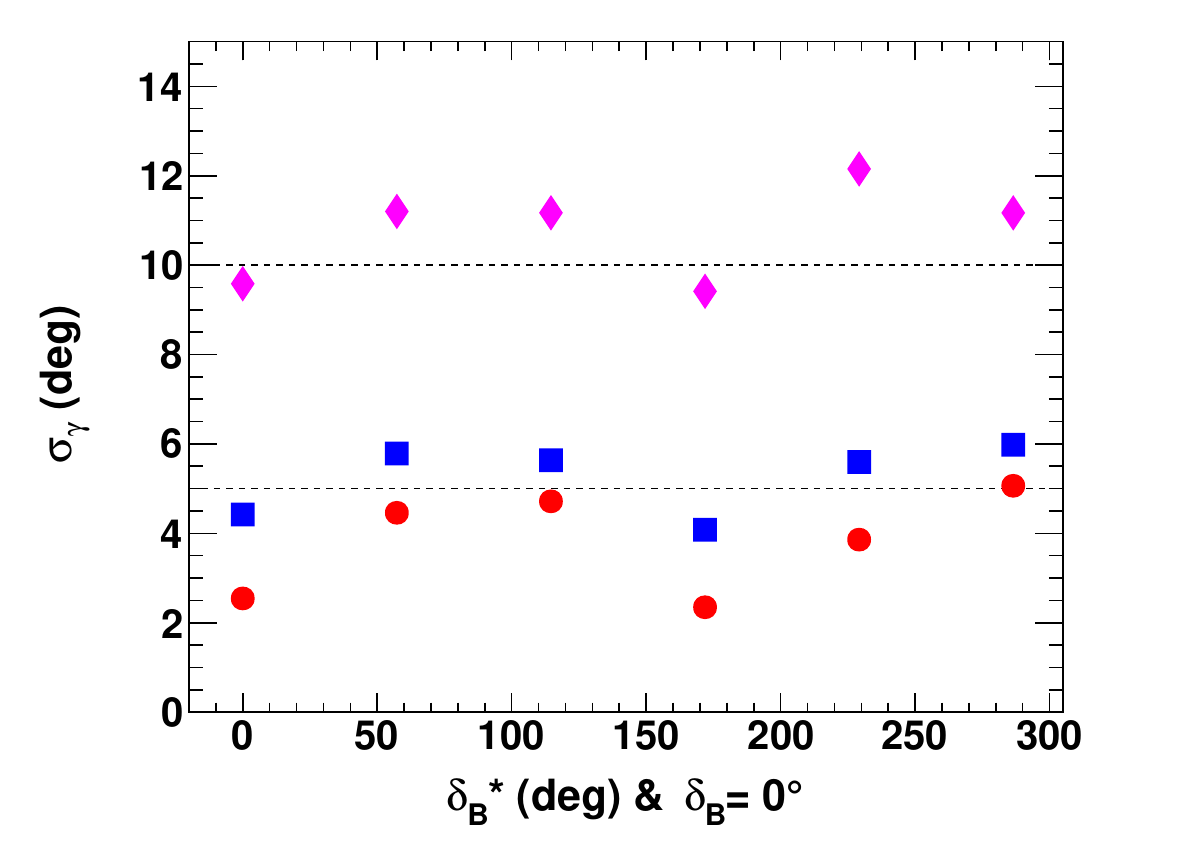}
\includegraphics[width=0.425\textwidth]{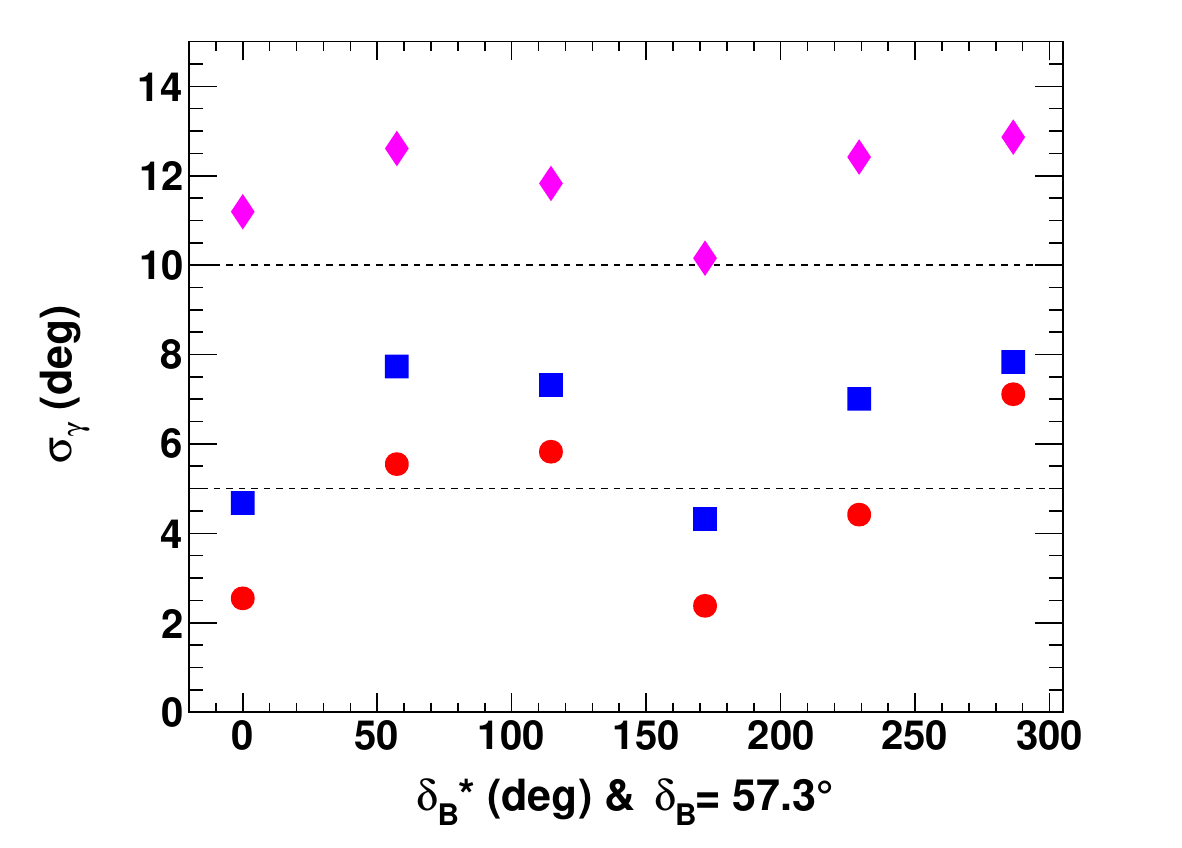} \\
\includegraphics[width=0.425\textwidth]{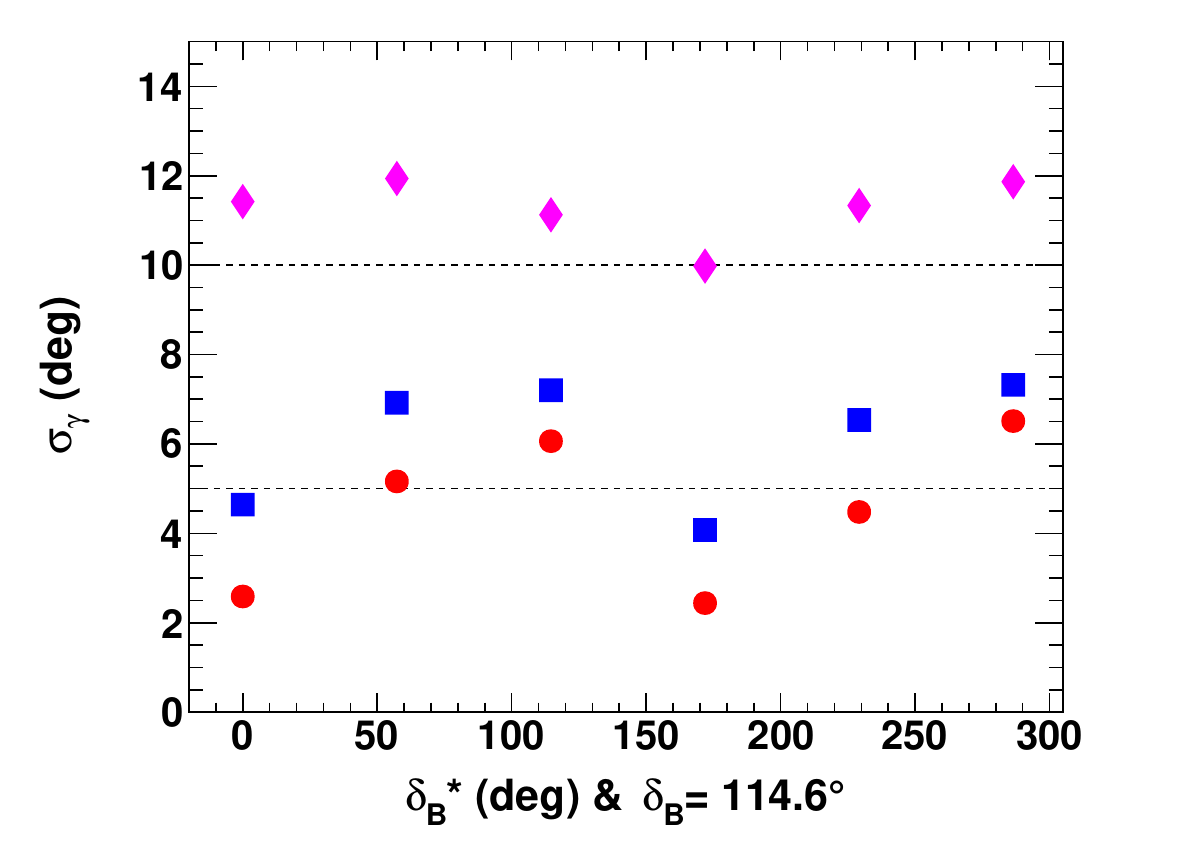}
\includegraphics[width=0.425\textwidth]{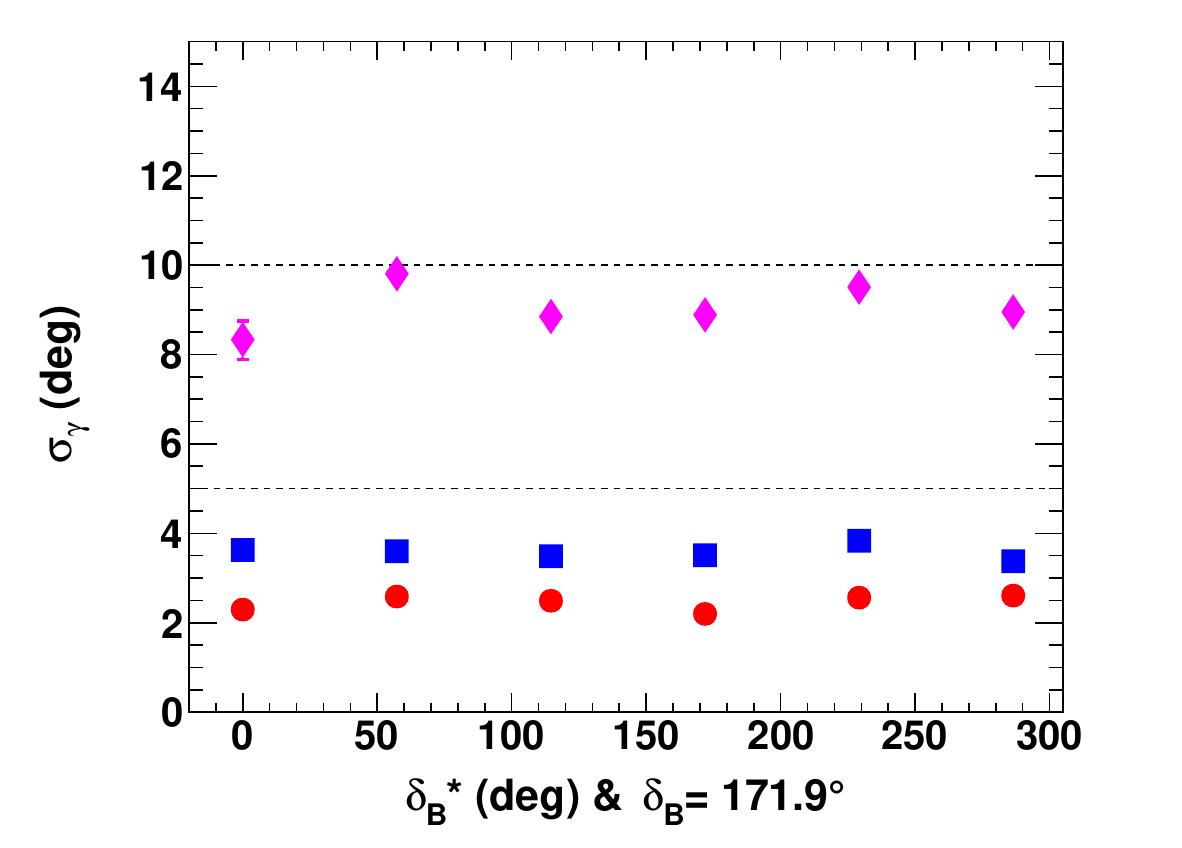} \\
\includegraphics[width=0.425\textwidth]{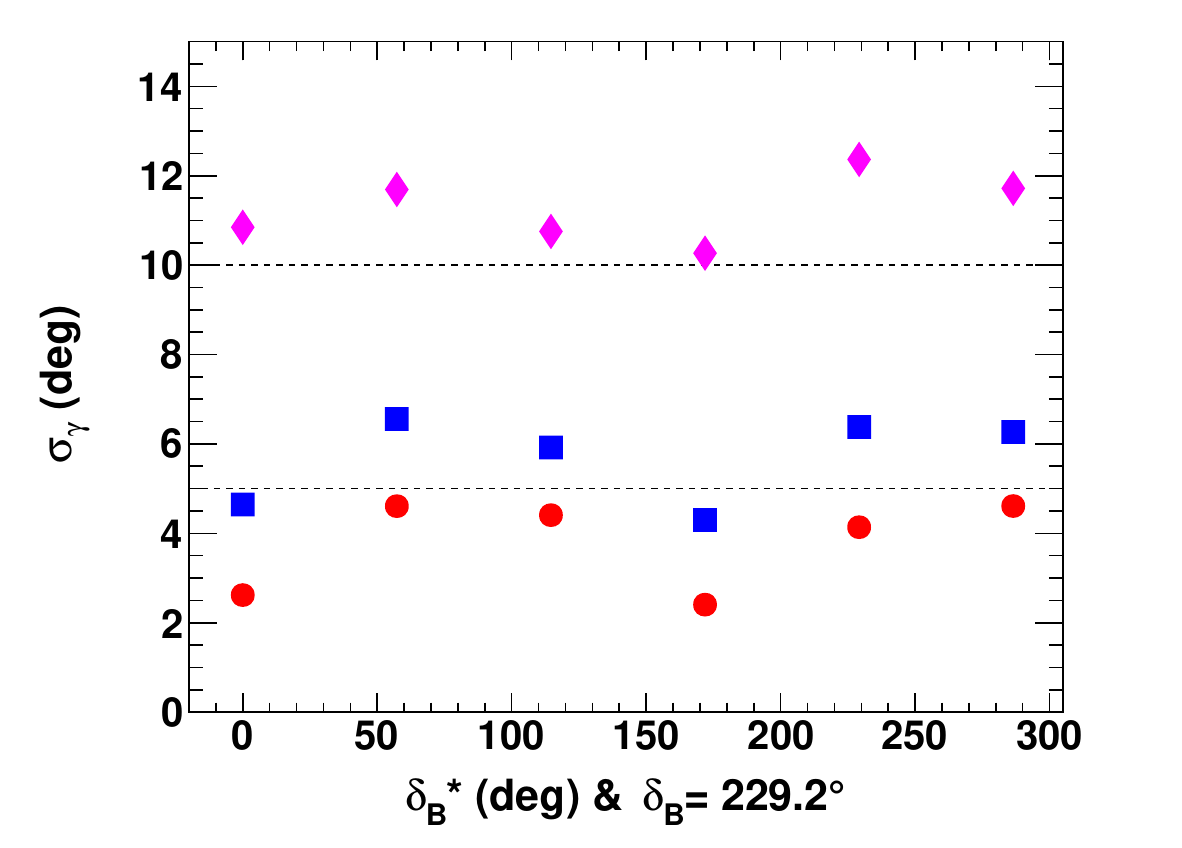}
\includegraphics[width=0.425\textwidth]{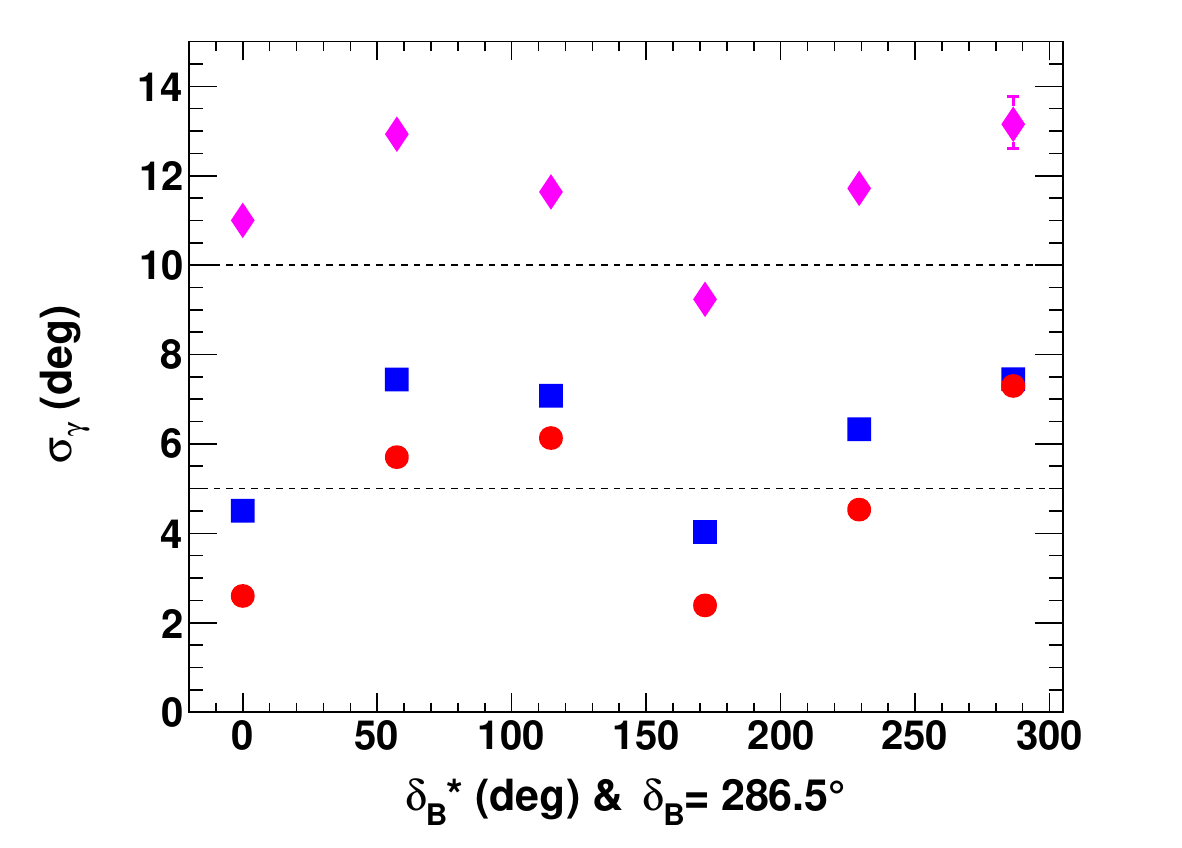}
\caption{\label{fig:AllLumi_rBst04_resol}Fitted mean value of $\gamma$ ($\sigma_{\gamma}$), for Run~1~\&~2 (pink lozenges), for Run $1-3$   (blue squares), and full HL-LHC  (red circles) LHCb dataset, as a function of $\deltaBst$, for $\rBst=0.4$, for an initial true value of  $65.66^\circ$ (1.146 rad).  On each figure, the horizontal dashed black  lines are guide for the eye at $\sigma_{\gamma}=5^\circ$ and  $10^\circ$. }
\end{figure}

\begin{figure}[h]
\centering
\includegraphics[width=0.425\textwidth]{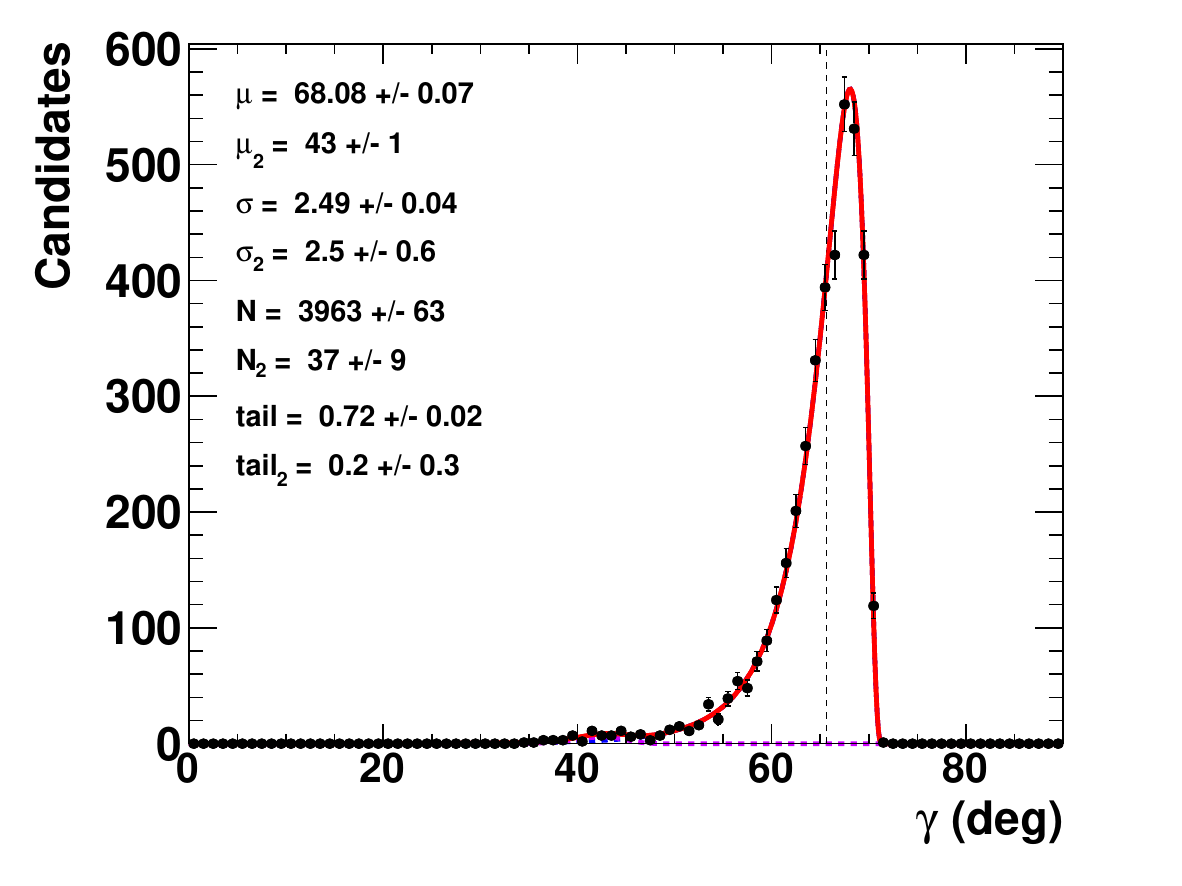}
\includegraphics[width=0.425\textwidth]{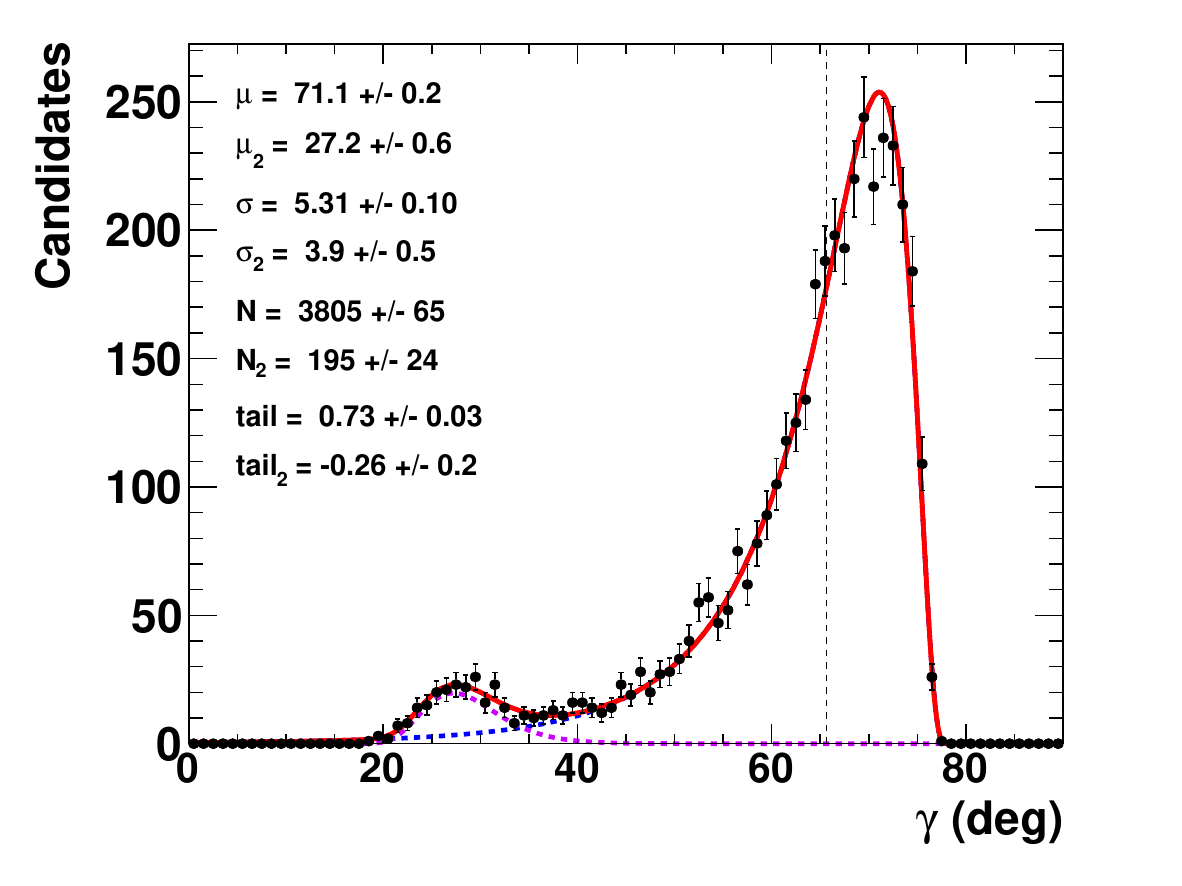}
\caption{\label{fig:HL-LHC_dB3dBst2} Fit to the distributions of  $\gamma$ obtained from 4000 pseudoexperiments, for an the expected full HL-LHC LHCb dataset. The initial configuration is $\gamma=65.66^{\circ}$, $\rBst=0.4$ (left) and 0.22 (right), $\deltaB=171.9^\circ$ (3 rad), and $\deltastB=114.6^\circ$ (2 rad). The purple dashed curve accounts for tails generated by the correlations with the nuisance parameters $\rBst$ and $\deltaBst$, while the blue dashed curve is the core part of the distribution, the plain red line is the sum of the two components of the fit.}
\includegraphics[width=0.425\textwidth]{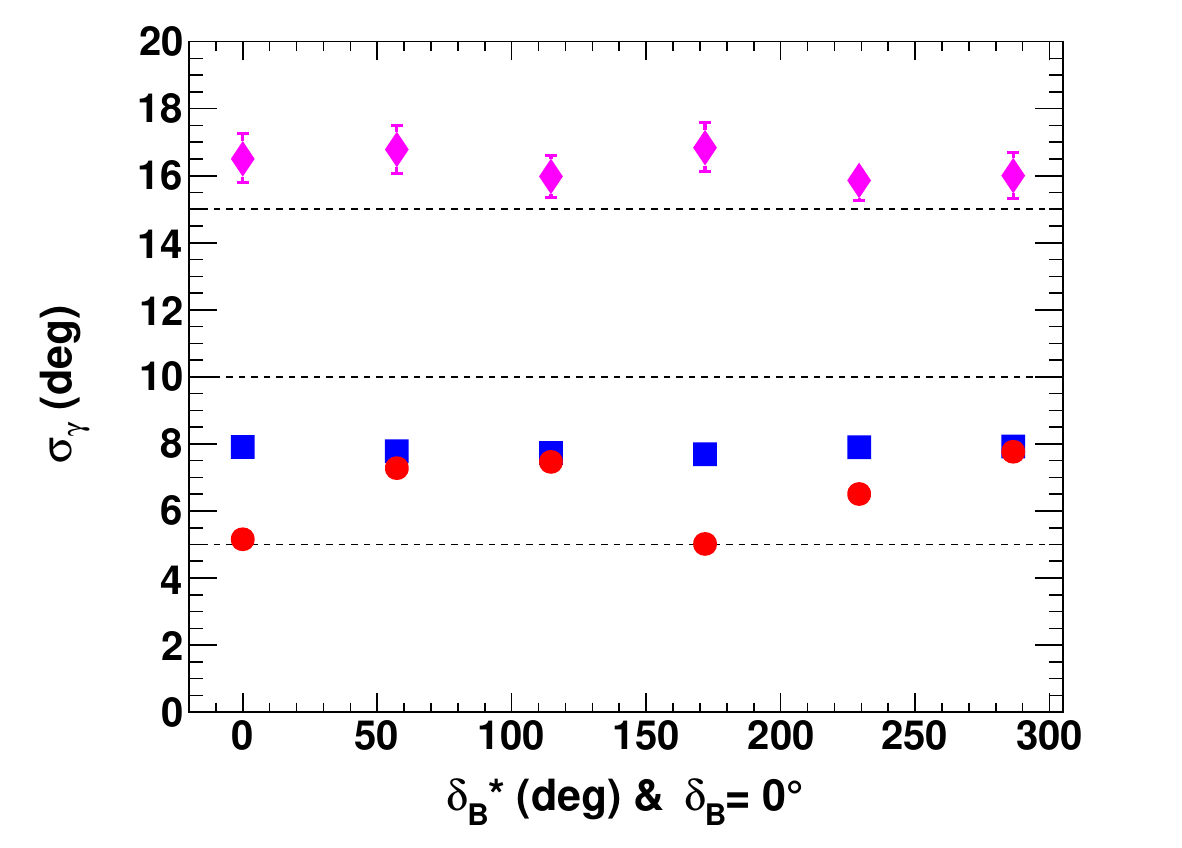}
\includegraphics[width=0.425\textwidth]{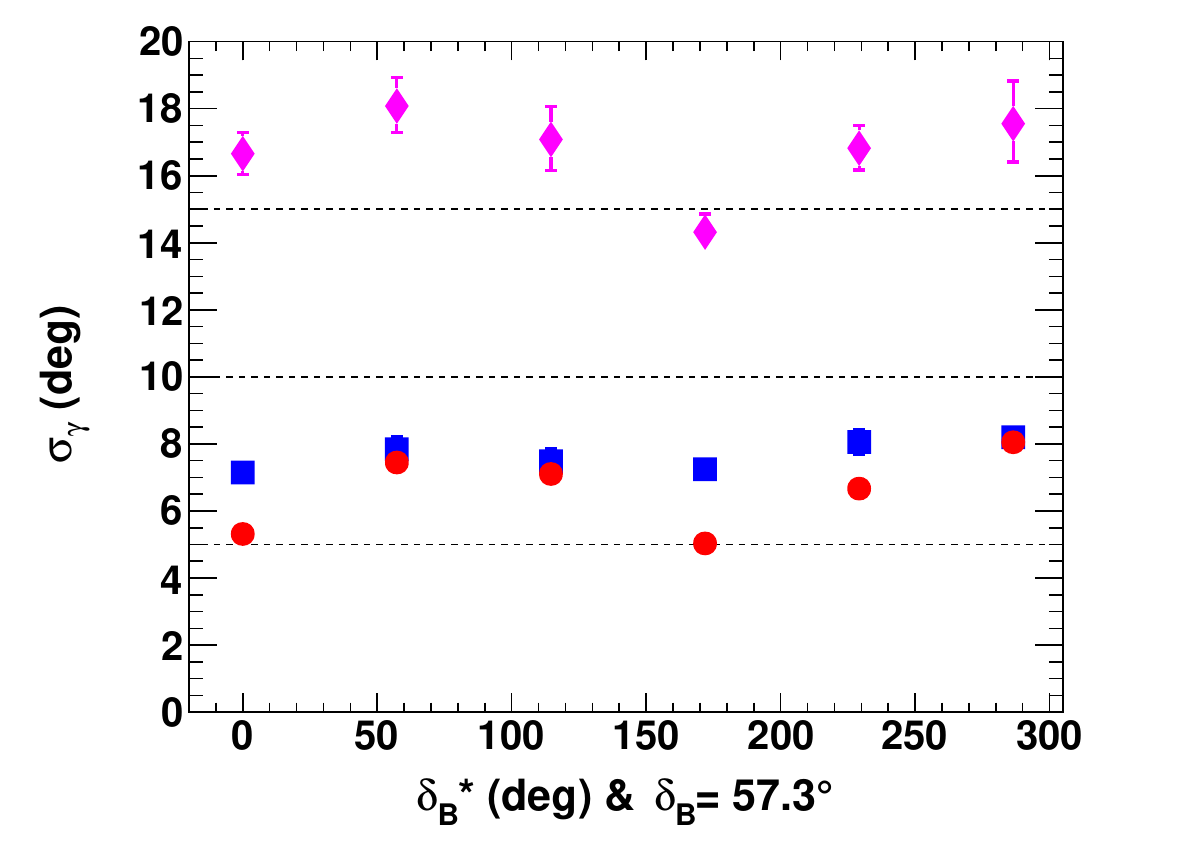} \\
\includegraphics[width=0.425\textwidth]{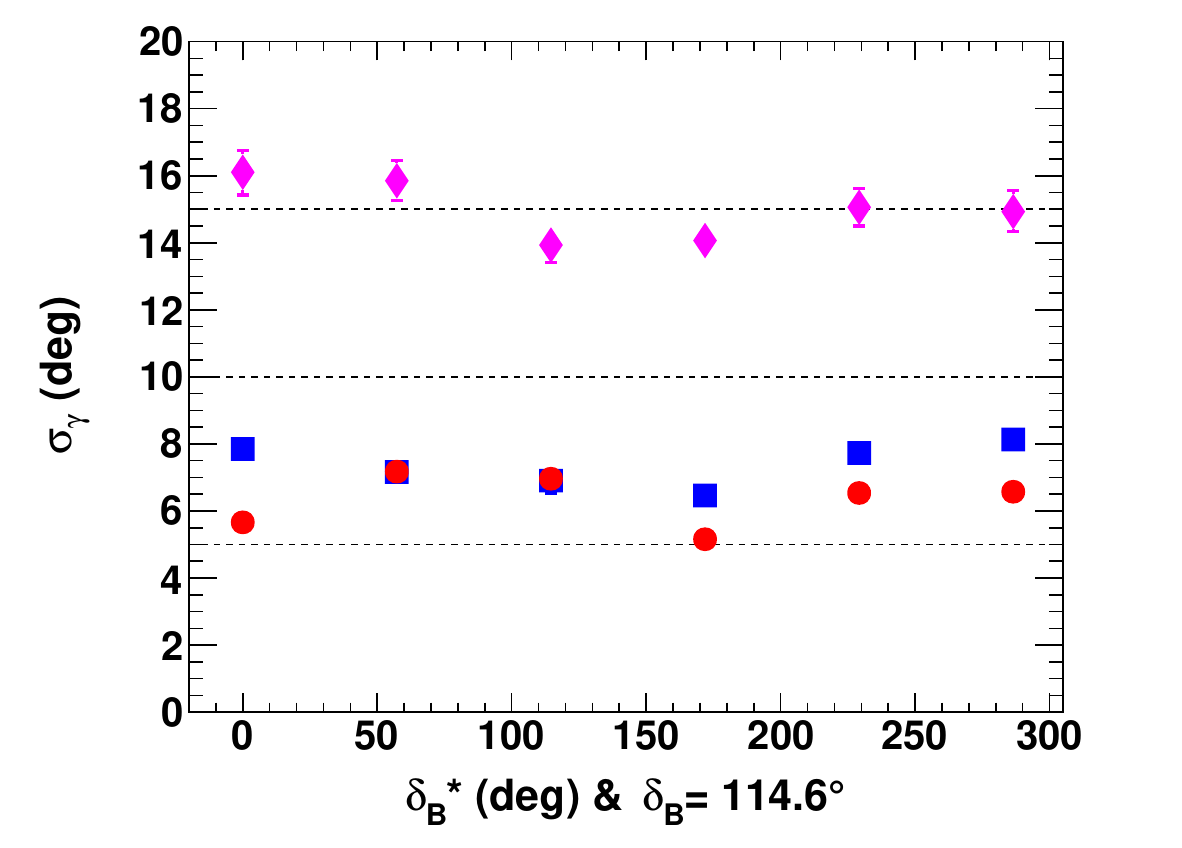}
\includegraphics[width=0.425\textwidth]{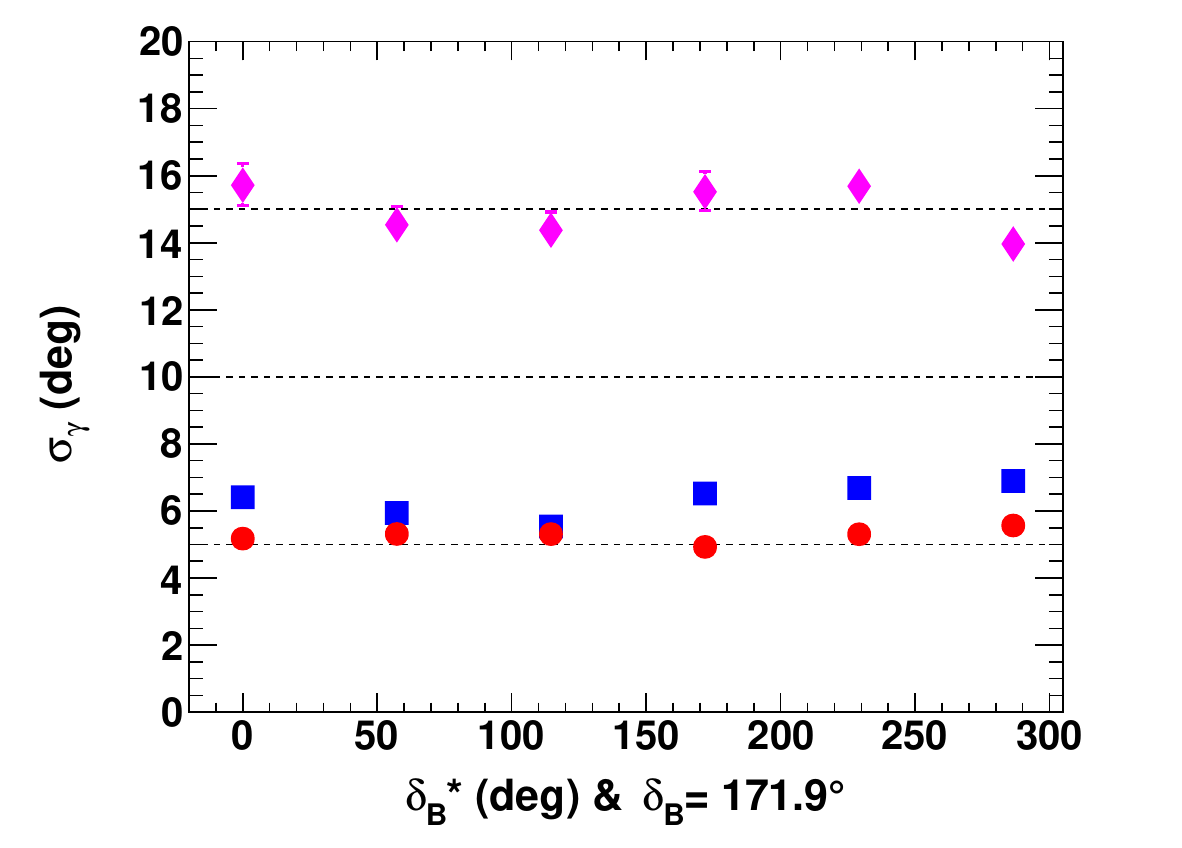} \\
\includegraphics[width=0.425\textwidth]{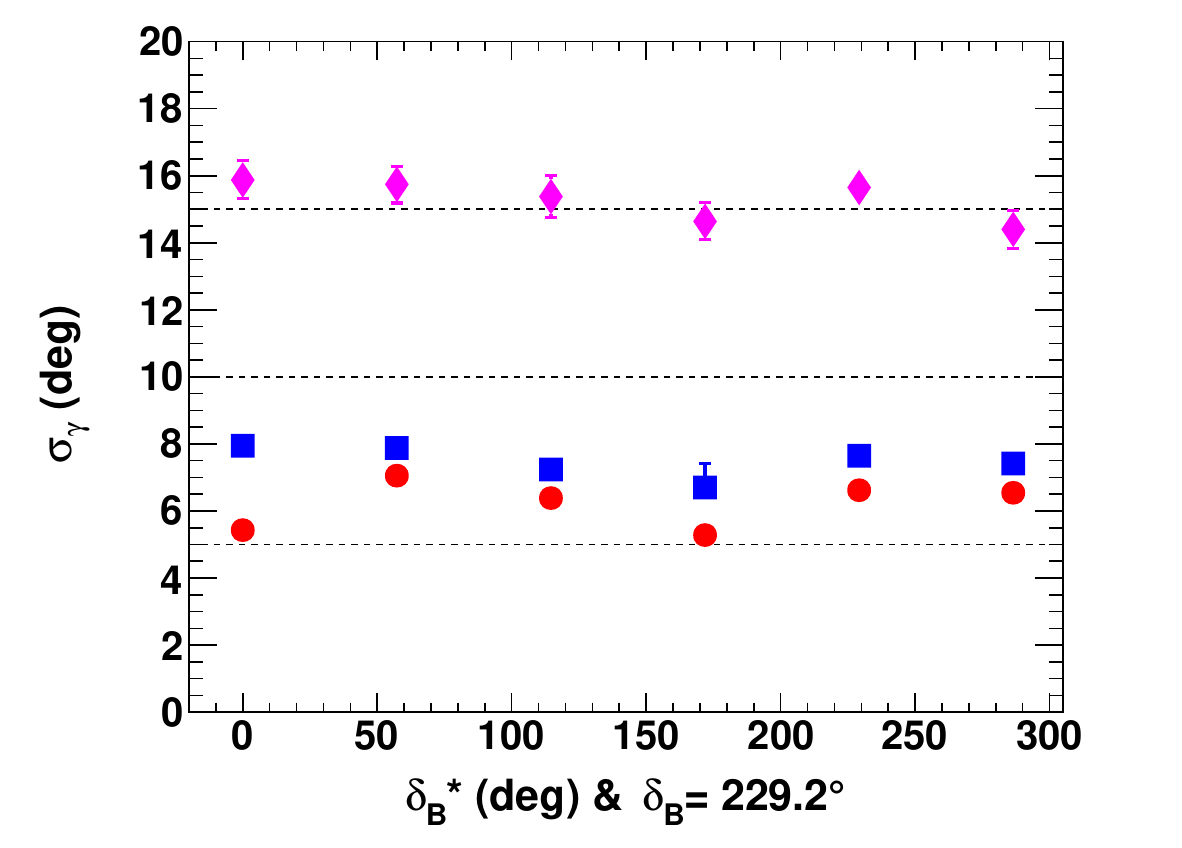}
\includegraphics[width=0.425\textwidth]{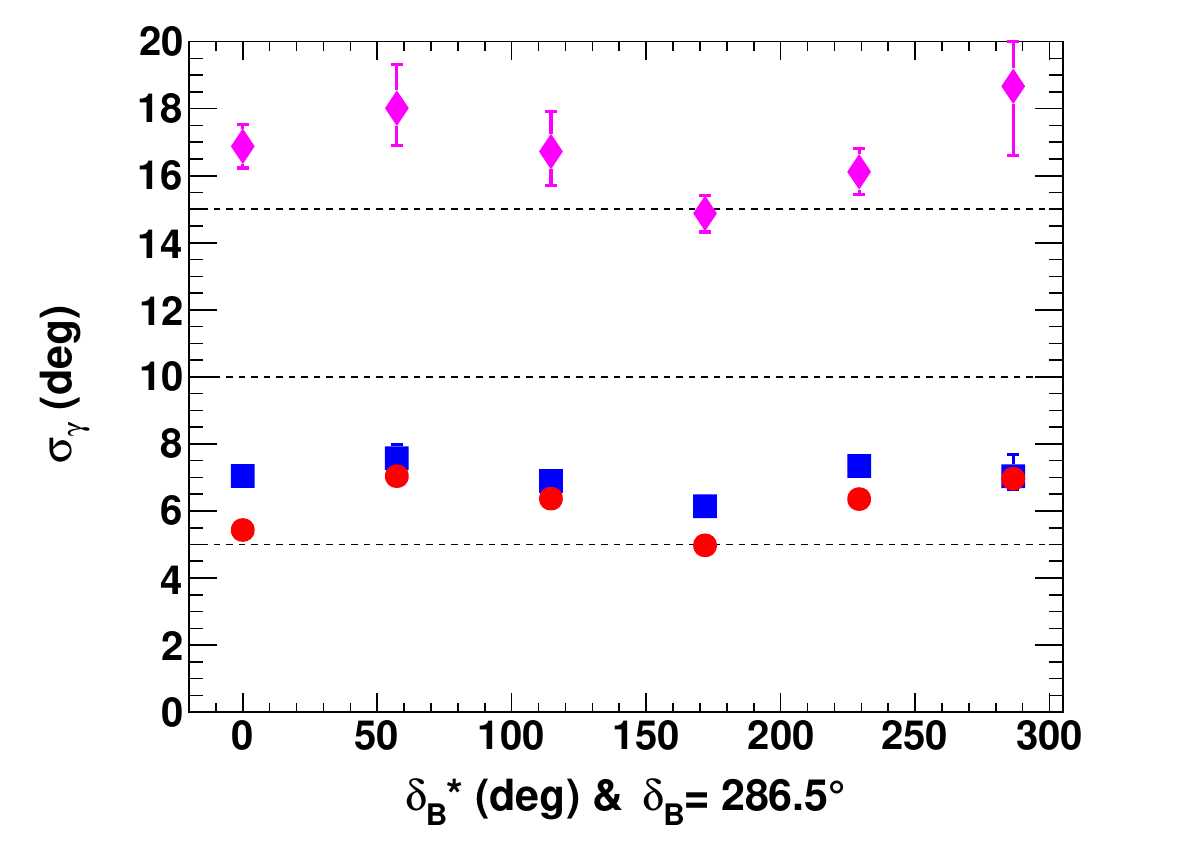}
\caption{\label{fig:AllLumi_rBst022_resol}Fitted resolution of $\gamma$ ($\sigma_{\gamma}$), for Run~1~\&~2 (pink lozenges), for Run $1-3$   (blue squares), and full HL-LHC  (red circles) LHCb dataset, as a function of $\deltaBst$, for $\rBst=0.22$, for an initial true value of  $65.66^\circ$ (1.146 rad). On each figure, the horizontal dashed black  lines are guide for the eye at $\sigma_{\gamma}=5^\circ, \ 10^\circ,$ and  $15^\circ$.  }
\end{figure}

\clearpage


For the configuration $\gamma=65.66^\circ$ (1.146 rad), $\deltaB=171.9^\circ$ (3.0 rad), and $\deltastB=114.6^\circ$ (2.0 rad), and  $\rBst=0.4$ (0.22), Fig.~\ref{fig:1-3_dB3dBst2} shows the fitted $\gamma$ distribution obtained for 4000 pseudoexperiments, for the expected Run $1-3$ LHCb dataset. the fitted values are : $\mu_{\gamma}=\left( 67.7 \pm 0.1 \right)^\circ$ $\left( (73.5 \pm 0.2)^\circ \right)$ and $\sigma_{\gamma}=\left( 3.5 \pm 0.1 \right)^\circ$ $\left( (5.5 \pm 0.2)^\circ \right)$, for $\rBst=0.4$ (0.22). Respectively, for the expected full HL-LHC LHCb dataset, the fitted values presented in  Fig.~\ref{fig:HL-LHC_dB3dBst2}  are : $\mu_{\gamma}=\left( 68.1 \pm 0.1 \right)^\circ$ $\left( (71.1 \pm 0.1)^\circ\right)$ and $\sigma_{\gamma}=\left( 2.5 \pm 0.1 \right)^\circ$ $\left( (5.3 \pm 0.1)^\circ \right)$, for $\rBst=0.4$ (0.22). The fitted values are slightly shifted up with respect to the initial $\gamma$ true value, but compatible within one standard deviation.  When comparing with numbers listed in Table~\ref{tab:dbeffect_rBst04_022_resol}, one can see that the resolution improves as $8.8/3.5=2.5$ ($14.4/5.5=2.6)$, for $\rBst=0.4$ (0.22), when moving from the Run~1~\&~2  to the expected Run $1-3$ LHCb datasets, while a factor 2.2 is expected. But, when moving from the expected Run $1-3$ to the full expected HL-LHC LHCb datasets, the improvement is only $3.5/2.5=1.4$ ($5.5/5.3=1.040)$, for $\rBst=0.4$ (0.22), while one may naively expect an improvement $8.8/2.2=4.0$. Part of this is certainly coming from the strong correlations in between the nuisance parameters $\rBst$ and $\deltaBst$.  A more sophisticated simultaneous global fit to the nuisance parameters $\rBst$ and $\deltaBst$, and $\gamma$ may be useful.

\begin{figure}[h]
\centering
\includegraphics[width=0.425\textwidth]{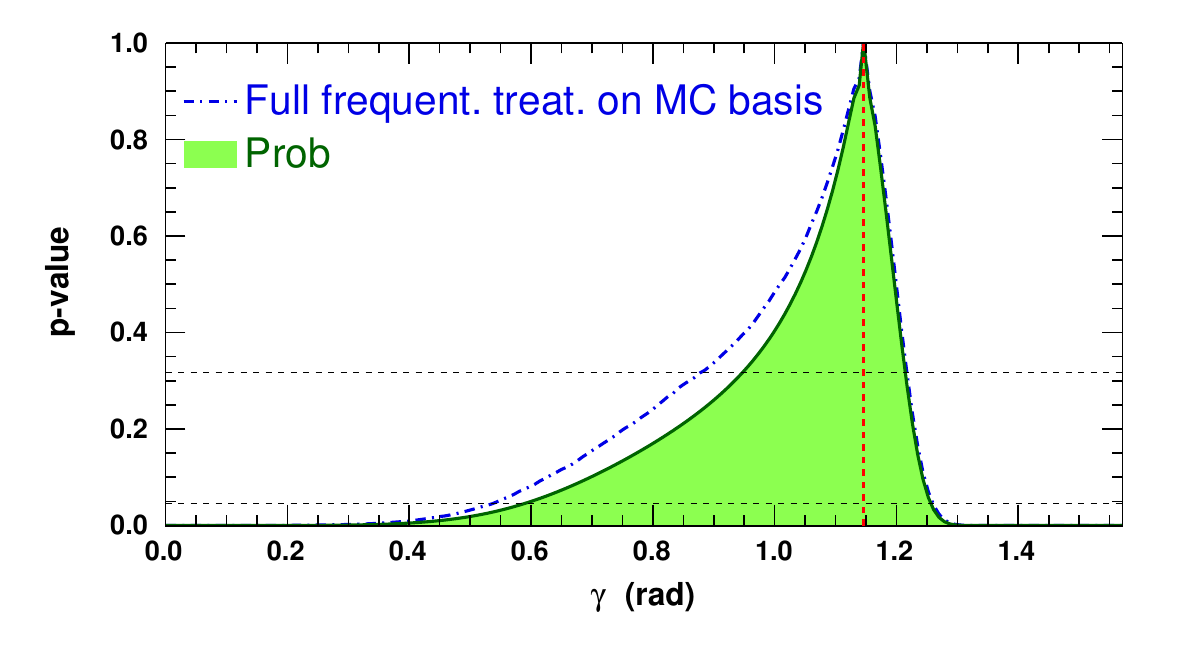}
\includegraphics[width=0.425\textwidth]{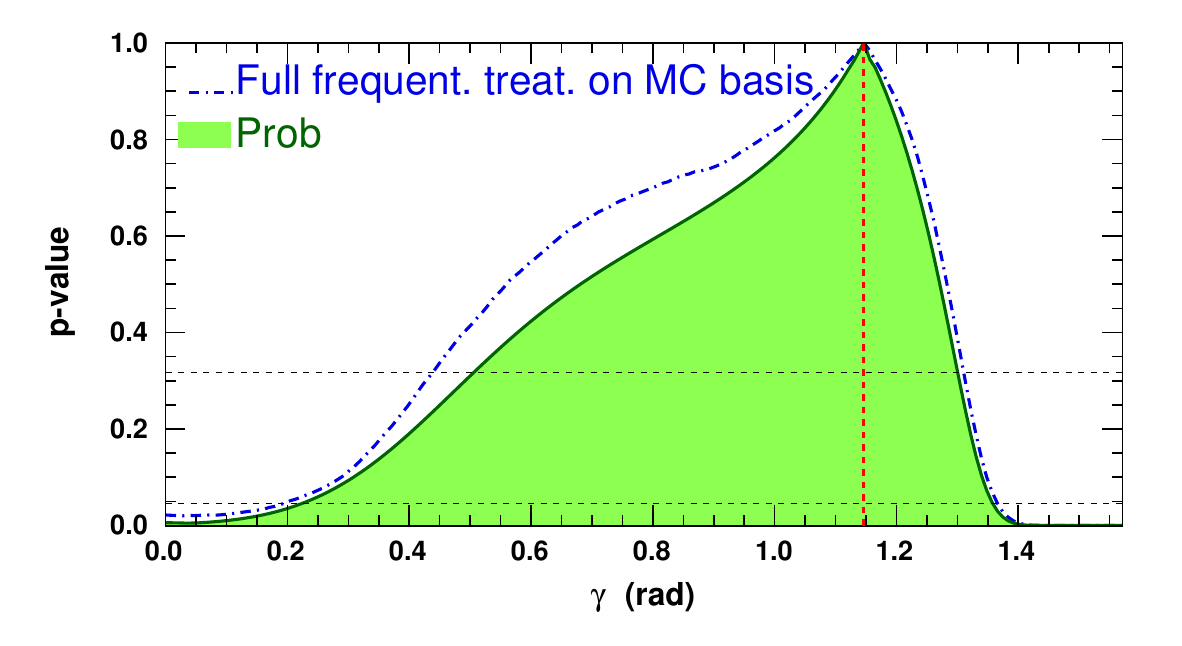} \\
\includegraphics[width=0.425\textwidth]{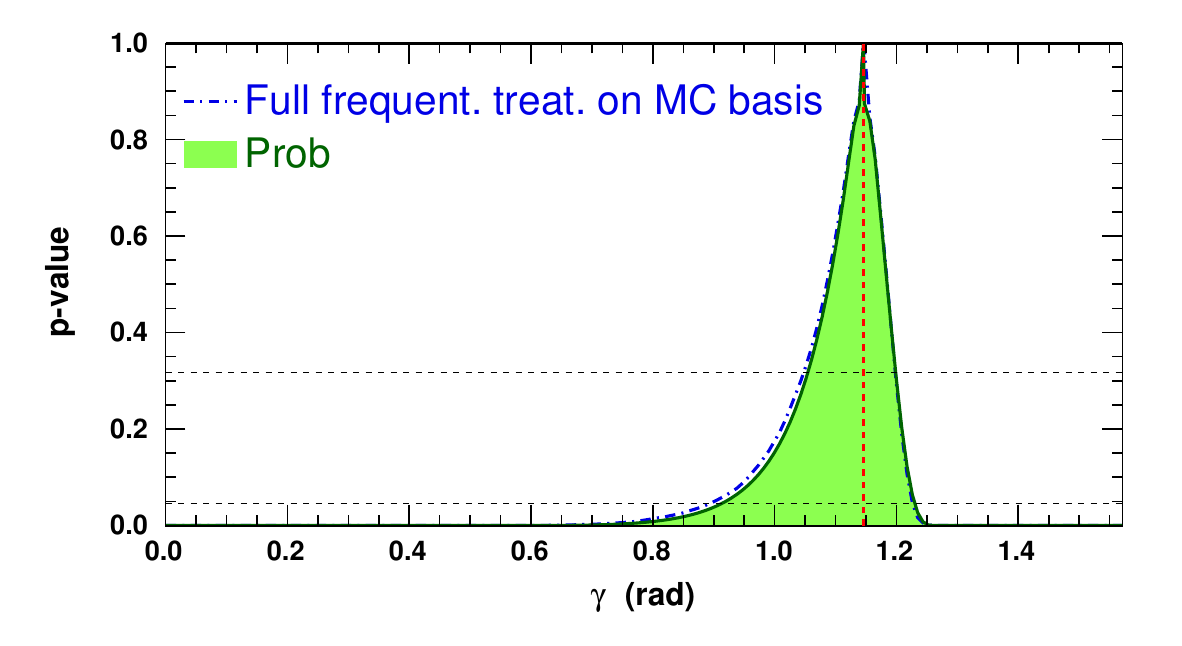}
\includegraphics[width=0.425\textwidth]{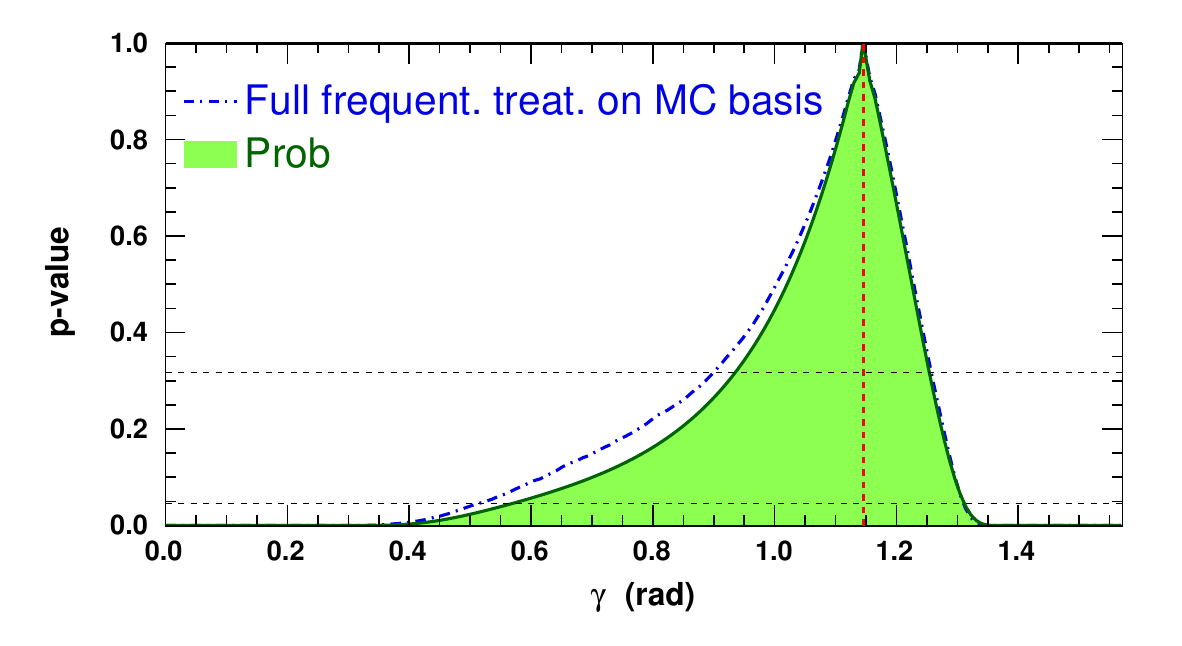}
\caption{\label{fig:gamma_1D_HL-LHC-plugin} Profile of the $p$-value distribution of the global $\chi^2$ fit to $\gamma$ for the set of true initial parameters: $\gamma=1.146$ rad, $\rBst=0.4$ (left) $0.22$ (right), $\deltaB=3.0$ rad, and $\deltastB=2.0$ rad. The integrated luminosity assumed here is that of LHCb data expected to be collected after the LHC Run~3 (top) and after the full HL-LHC (bottom) period. The corresponding distribution obtained from a full frequentist treatment on Monte-Carlo simulation basis~\cite{plugin} is superimposed to the profile obtained with {\ttfamily  TMath::Prob}.  On each figure the vertical dashed red line indicates the initial $\gamma$ true value, and the two horizontal dashed black lines, refer to 68.3 and 95.4~\% CL.}
\end{figure}

One has also to remember that the {\ttfamily  TMath::Prob} has still some under-coverage, {\it i.e.} 79 $(91)~\%$ and 94 $(89)~\%$) for $\rBst=0.4$ (0.22), with respect to the full frequentist treatment on Monte-Carlo simulation basis~\cite{plugin} as presented in Fig.~\ref{fig:gamma_1D_HL-LHC-plugin}, with the Run $1-3$ and the full expected HL-LHC LHCb datasets, respectively. The relative scale factors $F_{K3\pi}$, $F_{K\pi\piz}$, $F_{KK}$, and $F_{\pi\pi}$ used in this study have already a precision better than $2~\%$. The precision on the normalisation factors $C_{K\pi}$, $C_{K\pi,D\piz}$, and $C_{K\pi,D\gamma}$ may also benefit from another improved precision of the branching fraction of the decay modes $\BsDbphi$ and of the longitudinal polarisation fraction in the mode $\BsDstbphi$. But the normalisation factors are the same for all the set of Eqs.~\eqref{EQ__14_}-\eqref{EQ__34_} for $\Bs \rightarrow \Dtz \phi$ or $\Dtstz (\piz, \ \gamma)\phi$ their improved precision should be a second order effect. All of the above listed improvement are expected to happen to fully benefit from the total expected HL-LHC LHCb dataset.

The expected resolution on $\gamma$ for the other usual configuration ({\it i.e.} $\deltaBst=0$, 1, 2, 3, 4, 5 rad, and $\gamma=65.66^{\circ}$ (1.146 rad)) are presented in Fig.~\ref{fig:AllLumi_rBst04_resol}, for $\rBst=0.4$, and in Fig.~\ref{fig:AllLumi_rBst022_resol}, for $\rBst=0.22$ and for Run~1~\&~2, Run $1-3$, and full HL-LHC LHCb datasets. For  $\rBst=0.4$, the resolution ranges from $3.4^\circ$ to $7.8^\circ$ mostly, for Run $1-3$ and from  $2.2^\circ$ to $7.1^\circ$, or better, for the full HL-LHC dataset. For $\rBst=0.22$, the resolution ranges from $5.5^\circ$ to $8.2^\circ$, for Run $1-3$ and from  $3.3^\circ$ to $7.8^\circ$, or better, for the full HL-LHC dataset.

\begin{figure}[h]
\centering
\includegraphics[width=0.425\textwidth]{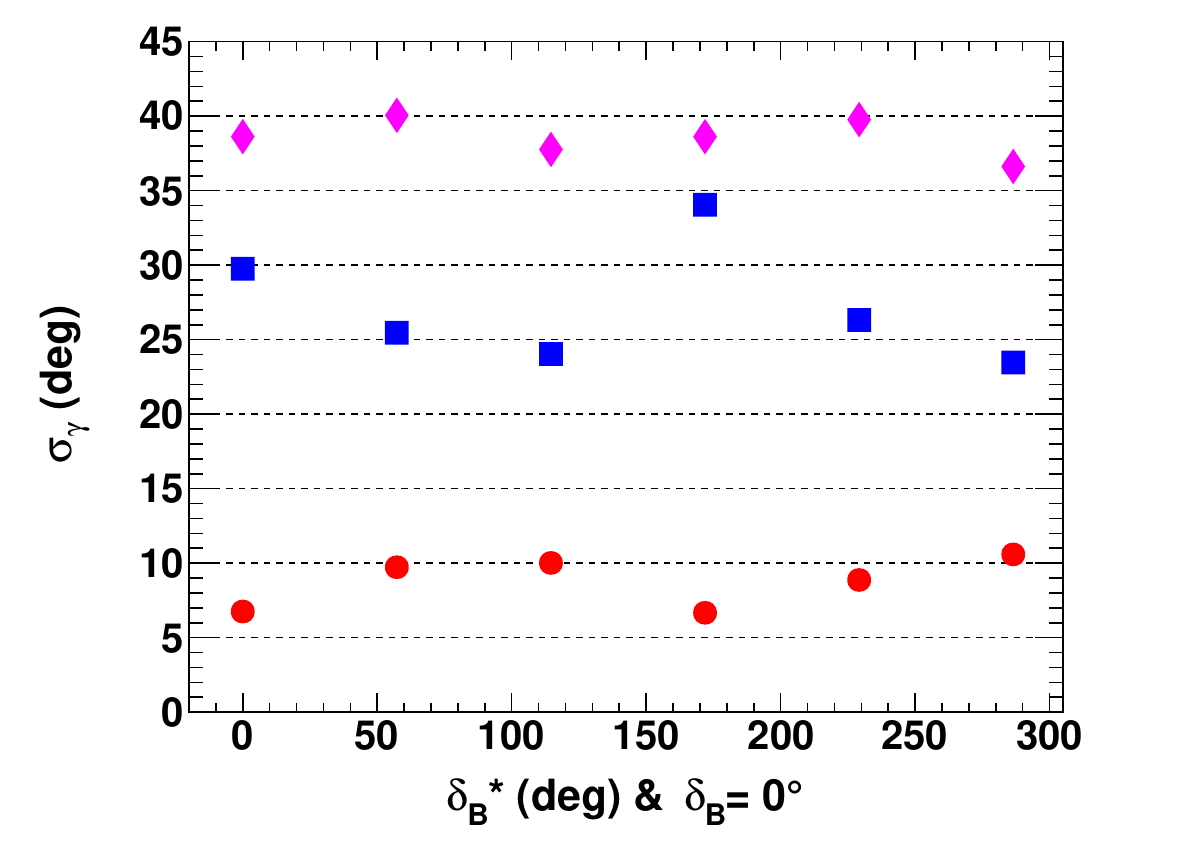}
\includegraphics[width=0.425\textwidth]{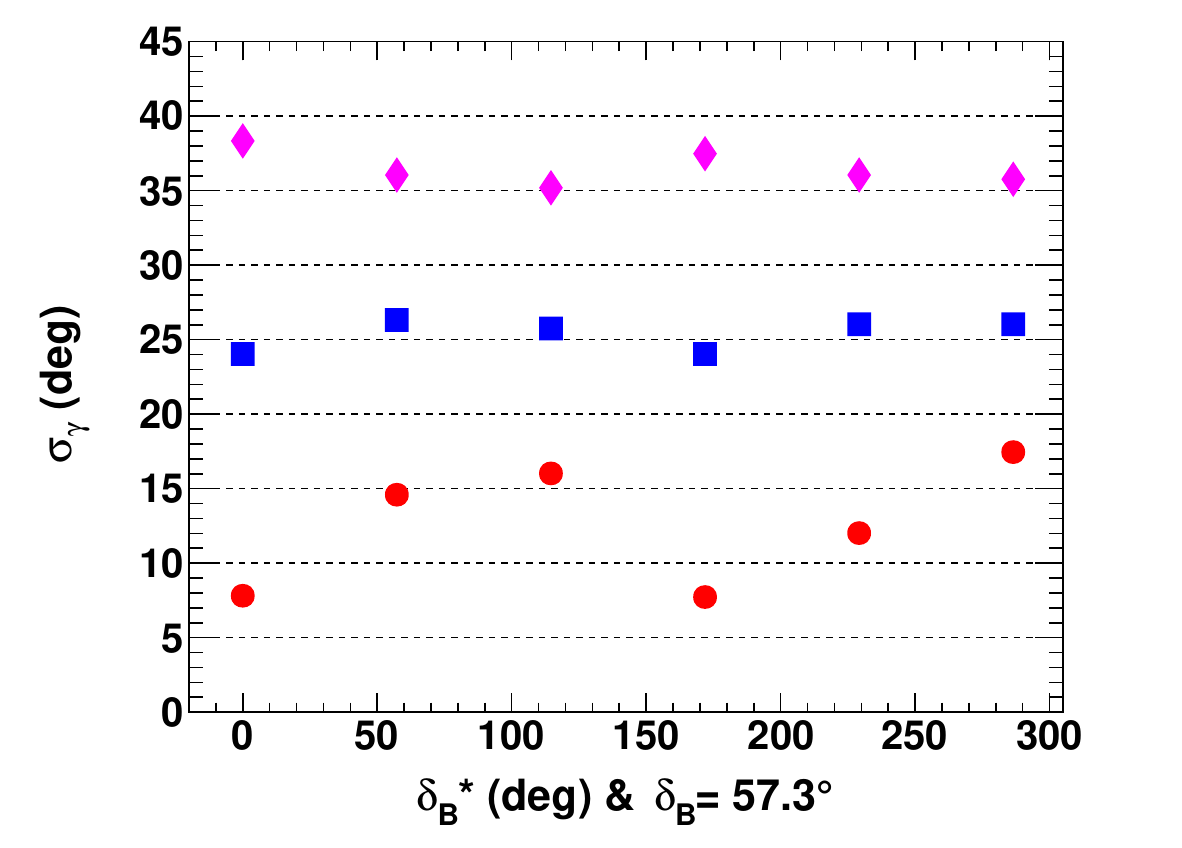} \\
\includegraphics[width=0.425\textwidth]{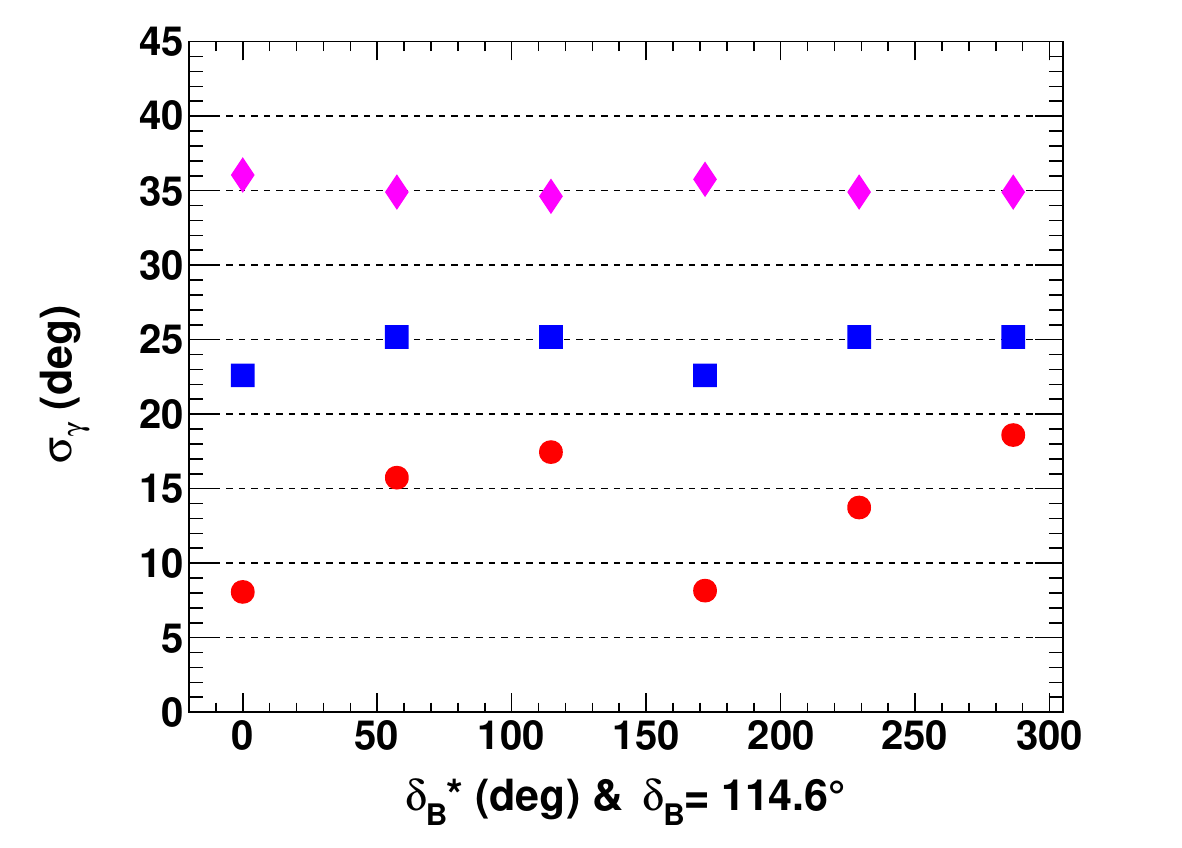}
\includegraphics[width=0.425\textwidth]{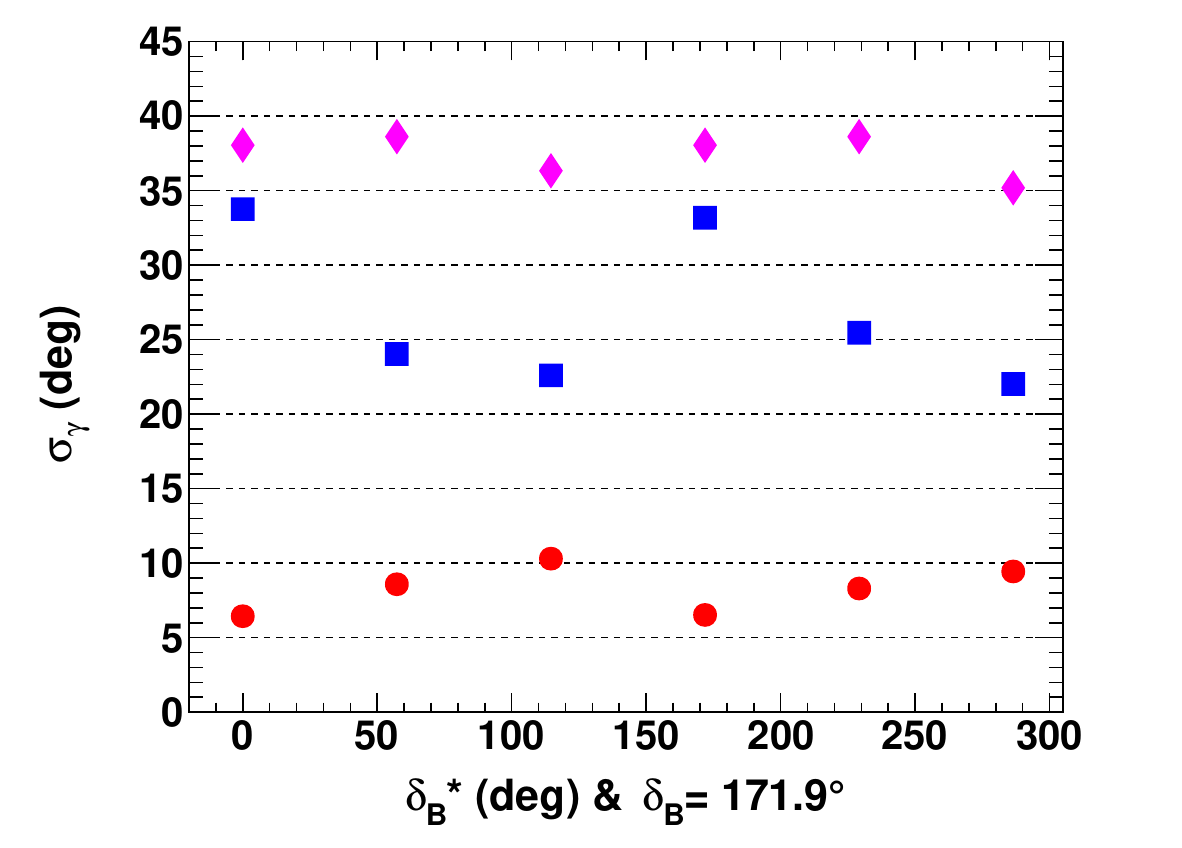} \\
\includegraphics[width=0.425\textwidth]{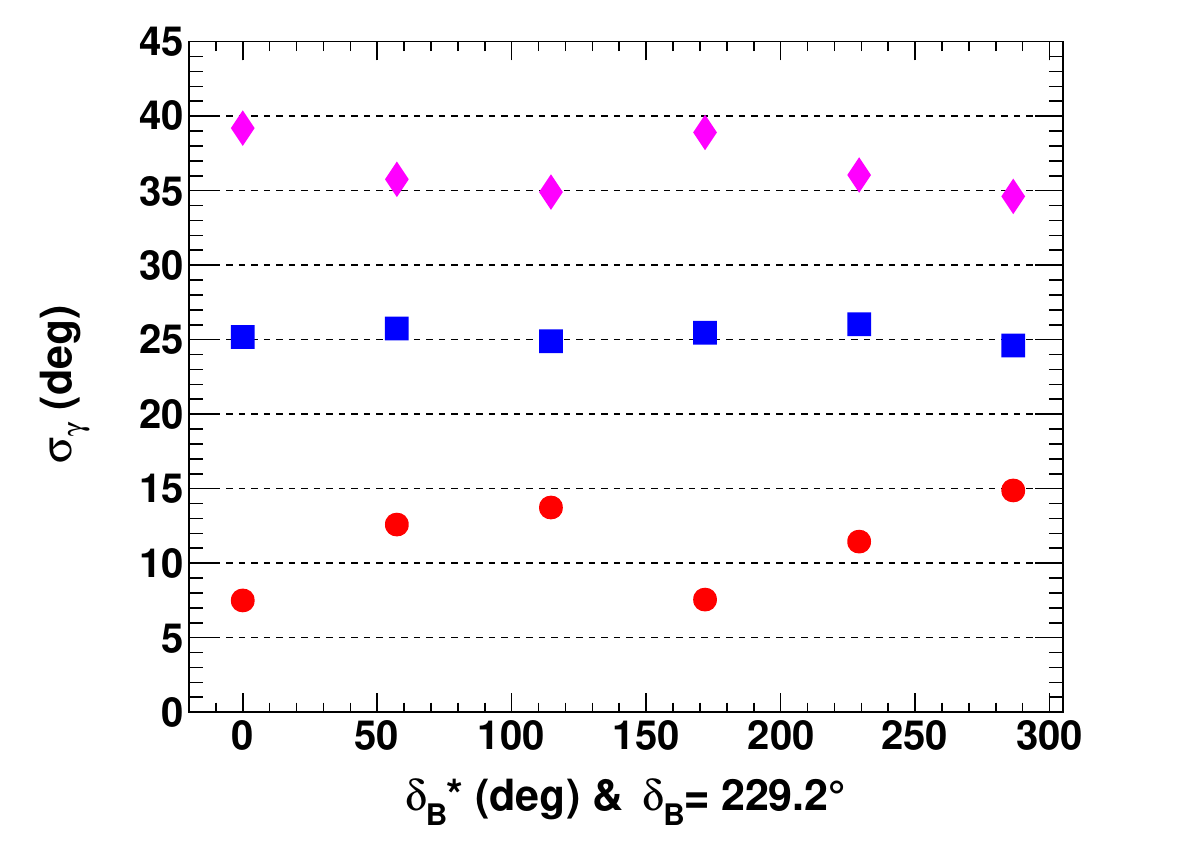}
\includegraphics[width=0.425\textwidth]{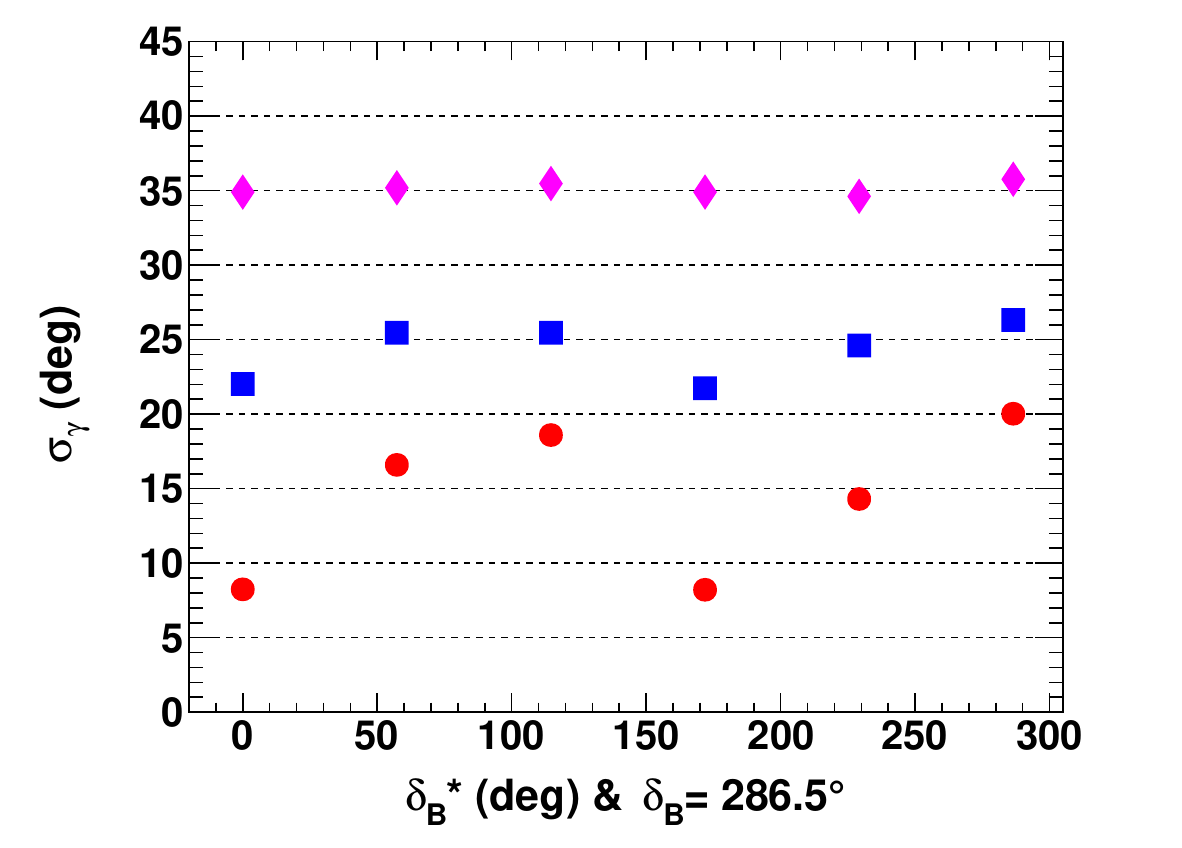} \\
\caption{\label{fig:resolpVal_022}
Half of the 68.3~\% CL intervals of the one-dimension $p$-value profiles of $\gamma$, for Run~1~\&~2 (pink lozenges), for Run $1-3$   (blue squares), and full HL-LHC  (red circles) LHCb dataset, as a function of $\deltaBst$, for $\rBst=0.22$, for an initial true value of  $65.66^\circ$. On each figure, the horizontal dashed black  lines are guide for the eye.}
\end{figure}

Another expected improvement could come from a time-dependent \CP Dalitz plane analysis of the decay $\Bs \rightarrow \tilde{D}^{(*)0}_{CP} \Kp \Km$ as anticipated in Ref.~\cite{Aaij:2018rol}. With the ultimate HL-LHC LHCb dataset, it should be possible to perform such an analysis, thus including the $\BsDtphi$ decay, to extract the CKM angle $\gamma$, as proposed a few years ago in~\cite{Nandi-London}.

or completeness, an alternate definition of the resolution as half of the 68.3~\% CL frequentist intervals of the one-dimension $p$-value profiles of 68.3~\% CL  is given in appendix~\ref{sec:appendE} in Figs.~\ref{fig:resolpVal_04} and~\ref{fig:resolpVal_022}. A better scaling of the performances with size of the datasets is observed, while relatively worse resolutions are obtained with respect to those displayed in Figs.~\ref{fig:AllLumi_rBst04_resol} and~\ref{fig:AllLumi_rBst022_resol}. However, the effects of the nuisance parameters  $\rBst$ and $\deltaBst$ are treated in a simplified way compared to the full treatment by the generated pseudoexperiments.

\subsection{Effect of the strong parameters from $D$-meson decays and of $y=\Delta\Gamma_s/2\Gamma_s$ }
\label{sec:strongDparams}
Most of the strong parameters of the $D$-meson decays to $K\pi$, $K3\pi$, $K\pi\piz$ are external parameters and are obtained from beauty- or charm-factories, such as BaBar, Belle, CLEO-c, LHCb~\cite{HFLAV}. Improvements on their determination are expected soon from  the updated BES-III experiment\cite{Asner:2008nq} or, later on, from future super $\tau$-charm factories~\cite{tau-char_fact}.
To check the impact of those improvements to the $\gamma$ sensitivity, a few scenarios have been tested. With the set of parameters $\gamma=1.146$ rad ($65.66^\circ$), $\rBst=0.4$, and $\deltaB=3.0$ rad and $\deltastB=2.0$, the uncertainties of present measurements of the $D$-meson parameters listed in Table~\ref{tab:para} have been scaled down and their impact on fitted $\gamma$ value from  pseudoexperiments is listed in Table~\ref{tab:Dstrong-params}. Since the uncertainties of the external parameters are presently  not yet dominant (Run~1~\&~2 data), the study is also performed for the expected full HL-LHC dataset. However, with much more data, future improvements on the measurement on the strong parameters from $D$-meson decays  don't seem to impact much the sensitivity to the CKM angle $\gamma$.

This exercise was repeated with the same initial configuration of the parameters $\gamma$, $\rBst$, and $\deltaBst$, for the uncertainty on  $y=\Delta\Gamma_s/2\Gamma_s$. The results of this study are listed in Table~\ref{y-params}. Here again, no obvious sensitivity to those changes is highlighted, neither for Run~1~\&~2, nor for the full HL-LHC dataset. In addition and to our knowledge, it should be stressed that the tested improvement on $y$ are not supported by any published prospective studies.

With the above studies one may conclude that the possibly large correlations of $\gamma$ with respect to the nuisances parameters $\rBst$  and $\deltaBst$ are definitely dominating the ultimate precision on $\gamma$ for the extraction with the $\BsDtphi$ modes.

\subsection{Effect of using or not the $\Bs \rightarrow \Dtstz\phi$ decays}
\label{sec:effectOfDstarphi}

It has been demonstrated in Ref.~\cite{Aaij:2018jqv} that the decays $\Bs \rightarrow \Dtstz\phi$  can be reconstructed in a clean way together with $\Bs \rightarrow \Dtz\phi$, with a similar rate and  a partial reconstruction method, where the $\gamma$ or the $\piz$ produced in the decay of the $\Dtstz$ are omitted. So far those modes were included in the sensitivity studies. Figures~\ref{fig:r_d_BstvsGamma_1-2_dB3dBst2_noDst}-\ref{fig:HL-LHC_dB3dBst2_noDst} in appendix~\ref{sec:appendF} show the 2-D $p$-value  profiles of the nuisance parameters $\rB$ and $\deltaB$ as a function of $\gamma$ and the fit to the distribution of $\gamma$ obtained from 4000 pseudoexperiments for the Run~1~\&~2, Run $1-3$, full HL-LHC  LHCb datasets, for the initial true values: $\gamma=65.66^\circ$ (1.146 rad), $\deltaB=171.9^\circ$ (3.0 rad), $\deltaB=114.6^\circ$ (2.0 rad) and,  $\rBst=0.4$ (0.22). For those figures the information from $\Bs \rightarrow \Dtstz\phi$ decays was not included. According to Figs.~\ref{fig:1-2_dB3dBst2_noDst}, \ref{fig:1-3_dB3dBst2_noDst}, and~\ref{fig:HL-LHC_dB3dBst2_noDst}, in appendix~\ref{sec:appendF}, there is a relative loss on precision to the unfolded value of $\gamma$ of about 20 (40~\%), when the $\Bs \rightarrow \Dtstz\phi$ decay are not used, for $\rBst=0.4$ (0.22). For future datasets the improvement obtained by including $\Bs \rightarrow \Dtstz\phi$ modes is less significant, but not negligible and helps to improve the measurement of $\gamma$.

\section{Conclusions}\label{sec:con}
Untagged $\BsDtphi$ decays provide another theoretically clean path to the measurement of the CKM-angle $\gamma$. By using the expected event yields for $D$ decay to $K\pi$, $K3\pi$, $K\pi\piz$, $KK$, and $\pi\pi$. We  have shown that a precision on $\gamma$ of about $8$ to $19^{\circ}$ can be achieved with LHCb Run~1~\&~2 data. With more data, a precision on  $\gamma$ of $3-8^{\circ}$ can be achieved with the LHCb Run $1-3$ dataset (23\invfb in 2025). Ultimately a precision of the order of $2-7^{\circ}$ has to be expected with the full expected HL-LHC LHCb dataset (300\invfb in 2038). The asymptotic  sensitivity is anyway dominated by the possibly large correlations of $\gamma$ with respect to the nuisances parameters $\rBst$  and $\deltaBst$.  The use of this method will improve our knowledge of $\gamma$ from $\Bs$ decays and help understand the discrepancy of $\gamma$ between measurements with $\Bp$ and $\Bs$ modes.

\begin{table}[h]
\centering
\caption{\label{tab:Dstrong-params}  Fitted resolution of $\gamma$ ($\sigma_{\gamma}$) in [deg] obtained from 4000 pseudoexperiments, as a function of decreasing uncertainties of the strong $D$-meson parameters (see Table~\ref{tab:para}). }
\begin{tabular}{l|rrrrr}
\hline \hline
uncertainties on $D$-meson params.  &  Now & $\times 1/2$   & $\times 1/5$  & $\times 1/10$ \\
\hline
Run~1~\&~2 ($\rBst=0.4$) & $ 8.8 \pm 0.2$ & $ 8.1 \pm 0.3 $  & $ 8.0\pm 0.3$ & $7.8\pm0.2 $ \\
Run~1~\&~2 ($\rBst=0.22$) & $12.9 \pm 0.3$ & $13.2\pm 0.5$ & $13.1\pm0.5$ & $12.8\pm0.9$ \\
full HL-LHC ($\rBst=0.4$) & $2.6\pm0.1$ & $2.5\pm0.1$ & $2.5\pm0.1$ & $2.5\pm0.1$ \\
full HL-LHC ($\rBst=0.22$) & $5.4\pm0.1$ & $5.3\pm0.1$ & $5.2\pm0.1$ & $5.1\pm0.1$ \\
\end{tabular}
\end{table}

\begin{table}[h]
\centering
\caption{\label{y-params}  Fitted resolution of $\gamma$ ($\sigma_{\gamma}$) in [deg] obtained from 4000 pseudoexperiments, as a function of decreasing uncertainties of $y=\Delta\Gamma_s/2\Gamma_s$. For the full HL-LHC dataset uncertainties for the strong $D$-meson parameters are divided by 2 with respect to the present measurement (see Table~\ref{tab:para}). }
\begin{tabular}{l|rrrrr}
\hline \hline
uncertainty on  $y=\Delta\Gamma_s/2\Gamma_s$ &  Now & $\times 1/2$   & $\times 1/5$  & $\times 1/10$ \\
\hline
Run~1~\&~2 ($\rBst=0.4$) & $ 8.8 \pm 0.2$ & $ 8.3 \pm 0.2 $  & $ 8.2\pm 0.2$ & $8.1\pm0.3 $ \\
Run~1~\&~2 ($\rBst=0.22$) & $12.9 \pm 0.3$ & $12.6\pm 0.4$ & $12.5\pm0.5$ & $12.5\pm0.5$ \\
full HL-LHC ($\rBst=0.4$) & $2.5\pm0.1$ & $2.5\pm0.1$ & $2.5\pm0.1$ & $2.5\pm0.1$  \\
full HL-LHC ($\rBst=0.22$) & $5.3\pm0.1$ & $5.3\pm0.1$ & $5.2\pm0.1$  & $5.2\pm0.1$ \\
\end{tabular}
\end{table}

\vspace{-1mm}
\centerline{\rule{80mm}{0.1pt}}
\vspace{2mm}


\acknowledgments{
 We are grateful to all the members of the CKMfitter group for their comments and for providing us with their private software based on a frequentist approach, while computing the many pseudoexperiments performed for this study. We especially would like to thanks J.~Charles, for his helpful comments while starting this analysis. We acknowledge support from National Natural Science Foundation of China (NSFC) under Contracts Nos. 11925504, 11975015; the 65$^{th}$ batch of China Postdoctoral Fund; the Fundamental Research Funds for the Central Universities, CNRS/IN2P3 (France), and STFC (United Kingdom) national agencies. Part of this work was supported through exchanges between Annecy, Beijing, and Clermont-Ferrand, by the France China Particle Physics Laboratory ({\it {\it i.e.}} FCPPL).}


\vspace{-1mm}
\centerline{\rule{80mm}{0.1pt}}
\vspace{2mm}

\newpage
\section{Appendix A: Fitted nuisance parameters $\rBst$ and $\deltaBst$} \label{sec:appendA}

\begin{figure}[h]
\centering
\includegraphics[width=0.425\textwidth]{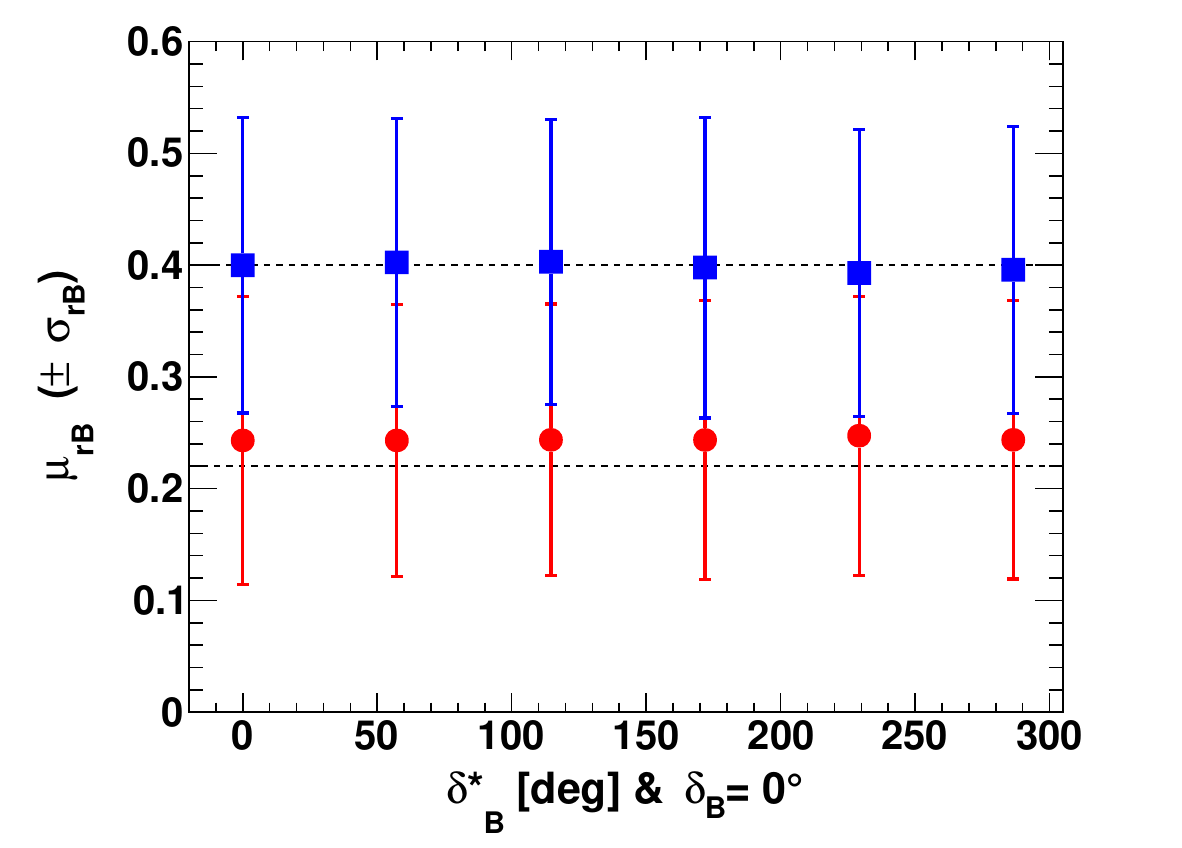}
\includegraphics[width=0.425\textwidth]{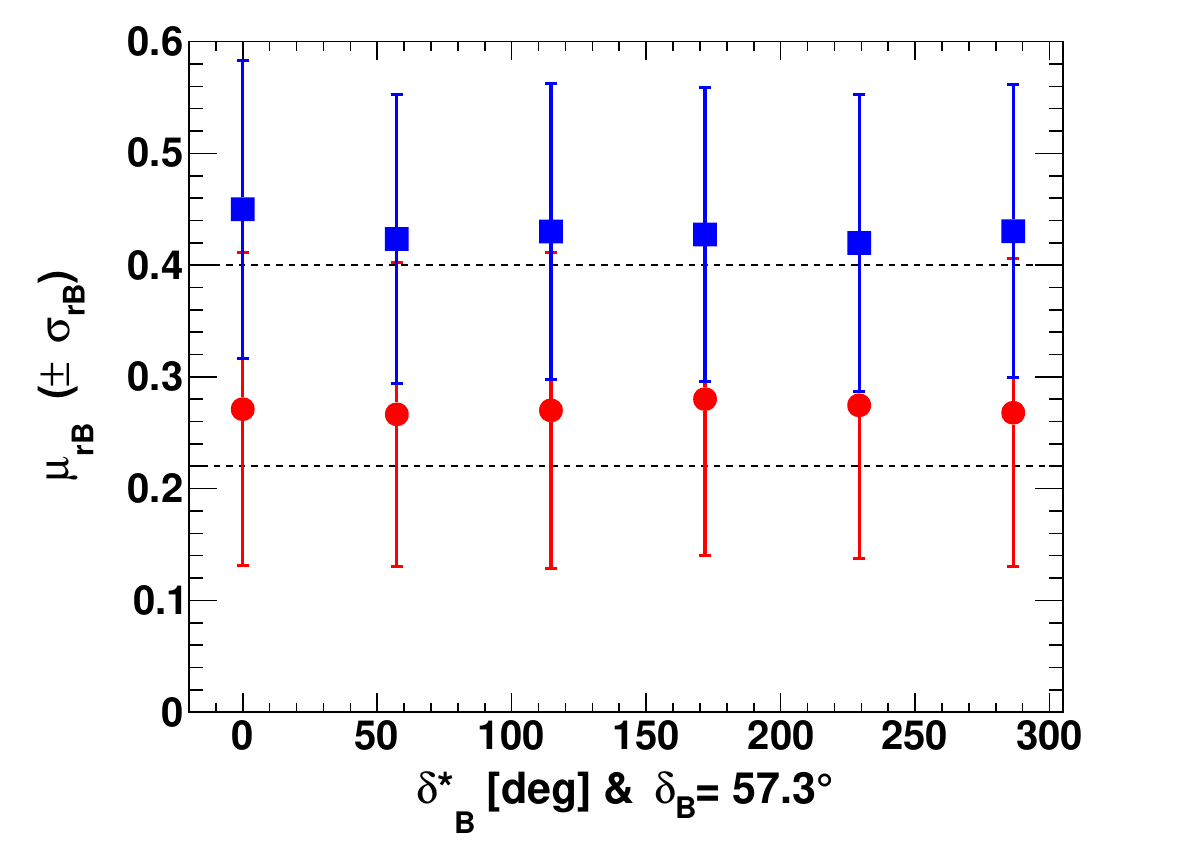} \\
\includegraphics[width=0.425\textwidth]{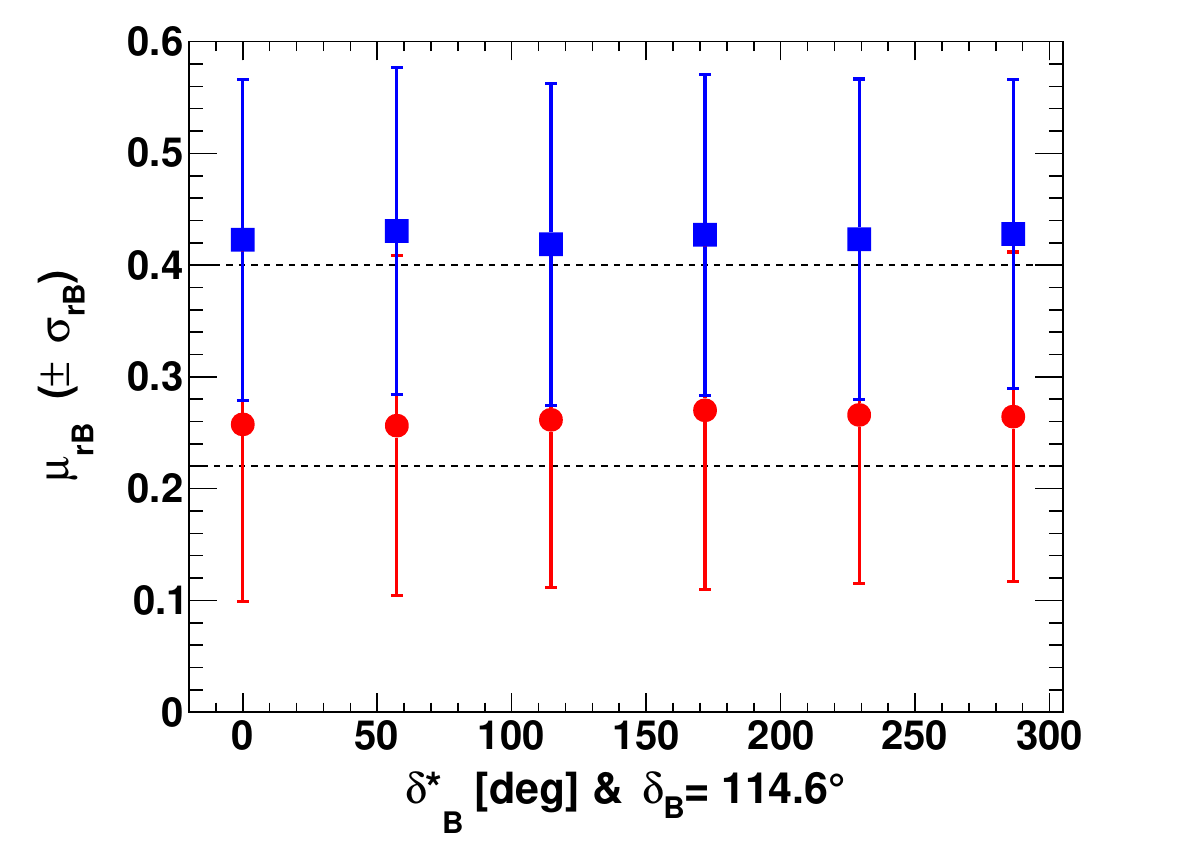}
\includegraphics[width=0.425\textwidth]{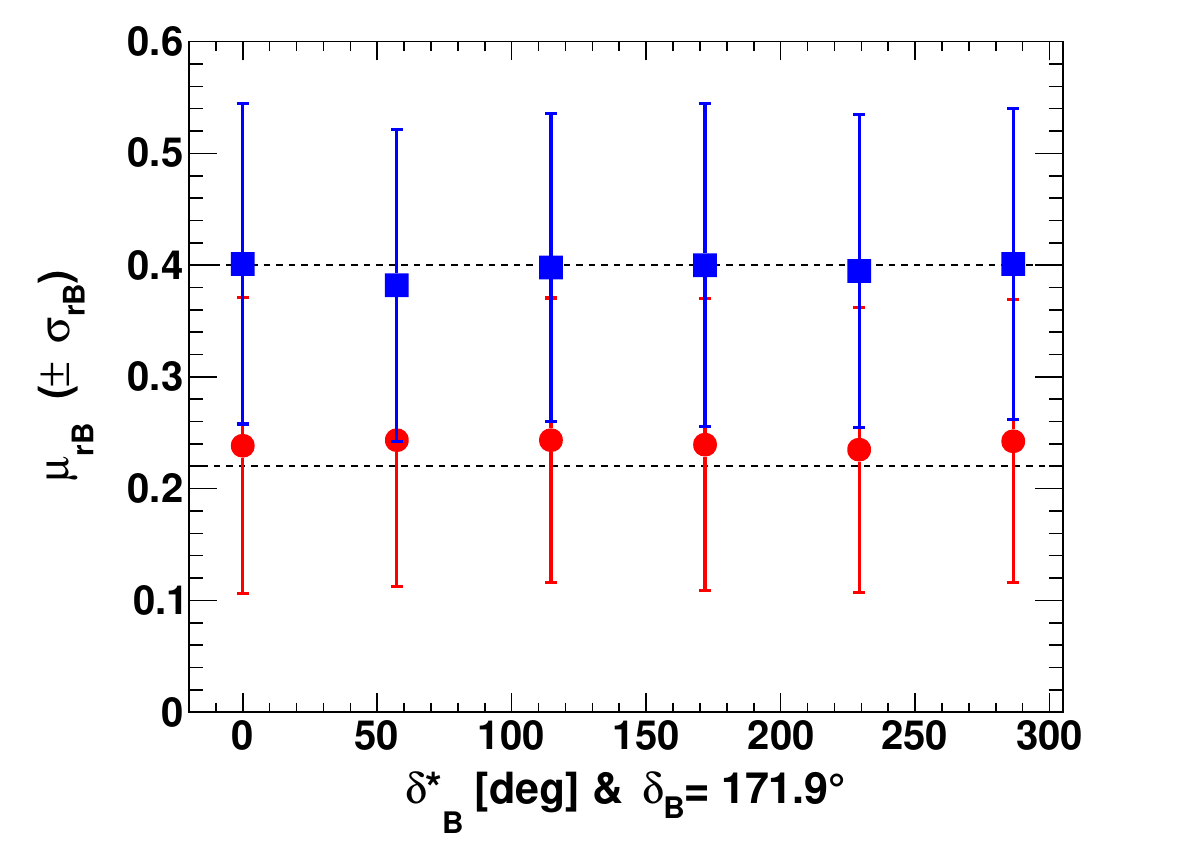} \\
\includegraphics[width=0.425\textwidth]{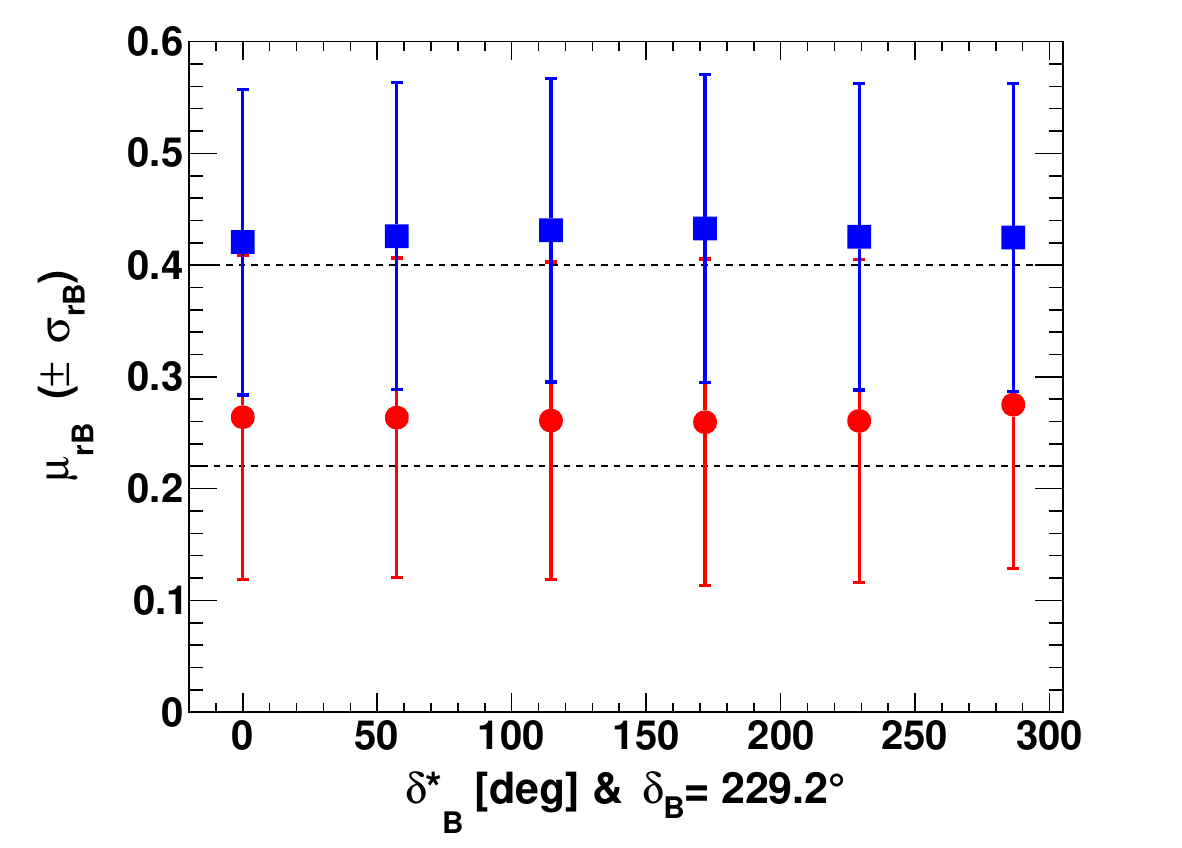}
\includegraphics[width=0.425\textwidth]{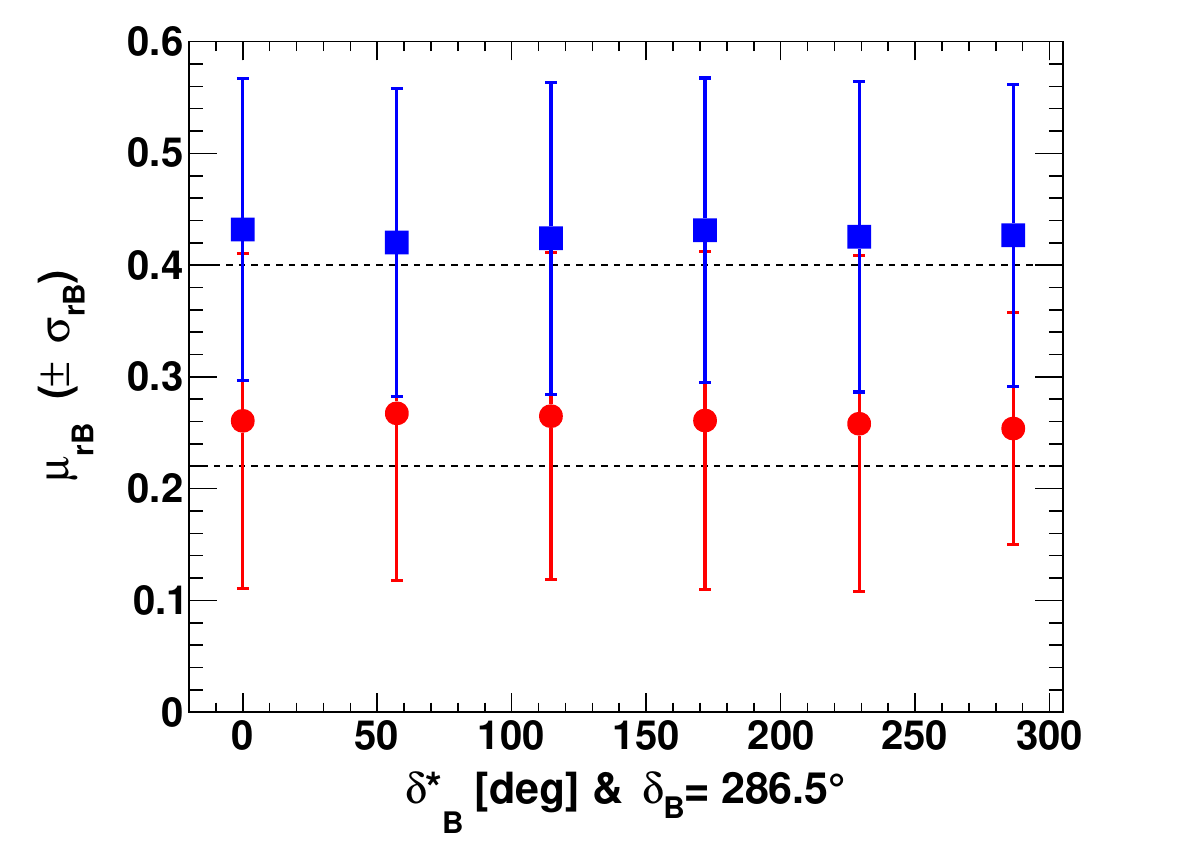}
\caption{\label{fig:dbeffect_rBst022_04_rB}Fitted mean value of $\rB$ ($\mu_{\rB}$), for $\rBst=0.22$ (red circles) and 0.4 (blue squares), as a function of $\deltaBst$, for an initial true value of $\gamma$ of $65.66^\circ$ (1.146 rad). On each figure, the horizontal dashed black  line indicates the initial  $\rB$ true value and the displayed uncertainties are the fitted resolutions on $\rB$ ({\it i.e.} $\sigma_{\rB}$). }
\end{figure}

\newpage
\begin{figure}[h]
\centering
\includegraphics[width=0.425\textwidth]{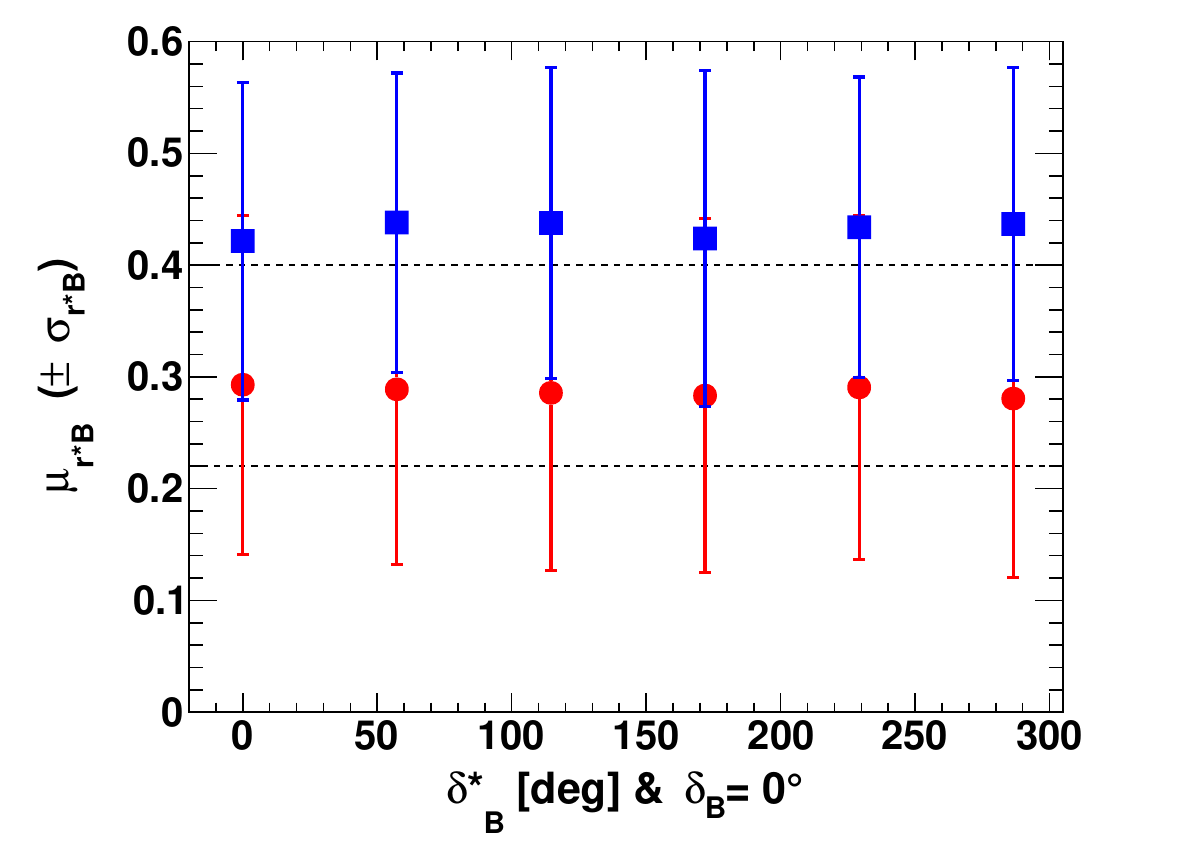}
\includegraphics[width=0.425\textwidth]{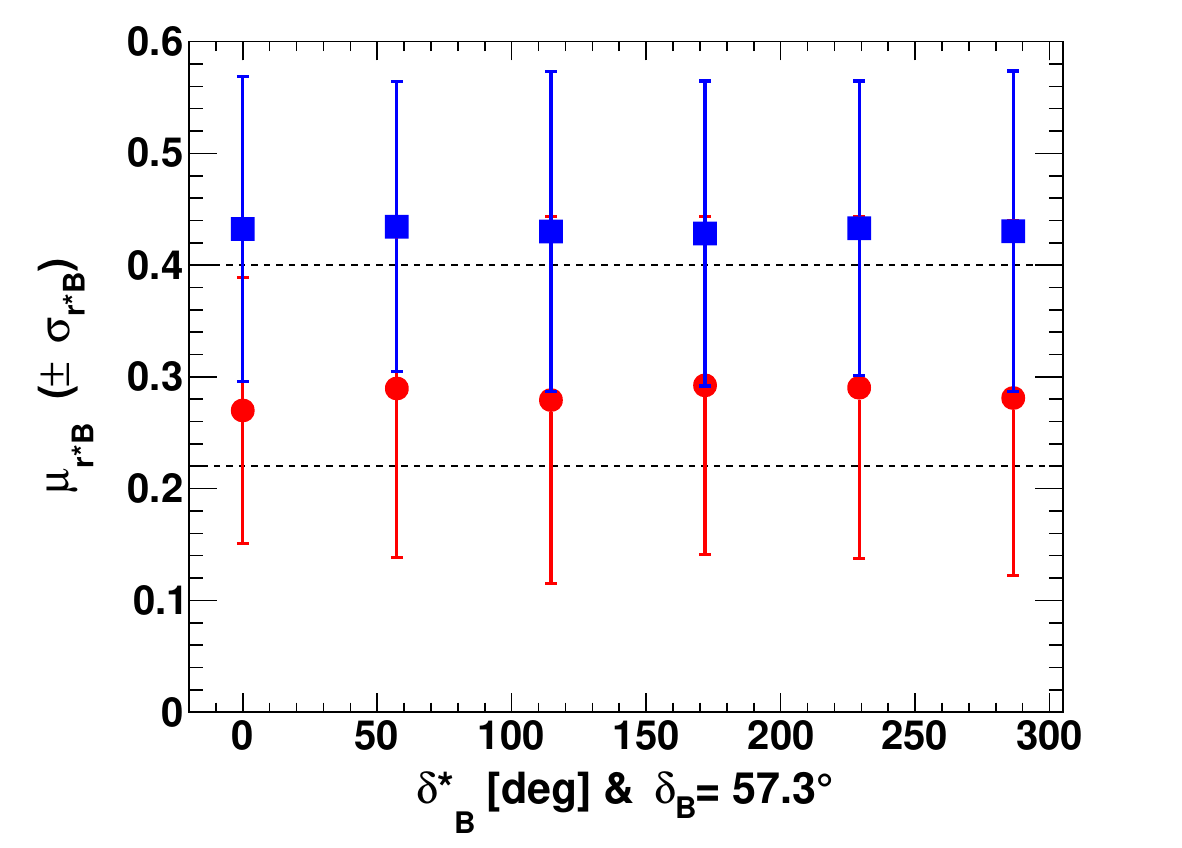} \\
\includegraphics[width=0.425\textwidth]{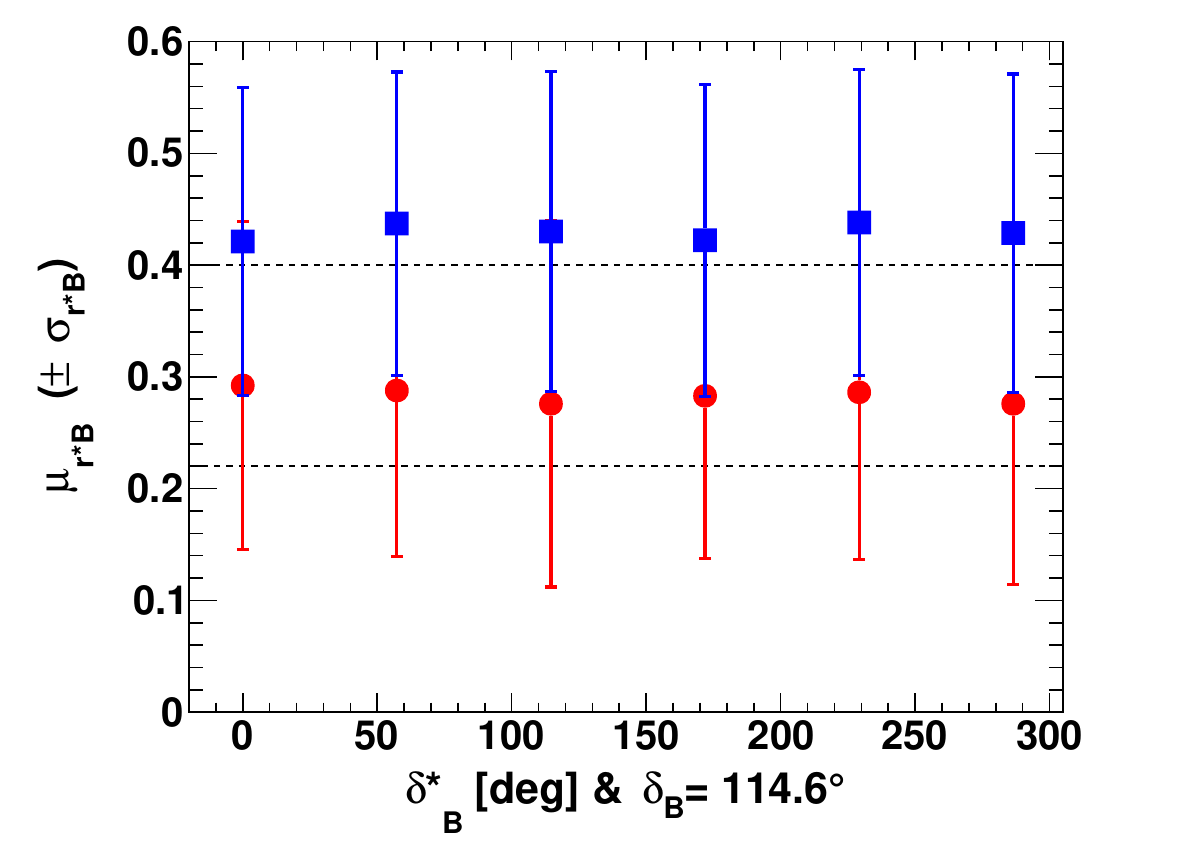}
\includegraphics[width=0.425\textwidth]{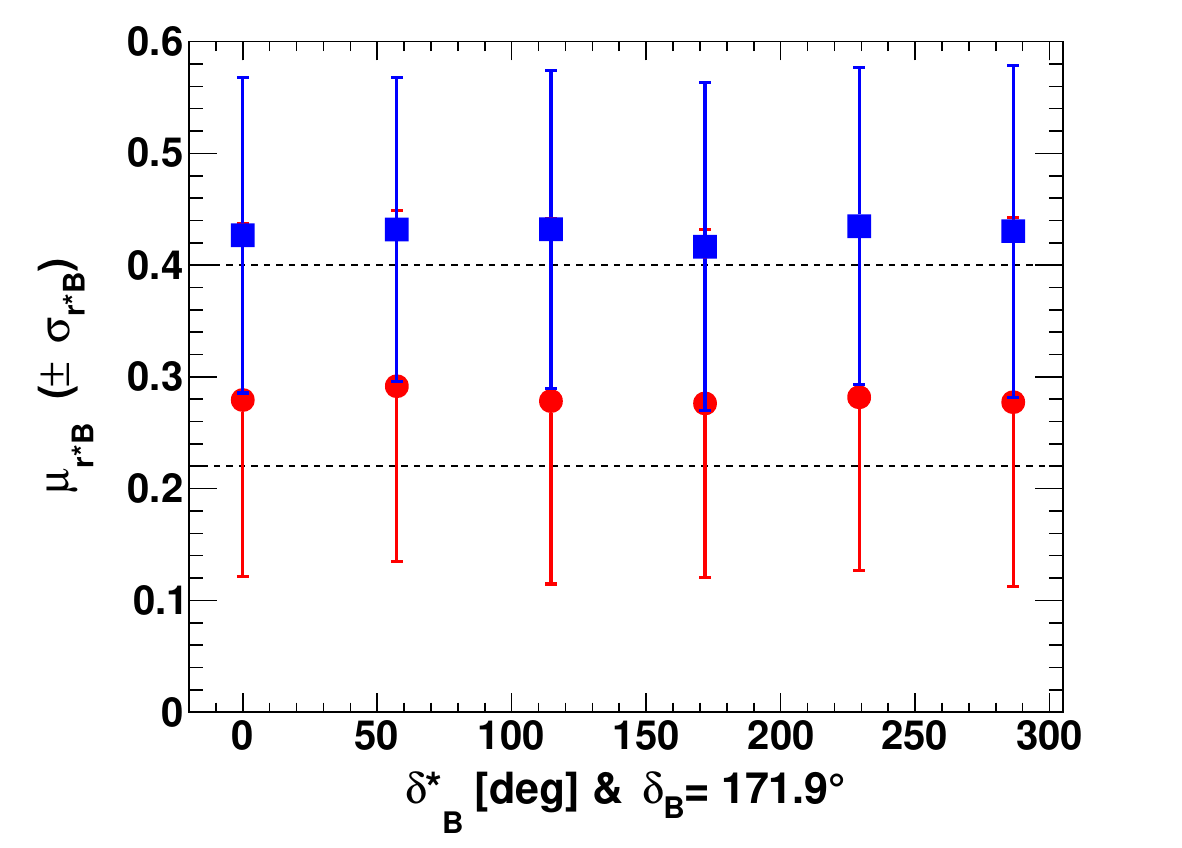} \\
\includegraphics[width=0.425\textwidth]{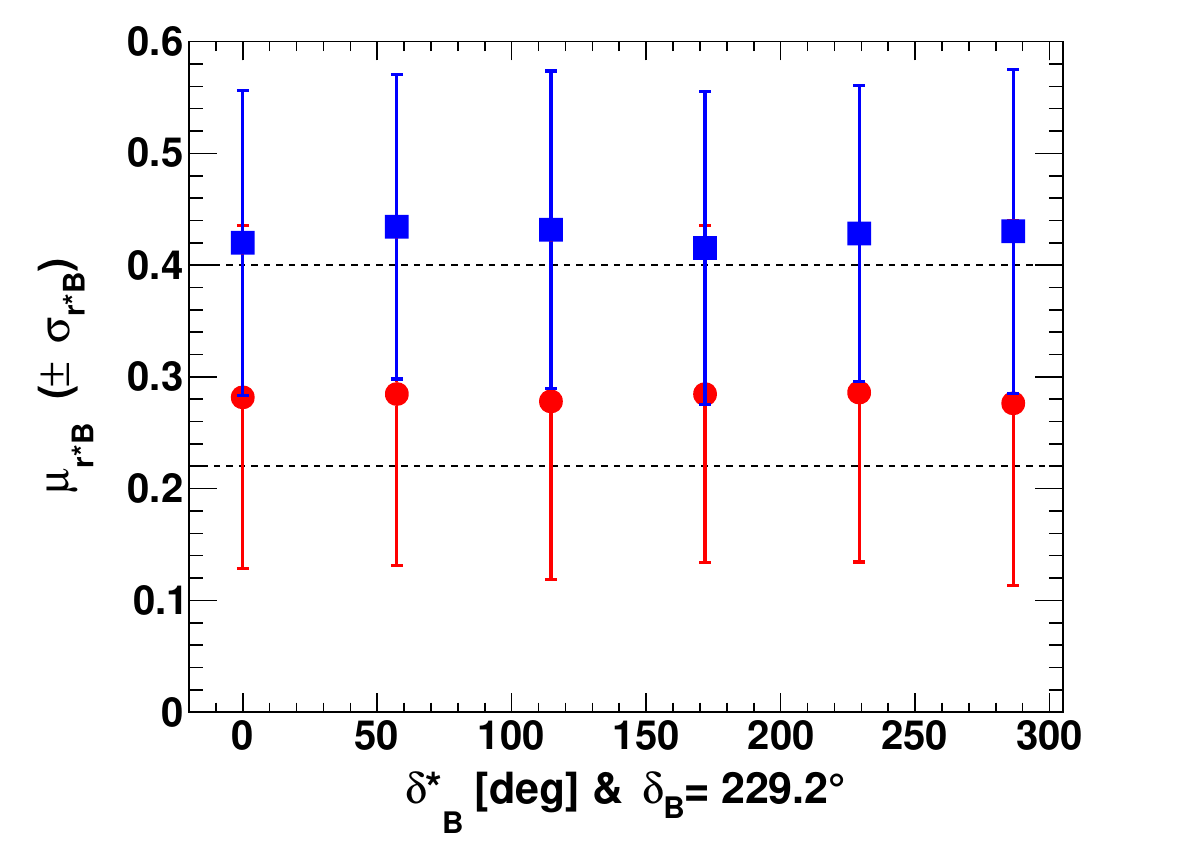}
\includegraphics[width=0.425\textwidth]{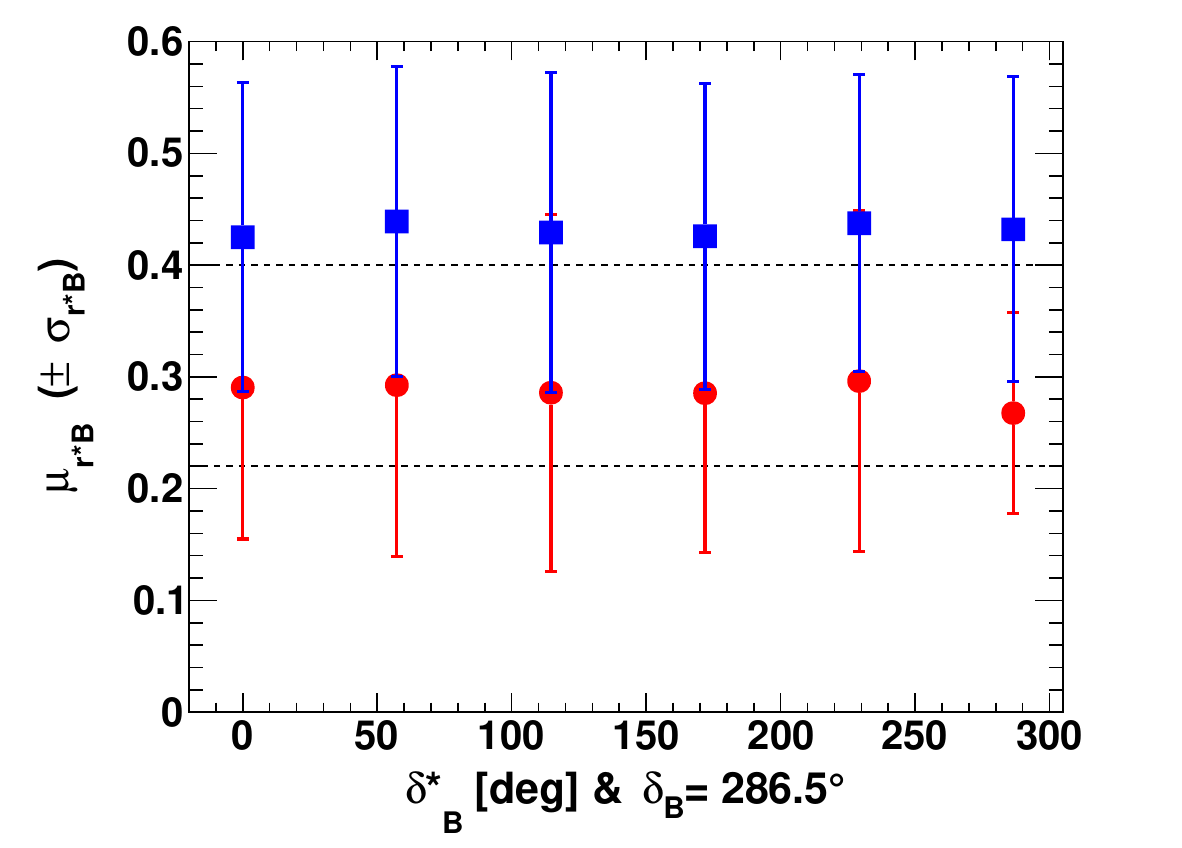}
\caption{\label{fig:dbeffect_rBst022_04_rBst}Fitted mean value of $\rstB$ ($\mu_{\rstB}$), for $\rBst=0.22$ (red circles) and 0.4 (blue squares), as a function of $\deltaBst$, for an initial true value of $\gamma$  of $65.66^\circ$ (1.146 rad). On each figure, the horizontal dashed black  line indicates the initial  $\rstB$ true value and the displayed uncertainties are the fitted resolutions on $\rstB$ ({\it i.e.} $\sigma_{\rstB}$).  }
\end{figure}

\newpage

\begin{figure}[h]
\centering
\includegraphics[width=0.425\textwidth]{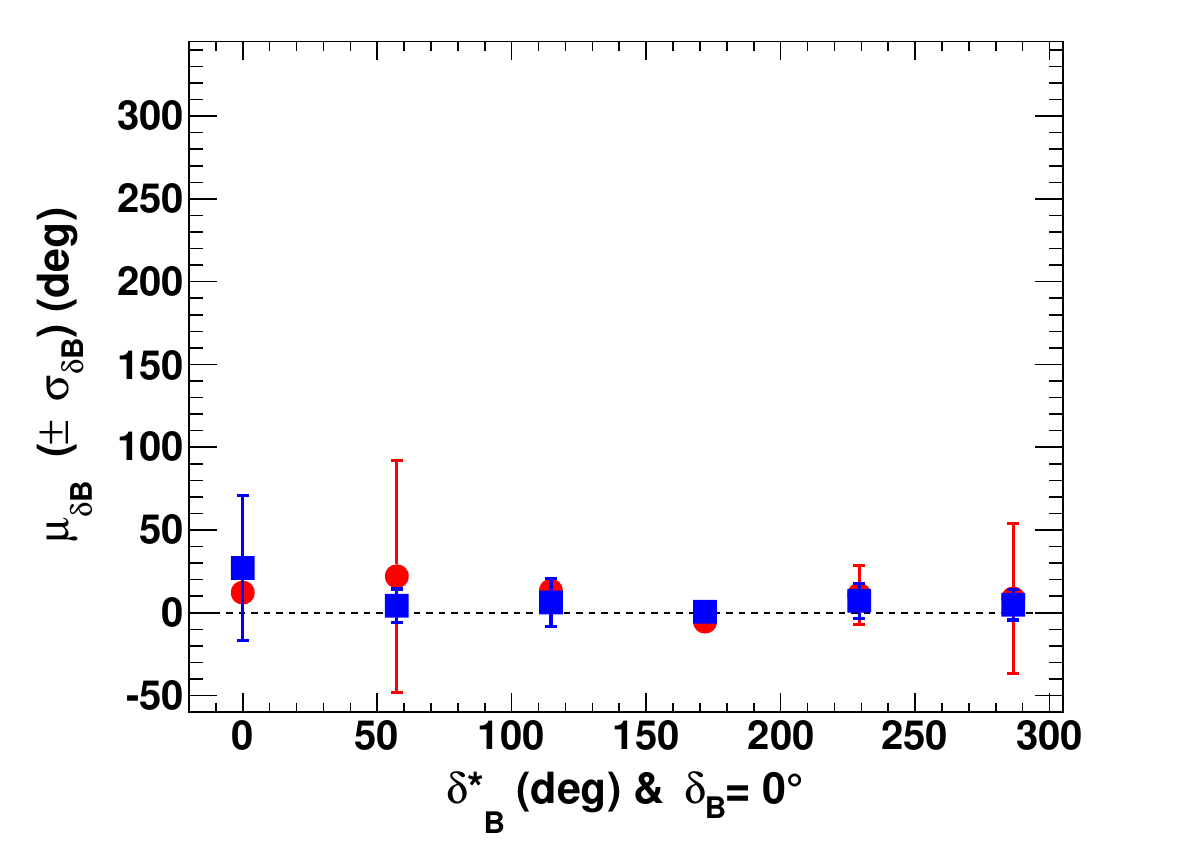}
\includegraphics[width=0.425\textwidth]{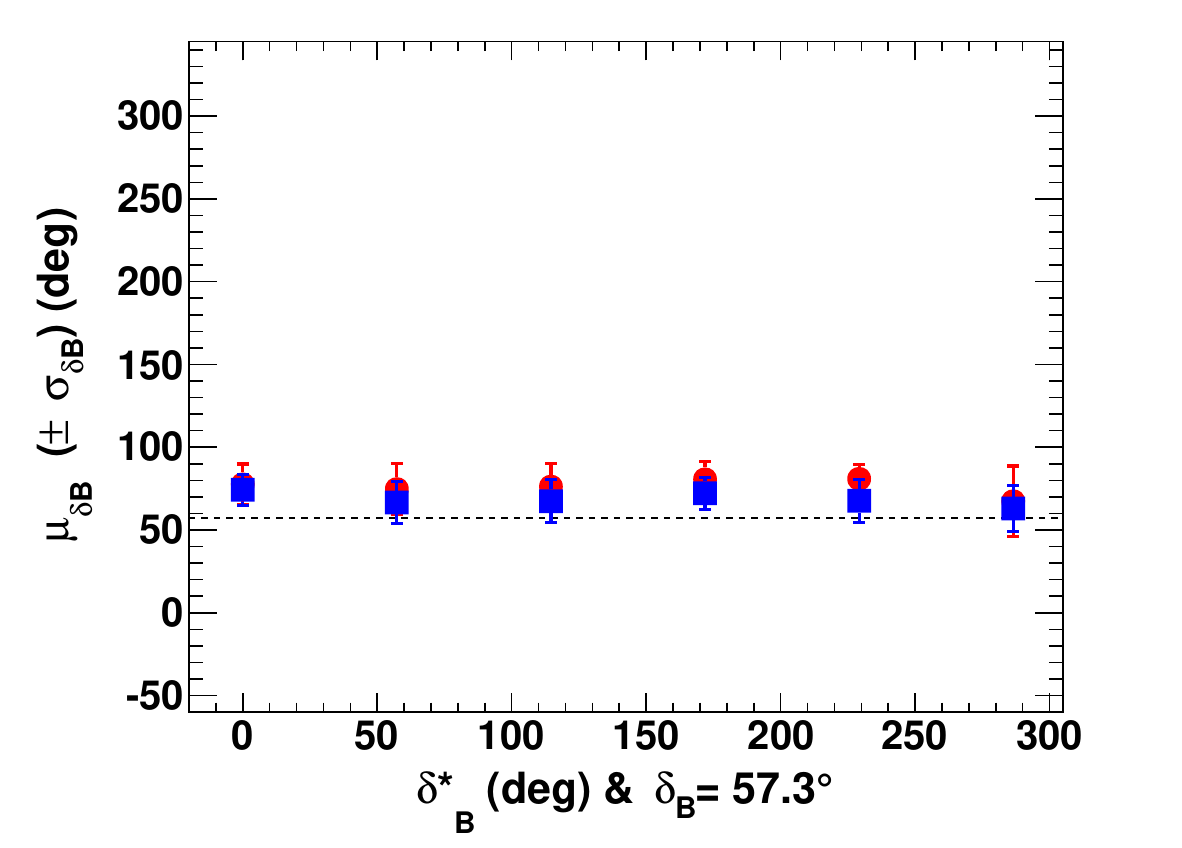} \\
\includegraphics[width=0.425\textwidth]{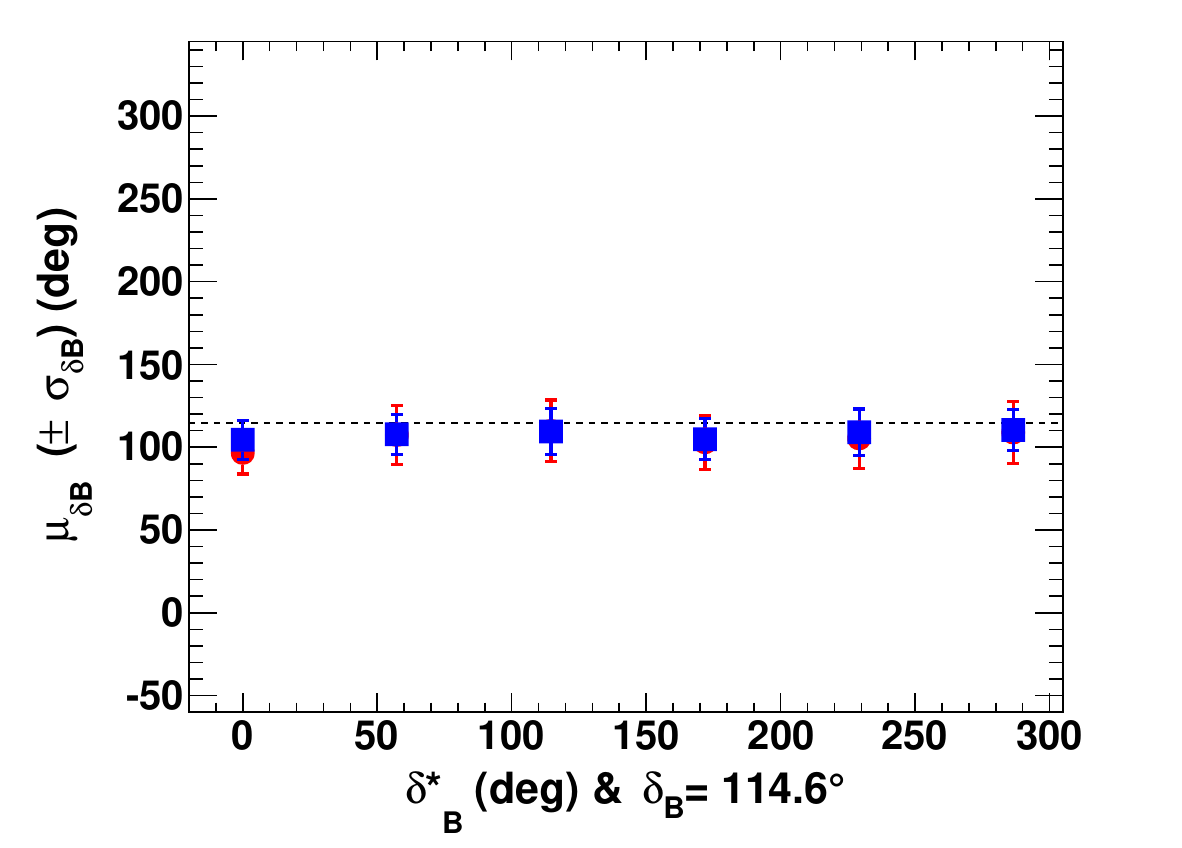}
\includegraphics[width=0.425\textwidth]{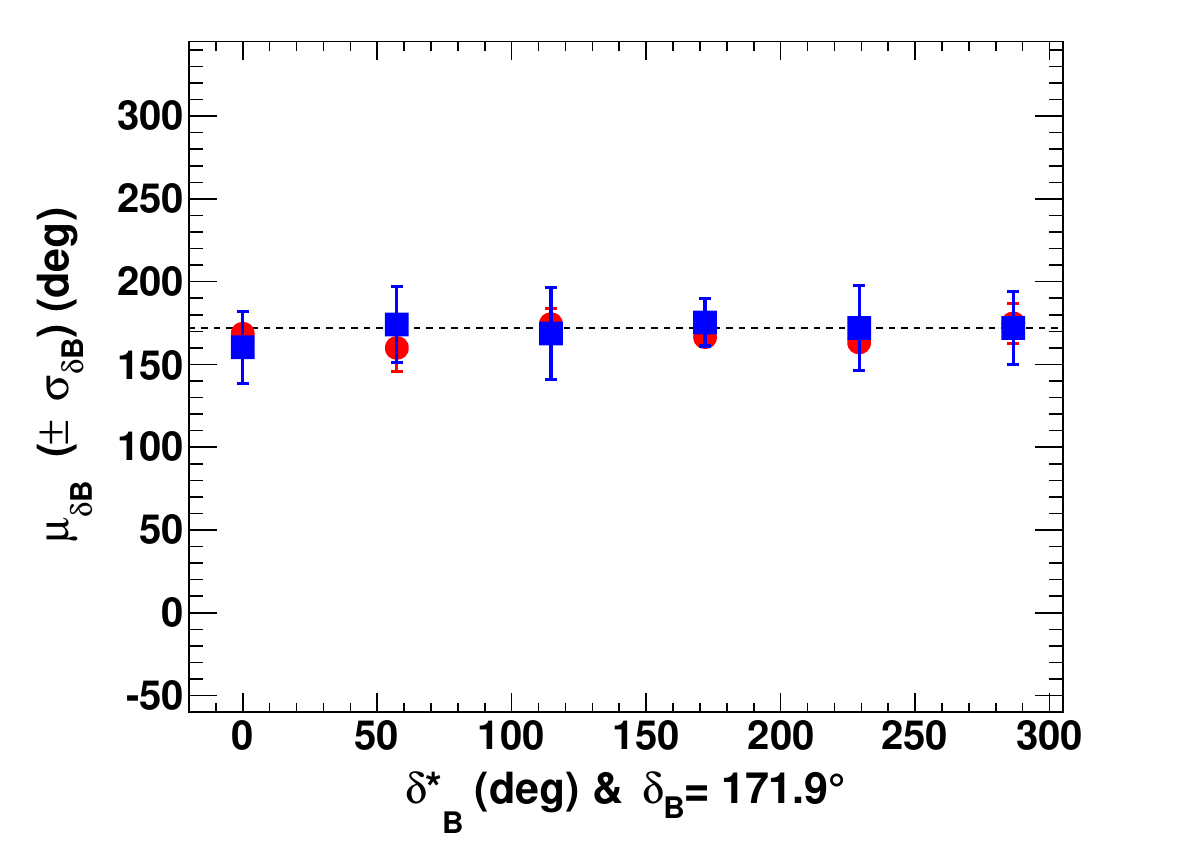} \\
\includegraphics[width=0.425\textwidth]{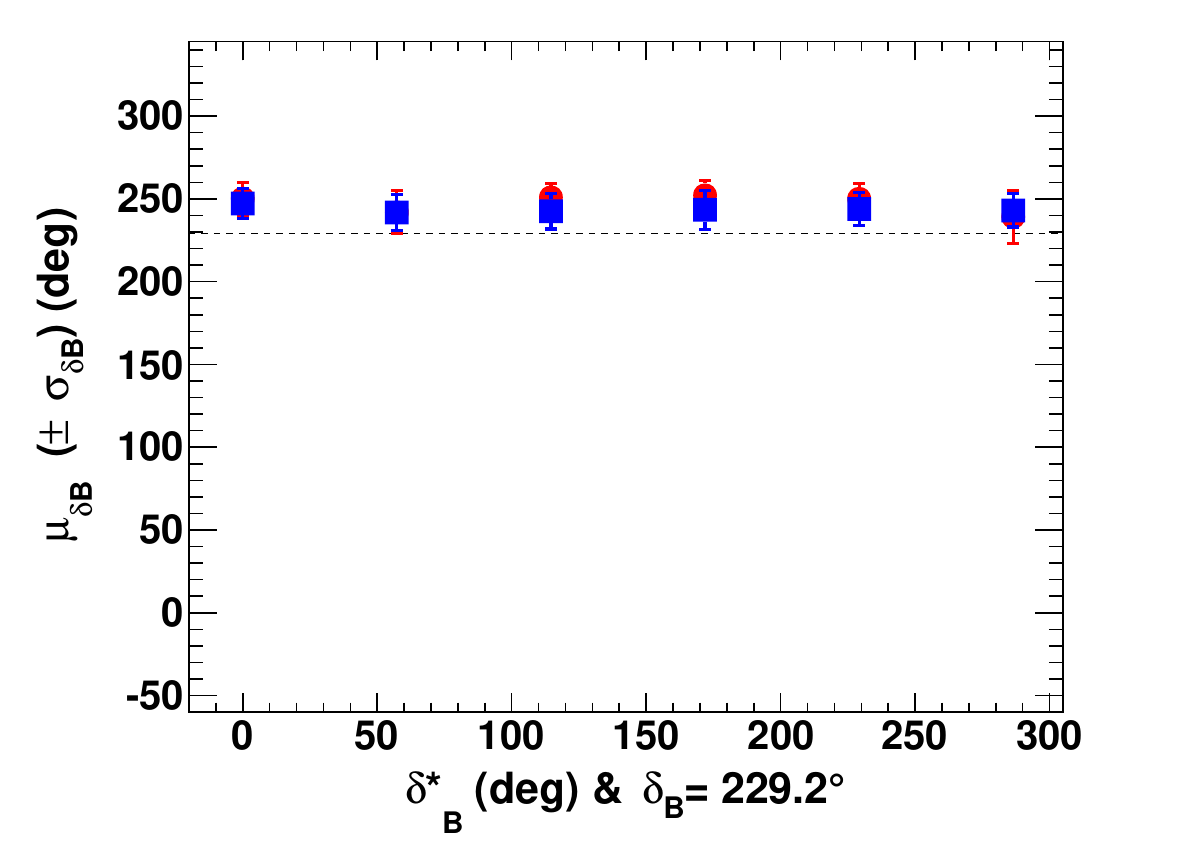}
\includegraphics[width=0.425\textwidth]{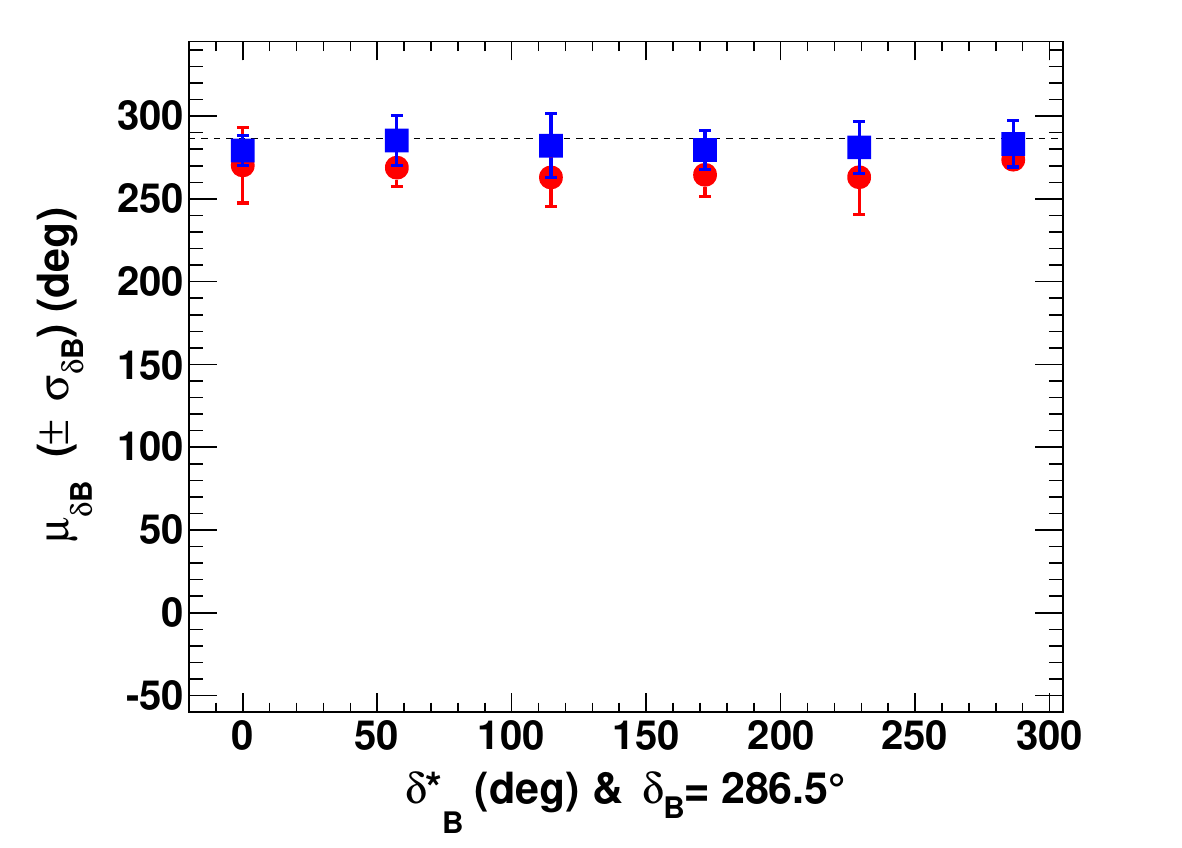}
\caption{\label{fig:dbeffect_rBst022_04_dB}Fitted mean value of $\deltaB$ ($\mu_{\deltaB}$), for $\rBst=0.22$ (red circles) and 0.4 (blue squares), as a function of $\deltaBst$, for an initial true value of $\gamma$ of  $65.66^\circ$ (1.146 rad). On each figure, the horizontal dashed black  line indicates the initial  $\deltaB$ true value and the displayed uncertainties are the fitted resolutions on $\deltaB$ ({\it i.e.} $\sigma_{\deltaB}$).}
\end{figure}

\newpage

\begin{figure}[h]
\centering
\includegraphics[width=0.425\textwidth]{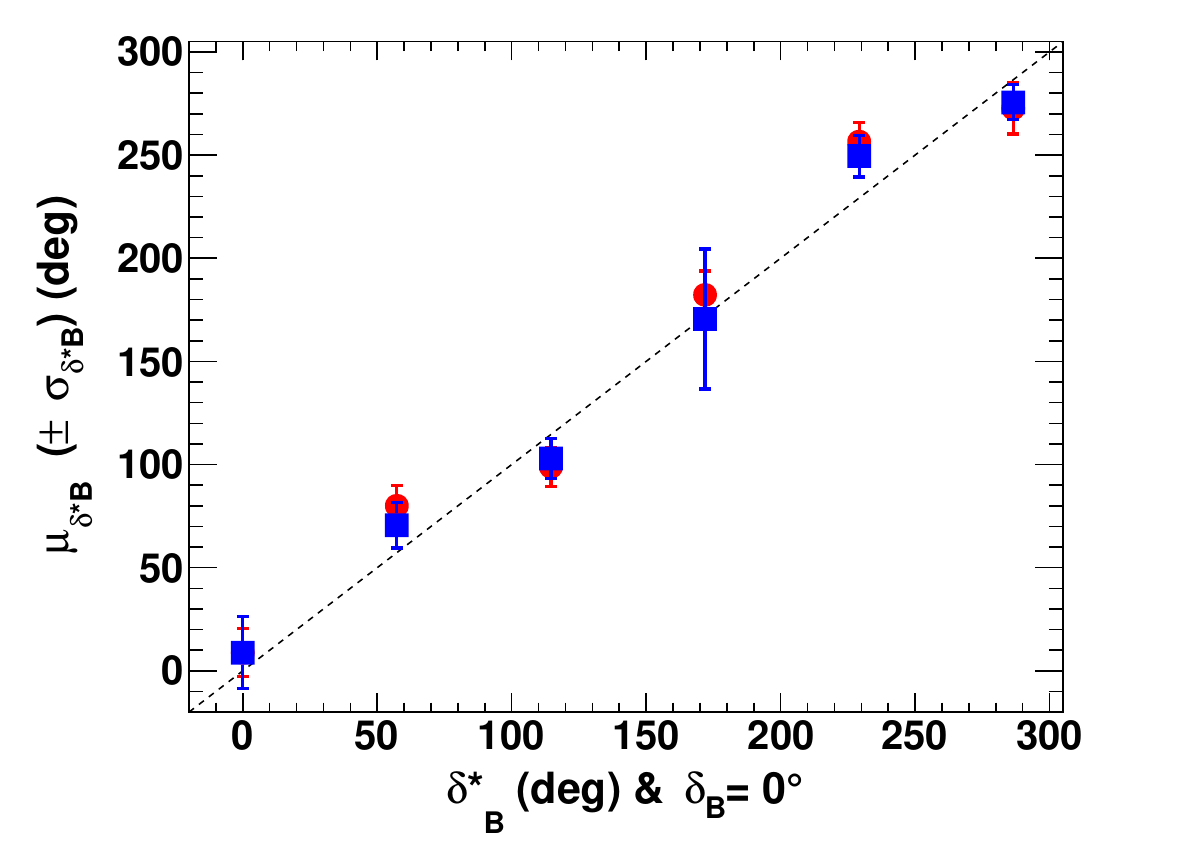}
\includegraphics[width=0.425\textwidth]{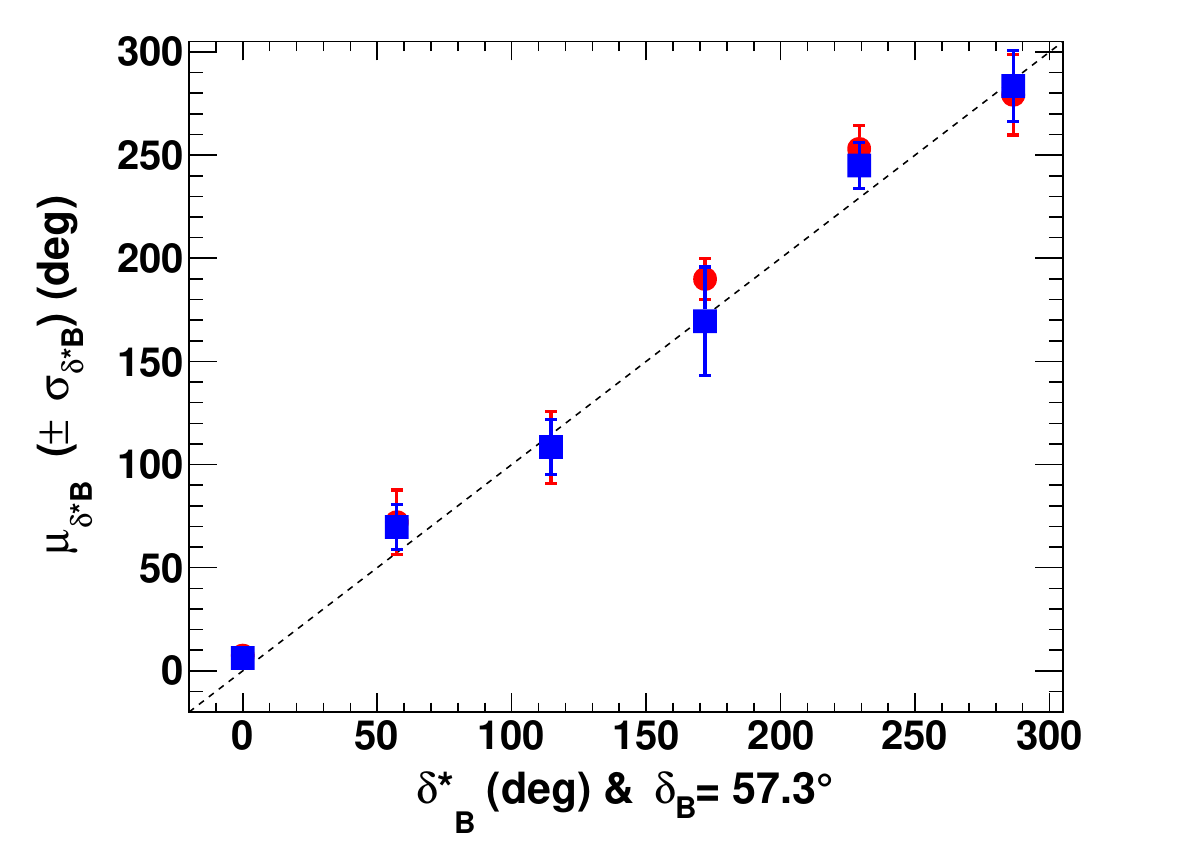} \\
\includegraphics[width=0.425\textwidth]{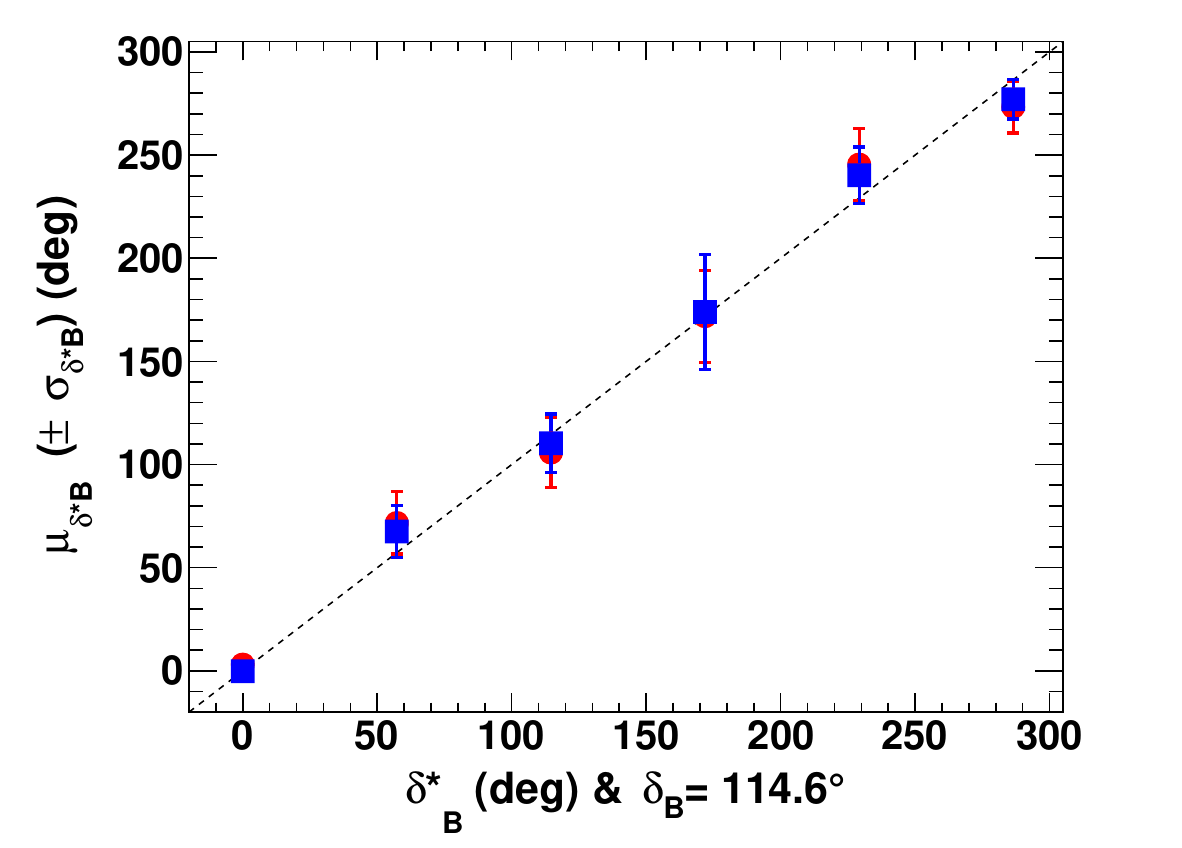}
\includegraphics[width=0.425\textwidth]{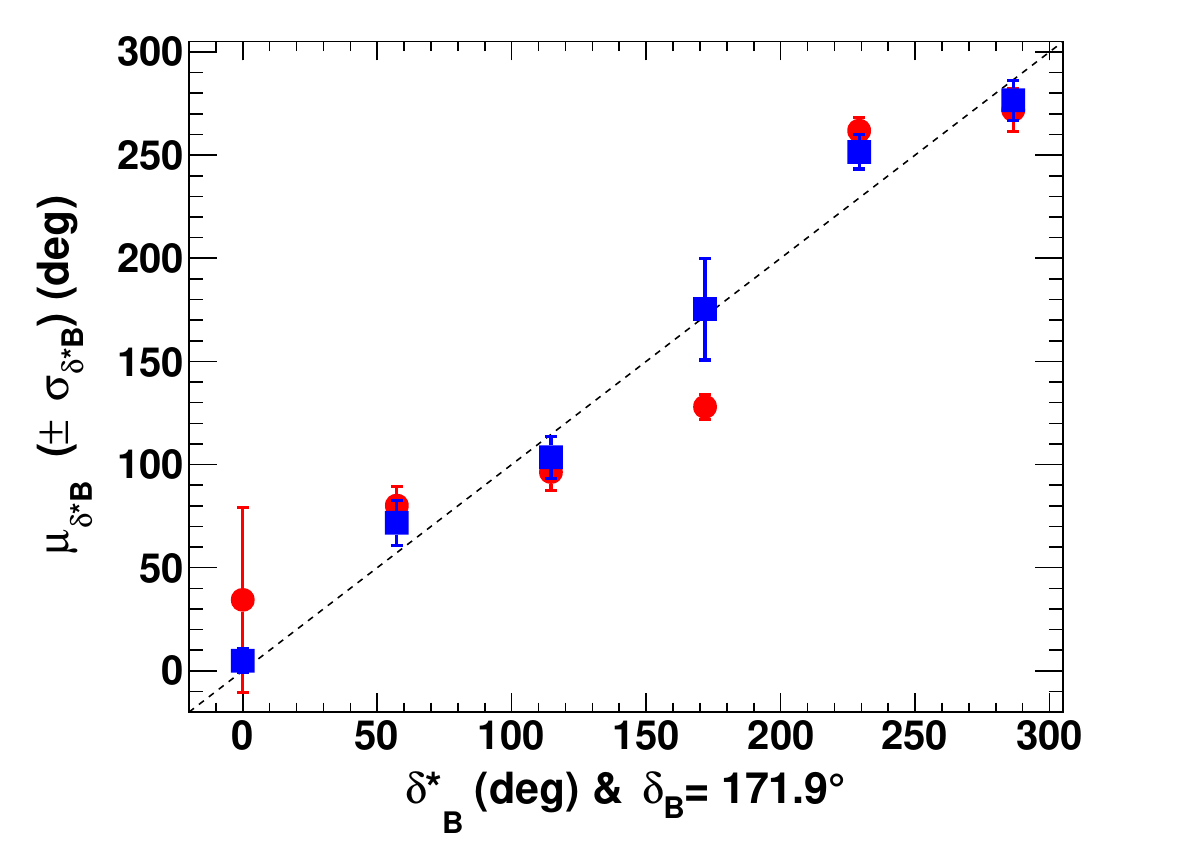} \\
\includegraphics[width=0.425\textwidth]{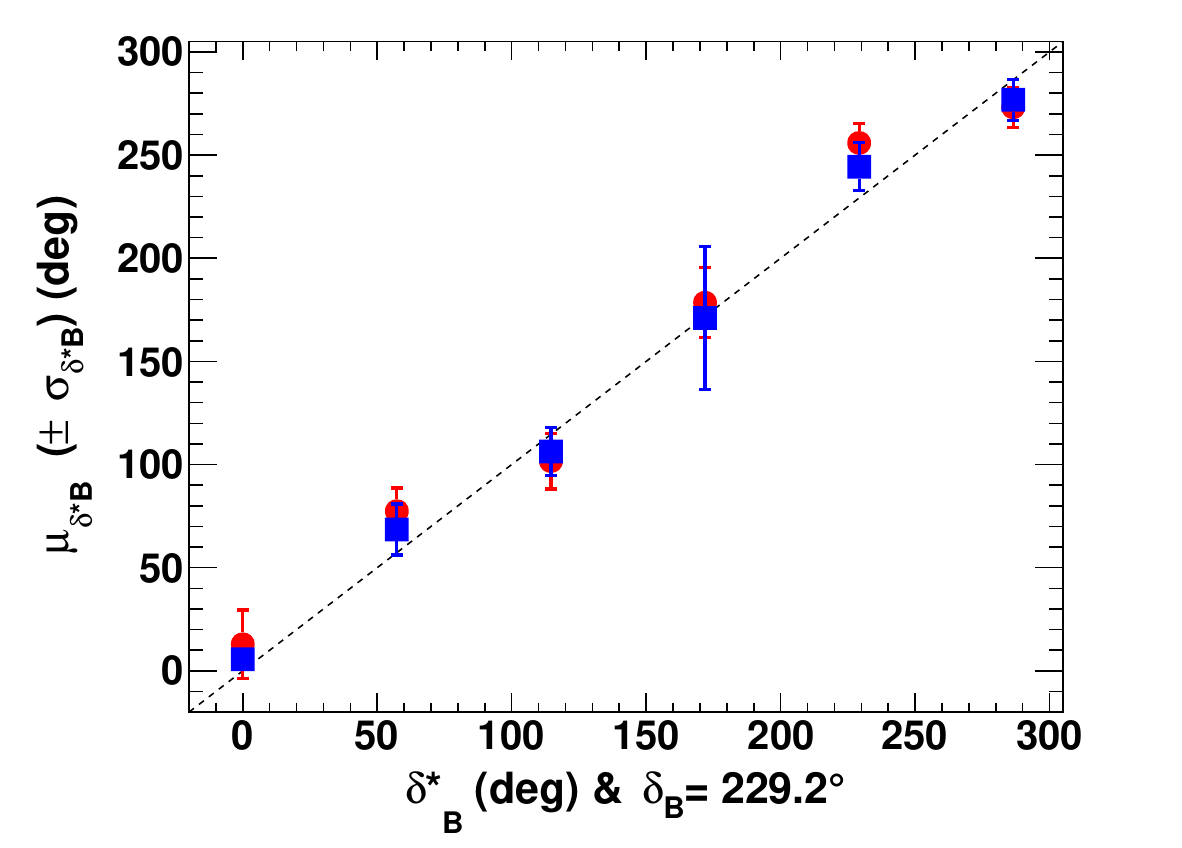}
\includegraphics[width=0.425\textwidth]{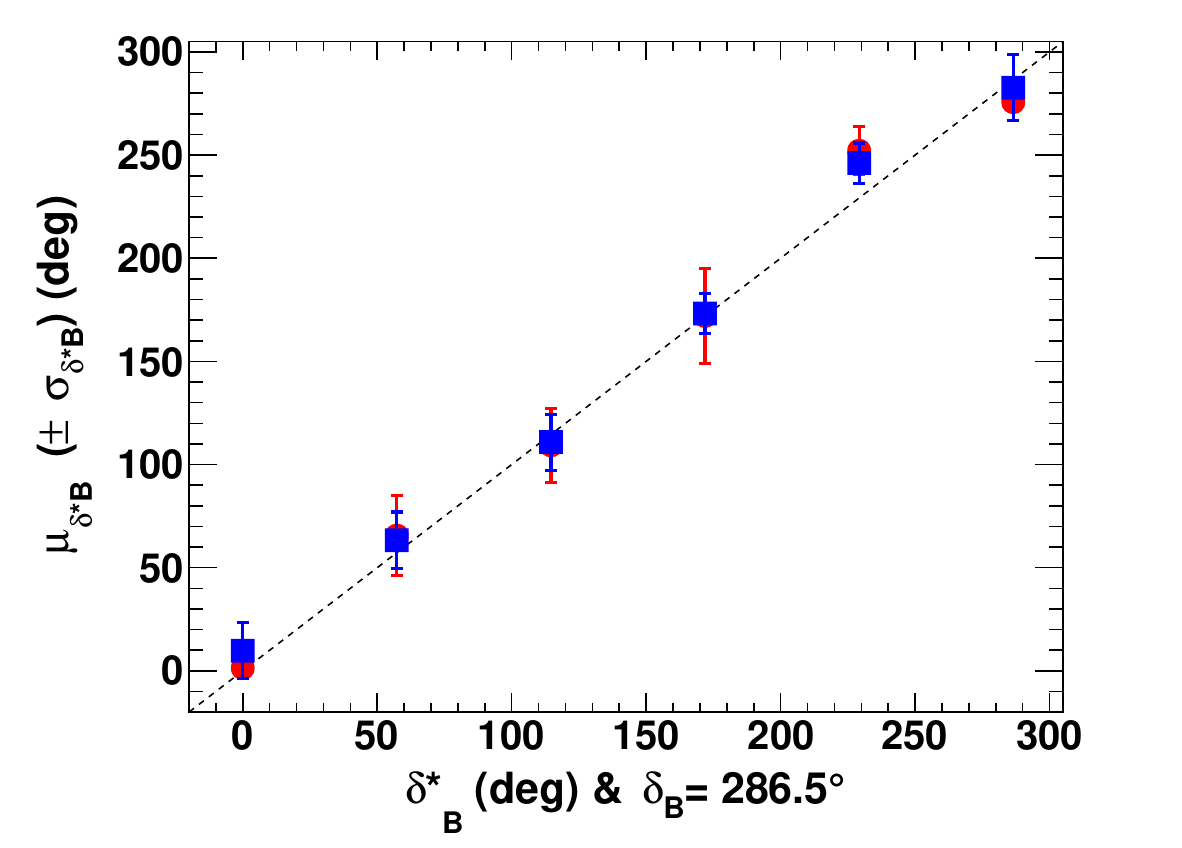}
\caption{\label{fig:dbeffect_rBst022_04_dBst}Fitted mean value of $\deltastB$ ($\mu_{\deltastB}$), for $\rBst=0.22$ (red circles) and 0.4 (blue squares), as a function of $\deltaBst$, for an initial true value of $\gamma$  of  $65.66^\circ$ (1.146 rad). On each figure, the dashed black  line indicates the initial  $\deltastB$ true value and the displayed uncertainties are the fitted resolutions on $\deltastB$ ({\it i.e.} $\sigma_{\deltastB}$).}
\end{figure}

\clearpage

\section{Appendix B: The case $\gamma$ equals $74^\circ$} \label{sec:appendB}
\label{sec:appendB}

\begin{figure}[h]
\centering
\includegraphics[width=0.425\textwidth]{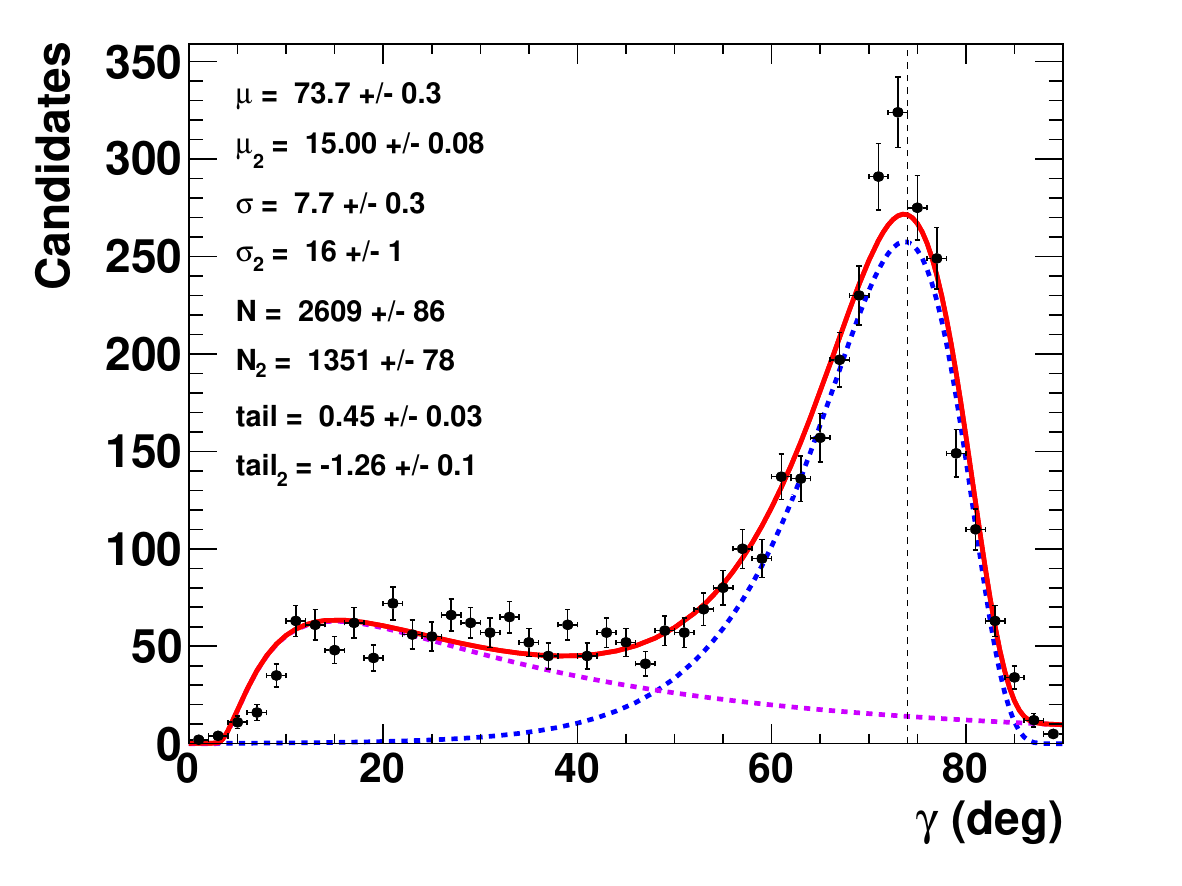}
\includegraphics[width=0.425\textwidth]{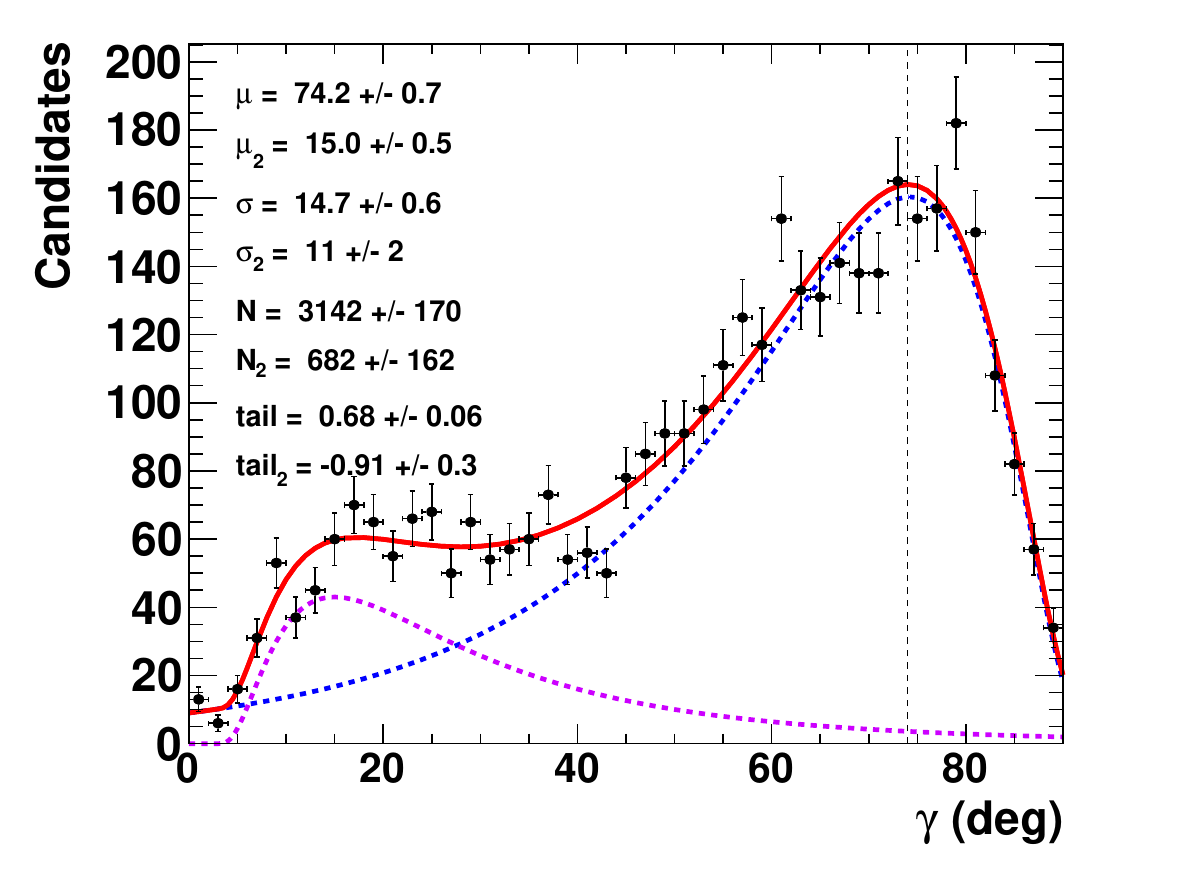}
\caption{\label{fig:74deg_dB3dBst2} Fit to the distributions of the  nuisance parameters  $\gamma$ obtained from 4000 pseudoexperiments. The initial configuration is $\gamma=74^{\circ}$, $\rBst=0.4$ (left) and 0.22 (right), $\deltaB=171.9^\circ$ (3 rad), and $\deltastB=114.6^\circ$ (2 rad). In the distributions, only the candidates with a value  $\gamma\  \in \ [0^\circ, \  90^\circ]$ are considered. The purple dashed curve accounts for tails generated by the correlations with the nuisance parameters $\rBst$ and $\deltaBst$, while the blue dashed curve is the core part of the distribution, the plain red line is the sum of the two components of the fit.}
\includegraphics[width=0.425\textwidth]{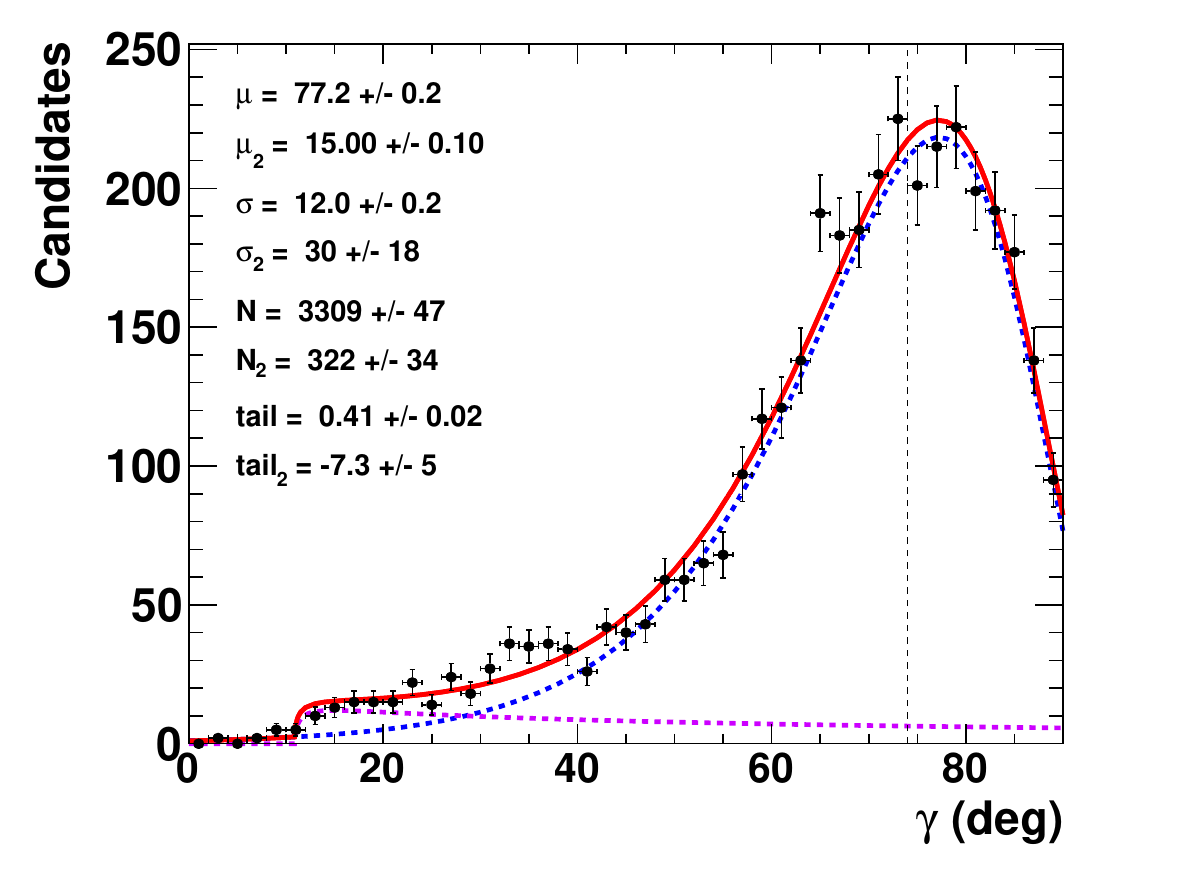}
\includegraphics[width=0.425\textwidth]{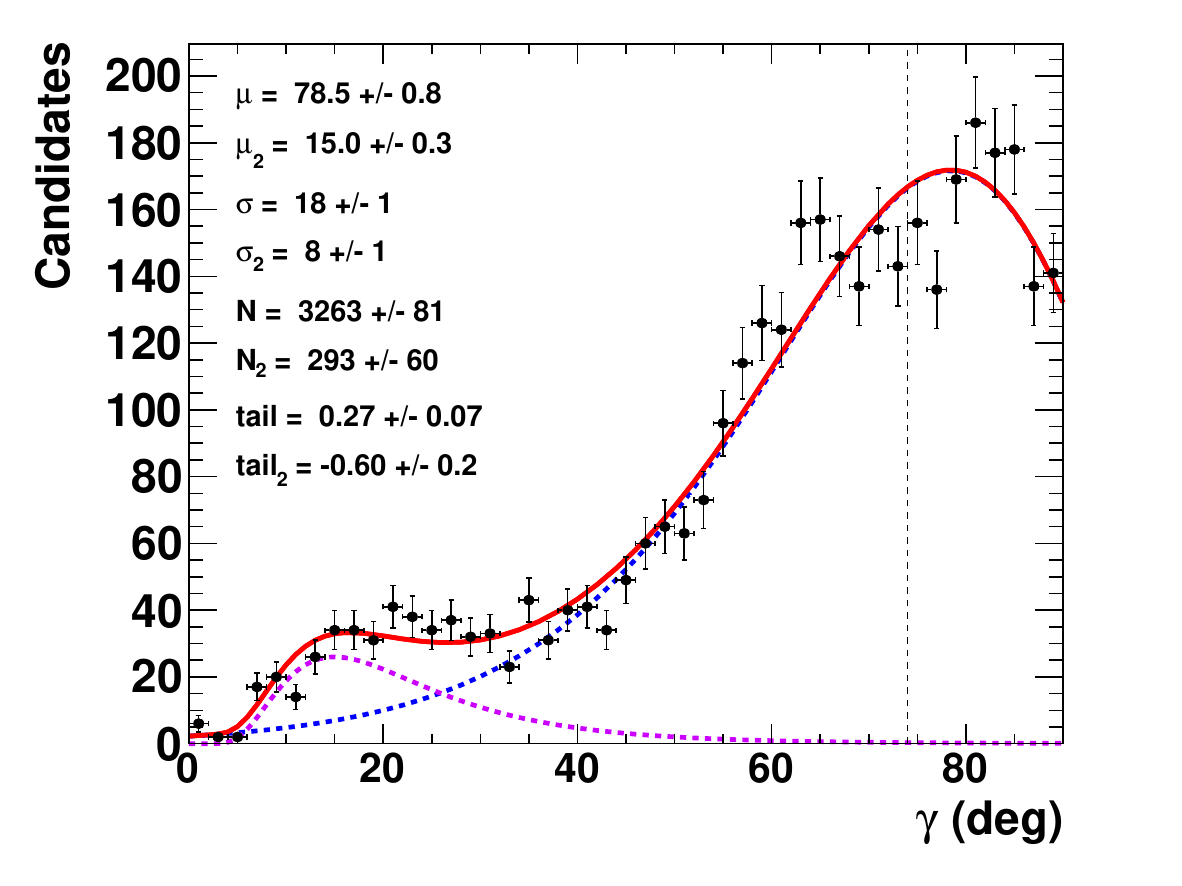}
\caption{\label{fig:74deg_dB1dBst5}Fit to the distributions of the  nuisance parameters  $\gamma$ obtained from 4000 pseudoexperiments. The initial configuration is $\gamma=74^{\circ}$, $\rBst=0.4$ (left) and 0.22 (right), $\deltaB=57.9^\circ$ (1 rad), and $\deltastB=286.5^\circ$ (5 rad). In the distributions, only the candidates with a value  $\gamma\  \in \ [0^\circ, \  90^\circ]$ are considered. The purple dashed curve accounts for tails generated by the correlations with the nuisance parameters $\rBst$ and $\deltaBst$, while the blue dashed curve is the core part of the distribution, the plain red line is the sum of the two components of the fit.}
\end{figure}

\clearpage

\section{Appendix C: Excluding the $\Bs\rightarrow \tilde{D}^{(*)0}(\pi\pi)\phi$ and $\Bs\rightarrow \tilde{D}^{(*)0}(K\pi\piz)\phi$ decays} \label{sec:appendC}

\begin{figure}[h]
\centering
\includegraphics[width=0.425\textwidth,height=0.125\textheight]{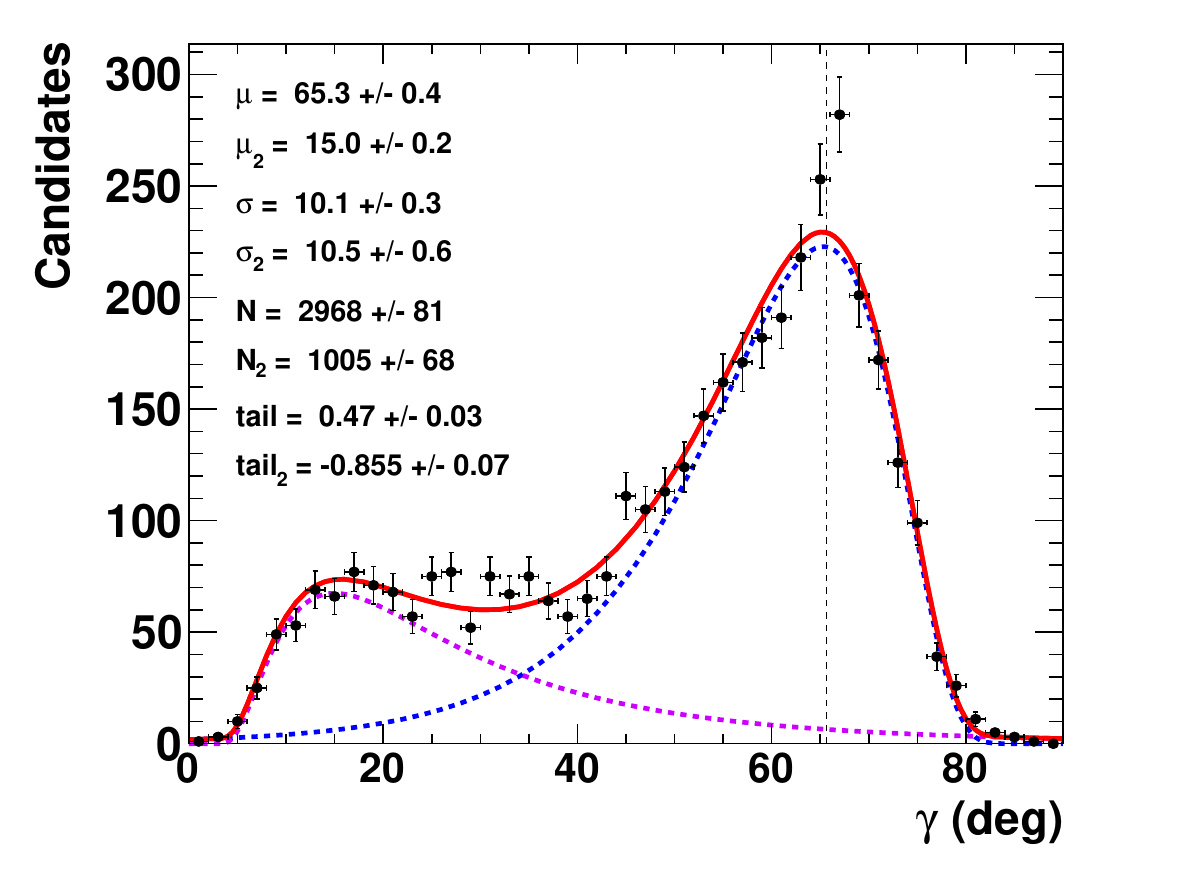}
\includegraphics[width=0.425\textwidth,height=0.125\textheight]{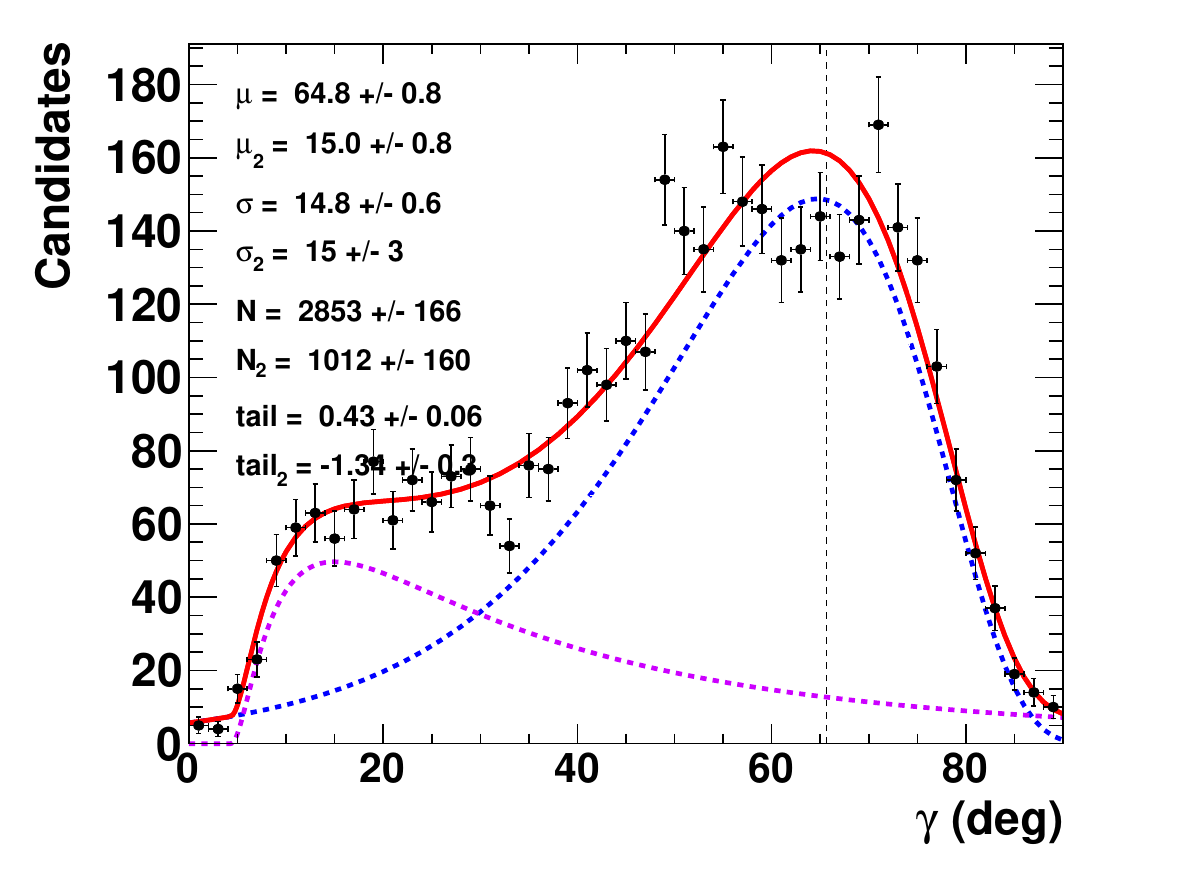} \\
\includegraphics[width=0.425\textwidth,height=0.125\textheight]{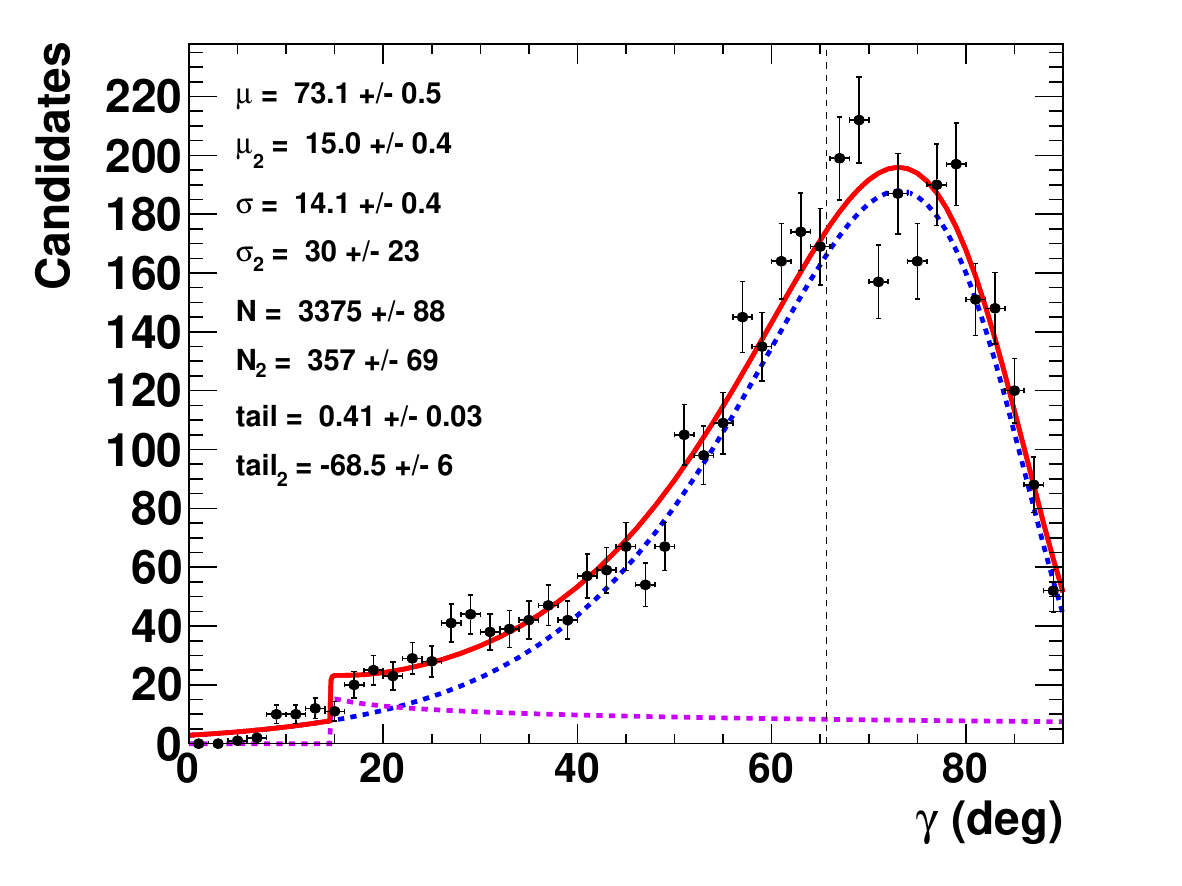}
\includegraphics[width=0.425\textwidth,height=0.125\textheight]{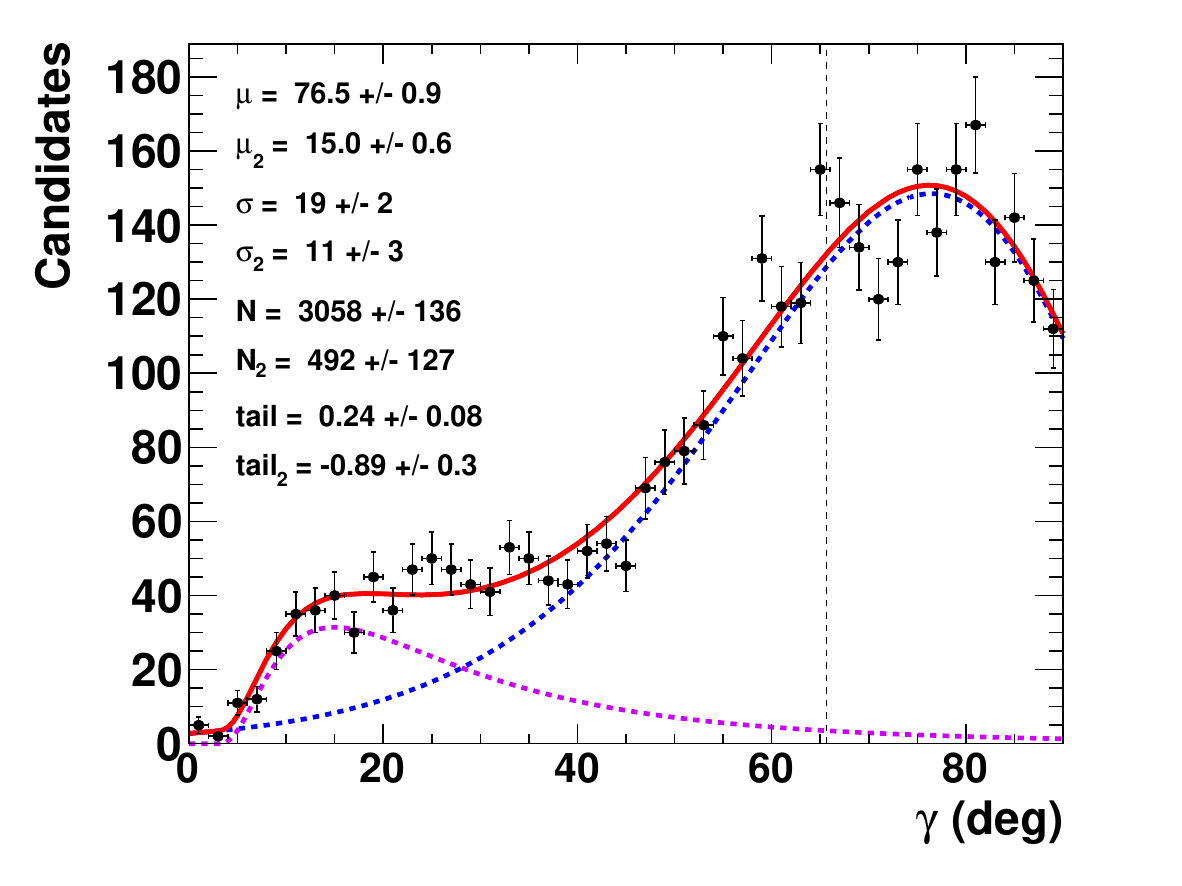}
\caption{\label{fig:1-2_nopipi} Fit to the distribution of   $\gamma$ obtained from 4000 pseudoexperiments, for an the Run~1~\&~2 LHCb dataset. The initial configuration is $\gamma=65.66^{\circ}$, $\rB=0.4$ (left) and 0.22 (right), and (top) $\deltaB=171.9^\circ$ (3 rad) and $\deltastB=114.6^\circ$ (2 rad) or (bottom) $\deltaB=57.3^\circ$ (1 rad) and $\deltastB=286.5^\circ$ (5 rad) (w/o $\Bs \rightarrow \tilde{D}^{(*)0}(\pi\pi)\phi$). The purple dashed curve accounts for tails generated by the correlations with the nuisance parameters $\rBst$ and $\deltaBst$, while the blue dashed curve is the core part of the distribution, the plain red line is the sum of the two components of the fit.}
\includegraphics[width=0.425\textwidth,height=0.125\textheight]{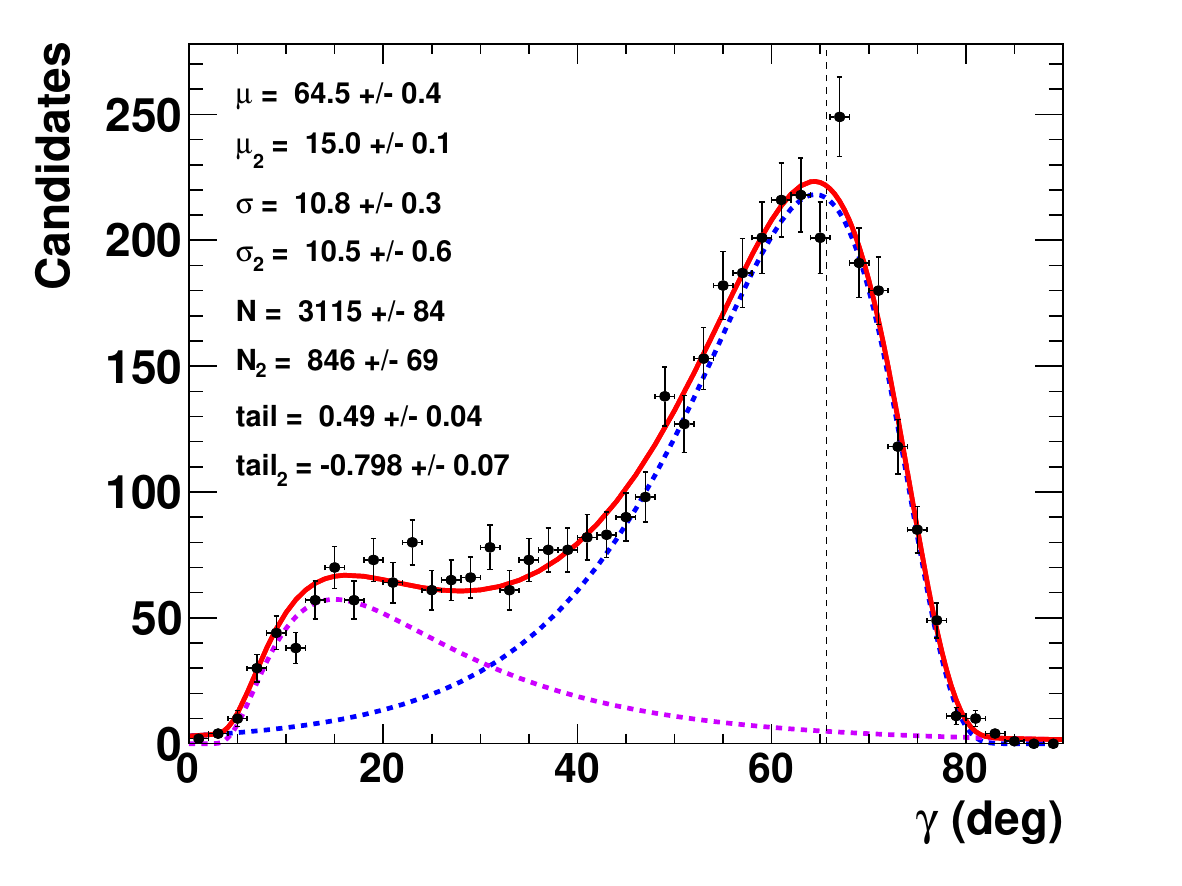}
\includegraphics[width=0.425\textwidth,height=0.125\textheight]{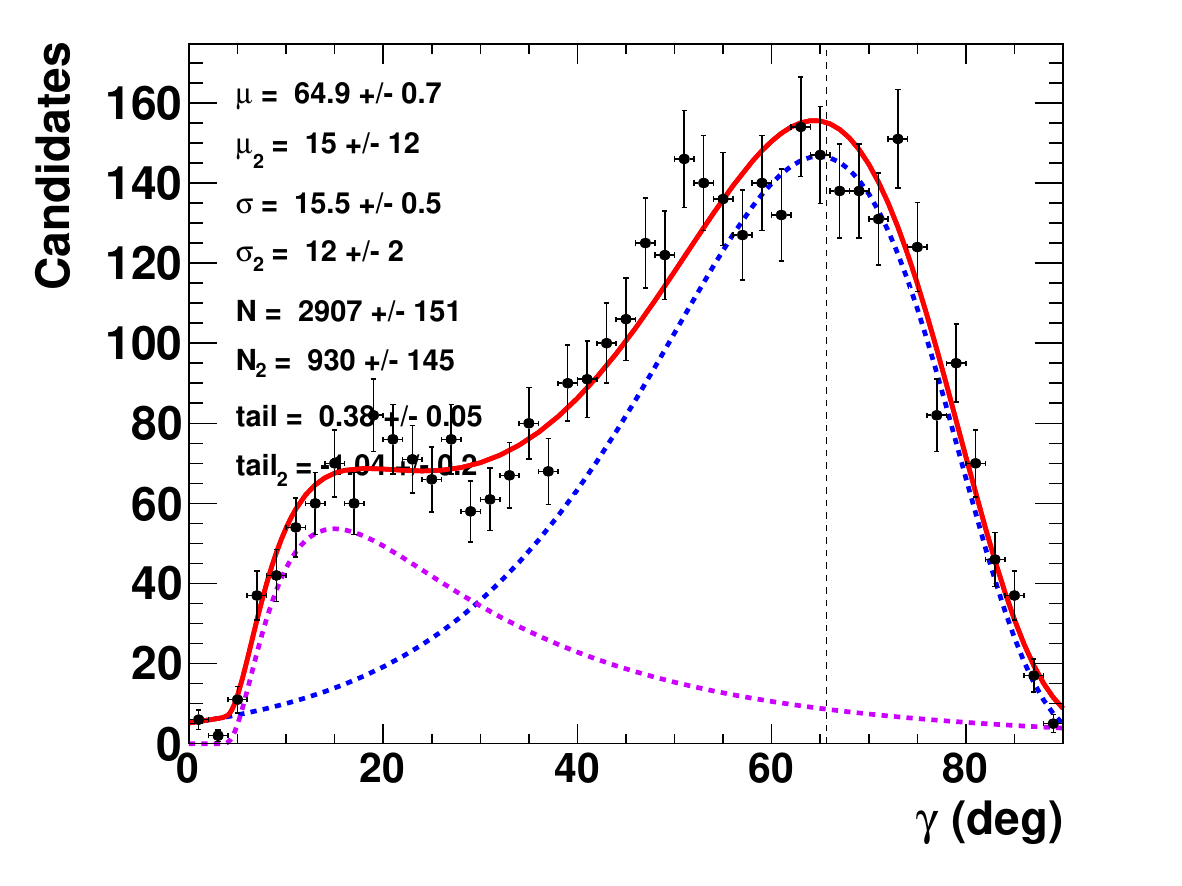} \\
\includegraphics[width=0.425\textwidth,height=0.125\textheight]{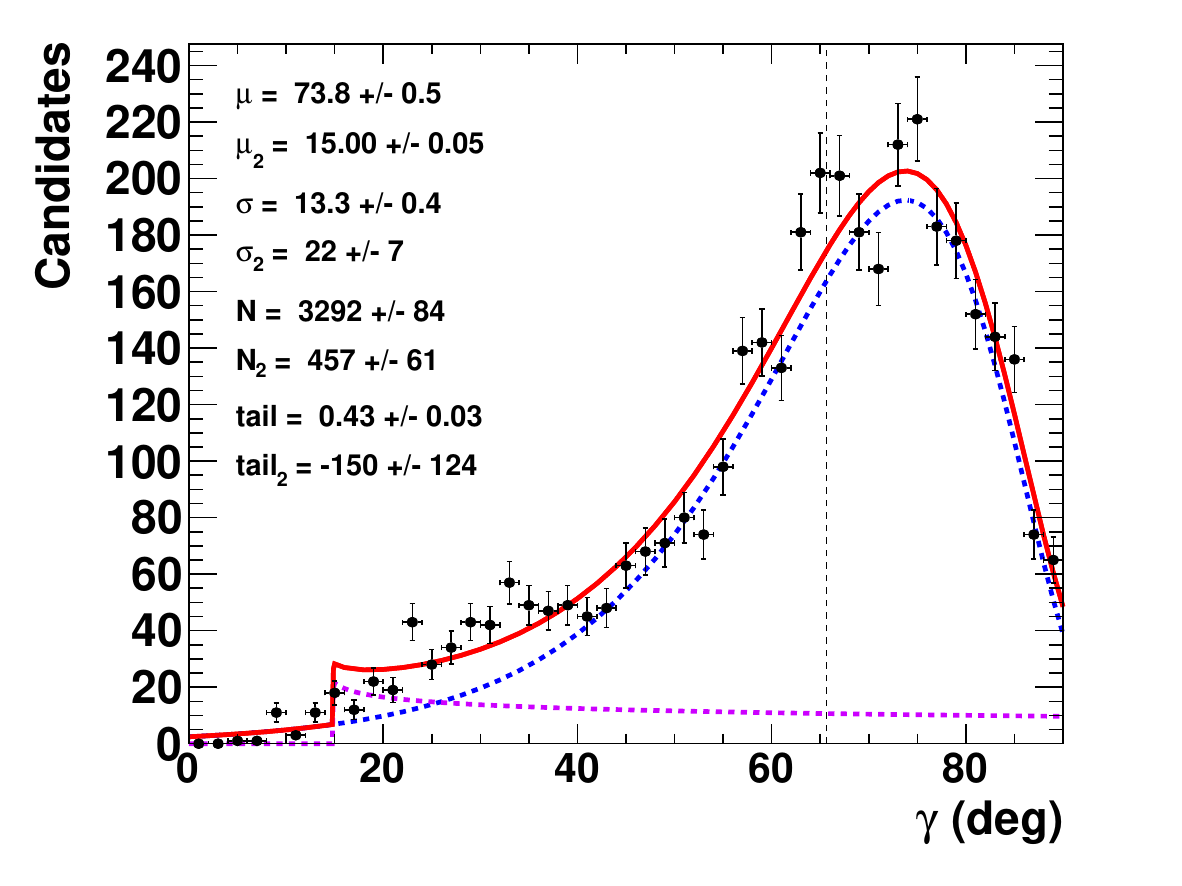}
\includegraphics[width=0.425\textwidth,height=0.125\textheight]{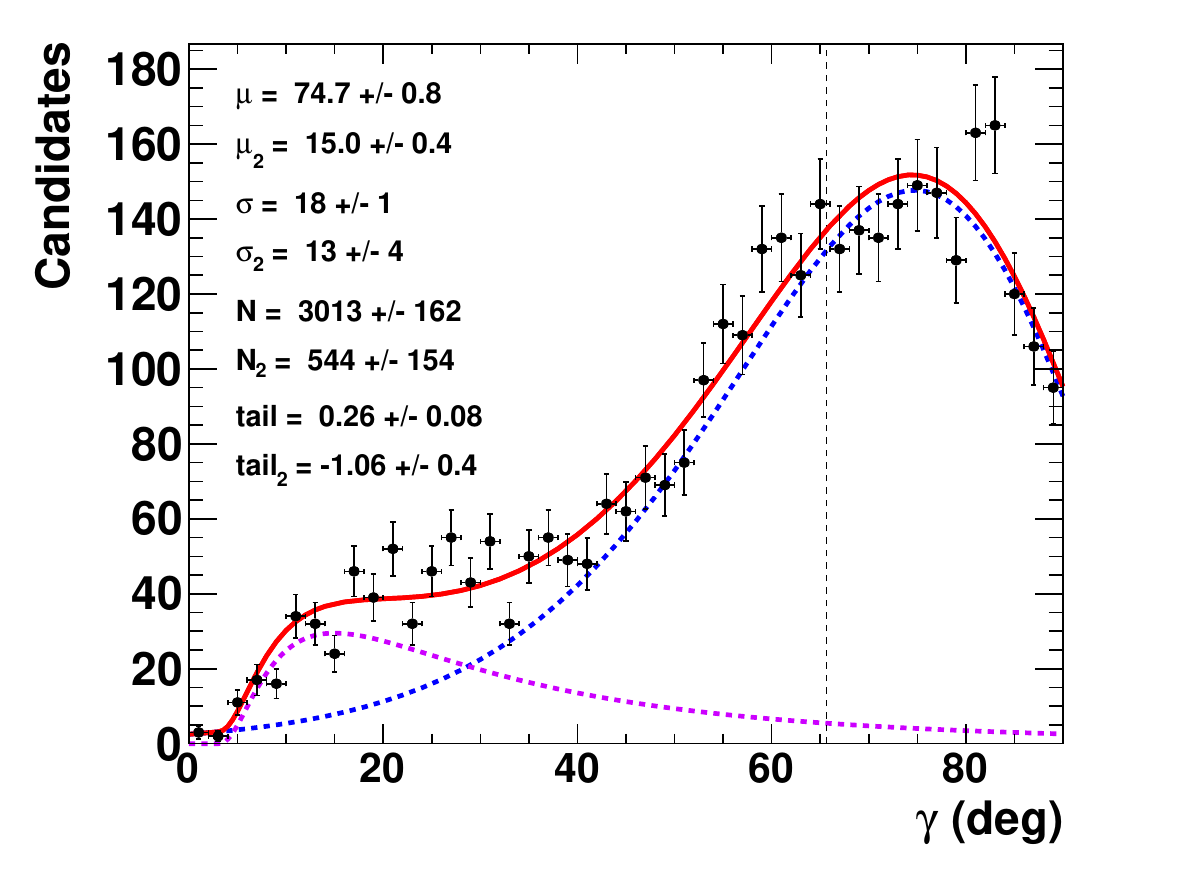}
\caption{\label{fig:1-2_noKpipi0} Fit to the distribution of   $\gamma$ obtained from 4000 pseudoexperiments, for an the Run~1~\&~2 LHCb dataset. The initial configuration is $\gamma=65.66^{\circ}$, $\rB=0.4$ (left) and 0.22 (right), and (top) $\deltaB=171.9^\circ$ (3 rad) and $\deltastB=114.6^\circ$ (2 rad) or (bottom) $\deltaB=57.3^\circ$ (1 rad) and $\deltastB=286.5^\circ$ (5 rad) (w/o $\Bs \rightarrow \tilde{D}^{(*)0}(K\pi\piz)\phi$). The purple dashed curve accounts for tails generated by the correlations with the nuisance parameters $\rBst$ and $\deltaBst$, while the blue dashed curve is the core part of the distribution, the plain red line is the sum of the two components of the fit.}
\end{figure}

\clearpage

\section{Appendix D: Other examples of two-dimension $p$-value  profiles for the full HL-LHC LHCb dataset}
\label{sec:appendD}

\begin{figure}[h]
\centering
\includegraphics[width=0.425\textwidth,height=0.15\textheight]{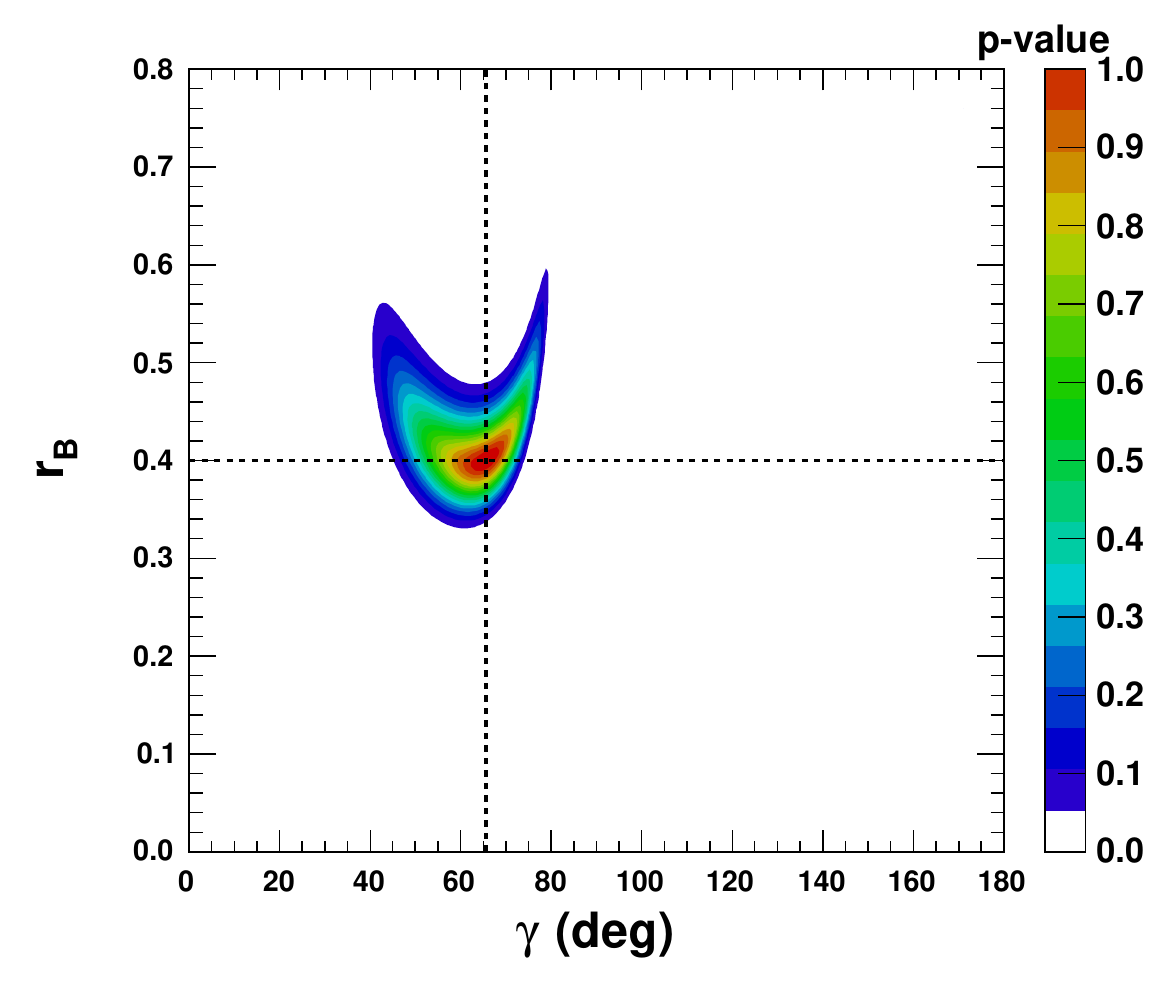}
\includegraphics[width=0.425\textwidth,height=0.15\textheight]{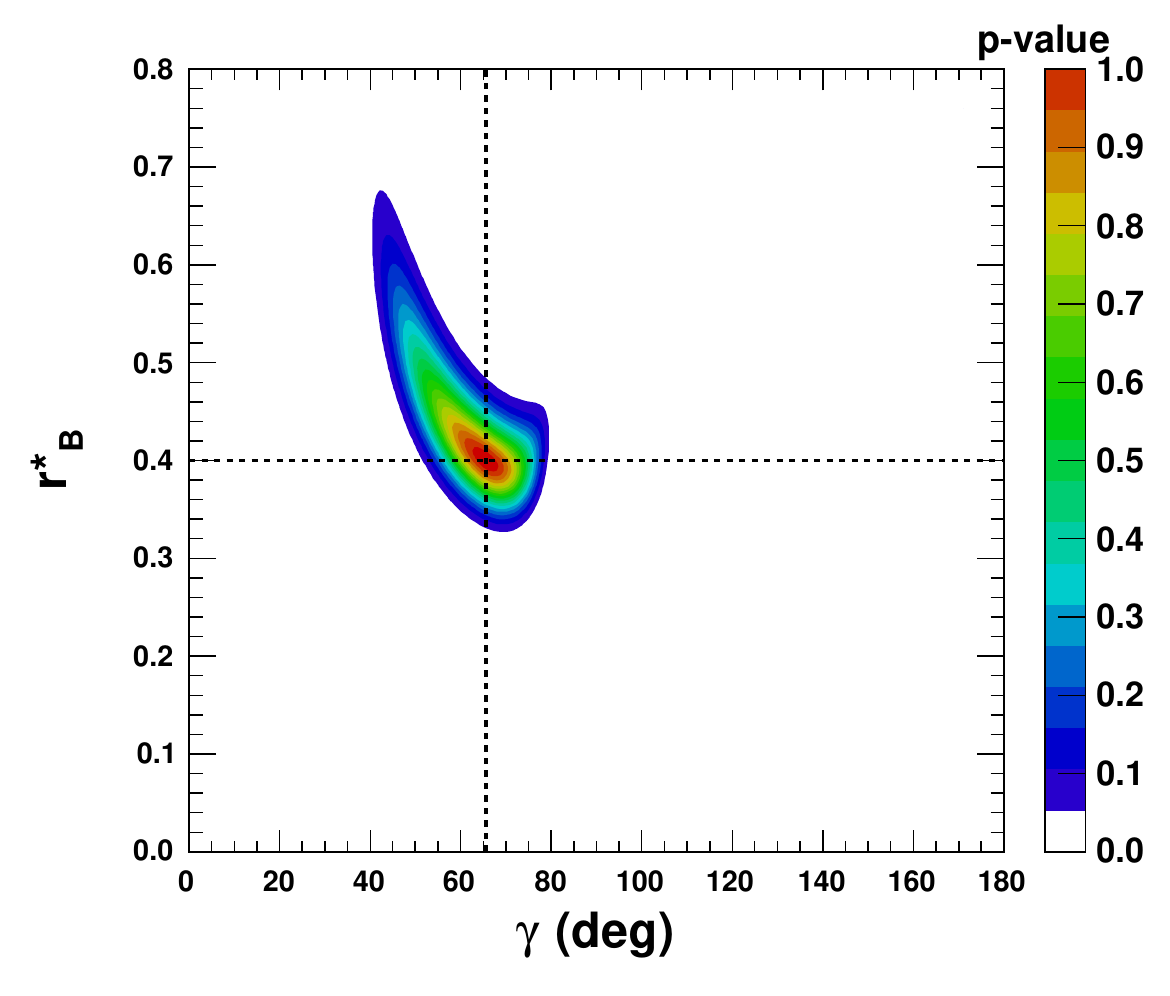} \\
\includegraphics[width=0.425\textwidth,height=0.15\textheight]{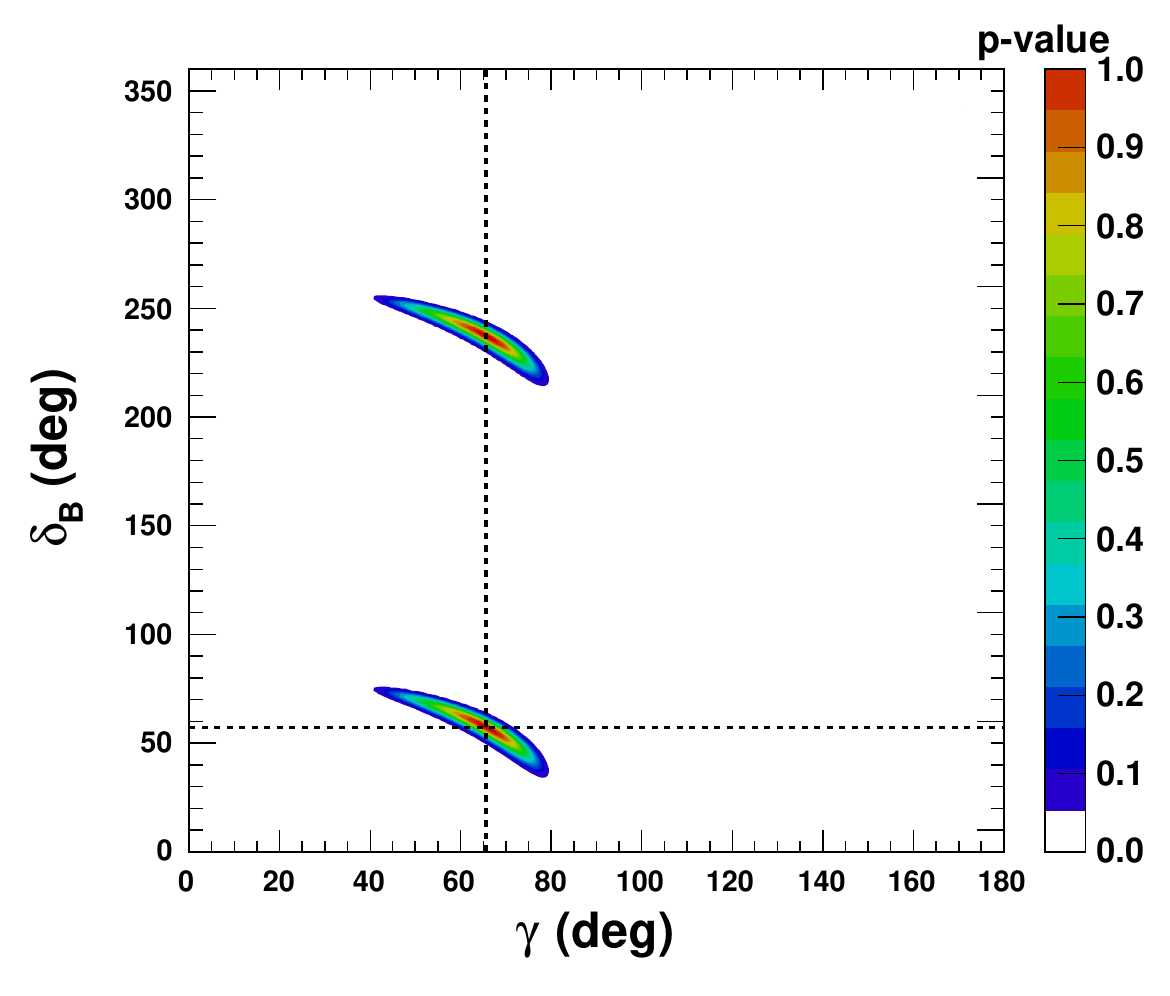}
\includegraphics[width=0.425\textwidth,height=0.15\textheight]{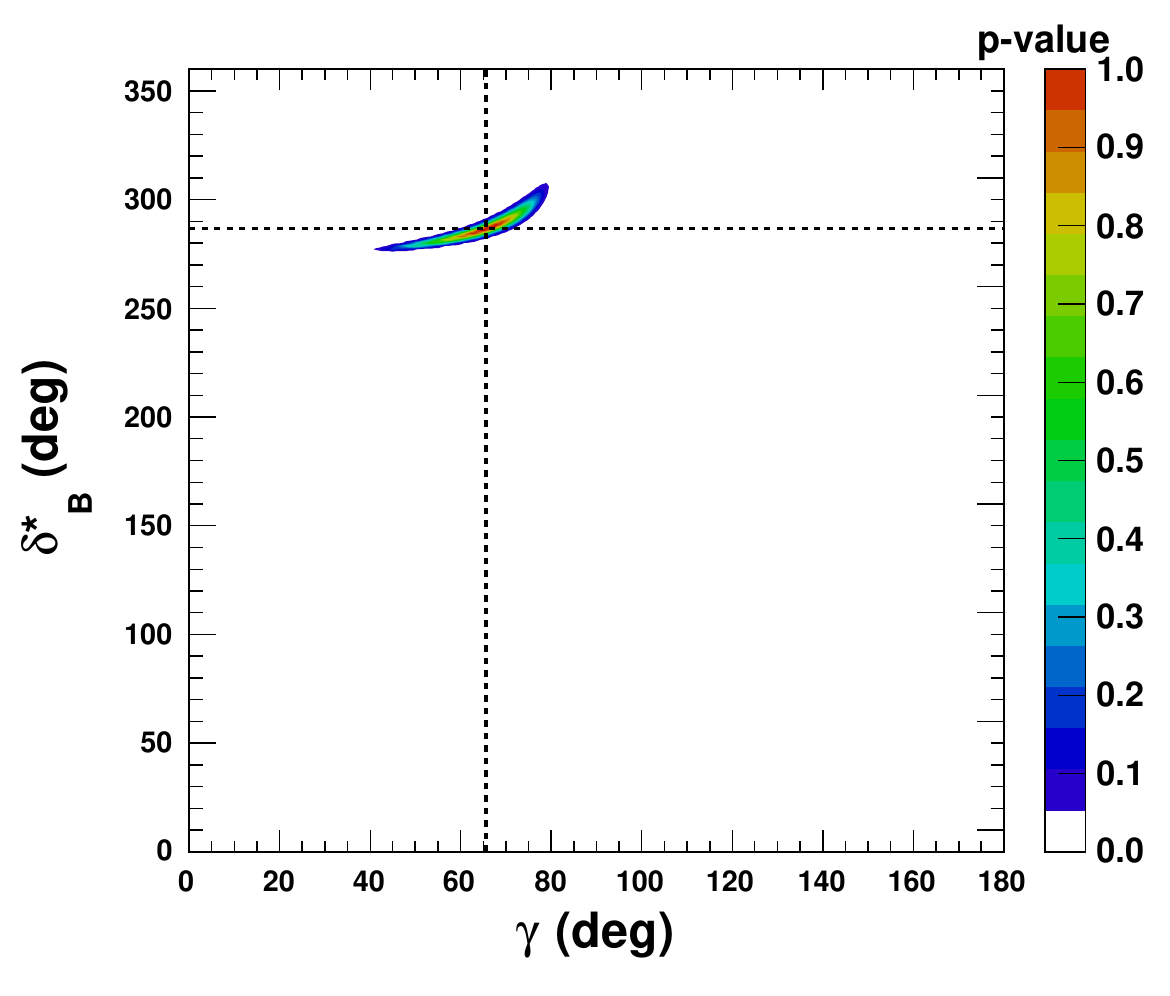}
\caption{\label{fig:2D-RunHL-LHC_rBst04_15} Two-dimension $p$-value  profile of the nuisance parameters $\rBst$ and $\deltaBst$, for the full HL-LHC LHCb dataset,  as a function of $\gamma$. On each figure the dashed black  lines indicate the initial  $\rBst$ and $\deltaBst$ ($\gamma$) true values: $\gamma=65.66^\circ$ (1.146 rad), $\deltaB=57.3^\circ$ (1.0 rad), and $\deltastB=286.5^\circ$ (5.0 rad), and  $\rBst=0.4$.}
\end{figure}

\begin{figure}[h]
\centering
\includegraphics[width=0.425\textwidth,height=0.15\textheight]{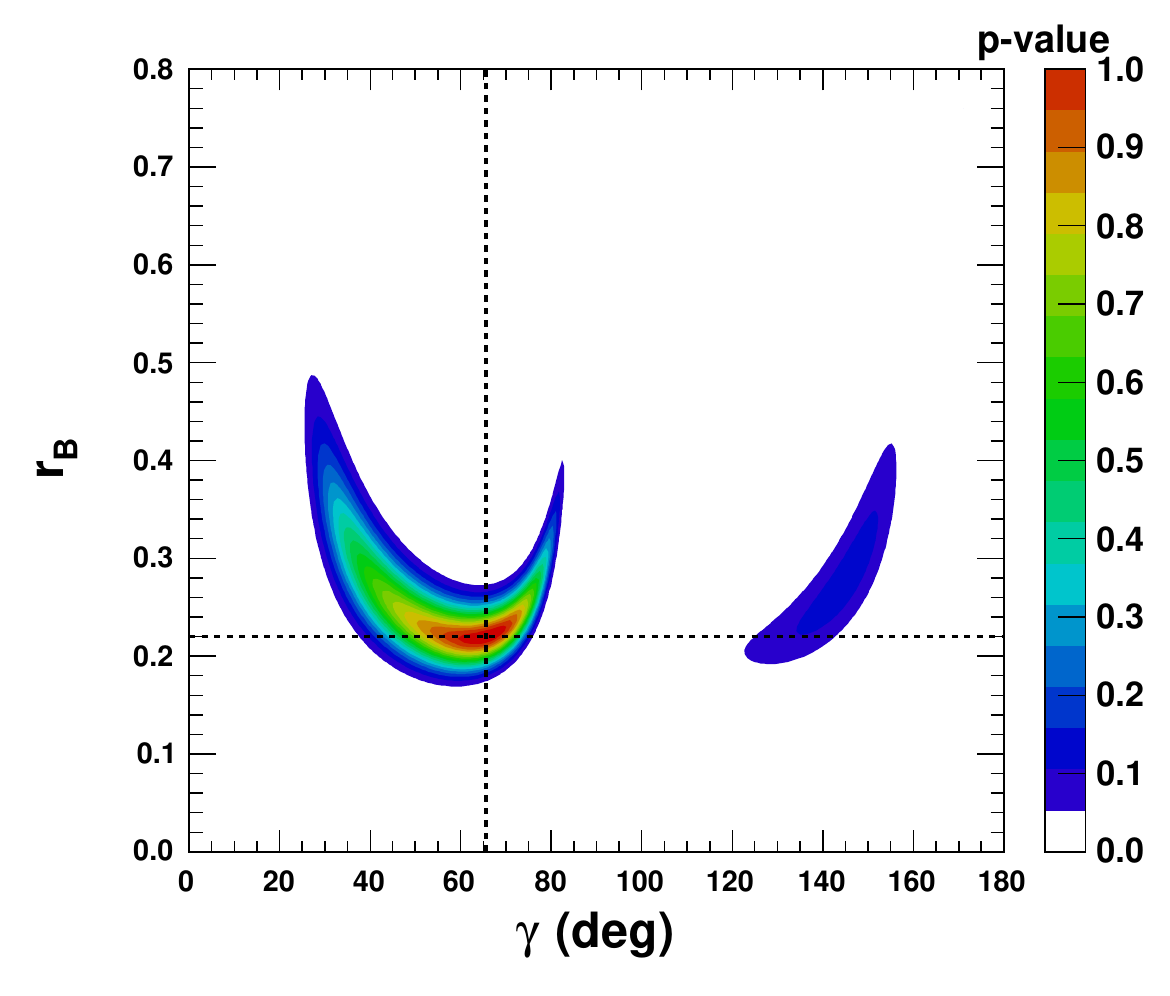}
\includegraphics[width=0.425\textwidth,height=0.15\textheight]{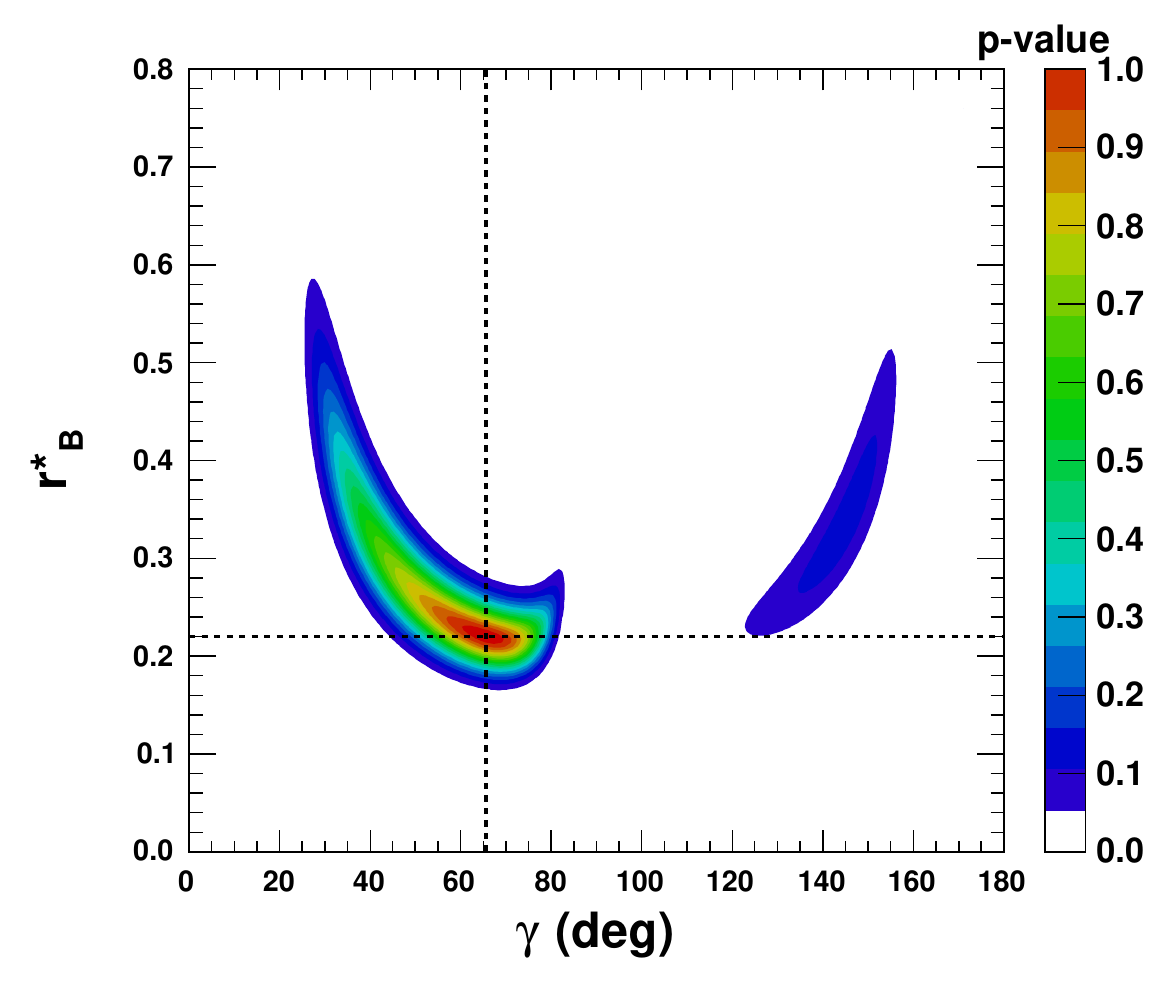} \\
\includegraphics[width=0.425\textwidth,height=0.15\textheight]{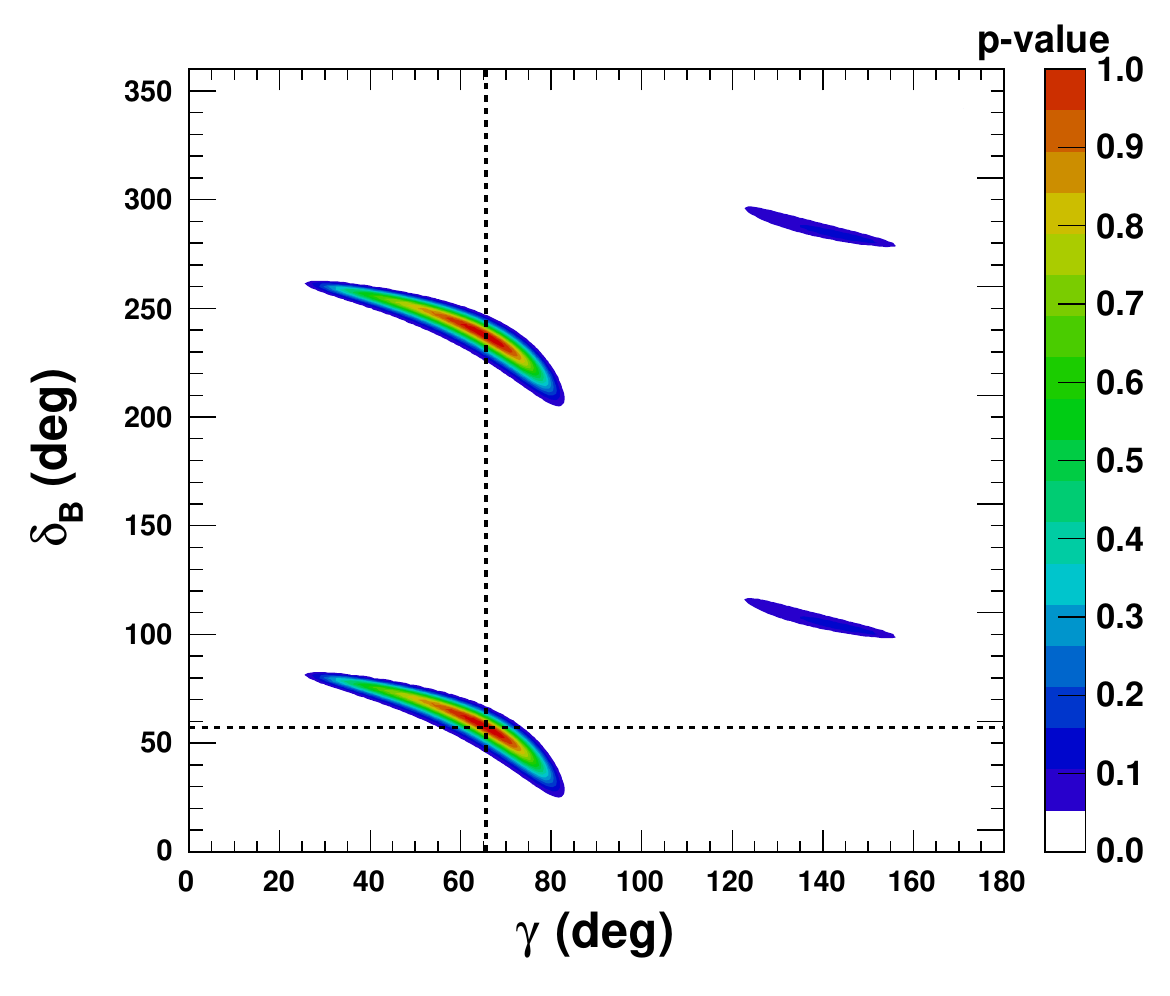}
\includegraphics[width=0.425\textwidth,height=0.15\textheight]{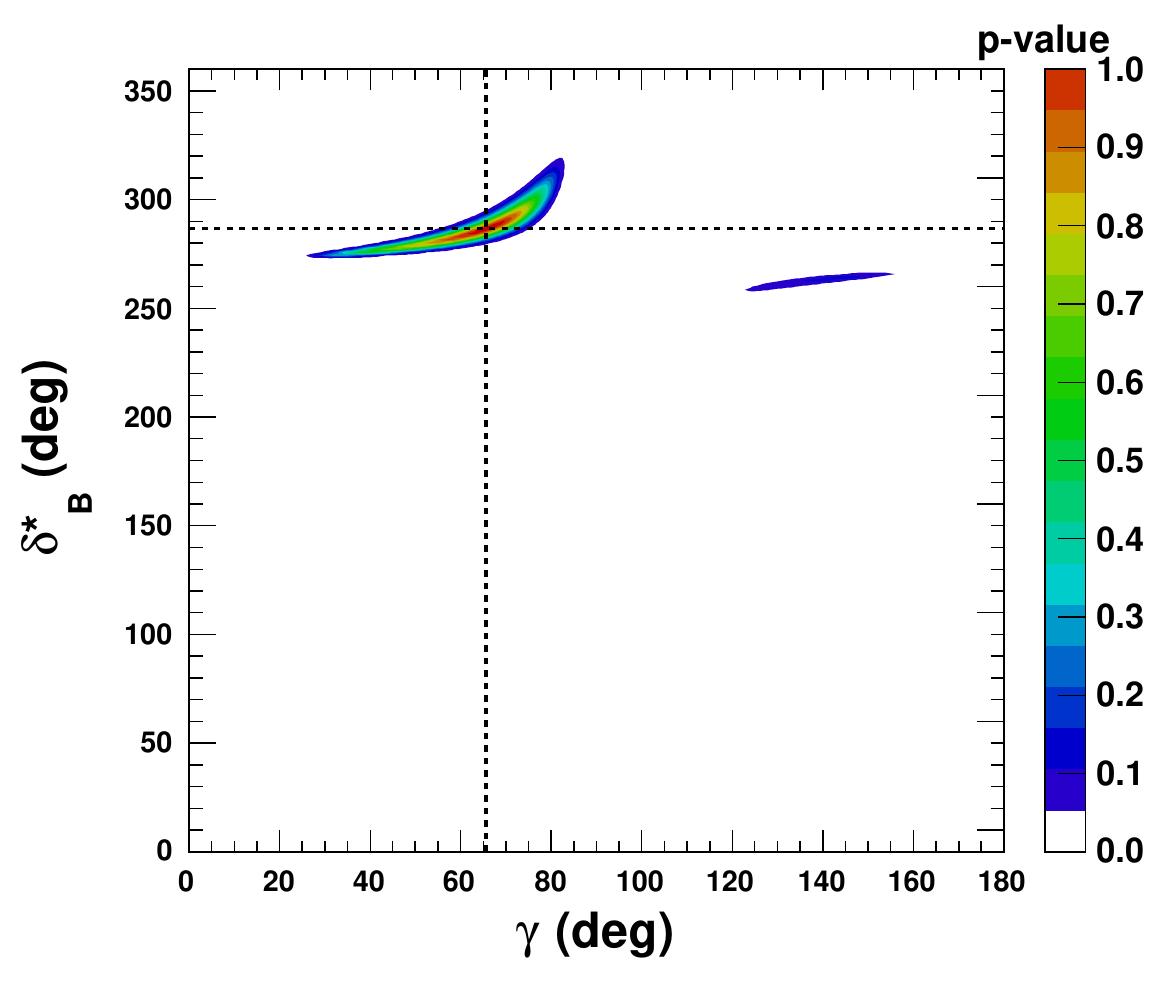}
\caption{\label{fig:2D-RunHL-LHC_rBst022_15} Two-dimension $p$-value  profile of the nuisance parameters $\rBst$ and $\deltaBst$, for the full HL-LHC LHCb dataset,  as a function of $\gamma$. On each figure the dashed black  lines indicate the initial true values: $\gamma=65.66^\circ$ (1.146 rad),  $\deltaB=57.3^\circ$ (1.0 rad), and $\deltastB=286.5^\circ$ (5.0 rad), and  $\rBst=0.22$.}
\end{figure}

\clearpage

\section{Appendix E: Half of the 68.3~\% CL intervals of the one-dimension $p$-value profiles of $\gamma$}
\label{sec:appendE}

\begin{figure}[h]
\centering
\includegraphics[width=0.425\textwidth]{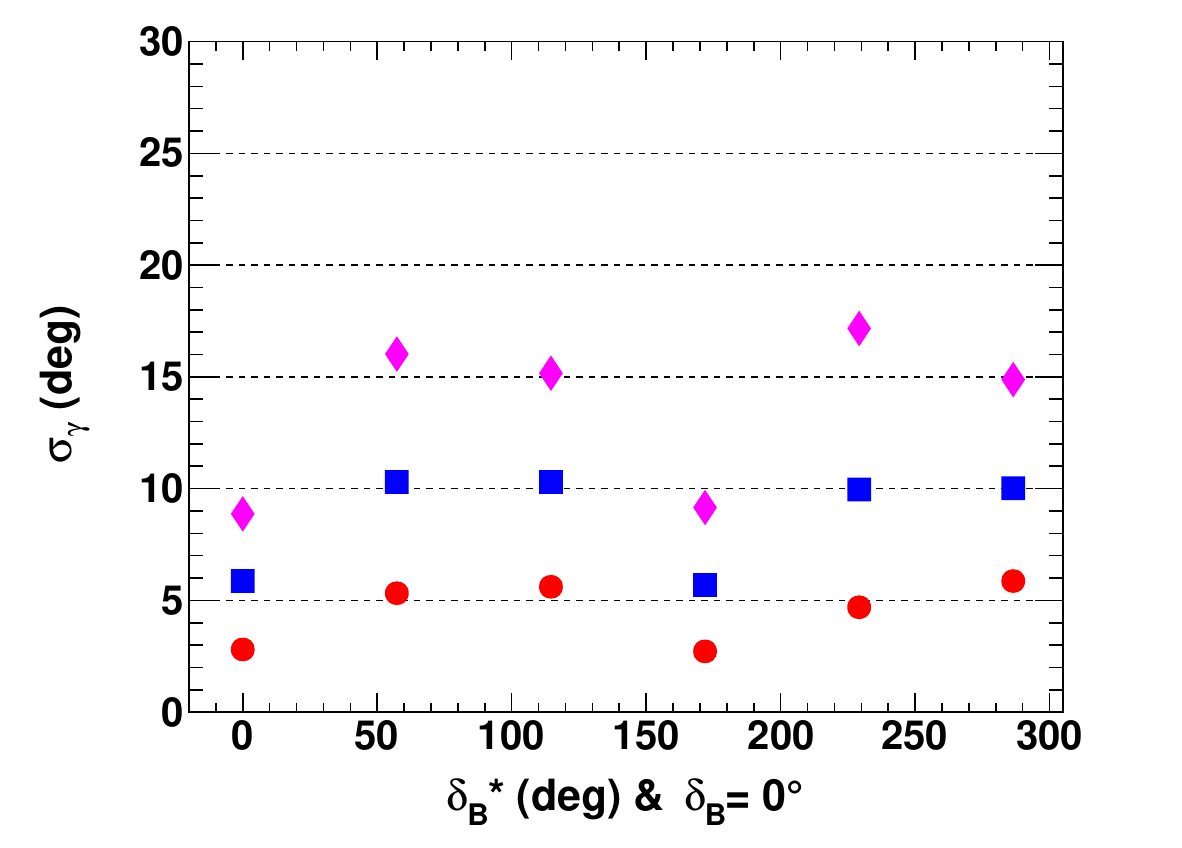}
\includegraphics[width=0.425\textwidth]{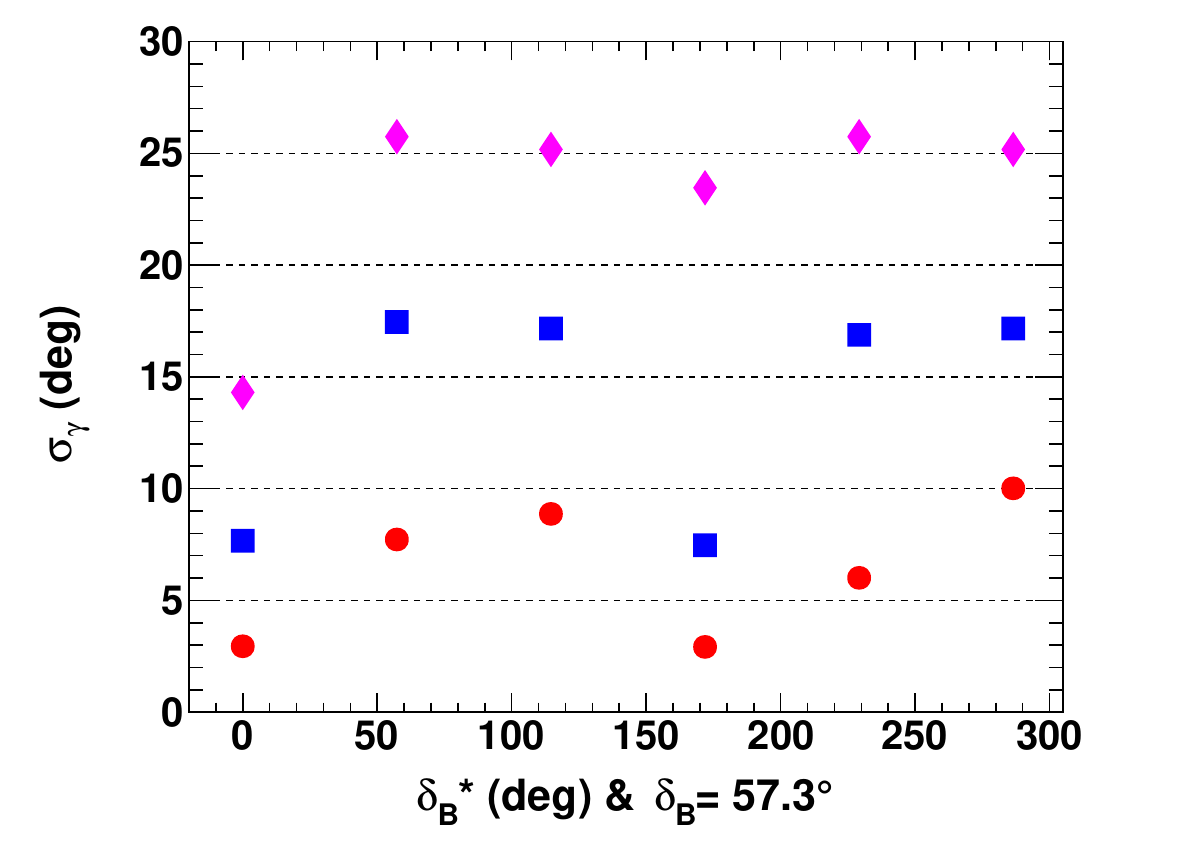} \\
\includegraphics[width=0.425\textwidth]{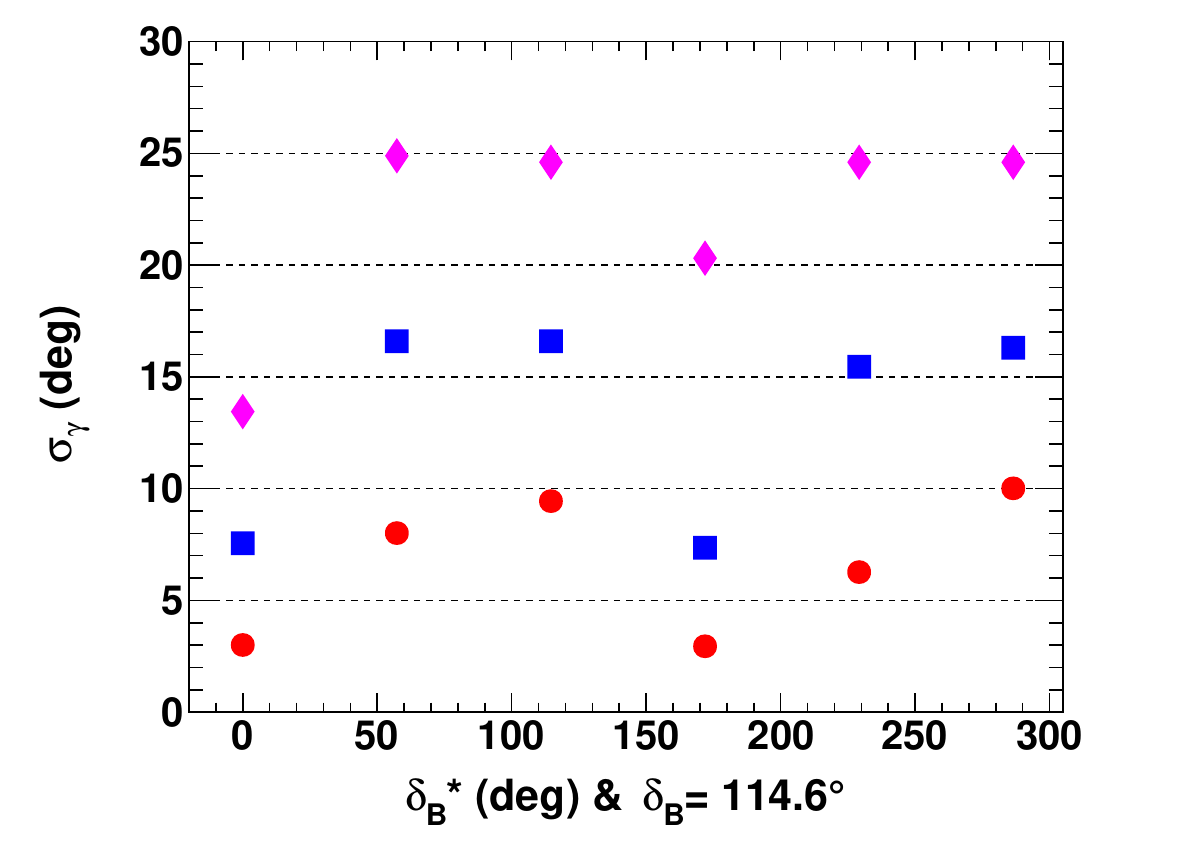}
\includegraphics[width=0.425\textwidth]{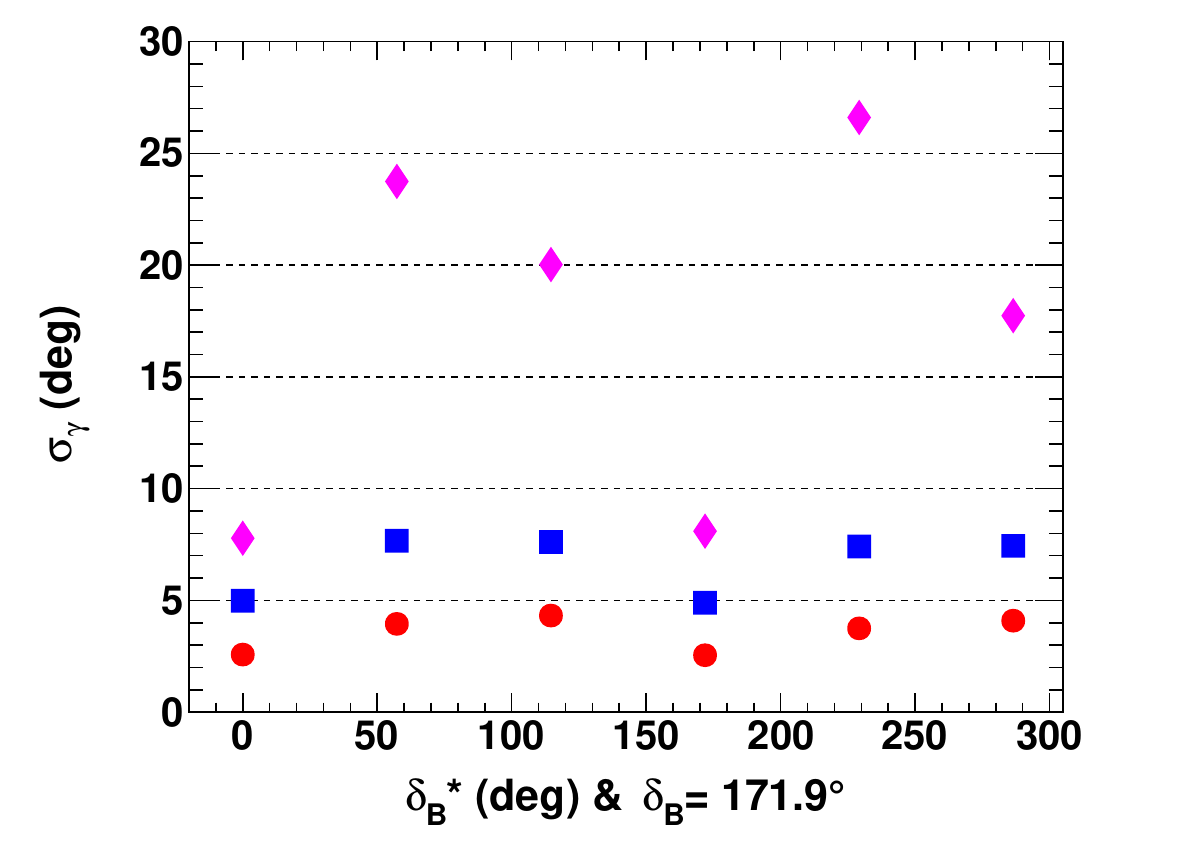} \\
\includegraphics[width=0.425\textwidth]{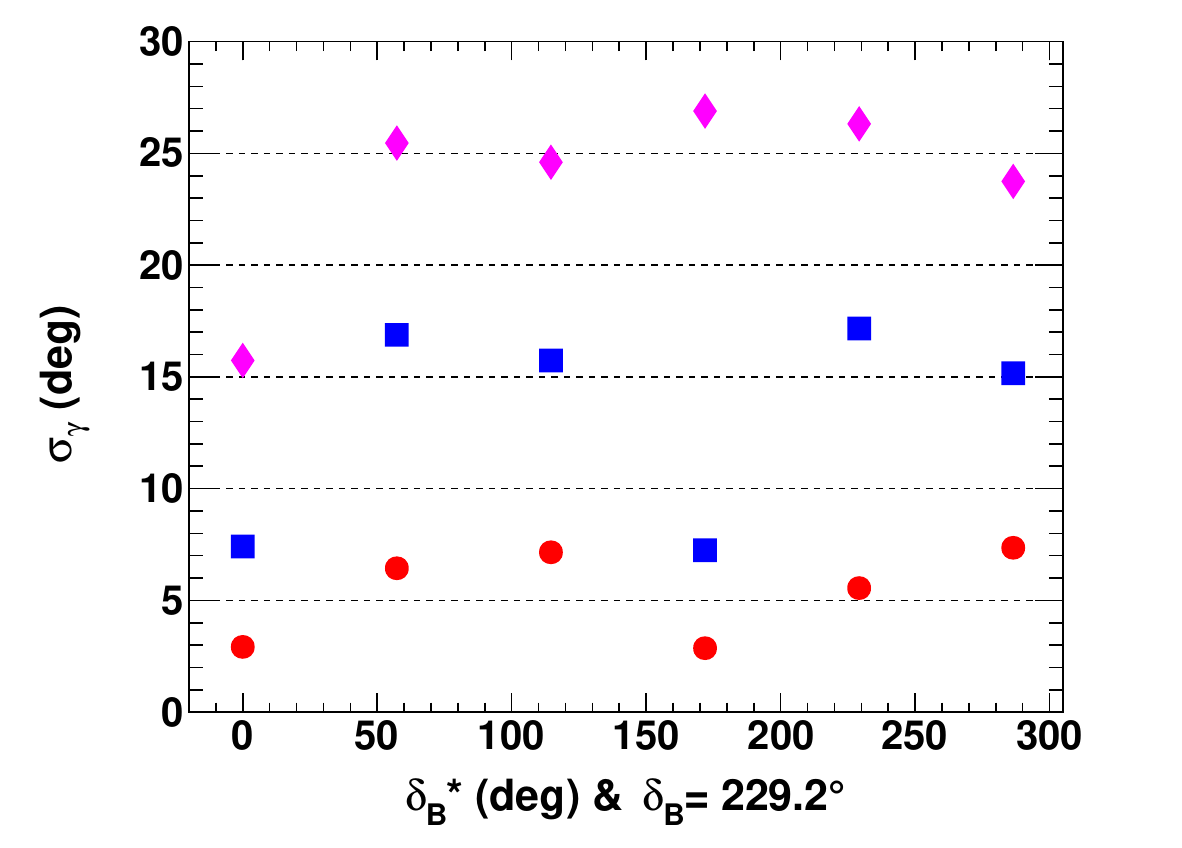}
\includegraphics[width=0.425\textwidth]{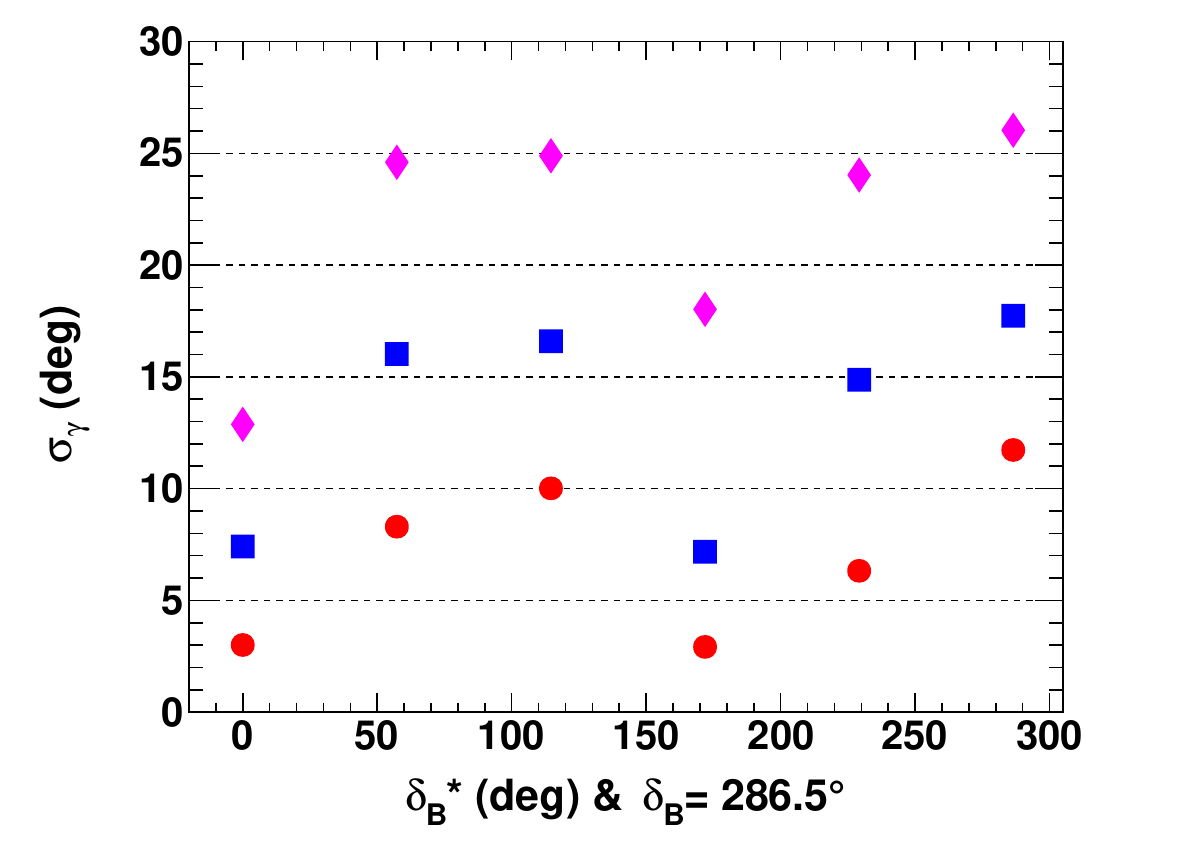} \\
\caption{\label{fig:resolpVal_04}
  Half of the 68.3~\% CL intervals of the one-dimension $p$-value profiles of $\gamma$, for Run~1~\&~2 (pink lozenges), for Run $1-3$   (blue squares), and full HL-LHC  (red circles) LHCb dataset, as a function of $\deltaBst$, for $\rBst=0.4$, for an initial true value of  $65.66^\circ$. On each figure, the horizontal dashed black  lines are guide for the eye.}
\end{figure}

\clearpage

\clearpage

\section{Appendix F: Excluding the $\Bs \rightarrow \Dtstz\phi$ decays}
\label{sec:appendF}
\begin{figure}[h]
\centering
\includegraphics[width=0.425\textwidth,height=0.175\textheight]{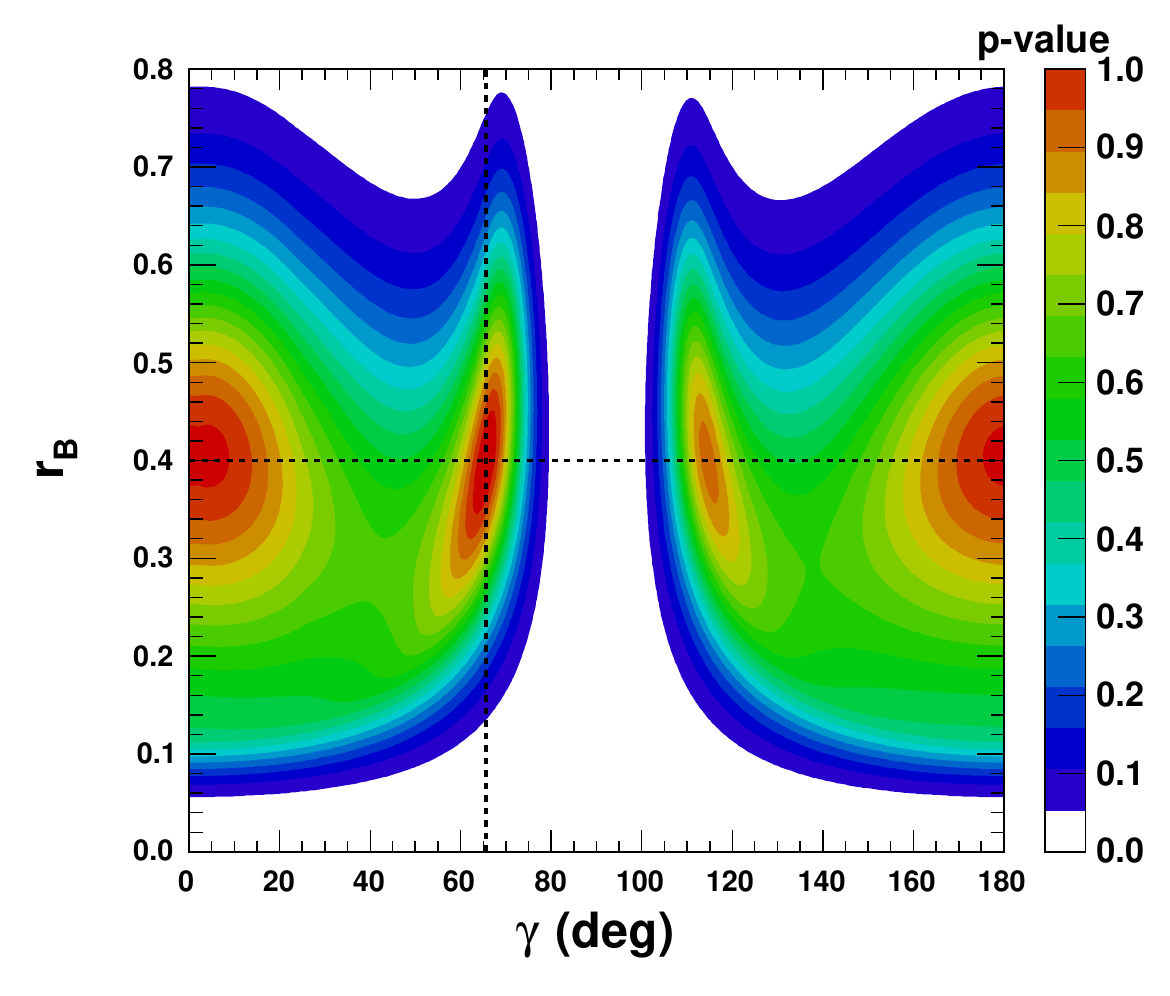}
\includegraphics[width=0.425\textwidth,height=0.175\textheight]{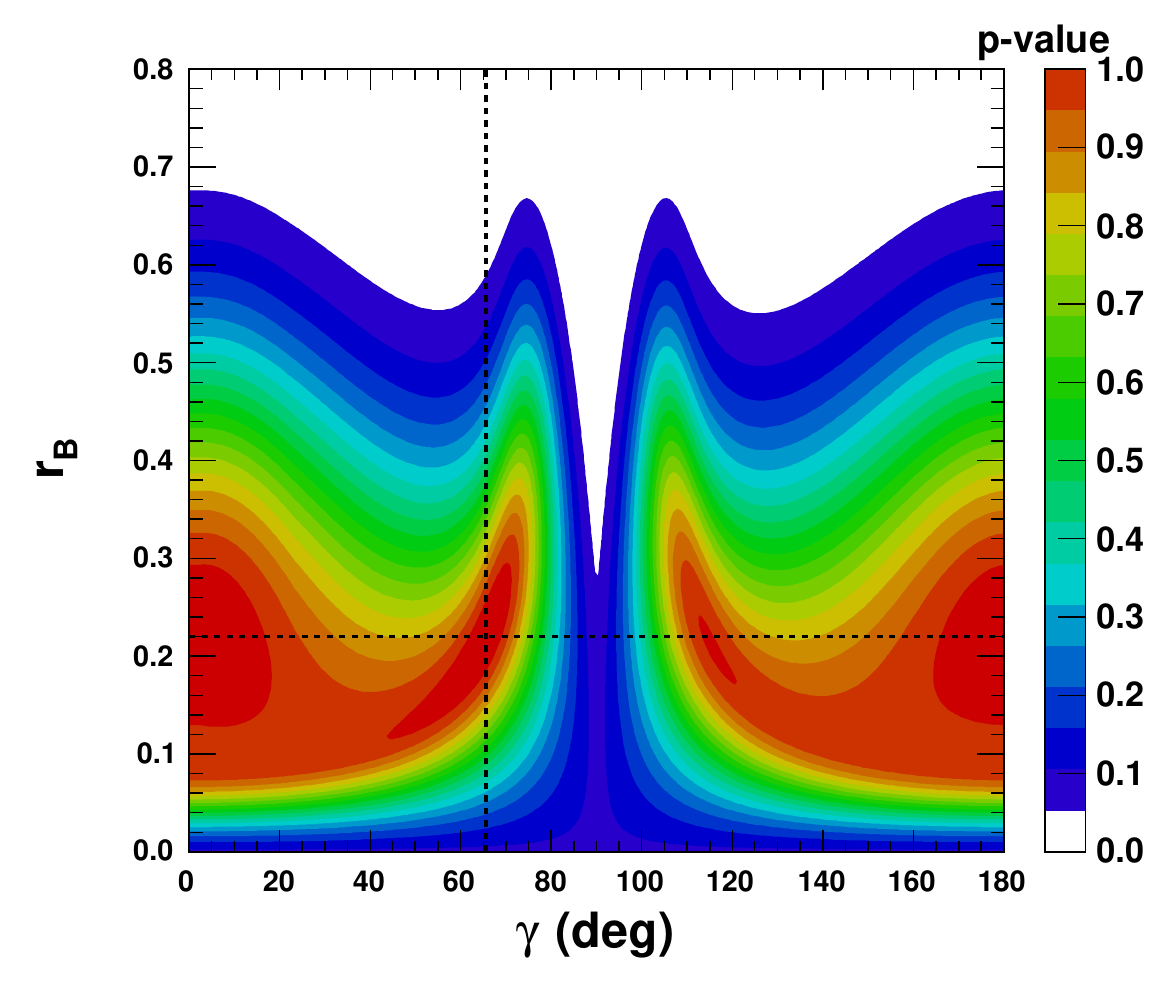} \\
\includegraphics[width=0.425\textwidth,height=0.175\textheight]{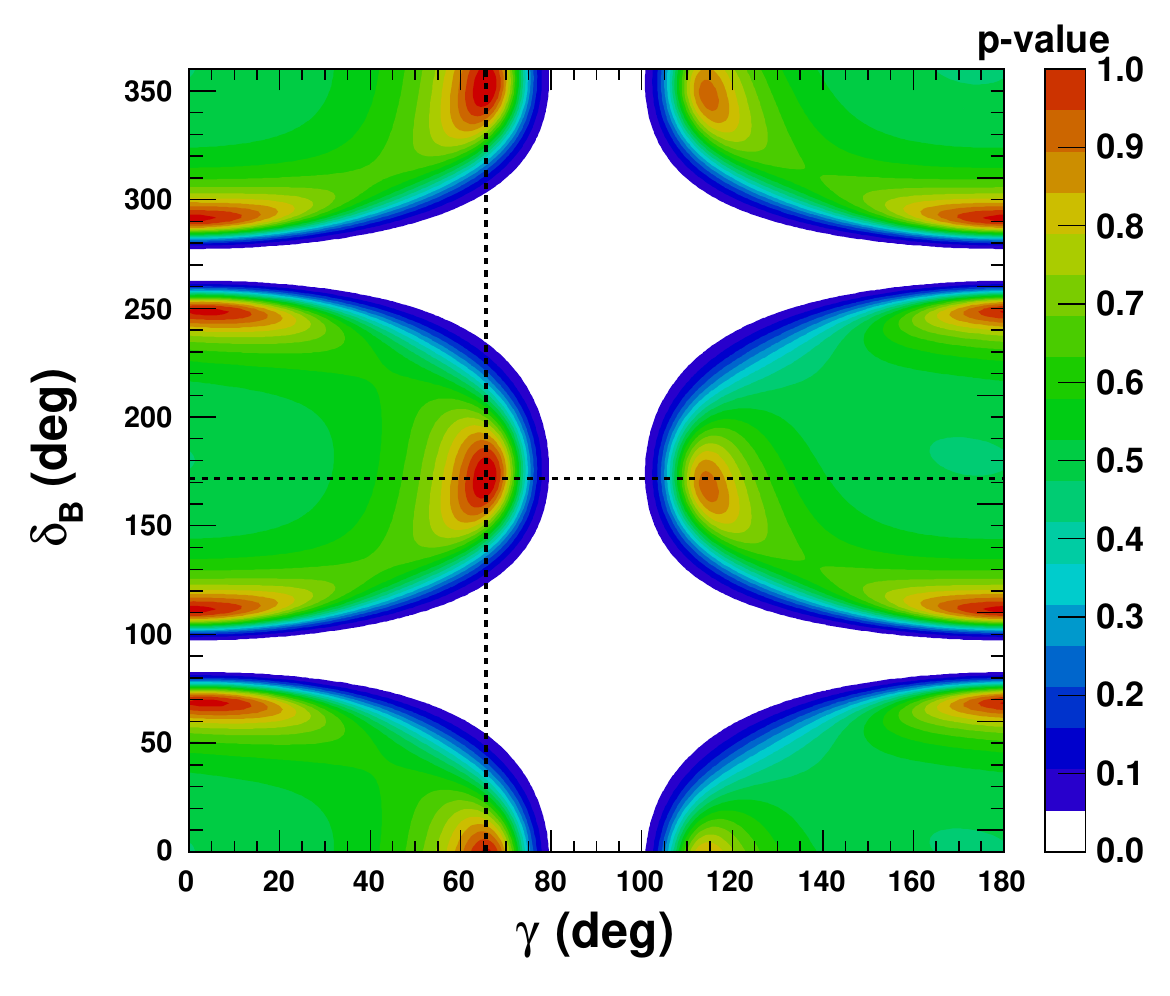}
\includegraphics[width=0.425\textwidth,height=0.175\textheight]{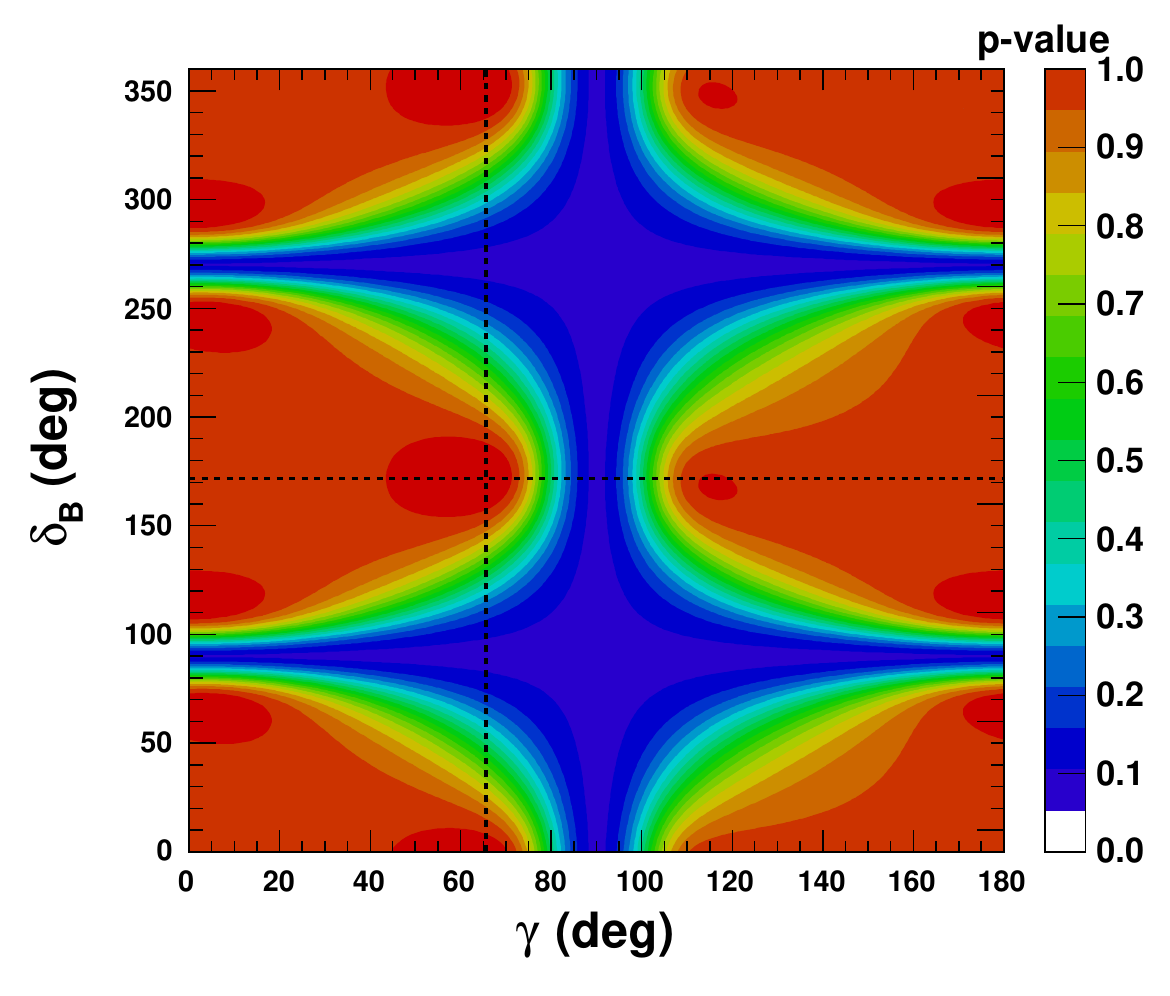} \\
\caption{\label{fig:r_d_BstvsGamma_1-2_dB3dBst2_noDst}
Two-dimension $p$-value  profile of the nuisance parameters $\rB$ and $\deltaB$, for the Run~1~\&~2 LHCb dataset, as a function of $\gamma$. On each figure the dashed black  lines indicate the initial true values: $\gamma=65.66^\circ$ (1.146 rad), $\deltaB=171.9^\circ$ (3.0 rad), and  $\rB=0.4$ (left) and 0.22 (right) (w/o $\Bs \rightarrow \Dstz\phi$).}
\includegraphics[width=0.425\textwidth]{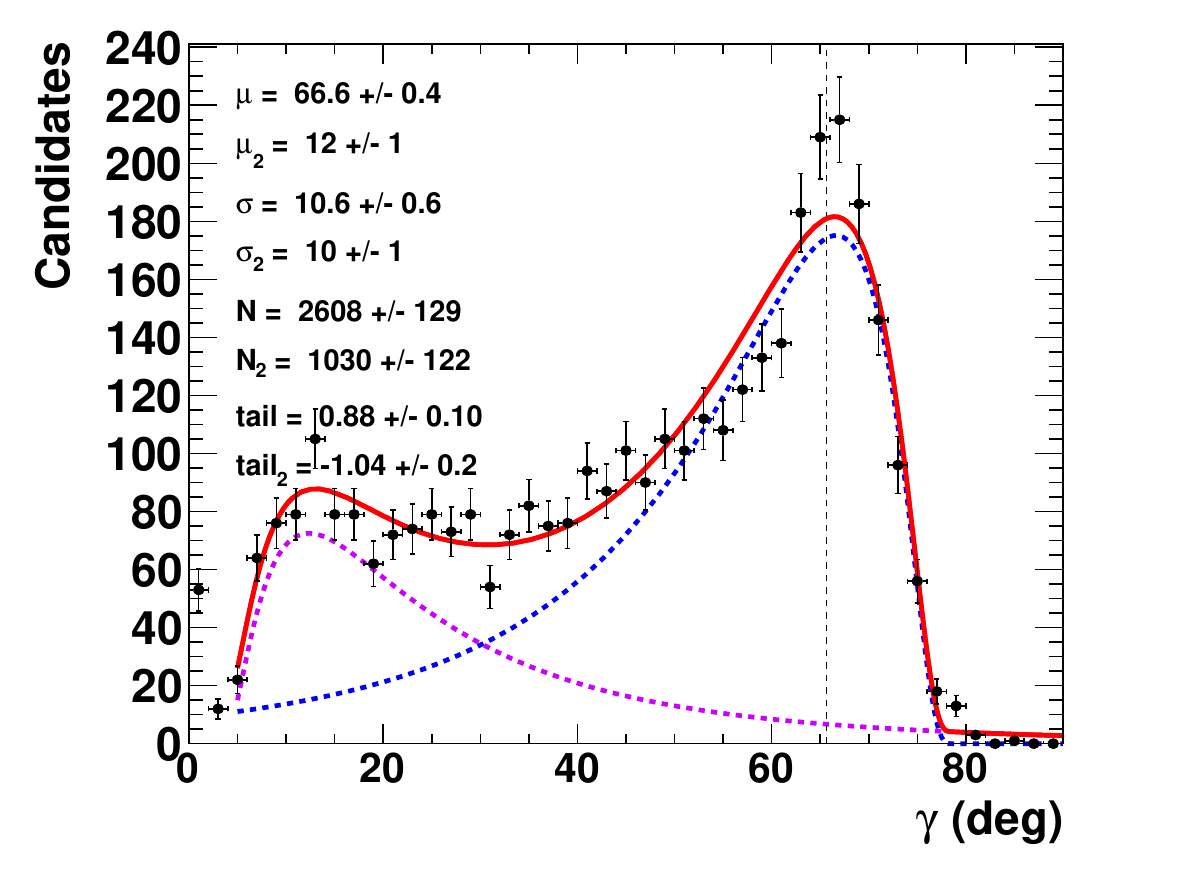}
\includegraphics[width=0.425\textwidth]{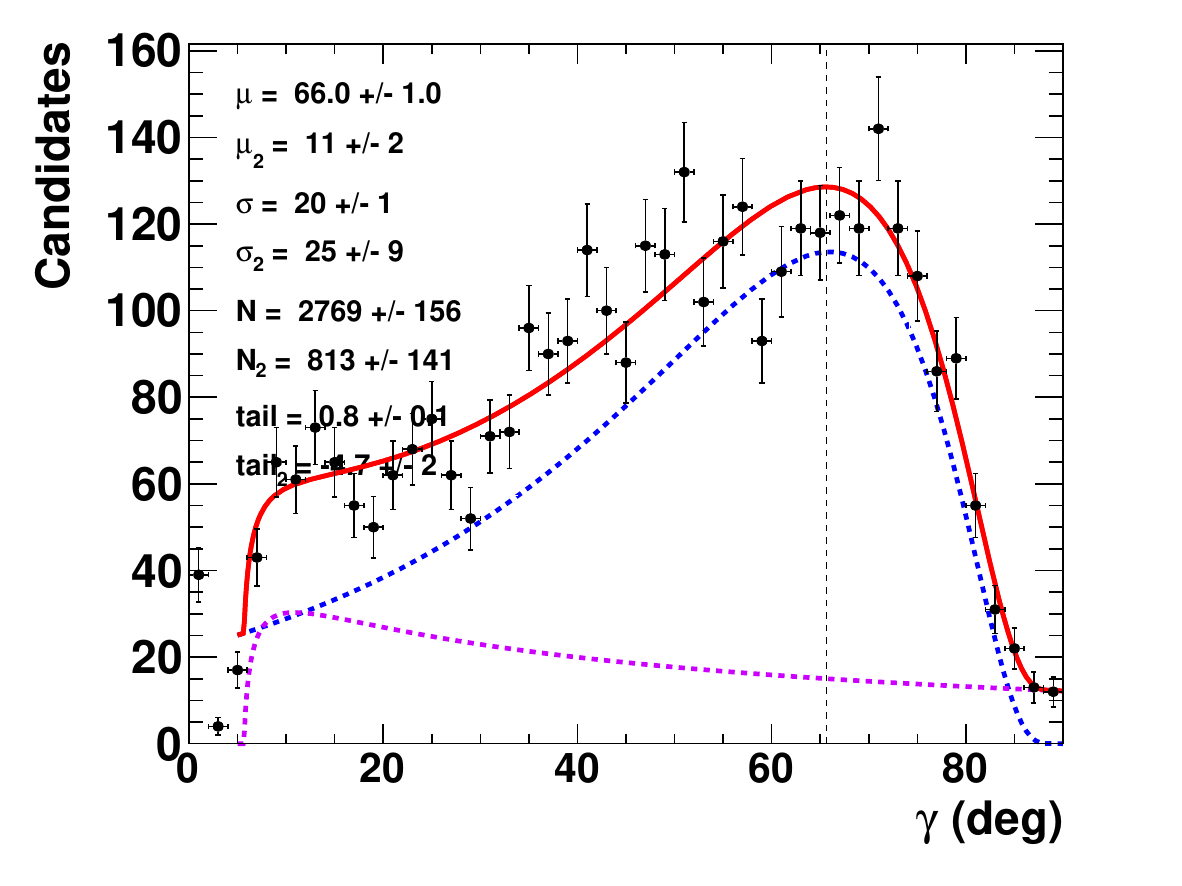}
\caption{\label{fig:1-2_dB3dBst2_noDst} Fit to the distribution of   $\gamma$ obtained from 4000 pseudoexperiments, for an the Run~1~\&~2 LHCb dataset. The initial configuration is $\gamma=65.66^{\circ}$, $\rB=0.4$ (left) and 0.22 (right), and $\deltaB=171.9^\circ$ (3 rad) (w/o $\Bs \rightarrow \Dstz\phi$). The purple dashed curve accounts for tails generated by the correlations with the nuisance parameters $\rBst$ and $\deltaBst$, while the blue dashed curve is the core part of the distribution, the plain red line is the sum of the two components of the fit.}
\end{figure}

\clearpage

\begin{figure}[h]
\centering
\includegraphics[width=0.425\textwidth,height=0.175\textheight]{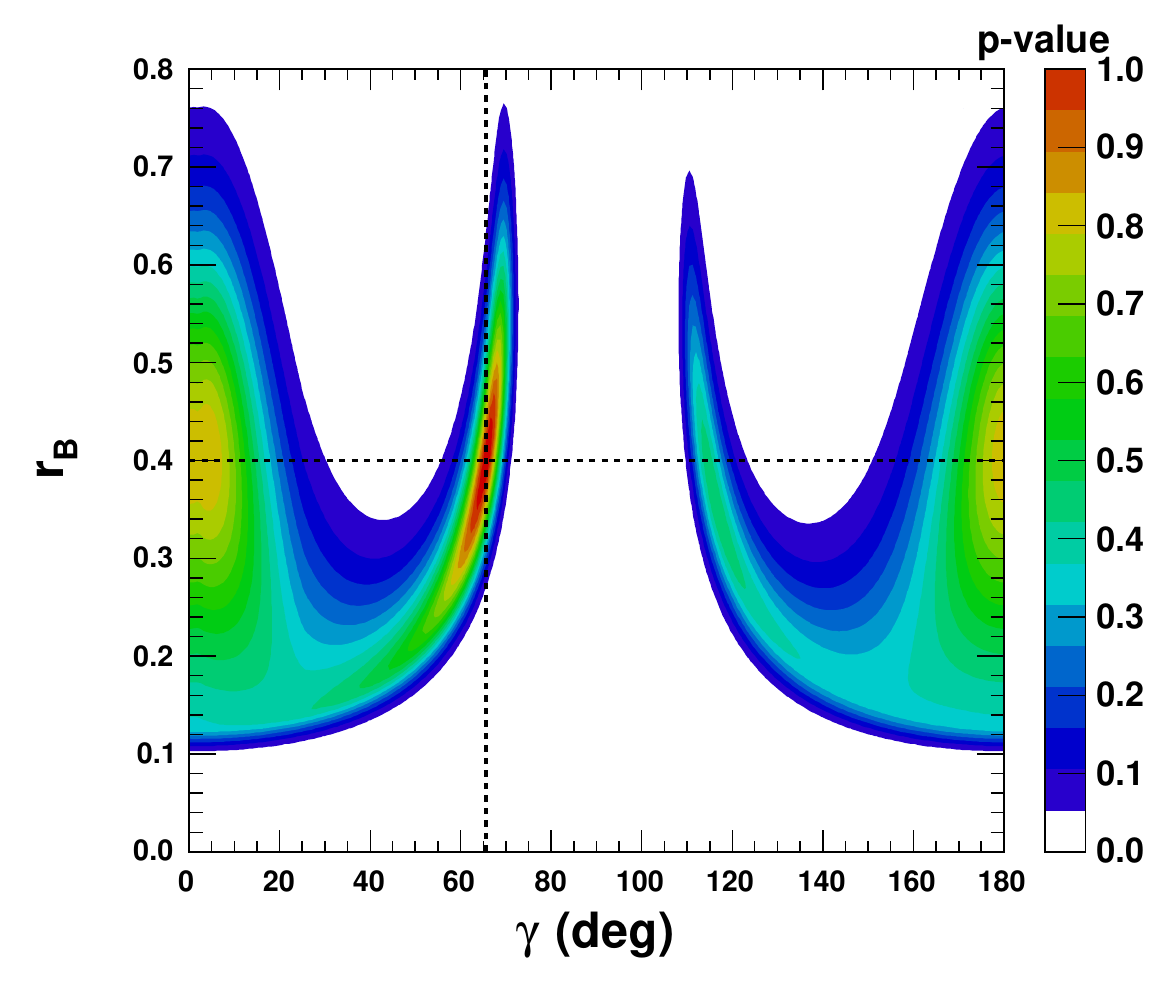}
\includegraphics[width=0.425\textwidth,height=0.175\textheight]{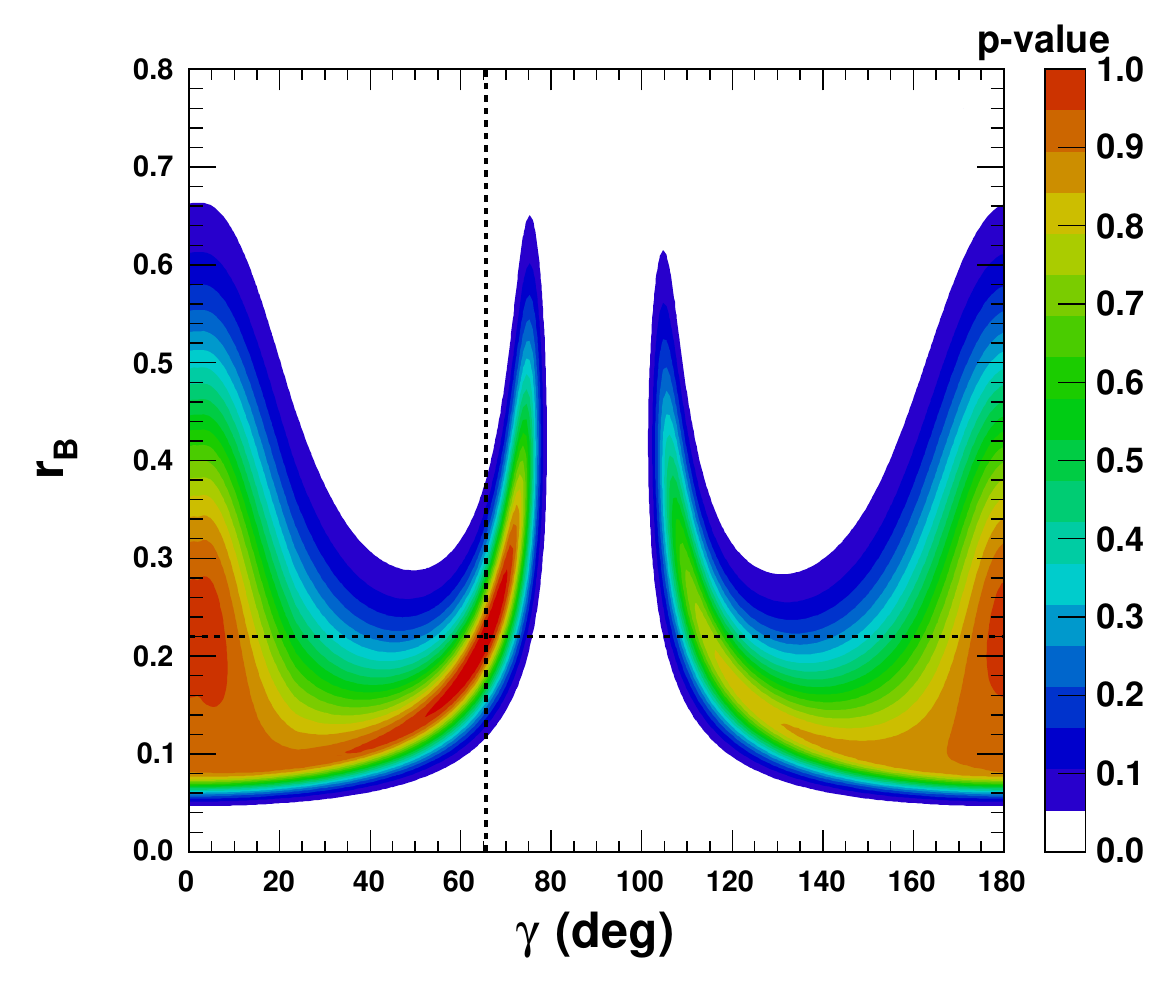} \\
\includegraphics[width=0.425\textwidth,height=0.175\textheight]{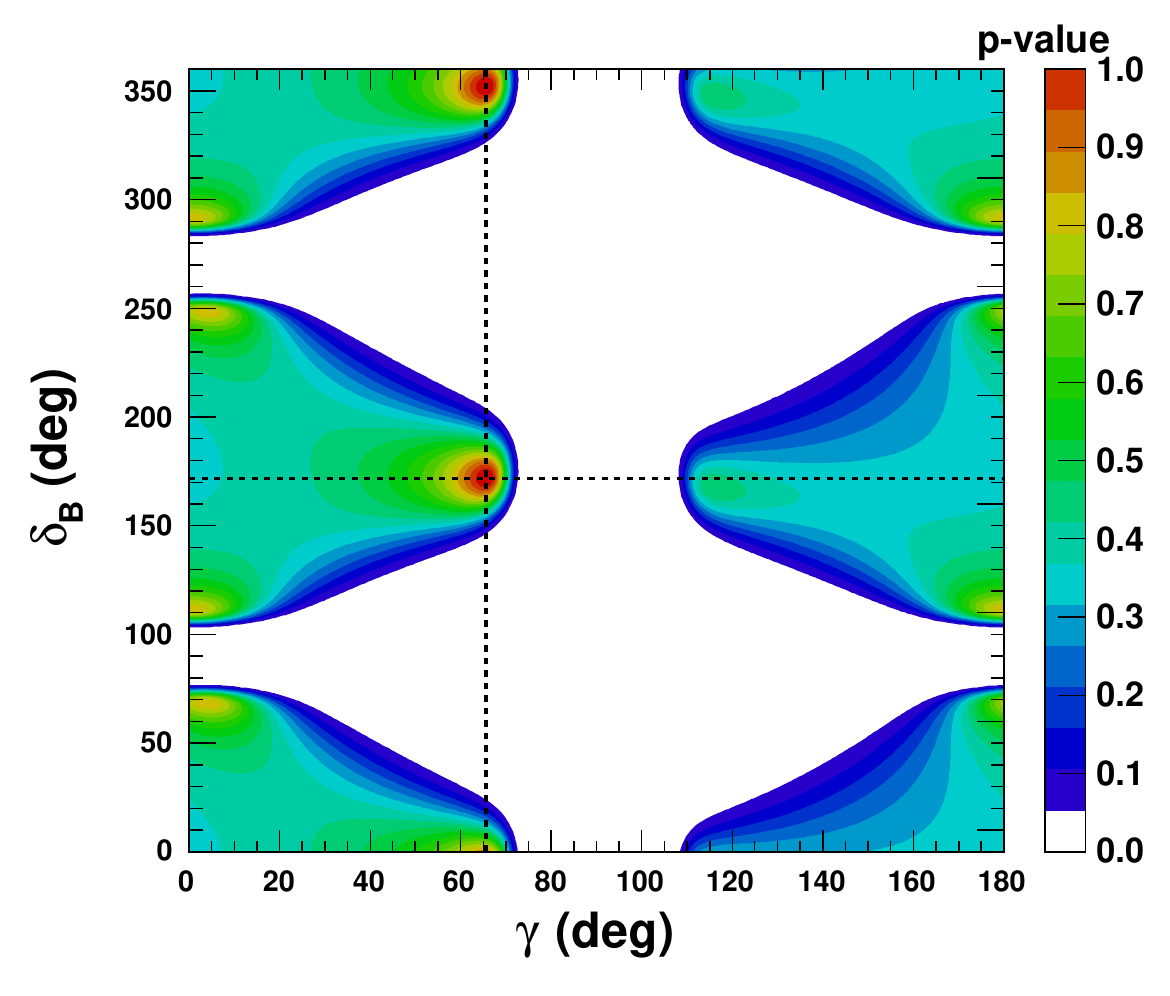}
\includegraphics[width=0.425\textwidth,height=0.175\textheight]{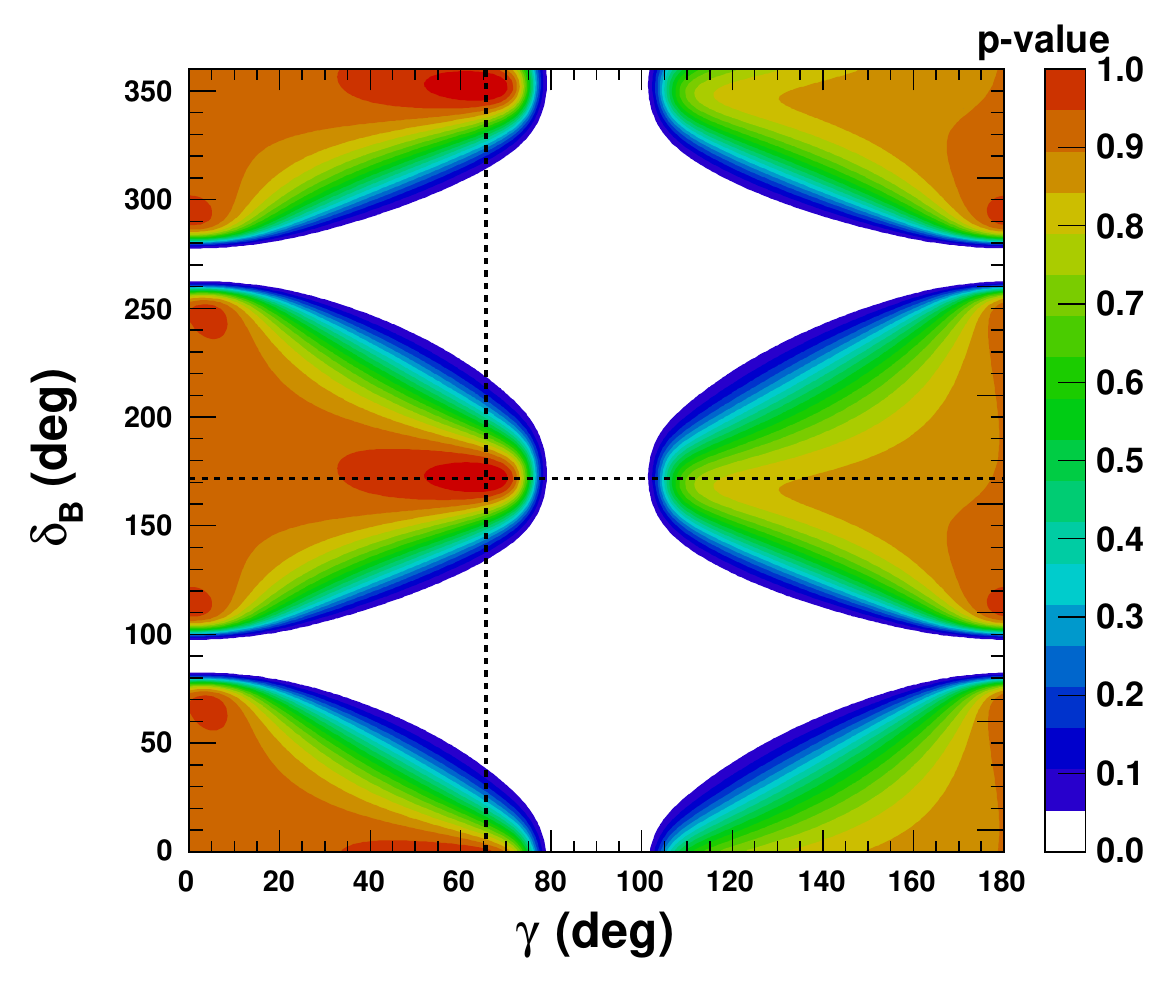} \\
\caption{\label{fig:r_d_BstvsGamma_1-3_dB3dBst2_noDst}
Two-dimension $p$-value  profile of the nuisance parameters $\rB$ and $\deltaB$  (Run $1-3$ LHCb),  as a function of $\gamma$. On each figure the dashed black  lines indicate the initial true values: $\gamma=65.66^\circ$ (1.146 rad), $\deltaB=171.9^\circ$ (3.0 rad), and $\deltastB=114.6^\circ$ (2.0 rad), and  $\rB=0.4$ (left) and 0.22 (right) (w/o $\Bs \rightarrow \Dstz\phi$).}
\includegraphics[width=0.425\textwidth]{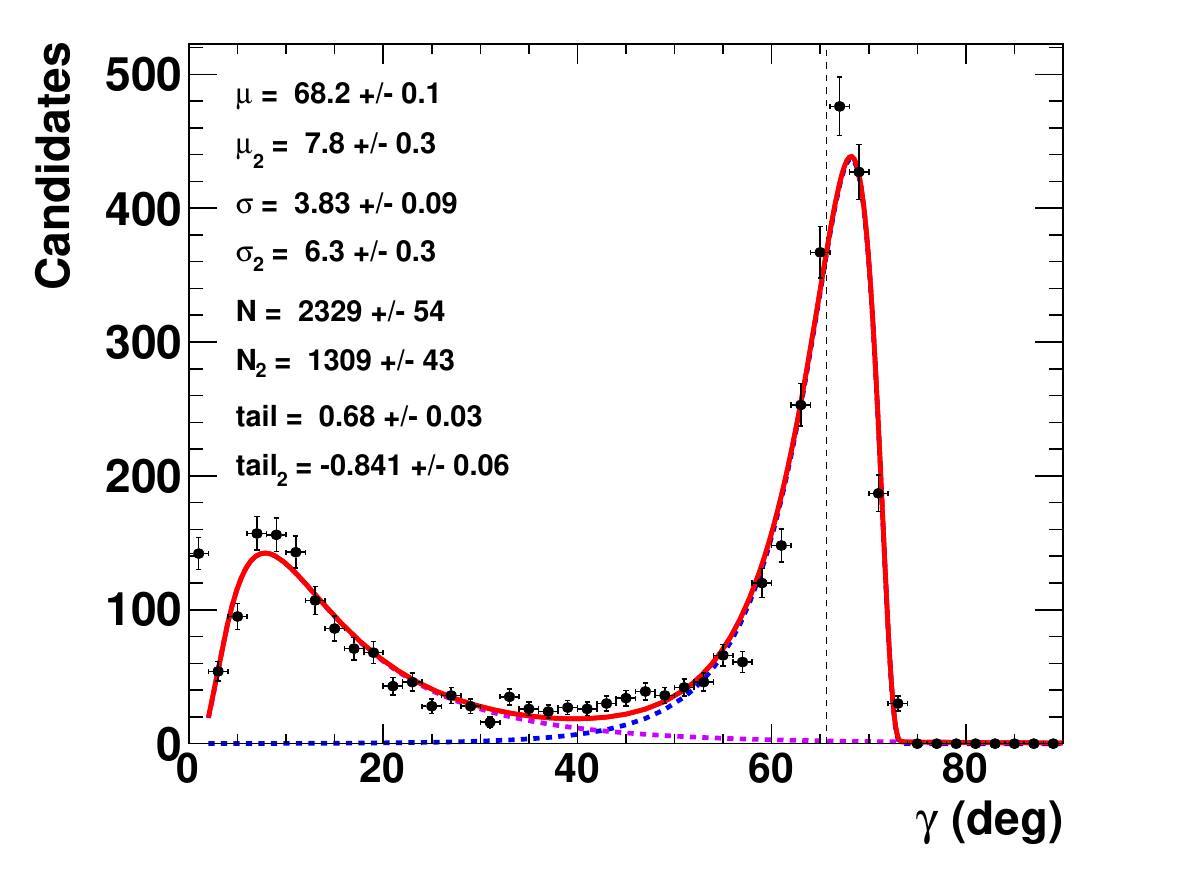}
\includegraphics[width=0.425\textwidth]{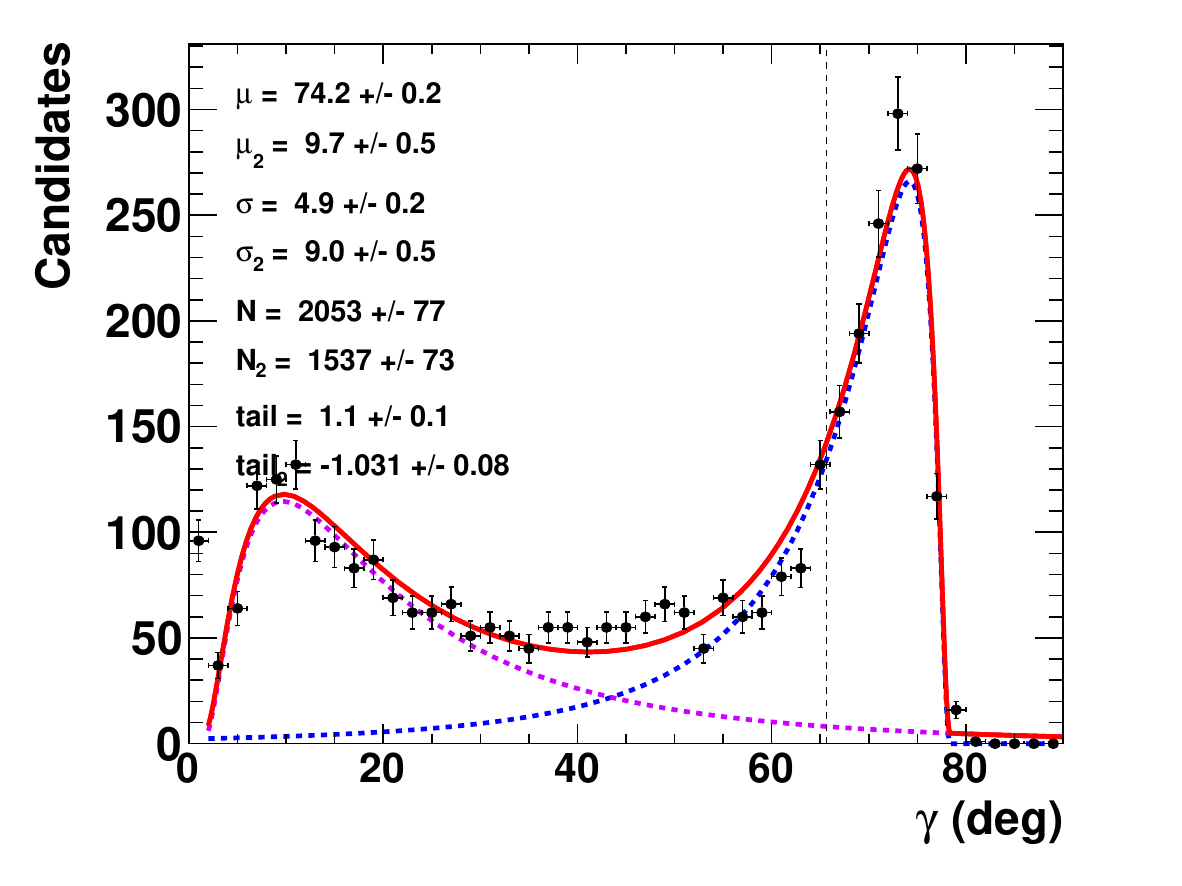}
\caption{\label{fig:1-3_dB3dBst2_noDst} Fit to the distribution of $\gamma$ obtained from 4000 pseudoexperiments (Run $1-3$ LHCb). The initial configuration is $\gamma=65.66^{\circ}$, $\rB=0.4$ (left) and 0.22 (right), and $\deltaB=171.9^\circ$ (3 rad) (w/o $\Bs \rightarrow \Dstz\phi$). The purple dashed curve accounts for tails generated by the correlations with the nuisance parameters $\rBst$ and $\deltaBst$, while the blue dashed curve is the core part of the distribution, the plain red line is the sum of the two components of the fit.}
\end{figure}

\newpage

\begin{figure}[h]
\centering
\includegraphics[width=0.425\textwidth,height=0.175\textheight]{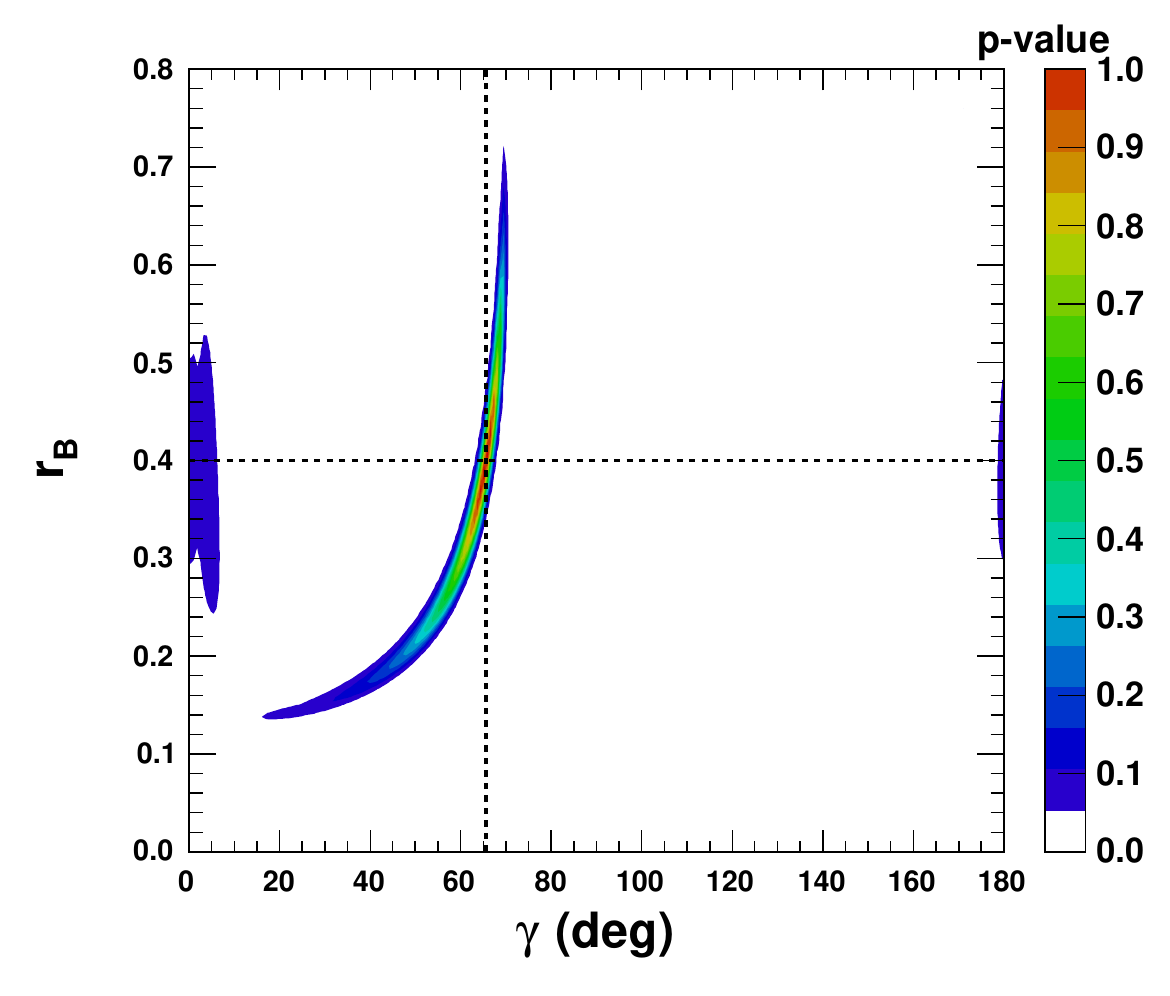}
\includegraphics[width=0.425\textwidth,height=0.175\textheight]{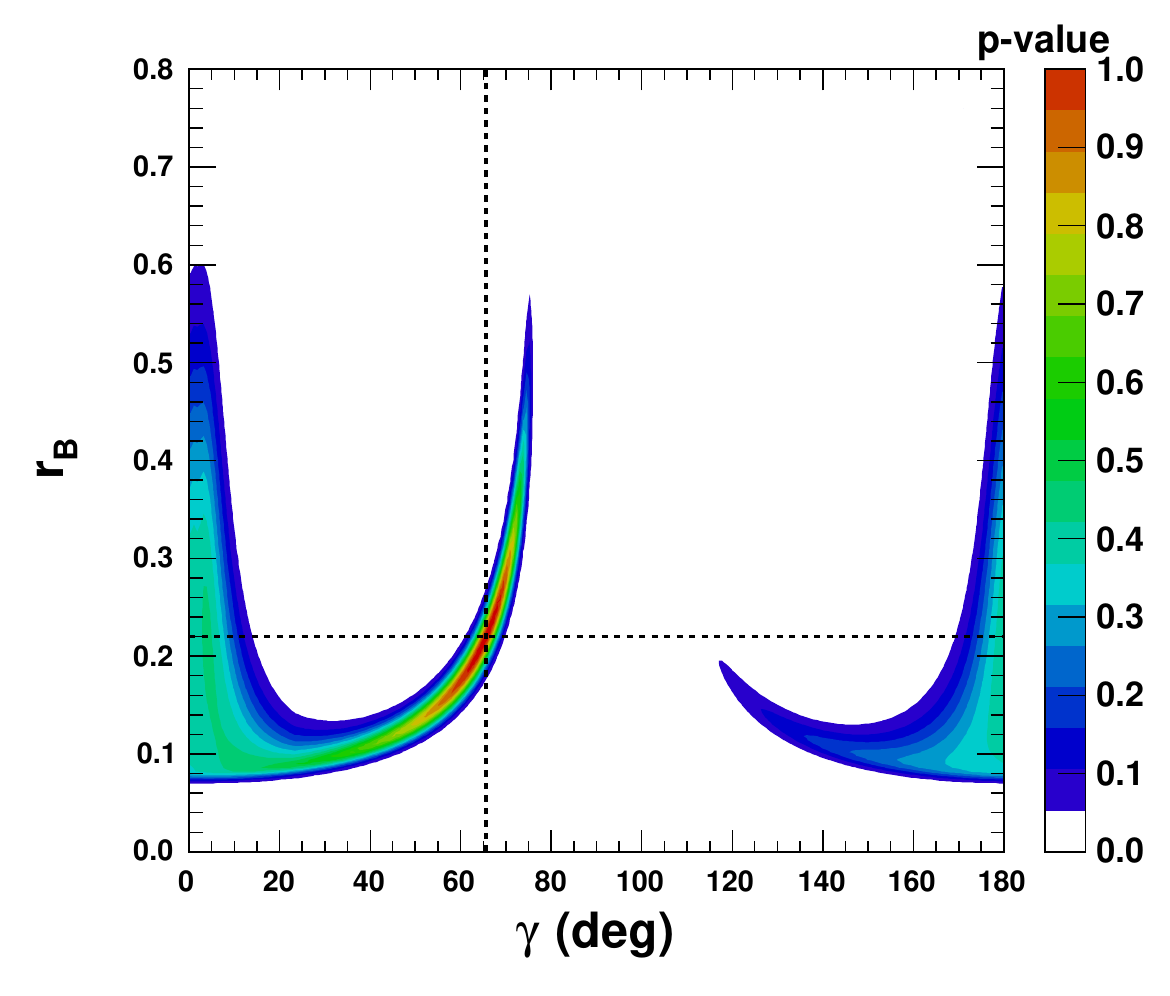} \\
\includegraphics[width=0.425\textwidth,height=0.175\textheight]{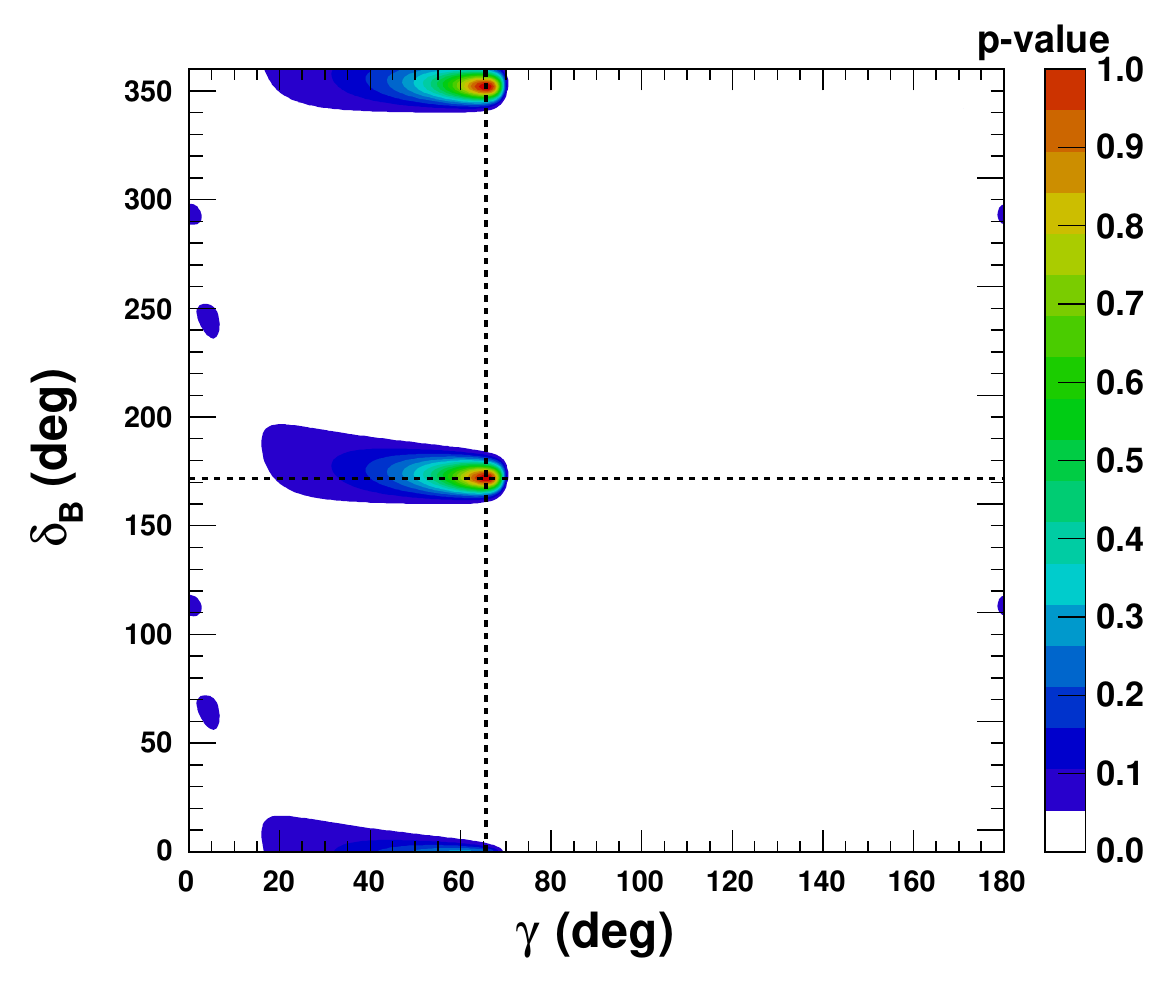}
\includegraphics[width=0.425\textwidth,height=0.175\textheight]{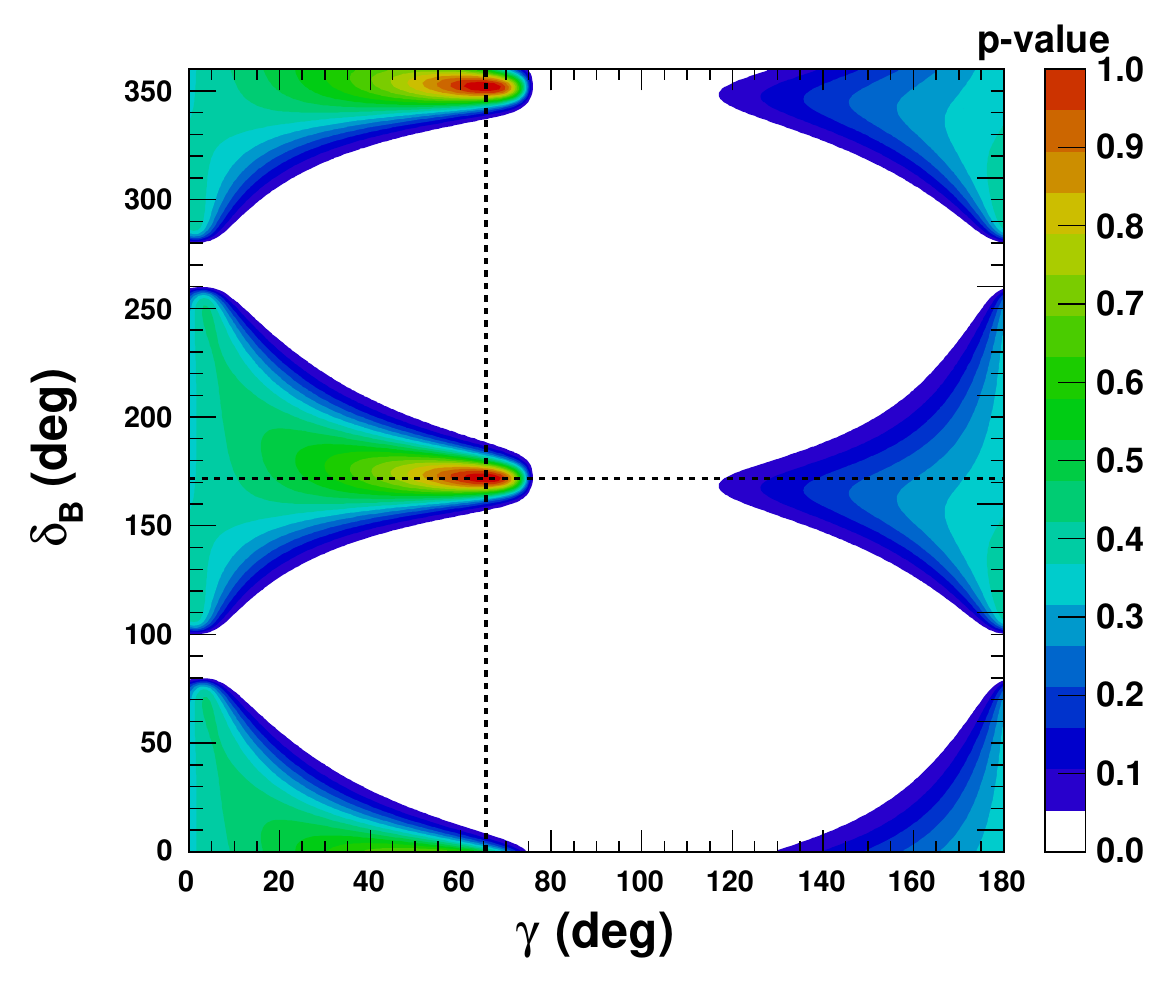} \\
\caption{\label{fig:r_d_BstvsGamma_HL-LHC_dB3dBst2_noDst}
Two-dimension $p$-value  profile of the nuisance parameters $\rB$ and $\deltaB$ (full HL-LHC LHCb),  as a function of $\gamma$. On each figure the dashed black  lines indicate the initial true values: $\gamma=65.66^\circ$ (1.146 rad), $\deltaB=171.9^\circ$ (3.0 rad), and  $\rB=0.4$ (left) and 0.22 (right) (w/o $\Bs \rightarrow \Dstz\phi$).}
\includegraphics[width=0.425\textwidth]{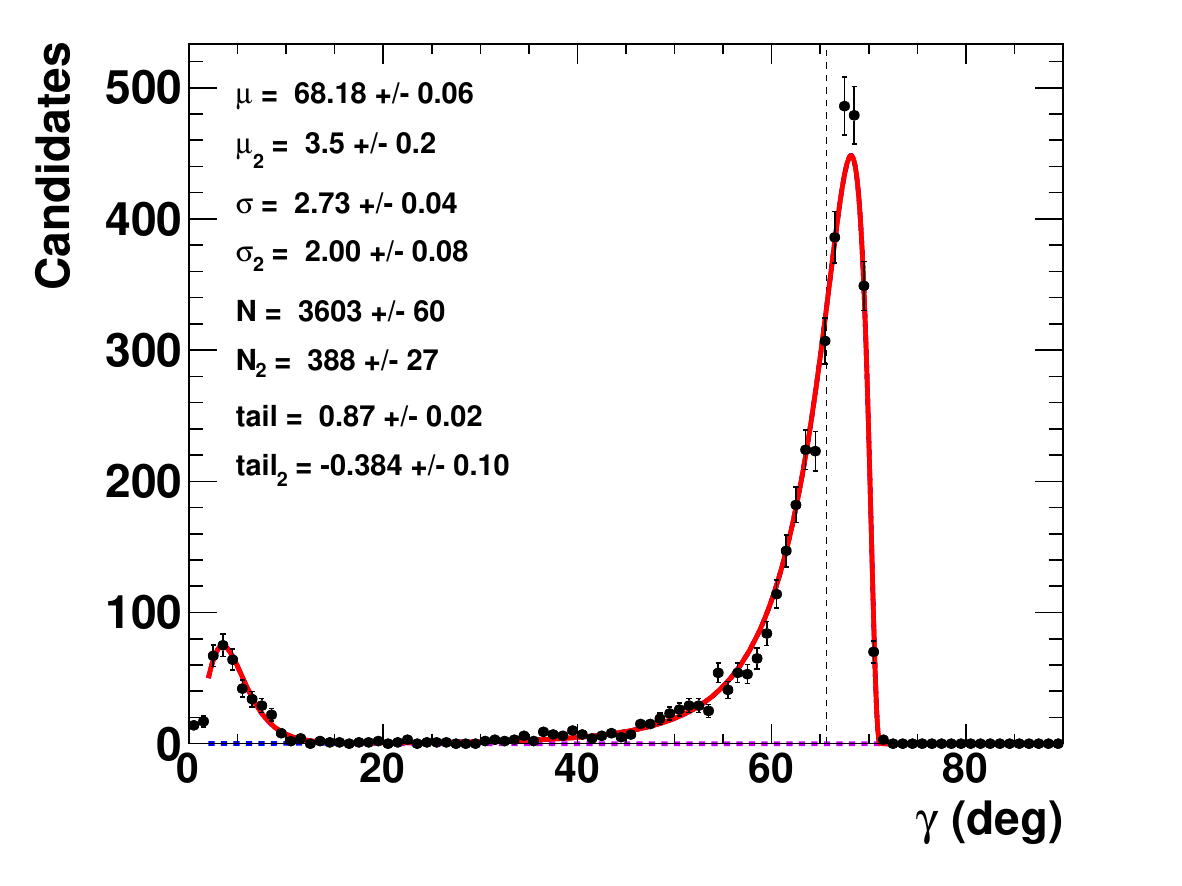}
\includegraphics[width=0.425\textwidth]{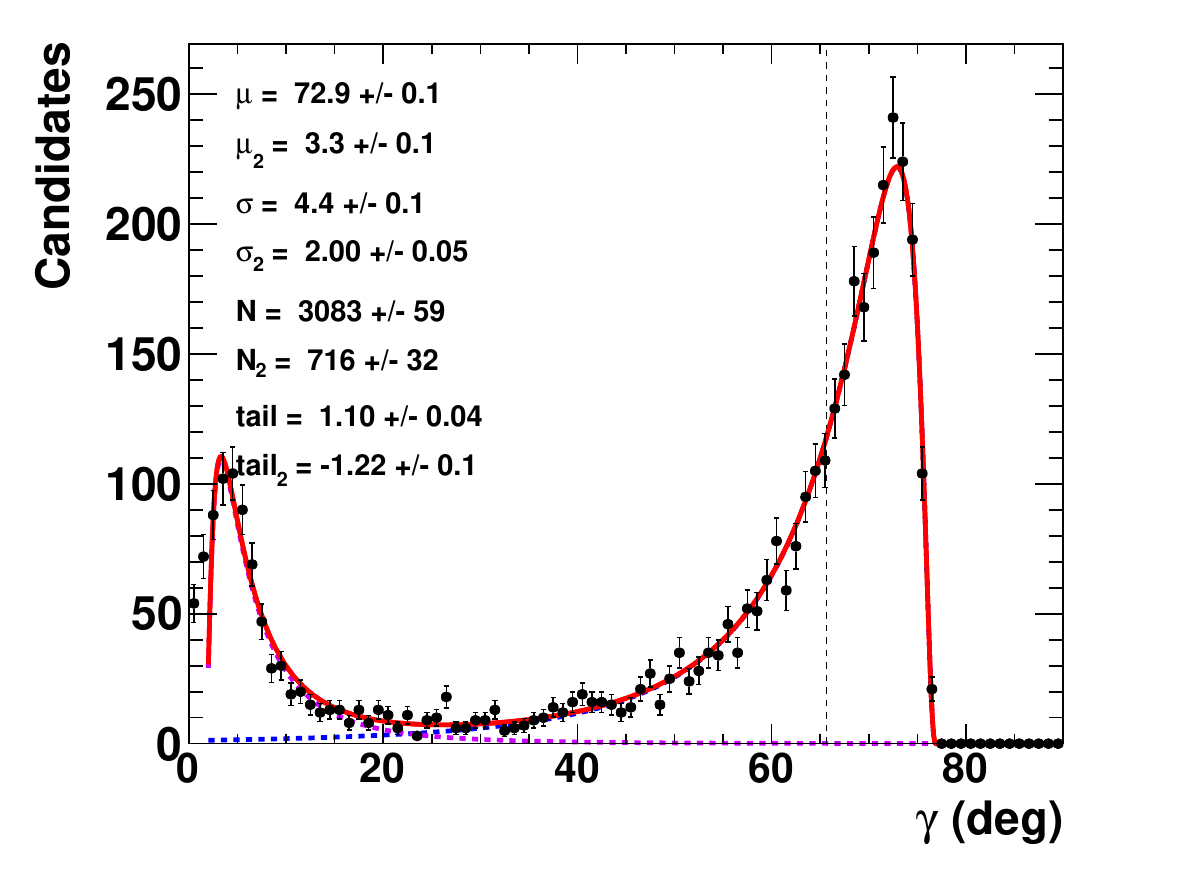}
\caption{\label{fig:HL-LHC_dB3dBst2_noDst} Fit to the distributions of  $\gamma$ obtained from 4000 pseudoexperiments (full HL-LHC LHCb). The initial configuration is $\gamma=65.66^{\circ}$, $\rB=0.4$ (left) and 0.22 (right), and $\deltaB=171.9^\circ$ (3 rad) (w/o $\Bs \rightarrow \Dstz\phi$). The purple dashed curve accounts for tails generated by the correlations with the nuisance parameters $\rBst$ and $\deltaBst$, while the blue dashed curve is the core part of the distribution, the plain red line is the sum of the two components of the fit.}
\end{figure}

\clearpage
\newpage

\clearpage
\end{CJK*}

\end{document}